\documentclass[a4paper,12pt]{report}

\usepackage{amsmath,graphicx}
\usepackage{multirow}
\usepackage{amsfonts}
\usepackage{amssymb}
\usepackage{amscd}
\usepackage{cite}
\usepackage{amsmath}
\usepackage{bbm}

\usepackage[latin1]{inputenc}
\usepackage{fancyhdr}
\def\hybrid{\topmargin -20pt    \oddsidemargin 0pt
        \headheight 0pt \headsep 0pt
        \textwidth 6.25in       % A4 paper
        \textheight 9.5in       % A4 paper
        \marginparwidth .875in
        \parskip 5pt plus 1pt   \jot = 1.5ex}
\hybrid

\def\cN{{\cal N}}
\renewcommand{\Re}{\operatorname{Re}}
\renewcommand{\Im}{\operatorname{Im}}
\newcommand\e{\mathrm{e}}
\newcommand\iu{\operatorname{i}}
\newcommand\diff{\mathrm{d}}

\newcommand{\pt}{\partial}

\def\a{\alpha}
\def\L{I}

\def\ax{{\tilde \phi}} % Axion
\def\Kdil{{k_\phi}}    %Killing vector dilaton
\def\Kax{{k_{\tilde \phi}}}  % Killing vector axion
\def\Kxi{{k}}                % Killing vector \xi
\def\Ktxi{{\tilde k}}        % Killing vector \tilde \xi
\def\kk{{k}}          % Killing vector contracted with symplectic vector

\def\Proj{{\Pi}}

\def\PQ{\tilde \pi}
\def\K{\hat{K}}

\def\P{{r}}  % \xi direction
\def\Q{{s}}  % \tilde \xi direction
\def\A{{t}}  % Axion direction
\def\R{{R}}  % real coefficient
\def\C{C}    % vector on Mv
\def\D{D}    % vector on Mh
\def\E{\rho}
\def\PQ{\tilde \pi}

\def\mino{m_{3/2}}

\newcommand\admu{\mu}                   %E7 adjoint pure
\newcommand\admui{\hat{\mu}}            %Sl(2,R) adjoint pure
\newcommand\admuA{\mu}                  %SO(6,6) adjoint pure
\newcommand\admuspin{\mu}               % spinor in the adjoint pure

\newcommand\U{\mu}                      %highest weight SU(2) in E7 formalism

\begin{document}

\begin{titlepage}
\begin{center}
%\rightline{\small ZMP-HH/10-5}
\vskip 2cm

{\Large \bf
Generalized Geometry and\\ Partial Supersymmetry Breaking}
\vskip 1.2cm
{\bf Hagen Triendl}

\vskip 0.8cm

{\em Institut de Physique Th\'eorique, CEA Saclay\\
Orme des Merisiers, F-91191 Gif-sur-Yvette, France}
\vskip 0.8cm

{\tt hagen.triendl@cea.fr}

\end{center}

\vskip 3.5cm

\begin{center} {\bf ABSTRACT } \end{center}
%\vspace{-2mm}

\noindent

This review article consists of two parts.\footnote{Based on the author's Ph.D.\ thesis, defended on 14th July 2010.}
In the first part we use the formalism of (exceptional) generalized geometry to derive the scalar field space of $SU(2)\times SU(2)$-structure compactifications. We show that in contrast to $SU(3)\times SU(3)$ structures, there is no dynamical $SU(2)\times SU(2)$ structure interpolating between an $SU(2)$ structure and an identity structure. Furthermore, we derive the scalar manifold of the low-energy effective action for consistent Kaluza-Klein truncations as expected from $\cN=4$ supergravity.

In the second part we then determine the general conditions for the existence of stable Minkowski and AdS $\cN=1$ vacua in spontaneously broken gauged $\cN=2$ supergravities and construct the general solution under the assumption that two appropriate commuting isometries exist in the hypermultiplet sector.
Furthermore, we derive the low-energy effective action below the scale of partial
supersymmetry breaking and show that it
satisfies the constraints of $\cN = 1 $ supergravity.
We then apply the discussion to special quaternionic-K\"ahler geometries which appear in the low-energy limit of $SU(3)\times SU(3)$-structure compactifications and construct Killing vectors with the right properties. Finally we discuss the string theory realizations for these solutions.

\bigskip

\vfill
October, 2010

%\today

\end{titlepage}

%%%%%%%%%%%%%%%%%%%%%%%%%%%%%% Acknowledgments %%%%%%%%%%%%%%%%%%%%%%%%
\thispagestyle{empty}
$\phantom{A}$
\vfill
\parbox{14.1cm}{
{\Large \bf Acknowledgments}\\ \\
This review article is based on my Ph.D.\ thesis. First of all I am deeply grateful to my supervisor Jan Louis for his guidance and encouragement, in particular for sharing his insights and intuition during the last three years.
I am very thankful to Paul Smyth for a fruitful and enjoyable collaboration.
Furthermore, I am thankful to Sarah Andreas, Fatih Argin, Michele Cicoli, Thomas Danckaert, Christian Gro\ss, Christian Hambrock, Cecilie Hector, Martin Hentschinski, Michael Herbst, Manuel Hohmann, Daniel Koch, Torben Kneesch, Danny Mart\'inez-Pedrera, Sebastian Mendizabal, Jan M\"oller, Chlo\'e Papineau, Ron Reid-Edwards, Martin Schasny, Jonas Schmidt, Vid Stojevic, Alex Thio, Roberto Valandro and Mattias Wohlfarth for creating a warm and inspiring atmosphere in Hamburg and for many lifely discussions.
I am also indepted to Davide Cassani, Vicente Cort\'es, Gianguido Dall'Agata, Bernard de Wit, Luca Martucci, Paul-Andy Nagy and Fabian Schulte-Hengesbach for interesting and stimulating discussions.

Finally I want to thank the II. Institute for Theoretical Physics, University of Hamburg, for hospitality during my Ph.D.. This work was supported by the German Science Foundation (DFG) under the Collaborative Research Center (SFB) 676.
}

\cleardoublepage

\tableofcontents
\cleardoublepage

%%%%%%%%%%%%%%%%%%%%%%%%%%%%%%%%%%%%%%%%%%%%%%%
\chapter{Introduction} \label{section:intro}
%%%%%%%%%%%%%%%%%%%%%%%%%%%%%%%%%%%%%%%%%%%%%%%

%%%%%%%%%%%%%%%%%%%%%%%%%%%%%%%%%%%%%%%%%%%%%%%
\section{String theory and flux compactifications} \label{section:flux_comp}
%%%%%%%%%%%%%%%%%%%%%%%%%%%%%%%%%%%%%%%%%%%%%%%
String theory provides a promising way to construct theories of particle physics and gravity that are well-defined in the UV limit (see for example \cite{Green:1987sp,Lust:1989tj,Polchinski:1998rq,Becker:2007zj} for a comprehensive introduction to string theory).
In contrast to theories of point particles, string theory considers the string, an object extended in one space dimension, as the fundamental constituent of the theory. Accordingly, the worldline of a point particle in spacetime is replaced by the string ``worldsheet'', which is a surface with one timelike and one spacelike direction embedded in spacetime.
This replacement has a deep impact on pertubation theory in the corresponding quantum theory:
In Feynman diagrams the propagator lines of point particles are replaced by surfaces representing propagating strings and therefore interaction vertices are smoothed out. As a consequence, all scattering diagrams of strings turn out to be finite, which suggests that string theory is UV finite.

In order to actually compute the scattering amplitudes, one considers the theory on the two-dimensional worldsheet of a closed string with only one dimensionful, free parameter given by the string tension $\alpha'$.\footnote{The other free parameter, the string coupling constant $g_s$, is dimensionless and determined by the expectation value of the ten-dimensional dilaton $\phi^{(10)}$, a massless scalar.}
More precisely, it is a sigma model, i.e.\ a model of scalar fields parameterizing a manifold, consisting of the $D$ bosonic coordinate fields that describe the embedding of the string into $D$-dimensional spacetime and a number of additional fermionic and bosonic scalar fields introduced for consistency.
From the dynamics it then follows that left- and right-moving modes can be treated independently, whose quantization gives the spectrum of the string theory.
Stability of the string theory requires the absence of tachyonic modes in the spectrum and of tadpoles in the perturbation theory. It turns out that all string theories that meet these conditions admit some amount of supersymmetry in the two-dimensional worldsheet theory, so-called ``worldsheet supersymmetry'', and are called superstring theories.
Furthermore, since one wants to recover a theory of gravity, a string theory should admit a massless spectrum (including the gravitational interaction). Such string theories are called critical.

Of particular interest are stable, critical superstring theories that can be realized in a flat background.
It turns out that such there are five different superstring theories of this type, and all of them require a ten-dimensional spacetime.
If both left- and right-moving sectors are supersymmetric, the theory is called type II string theory. Depending on the relative chirality (under the ten-dimensional Lorentz group) of left- and right-moving states, we distinguish between type IIA (non-chiral) and type IIB (chiral) string theories.
By identifying left- and right-movers of the type II superstring, i.e.\ by modding out its orientation, one can define type I string theory analogously. However, the type I string needs further field content in order to be a consistent quantum theory, which is realized by adding open string states to the theory.
Heterotic string theories are defined in a different way. Here, one introduces fermionic fields only for left-movers, while the right-moving spectrum is completed by 16 additional bosonic scalar fields.

The quantization of the superstring leads to an infinite tower of string excitations that form multiplets of a spacetime supersymmetry group.
Its massless spectrum coincides with the spectrum of ten-dimensional supergravity theories, whose ten-dimensional Planck scale is set by $\alpha'$. Moreover, the masses of higher string states are of order $\alpha'$, which means that the masses of further string states are generically so large that the observable spectrum coincides with the massless spectrum of ten-dimensional supergravity. While type I and heterotic string theories descend to $\cN=1$ supergravities in ten dimensions with a number of vector multiplets resembling either the gauge group $SO(32)$ or $E_8\times E_8$, the low-energy limit of type II theories is given by the chiral (IIB) or non-chiral (IIA) version of $\cN=2$ supergravity. Chiral and non-chiral here means that the two supersymmetry generators are Majorana-Weyl spinors of the same or opposite chirality, respectively. Note that in ten dimensions $\cN=2$ supersymmetry together with the choice of chirality already determines the theory completely.

In this review we shall only consider type II string theories and the corresponding $\cN=2$ supergravities. Let us therefore briefly discuss the massless spectra of these theories from the worldsheet point of view.
The massless spectrum of a type II string theory consists of the tensor product of lowest-order left- and right-moving excitations, which form massless representations under the ten-dimensional Lorentz group.
More precisely, the lowest-order excitations consist of two representations: A Lorentz vector, the Neveu-Schwarz (NS) sector, and a spinor field, the Ramond (R) sector. As a consequence, the tensor product of left- and right-movers consists of the four combinations of NS and R sector, again forming representations of the ten-dimensional Lorentz group. For example, the NS-NS sector forms a Lorentz tensor that decomposes into the ten-dimensional (symmetric) metric $g_{MN}$, an anti-symmetric tensor field $B_{MN}$ and the dilaton $\phi^{(10)}$ that is a scalar corresponding to the trace component. Similarly, the R-R sector is a spinor bilinear that corresponds to a formal sum $C$ of form fields of odd (even) degree in type IIA (IIB). Finally, the NS-R and the R-NS sector give the two gravitini and dilatini of opposite (same) chirality in type IIA (IIB).

In order to make contact with observations, one needs to compactify superstring theories to four dimensions. This goes back to the idea of Kaluza and Klein \cite{Kaluza:1921tu,Klein:1926tv} to consider compact, spacelike extra-dimensions in a gravity theory in order to unify interactions.
More generally, a $D$-dimensional theory on a space that is a product of $d$-dimensional, infinitely extended spacetime with a $(D-d)$-dimensional compact space $Y_{D-d}$ corresponds to an effectively $d$-dimensional theory but with an infinite Kaluza-Klein tower of massive states. The masses are related to the size of $Y_{D-d}$, the scale of compactification. Furthermore, on all mass levels, the representations of the $D$-dimensional Lorentz group $SO(1,D-1)$ decompose into representations of $SO(1,d-1)$.
At energies below the compactification scale, massive states decouple and one can integrate them out of the theory. For convenience, one simply truncates the spectrum at a given energy scale and only considers the light spectrum. The light field content in turn is determined by the topological features of $Y_{D-d}$.

Applying the concept of dimensional reduction to the superstring in $D=10$, one should consider backgrounds where six space dimensions form a compact manifold whose size is considered to be at a scale not accessible for present-day experiments. The low-energy limit of string theory in this background should then reproduce the known Standard Model of Particle Physics.
More generally, it is interesting to ask what is the low-energy limit of string theory on a background of the form
\begin{equation} \label{stringbackground}
 M_{1,d-1} \times Y_{10-d} \ ,
\end{equation}
where in the following we will restrict to $M_{1,d-1}$ being some maximally-symmetric non-compact space, i.e.\ Minkowski, AdS or de Sitter (dS) space, while $Y_{10-d}$ is a compact manifold of
dimension $(10-d)$.\footnote{In general, one can also allow for some warp factor between the two factors in \eqref{stringbackground}, in other words the metric of $M_{1,d-1}$ can carry a dependence on the coordinates of $Y_{10-d}$ through a prefactor $\e^A$ where $A$ is a function on $Y_{10-d}$. This warp factor can have dramatic effects: Strong warping can even lead to throat-like geometries where the local modes decouple from the rest of the geometry \cite{Klebanov:2000hb}. We will not analyze the effect of warping in this review even though most of the result carry over to the case of mildly warped geometries, i.e.\ where the warp factor $\e^A$ stays finite and non-zero at every point of $Y$. For a discussion of warped compactifications, see for example \cite{Koerber:2008sx,Martucci:2009sf}.}

The resulting low-energy effective actions can be very complicated, the scalars for instance naturally form a non-linear sigma model. In general it is unclear how the corresponding string theory might behave on this background, since there is no control over string corrections in the theory.
An exception are low-energy effective theories that are supersymmetric: The renormalization properties of supersymmetric theories limit the influence of quantum effects and string corrections. For instance, in an $\cN=1$ supergravity only the K\"ahler potential on the target space of scalars gets corrected by (string) loop effects.
Therefore, it is favorable to focus on string compactifications where the low-energy effective action is supersymmetric.
Such supersymmetric string compactifications lead to lower-dimensional supergravity theories in the low-energy limit, whose phenomenological properties in turn can be studied.
In these theories, supersymmetry then can be broken in a soft way at low scales. This not only enables one to retain control over the theory after supersymmetry breaking but, as a side effect, yields many phenomenologically desirable features for the theory.
For instance, the minimal supersymmetric extension of the Standard Model of Particle Physics unifies the gauge couplings indicating that the Standard Model gauge group might originate from a simple Lie group (see for instance \cite{Louis:1998rx,Bustamante:2009us} and references therein). This may also explain the hypercharge pattern of Standard Model particles.
Furthermore, in a theory with spontaneously broken supersymmetry, the loop corrections to the Higgs mass are limited by the scale of supersymmetry breaking. This solves the Standard Model hierarchy problem, by giving an alternative scenario to the cancellation of a bare Higgs mass and its loop corrections of order Planck (or GUT) scale to the much lower scale of electroweak symmetry breaking.

Backgrounds of the form \eqref{stringbackground} allow for a supersymmetric description at low energies if one can find a $d$-dimensional supersymmetry generator acting on the light spectrum.
In other words, the backgrounds has to admit a $d$-dimensional supersymmetry generator inside the ten-dimensional supersymmetry algebra whose restriction to the low-energy spectrum is well-defined, i.e.\ it maps light modes to other light modes.
Generically, this corresponds to the requirement that $Y_{10-d}$ admits globally defined and nowhere-vanishing spinors. This in turn limits the possible transformations that can be used to glue charts on $Y_{10-d}$ together in that they leave these spinors invariant. In turn, the group of such transformations, called the structure group, is reduced from all metric-preserving transformations on $Y_{10-d}$ to some subgroup $G$ such that the nowhere-vanishing spinors on $Y_{10-d}$ become singlets of the structure group. Correspondingly, the manifold $Y_{10-d}$ is called a $G$-structure manifold \cite{Gates:1984nk,Strominger:1986uh,Hull:1986kz,Hitchin:2000jd,Hitchin:2001rw,Gauntlett:2001ur,Gauntlett:2002sc,Chiossi:2002tw,Gauntlett:2003cy}.

A subclass of such backgrounds are
Calabi-Yau manifolds, where the globally
defined spinors are also covariantly constant with respect to the
Levi-Civita connection. This reduces the group of linear transformations used in the Levi-Civita connection, the so-called holonomy group, to $G\subset SU(n),n\le (10-d)/2,$ and the Ricci-tensor vanishes.\footnote{There is only a single four-dimensional $SU(2)$-holonomy manifold, which is called K3. For a review, see \cite{Aspinwall:1996mn}. Note that for seven- and eight-dimensional manifolds $Y$ the holonomy group can be $G_2$ and $Spin(7)$, respectively. Since we restrict to the case $d \ge 4$, we do not consider such manifolds in the following.} Therefore, Calabi-Yau compactifications lead to vacuum solutions to the equations of motion (Einstein equation) in the absence of energy sources. For a review on Calabi-Yau manifolds see for instance \cite{Greene:1996cy}.

String theory in addition allows for a number of supersymmetric energy sources, which need to be studied in the context of compactifications in order to understand the generic low-energy behavior of string theory and furthermore offer attractive features for model building. For instance, the background can exhibit topological twists in the gauge and form fields of the ten-dimensional theory, so-called fluxes, which yield field strengths for the corresponding fields that cannot be turned off dynamically. These fluxes lead to additional couplings in the low-energy effective action that e.g.\ enables one to reduce the number of unwanted massless degrees of freedom.
Furthermore, in compactifications of the type II string one can orientifold the string background in order to reduce the amount of supersymmetry of the compactification. This amounts to modding out the combined action of a $\mathbb{Z}_2$ symmetry of $Y$ and the exchange of left- and right-movers on the worldsheet. At the $\mathbb{Z}_2$ fixed-point locus in $Y$, the so-called orientifold plane, the string is unoriented and feels a negative energy density and a R-R charge. This gives the possibility to include supersymmetric D-branes in the setup, which have positive energy density and R-R charge and thereby cancel the charge of the orientifold plane. These D-branes are defined as the boundary of open strings and thereby can introduce non-Abelian gauge groups.
Both fluxes and localized energy sources such as
D-branes and orientifold planes contribute to the energy-momentum tensor and therefore may demand a different (non-Calabi-Yau) geometry (for reviews see for example \cite{Grana:2005jc,Douglas:2006es,Blumenhagen:2006ci,Wecht:2007wu,Samtleben:2008pe}).
Therefore it is necessary to study general $G$-structure manifolds and compactifications thereon.

In contrast to Calabi-Yau manifolds, the covariant derivative for the nowhere-vanishing spinors on such manifolds is non-vanishing and can be decomposed into irreducible representations under the structure group $G$. These components are called ``intrinsic torsion classes'' and can be used to classify $G$-structure manifolds (see for instance \cite{Joyce:2000,Chiossi:2002tw,Gauntlett:2002sc,Gurrieri:2002wz}).

In type II theories a slightly more general setup is possible. The ten-dimensional action admits two supersymmetries, acting on the left- and right-movers, respectively. Each of them descends to a supersymmetry generator of the low-energy theory via a nowhere-vanishing spinor on $Y$. If these spinors are different, they define different ``structure
groups'' and the corresponding backgrounds are defined by a $G \times G$ structure \cite{Jeschek:2004wy,Grana:2004bg,Grana:2005sn,Grana:2005ny,Grana:2006hr}.

In this article we will review $SU(3)\times SU(3)$-structure backgrounds of type II theories, which admit $8$ real supercharges, following \cite{Grana:2005ny,Grana:2006hr}. Furthermore, we shall discuss type II backgrounds that allow for $16$ supercharges, corresponding to compactifications on manifolds with $SU(2)\times SU(2)$ structure. Aspects of such backgrounds were previously discussed for example in \cite{Gauntlett:2003cy,Bovy:2005qq,Grana:2005sn,ReidEdwards:2008rd,Spanjaard:2008zz,Lust:2009zb,Louis:2009dq}.
It turns out that $SU(n)\times SU(n)$ structures can be discussed rather conveniently in the framework of generalized geometry \cite{Hitchin:2004ut,Gualtieri:2003dx,Witt:2004vr,Witt:2005sk}. Let us introduce these concepts next.

%%%%%%%%%%%%%%%%%%%%%%%%%%%%%%%%%%%%%%%%%%%%%%%
\section{Covariant formulations and parameter spaces}
%%%%%%%%%%%%%%%%%%%%%%%%%%%%%%%%%%%%%%%%%%%%%%%
In the last ten years, much progress has been made in finding geometric formulations that not only describe geometrical degrees of freedom in string theory compactifications but covariantize as many symmetries of string theory as possible. The advantage of this approach is that the understanding of some subsector of fields, for instance the subsector of fields coming from the metric, enables one to determine the geometry of the complete moduli space and to make symmetries and dualities of the theory manifest.
Furthermore, such formulations are convenient for describing more general string backgrounds that differ from the class of Calabi-Yau compactifications by the inclusion of intrinsic torsion and backgrounds that distinguish between left- and right-movers. These formalisms enables one to describe viable string backgrounds that may lack a description in terms of ordinary geometry.\footnote{
Throughout this paper we do not specify
if $Y_{10-d}$ is an honest manifold or a
generalization thereof.
For the discussion of $SU(n)\times SU(n)$ structures it is sufficient to consider backgrounds that admit a splitting of the ten-dimensional tangent bundle (and generalized version thereof) into a $d$-dimensional Minkowskian tangent bundle and the corresponding internal one.
Nevertheless we always call $Y$ the compactification
manifold and the analysis in this review just carries over to this more general case.
}

The motivation comes from string compactifications on an $n$-dimensional torus, which lead to effective theories that are not only highly supersymmetric but also possess a large bosonic symmetry group.
In all such string compactifications, the massless scalar degrees of freedom coming from the metric and the two-form $B$ in the Kaluza-Klein reduction form representations under the Lie group $SO(n,n)$, the so-called ``T-duality group'', which includes the geometrical symmetry group $Gl(n,\mathbb{R})$ as a subgroup (see \cite{Giveon:1994fu} and references therein). The additional generators of $SO(n,n)$ describe symmetries of the massless string spectrum that are not symmetries of the geometry.
Making this observation, it seems natural to build new string backgrounds by ``twisting'' the torus by some symmetry transformation in $SO(n,n)$. This means that different patches on the torus are not glued together by direct identification but by some $SO(n,n)$ transformation. If these twists cannot be interpreted in terms of fluxes or intrinsic torsion, one refers to them as ``non-geometric flux''.
This leads to so-called twisted tori and T-folds \cite{Dabholkar:2002sy,Hull:2004in,Dabholkar:2005ve,Kachru:2002sk,Flournoy:2004vn,Dall'Agata:2005ff,Hull:2005hk,Dall'Agata:2005mj,Shelton:2005cf}.
Note that the $SO(10-d,10-d)$ symmetry of the massless spectrum is not a symmetry of the massive string states. In particular, from the pattern of BPS states of torus compactifications, one sees that this symmetry group should be broken to the discrete group $SO(10-d,10-d,\mathbb{Z})$ \cite{Hull:1994ys}. In order to construct not only a viable background for the low-energy effective supergravity theory but also for the complete string theory, we should therefore only use $SO(10-d,10-d,\mathbb{Z})$ transformations to glue together patches on the torus.

One could think that this way of constructing new string backgrounds is limited to torus compactifications, where the T-duality group arises as the symmetry group, which reflects the high degree of supersymmetry in the compactification and cannot be expected to be present for more general compactifications on some $(10-d)$-dimensional manifold $Y_{10-d}$. However, if we just consider the theory on $M_{1,d-1}\times {y}$, where $y$ is some given point on $Y_{10-d}$, we find exactly those degrees of freedom that also appear in the massless spectrum of a $(10-d)$-torus compactification. Therefore, the T-duality group is also a symmetry of the theory at any point in internal space $Y$. This means that over any point of the internal space we can assemble the degrees of freedom into representations of the T-duality group and thereby find a formulation that is covariant with respect to $SO(10-d,10-d)$.
Covariance here means that we can apply local $SO(10-d,10-d)$ transformations to the formulation without changing the theory itself. This is in complete analogy to diffeomorphism invariance of the geometrical formulation.
The $SO(10-d,10-d)$-covariant formalism constructed in this way is called generalized geometry and has been formulated and discussed in the mathematical literature in \cite{Hitchin:2004ut,Gualtieri:2003dx,Witt:2004vr,Witt:2005sk,Cavalcanti:2005hq,Hitchin:2005cv}.

If we now vary over the point $y$ in $Y$, we see that the embedding of fields into $SO(10-d,10-d)$ representations can vary. In particular, when moving from one coordinate patch to another, the representations in these patches may be related by some $SO(10-d,10-d)$ transformation. Note that, analogous to the torus case, we may only use $SO(10-d,10-d,\mathbb{Z})$ to glue together patches.
Therefore, the $SO(10-d,10-d)$ representations form bundles over the space $Y$, with a quantized curvature form, where the transition functions take values in the T-duality group. In particular, one replaces the tangent bundle $TY$ on $Y$ by the generalized tangent bundle ${\cal T} Y$, which locally looks like $TY \oplus T^*Y$. In other words, ${\cal T} Y$ admits a canonical pairing of split signature and therefore transforms under the action of the group $SO(10-d,10-d)$.

The formalism of generalized geometry is ideally suited to describe $SU(n)\times SU(n)$ structures.
Since $SU(n)\times SU(n)$ forms a subgroup of $SO(2n,2n)$, we can describe such a background by a breaking of the structure group $SO(2n,2n)\to SU(n)\times SU(n)$, similarly to the breaking $SO(2n)\to SU(n)$.
It has been shown in \cite{Gualtieri:2003dx} that an $SU(n)\times SU(n)$ structure is parameterized in terms of two almost complex structures on ${\cal T} Y$, which in turn can be mapped to a pair of ``pure'' $SO(2n,2n)$ spinors \cite{Hitchin:2004ut}.

In this review we shall apply the pure-spinor formalism of generalized geometry to $SU(2)\times SU(2)$ structures and state the purity and compatibility conditions for a pair of $SO(2n,2n)$ spinors describing an $SU(2)\times SU(2)$ structure, following our work \cite{Triendl:2009ap}. From simple chirality arguments we find that generic $SU(2)\times SU(2)$ structures do not exist but instead only manifolds with a single $SU(2)$ structure or with an identity structure can occur. The latter correspond to backgrounds with $32$ supercharges (which we do not study any further in this review). As a consequence, all smooth type II compactifications with $16$ supercharges correspond to backgrounds with an honest $SU(2)$ structure. In the context of generalized geometry, we shall denote them by $SU(2)\times SU(2)$-structure backgrounds, since $SU(2)\times SU(2)$ is still the structure group of the generalized tangent bundle.

In type II theories one can even go beyond the pure-spinor approach. The inclusion of the Ramond-Ramond sector extends the symmetry group of the massless spectrum to the so-called ``U-duality group'' $E_{11-d(11-d)}$ and makes it possible to arrange the massless fields in $E_{11-d(11-d)}$ representations \cite{Hull:2007zu,Pacheco:2008ps,Sim:2008,Grana:2009im}. The resulting formalism is called ``exceptional generalized geometry'' and is maximally covariant with respect to the symmetries of type II theories. The representations of $E_{11-d(11-d)}$ again form bundles over $Y$ that are glued together by $E_{11-d(11-d)}(\mathbb{Z})$-transformations \cite{Hull:2007zu}. For instance, the generalized tangent bundle ${\cal T} Y$ is replaced by the exceptional tangent bundle ${\cal E} Y$ which locally has the properties of $TY \oplus T^*Y$ plus a $SO(10-d,10-d)$-spinor bundle and forms the fundamental representation of $E_{11-d(11-d)}$.
The pure $SO(10-d,10-d)$ spinors parameterizing an $SU(n)\times SU(n)$ structure can be embedded into representations of $E_{11-d(11-d)}$ \cite{Grana:2009im}. As expected, this incorporates also the R-R fields into the formalism. Therefore, deformations of these embedded spinors contain all massless internal degrees of freedom of a type II background of the form \eqref{stringbackground}.

These parametrizations turn out to be very useful. One can rewrite the ten-dimensional theory in such a way that only a subalgebra of the Lorentz algebra $so(1,9)$ and thus of the supersymmetry algebra is manifest in the theory \cite{deWit:1986mz}. More precisely, one can reorder the ten-dimensional field content such that it forms representations of the lower-dimensional Lorentz group $SO(1,d-1)$ and of the internal structure group $SU(n)\times SU(n)$ \cite{Grana:2006hr}. The $SO(1,d-1)$ scalars are then identified with the internal degrees of freedom of the theory, which over each point of the ten-dimensional space in \eqref{stringbackground} can be parameterized by the pure spinors and their embeddings in $E_{11-d(11-d)}$ representations. The parameter space of $SU(n)\times SU(n)$ structures then serves as the target space of $SO(1,d-1)$ scalars in the ten-dimensional theory. We shall refer to these spaces as \emph{parameter spaces} in the following.
Note that this rewriting of the theory is performed in ten dimensions, i.e. no dimensional reduction is performed so far. However, it enables us to derive the $d$-dimensional field content when carrying out a Kaluza-Klein reduction on the background \eqref{stringbackground}, as we discuss next.

%%%%%%%%%%%%%%%%%%%%%%%%%%%%%%%%%%%%%%%%%%%%%%%
\section{Consistent truncations and low-energy effective actions}
%%%%%%%%%%%%%%%%%%%%%%%%%%%%%%%%%%%%%%%%%%%%%%%
So far we just discussed different descriptions of the ten-dimensional theory. One still needs to carry out a Kaluza-Klein reduction in order to find the $d$-dimensional low-energy effective theory. In general this is done by truncation of the spectrum below the compactification scale. Such a truncation should be performed in a consistent way, which means that the solutions to the equations of motion of the truncated theory all lift to solutions of the full theory. In particular, a consistent truncation should respect the symmetries of the theory.
Finding a consistent truncation is in general a non-trivial procedure that might strongly depend on the model under consideration.

In order to make general statements about the resulting effective theory, in this review we shall -- after stating the corresponding conditions -- simply assume the existence of a consistent truncation, in the spirit of \cite{Grana:2005ny,Grana:2006hr}. This will enable us to analyze general properties of the low-energy effective action for $SU(n)\times SU(n)$-structure backgrounds and vacua thereof.
In particular, we assume the existence of a finite basis of light modes, in which we then expand the pair of pure spinors describing the $SU(n)\times SU(n)$ structure (and their embeddings into $E_{11-d(11-d)}$ representations) and in this way parameterize the scalar sector of the theory in $d$ dimensions. The space spanned by these pure spinors consists of the light deformations of the theory and therefore gives the target space of $d$-dimensional scalar fields. We shall denote this space by \emph{scalar field space} in the following.

For a consistent truncation, the scalar field space inherits many properties from the parameter space of the ten-dimensional theory.
For example, by use of the pure spinor parametrization and its $E_{7(7)}$ embedding one can determine the scalar field space ${\cal M}$ of $SU(3)\times SU(3)$-structure compactifications to be a direct product of a special-K\"ahler and a quaternionic-K\"ahler manifold \cite{Hitchin:2004ut,Grana:2005ny,Grana:2006hr,Cassani:2007pq,Cassani:2008rb,Grana:2009im}
\begin{equation}\label{N=2product}
{\cal M} = {\cal M}_{\rm v} \times{\cal M}_{\rm h} \ .
\end{equation}
The low-energy effective action of $SU(3)\times SU(3)$-structure compactifications is an $\cN=2$ supergravity, in agreement with \eqref{N=2product}. Moreover, the quaternionic-K\"ahler manifold ${\cal M}_{\rm h}$ can be shown to be in the image of the c-map, i.e.\ it is a principal fibre bundle over a special-K\"ahler manifold \cite{Cecotti:1988qn,Ferrara:1989ik}.
The fibre of such a special quaternionic-K\"ahler manifolds admits a Heisenberg algebra of isometries \cite{deWit:1990na,deWit:1992wf}.
In $SU(3)\times SU(3)$-structure compactifications possible electric and magnetic gaugings of these isometries originate from fluxes, intrinsic torsion and non-geometric fluxes on $Y$ \cite{Polchinski:1995sm,Michelson:1996pn,D'Auria:2004tr,D'Auria:2004wd,Grana:2006hr}.

Before discussing $SU(2)\times SU(2)$-structure backgrounds, let us make one more remark on the motivation to study fluxes, intrinsic torsion and non-geometric fluxes in compactification setups.
One generic feature of string compactifications is the large number of scalars that are light compared to the compactification scale and therefore appear in the low-energy effective action. The precise number of such scalar fields and their masses depend on the chosen truncation of the theory but are usually linked to the topology of the background, e.g.\ for geometric backgrounds the topology of the compact space $Y$ and fluxes for the form fields.
In particular, massless scalar fields correspond to supersymmetry-preserving deformations of the background. The vacuum expectation values of these fields, so-called moduli, parameterize a family of supersymmetric backgrounds that have the same topology. The corresponding low-energy effective actions differ in their couplings and masses, i.e.\ couplings and masses are moduli-dependent. As a consequence, the presence of moduli eliminates the predictability of the considered theory. Viable models therefore should include additional effects that lift the masses of all scalars to a non-zero value, thereby stabilizing all moduli.
In $\cN=2$ gauged supergravity, an adequate potential to lift the scalar field sector is induced by electric and magnetic gaugings, which enables one to address the moduli problem in the context of flux compactifications \cite{Louis:2002ny,Dall'Agata:2003yr,Sommovigo:2004vj,D'Auria:2004yi}. We shall come back to this issue when discussing supersymmetric minima of $\cN=2$ supergravities.

For $SU(2)\times SU(2)$-structure compactifications the assumption of a consistent truncation leads even further. The low-energy effective theory should be a gauged $\cN=4$ supergravity, whose couplings are highly constrained by the large amount of supersymmetry, see for example \cite{Haack:2001iz,D'Auria:2002tc,D'Auria:2003jk,Derendinger:2004jn,Derendinger:2005ph,Schon:2006kz}.
For example, the light scalar fields of type IIA compactifications have to parameterize cosets
of the form \cite{de Roo:1984gd}
\begin{equation}\label{N=4coset}
\mathcal{M}_{\cN=4} = \frac{SO(10-d, n_{\rm v})}{SO(10-d)\times
  SO(n_{\rm v})}\times \mathbb{R}^+\ ,
\end{equation}
where $n_{\rm v}$ counts the number of vector multiplets and the
$\mathbb{R}^+$ factor corresponds to the dilaton.\footnote{In $d=4$
  the $\mathbb{R}^+$ factor is enlarged to the coset $Sl(2,\mathbb{R})/SO(2)$
  since the antisymmetric tensor of the NS-NS sector is dual to an axion
  and contributes to the scalar couplings.}
As we shall show, based on \cite{Triendl:2009ap}, we indeed can identify $\mathcal{M}_{\cN=4}$ as the deformation space of
$SU(2) \times SU(2)$-structure manifolds.
For a special class of $SU(2)$ structure backgrounds in $d=4$ this was already
discussed in \cite{ReidEdwards:2008rd}.
Here we analyze the
generic situation and concentrate on the scalar field space, which
corresponds to the kinetic terms of the scalars in the low-energy
effective Lagrangian.
A more detailed derivation of this Lagrangian
including the possible gaugings and the potentials will be presented
elsewhere \cite{LST}.

%%%%%%%%%%%%%%%%%%%%%%%%%%%%%%%%%%%%%%%%%%%%%%%
\section[Partial supersymmetry breaking and $\cN=1$ theories]{Partial supersymmetry breaking and $\cN=1$ effective theories}
%%%%%%%%%%%%%%%%%%%%%%%%%%%%%%%%%%%%%%%%%%%%%%%
A central question when analyzing the low-energy effective action of flux compactifications is the amount of supersymmetry that is preserved by the vacuum of the theory.
If symmetries of the theory are spontaneously broken in the vacuum, the scale of symmetry breaking naturally coincides with the compactification scale, which is usually high compared to scales accessible for phenomenology.
On the other hand, phenomenological features of supersymmetry and its good renormalization properties suggest that models should be preferred that allow supersymmetry to be unbroken down to considerably low energies. Due to this separation of the supersymmetry breaking and the compactification scale, it is natural to first find effective theories that allow for supersymmetric vacua and subsequently modify these theories in such a way that supersymmetry is broken by some low-energy effect.
We will concentrate on the first step and discuss the existence of $(\cN=1)$-supersymmetric vacua in low-energy effective theories.

The effective action arising from supersymmetric compactifications of type II theories usually admits a non-minimal amount of supersymmetry. As discussed in Section \ref{section:flux_comp}, globally defined nowhere-vanishing spinors on manifold $Y$ give rise to supersymmetries of the effective action. The number of such spinors determines the amount of supersymmetry in the action. On the other hand, imposing the existence of these spinors reduces the structure group of $Y$. For instance $SU(3)\times SU(3)$ structure compactifications allow for one nowhere-vanishing spinors on $Y_6$, leading to $\cN=2$ supergravity theories in $d=4$ in the reduction. Therefore, supersymmetric type II compactifications admit usually at least $\cN=2$ supersymmetry.
However, $\cN=1$ vacua are highly preferred because they are much closer to phenomenologically viable models.

One way to achieve an $\cN=1$ theory is given by truncating the $\cN=2$ supergravity such that the
surviving fields give only rise to an $\cN=1$-supersymmetric theory \cite{Andrianopoli:2001zh,Andrianopoli:2001gm,Andrianopoli:2002rm,Andrianopoli:2002vq}, corresponding to some orientifold projection in the type II compactification \cite{Grimm:2004uq,Grimm:2004ua,Benmachiche:2006df,Cassani:2007pq}.
An alternative scheme is provided by finding and classifying $\cN=1$ vacua of $\cN=2$ supergravities with a Minkowski or AdS geometry. It turns out that this is naturally related to gaugings in $\cN=2$ supergravity (see for example \cite{Andrianopoli:1996cm} and references therein for a discussion of gauged supergravity). If $\cN=2$ supergravity is for example ungauged, there are no $\cN=1$-supersymmetric vacua. Gaugings however might lead to spontaneous supersymmetry breaking, similar to the Higgs mechanism in the Standard Model. Analogously to electroweak symmetry breaking, the super-Higgs mechanism induces a mass for the ``gauge field'', which in the case of $\cN=2\to\cN=1$ breaking is one of the two gravitini, which therefore forms a massive $\cN=1$ gravitino multiplet.

It turns out that a crucial requirement for the appearance of $\cN=1$ vacua is the inclusion of magnetic charges in the theory. It has been shown already in \cite{Cecotti:1984rk,Cecotti:1984wn} that for a standard gauged $\cN=2$ supergravity with only electric charges no $\cN=1$ Minkowski vacua exist.
The possibility of partial supersymmetry breaking in globally $\cN=2$ supersymmetric theories in four spacetime dimensions was subsequently discovered in \cite{Bagger:1994vj,Antoniadis:1995vb}. In particular, it was observed in \cite{Antoniadis:1995vb} that the presence of a magnetic Fayet-Iliopoulos term spontaneously breaks $\cN=2$ to $\cN=1$.
The supergravity version of this situation was presented in \cite{Ferrara:1995gu,Ferrara:1995xi,Fre:1996js} for a specific class of gauged $\cN=2$ theories. There it was found that the no-go theorem of \cite{Cecotti:1984rk,Cecotti:1984wn} could be avoided in a specific basis for the scalar fields of the $\cN=2$ vector multiplets.
In this review we shall attempt a more systematic analysis of the problem, going beyond the few explicit examples mentioned above by finding and then solving the general conditions in $\cN=2$ supergravity for partial supersymmetry breaking in Minkowski and anti-de Sitter spacetimes (for an analysis in three-dimensional spacetime, see \cite{Hohm:2004rc}).

The situation becomes more delicate in the context of flux compactifications. In classical gravity one is faced with another no-go theorem, due to Gibbons \cite{Gibbons:1984kp}, de Wit et al.\ \cite{deWit:1986xg} and Maldacena and Nu\~{n}ez \cite{Maldacena:2000mw}, which forbids flux compactifications to Minkowski space in the absence of negative energy-density sources, regardless of the amount of supersymmetry preserved. Furthermore, in \cite{Mayr:2000hh} it was noted that even if one evades the various no-go theorems and finds an $\cN=1$ vacuum, worldsheet instanton corrections in $\cN=2$ flux compactifications could ruin the result and reinstate the no-go theorem forbidding partial supersymmetry breaking. We shall address these points in our analysis in the context of gauged $\cN=2$ supergravities coming from $SU(3)\times SU(3)$-structure compactifications.

An important tool for computations in $\cN=2$ supergravity is special geometry: The geometry of the special-K\"ahler component of the scalar field space \eqref{N=2product} is determined by a single holomorphic function, the so-called holomorphic prepotential.
In the known examples of partial supersymmetry breaking however the holomorphic prepotential does not exist, as one of the gauge bosons has been exchanged with its magnetic dual via a symplectic rotation \cite{Ceresole:1995jg}. The lack of a prepotential makes it difficult to generalize the discussion to arbitrary $\cN=2$ supergravities. Therefore, it is advantageous to reinstate the prepotential, which one can always do at the expense of having to introduce both electric and magnetic charges. It turns out that the embedding tensor formalism \cite{deWit:2002vt,deWit:2005ub} is ideally suited to address this problem. This formalism treats electric and magnetic gauge bosons on the same footing and the conditions for partial supersymmetry breaking can then be formulated as a condition on the embedding tensor itself.
Indeed, using this we shall construct a general solution for Minkowski and AdS vacua displaying  $\cN=1$ supersymmetry for a broad class of $\cN=2$ gauged supergravities, following our work \cite{Louis:2009xd}.
More precisely, the conditions for
partial supersymmetry breaking primarily determine the structure of
the embedding tensor, i.e.\ the spectrum of electric and magnetic
charges, but do not constrain the
scalar field space ${\cal M}_{\rm v}$ of the vector multiplets.
In the hypermultiplet sector on the other hand,
the scalar field space ${\cal M}_{\rm h}$ has to admit
at least two linearly independent, commuting
isometries. Gauging these isometries is necessary in order to induce masses for the two Abelian gauge
bosons which join the heavy gravitino in a massive $\cN=1$ gravitino
multiplet.  Partial supersymmetry breaking further demands that a
specific linear combination of the two Killing vectors generating the isometries
is holomorphic with respect to one of the three almost complex
structures which exist on ${\cal M}_{\rm h}$.

We explicitly identify two such Killing vectors for the specific class
of special quater\-nionic-K\"ahler manifolds, which are in the image of the c-map (cf.\ the discussion below \eqref{N=2product}).
More precisely, we give the construction of embedding tensors that lead to $\cN=1$ vacua for any moduli space that admits the Heisenberg algebra of Killing vectors naturally appearing in flux compactifications of type II string theory. Moreover, we find that by adjusting the charges one can realize $\cN=1$ vacua at any point of the moduli space.

We discuss the `uplift' of our solutions to flux compactifications. We show that by rewriting the conditions for partial supersymmetry breaking in terms of the embedding tensor, the compactification no-go theorem of \cite{Gibbons:1984kp,deWit:1986xg,Maldacena:2000mw} can be evaded by including non-geometric fluxes. As we are able to phrase the conditions for an $\cN=1$ vacuum in terms of a general holomorphic prepotential, this also opens up the possibility of finding solutions in the presence of instanton corrections. Finally, the flux quantization condition forces the embedding tensor to have integer entries only, leading to a lattice in the moduli space where $\cN=1$ vacua can be realized. More importantly, this might restrict the possibility of $\cN=1$ vacua to a subclass of moduli spaces.

After finding the conditions on the gauged $\cN=2$ supergravity, we go one step further and
derive the $\cN=1$ low-energy effective
action that is valid below the scale of partial supersymmetry breaking $\mino$
or, in other words, below the scale set by the heavy gravitino, based on \cite{Louis:2010ui}.
In order to achieve this we integrate out the entire massive
$\cN=1$ gravitino multiplet (containing fields with spin $s=(3/2,1,1,1/2)$)
together with
all other multiplets which, due to the symmetry breaking, acquire masses
of ${\cal O}(\mino)$.  This results in
an effective $\cN=1$ theory whose couplings are determined by the
couplings of the `parent' $\cN=2$ theory. Note that some aspects of this analysis have already been discussed in \cite{Louis:2002vy,Gunara:2003td}.

An interesting aspect of the $\cN=1$ effective theory is the structure of its
scalar field space ${\cal M}^{\cN=1}$, which is fixed by $\cN=1$ supersymmetry to be K\"ahler. In $\cN=2$ supergravities the scalar field space ${\cal M}$ is a direct
product of the form \eqref{N=2product}. In particular, ${\cal M}_{\rm h}$ is quaternionic-K\"ahler but not K\"ahler.
We shall see that the process of integrating out the two heavy
gauge bosons corresponds to taking the quotient of
%the $4 n_{\rm  h}$-dimensional quaternionic-K\"ahler manifold
${\cal M}_{\rm h}$ with respect to the two isometries generating the partial supersymmetry breaking.  This leaves a manifold $\hat{\cal M}_{\rm h}={\cal M}_{\rm h}/\mathbb{R}^2$ where the two
`missing' scalar fields are the Goldstone bosons eaten by the heavy
gauge bosons. We shall show that $\hat{\cal M}_{\rm h}$ is equipped with a
K\"ahler metric consistent with the $\cN=1$ supersymmetry of the low-energy effective theory \cite{Cortes}. It is also possible that, apart from the two gauge bosons, other scalar fields (from both vector and
hypermultiplets) acquire a mass of ${\cal O}(m_{3/2})$ and thus
have to be integrated out, leading to a further reduction of the scalar field space. However, as such scalars are not Goldstone bosons this process simply amounts to projecting to a K\"ahler submanifold of
$\hat{\cal M}_{\rm h}\times {\cal M}_{\rm v}$, rather than taking a quotient.  The resulting $\cN=1$ scalar field
space is then given by
\begin{equation}
  \label{N=1product}
 {\cal M}^{\cN=1}\ =\ \hat{\cal M}_{\rm h} \times \hat{\cal M}_{\rm v}\ ,
 \end{equation}
where $\hat{\cal M}_{\rm v}$ is a submanifold of ${\cal M}_{\rm v}$.\footnote{For
notational simplicity we did not introduce a new symbol for the
submanifold of $\hat{\cal M}_{\rm h}$.}

The dimension of ${\cal M}^{\cN=1}$ can be as large as $\operatorname{dim}{\cal M}-2$, if only the two Goldstone bosons have been removed from the scalar field space by integrating out the heavy gauge bosons. However, depending on the specific couplings, the dimension of ${\cal M}^{\cN=1}$
can be much smaller if most scalars are stabilized at the scale $\mino$.
Indeed we shall see that generically all scalars coming from vector multiplets are stabilized. For the class of special quaternionic-K\"ahler manifolds, a similar conclusion can be found for the scalars coming from the hypermultiplets.

%%%%%%%%%%%%%%%%%%%%%%%%%%%%%%%%%%%%%%%%%%%%%%%
\section{Organization of this review article}
%%%%%%%%%%%%%%%%%%%%%%%%%%%%%%%%%%%%%%%%%%%%%%%
This review is organized as follows. In Chapter \ref{section:Generalized} we introduce basic concepts for $SU(n)\times SU(n)$-structure backgrounds. In Section \ref{section:basics} we define them by imposing the existence of nowhere-vanishing spinors on $Y$. In Section \ref{section:spectrum} we decompose the field content of ten-dimensional type II supergravity into representations of the $d$-dimensional Lorentz group and the structure group and identify them with multiplets of the $d$-dimensional supersymmetry algebra. In particular, we discover that the gravitino multiplets that become massive in the compactification process sit in the $n$-plet representations of $SU(n)\times SU(n)$. In Section \ref{section:SUn} we take a first step to find the parameter space of the scalars of the $d$-dimensional Lorentz group by analyzing geometric $SU(n)$ structures and deriving the parameter space of geometric deformations.

In Chapter \ref{section:covariant} we then use the techniques of generalized geometry to derive the parameter space of all $d$-dimensional scalars for $SU(n)\times SU(n)$-structure backgrounds. In Section \ref{section:pure_spinors} we introduce the formalism of generalized geometry and give a description of $SU(n)\times SU(n)$ backgrounds in terms of pure spinors. Using this, we derive the parameter space of the NS-NS sector. In Section \ref{section:Exceptional} we then also incorporate the R-R scalars by introducing the $E_{11-d(11-d)}$-covariant formalism of exceptional generalized geometry and apply it to the case $d=6$ to derive the parameter space of the six-dimensional theory. After reviewing the embedding of pure spinor pairs that describe $SU(3)\times SU(3)$-structure backgrounds into representations of $E_{7(7)}$, we determine the parameter space of all scalars in $SU(2)\times SU(2)$-structure backgrounds for $d=4$.

In Chapter \ref{section:effective} we finally derive the scalar field spaces of low-energy effective supergravities of $SU(2)\times SU(2)$-structure compactifications and review the appearance of $\cN=2$ supergravity in $SU(3)\times SU(3)$-structure compactifications. In Section \ref{section:truncation} we discuss the consistency conditions for Kaluza-Klein truncations, which in Section \ref{section:moduliSU2} we assume to be satisfied in order to derive the scalar field spaces of the $d$-dimensional low-energy effective theory by use of the parameter spaces derived in Chapter \ref{section:covariant}. In Section \ref{section:EffActionN=2} we provide basic material for the following chapter and review gauged $\cN=2$ supergravities as they appear in $SU(3)\times SU(3)$-structure compactifications.

In Chapter \ref{section:SPSB} we discuss $\cN=1$ vacua of $\cN=2$ gauged supergravity and string theory. In Section \ref{section:vectors} we analyze the possibility of partial supersymmetry breaking in gauged $\cN=2$ supergravities. In particular, we show how to evade the classical no-go theorems for Minkowski vacua and construct the general solution. In Section \ref{section:LEEA} we then derive the effective $\cN=1$ supergravity by integrating out the massive fields in the spontaneously partially broken phase. In Section \ref{section:breaking_hyper} we focus on the class of special quaternionic-K\"ahler moduli spaces, which generally arise in $\cN=2$ compactifications of type II string theory. We apply our results of the preceding sections in order to construct the general solutions for both $\cN=1$ Minkowski and AdS vacua and to specify the properties of the $\cN=1$ effective action.
Finally, we comment on stringy effects in the corresponding flux compactifications in Section \ref{section:strings}.
We present our conclusions in Chapter \ref{section:conclusion}. Our conventions are given in the appendix.

The new results of Chapter \ref{section:Generalized} and Chapter \ref{section:covariant}, regarding $SU(2)\times SU(2)$ structures, have been published in \cite{Triendl:2009ap}, where also backgrounds of the form \eqref{stringbackground} with $d=5$ are discussed.
Furthermore, Chapter \ref{section:SPSB} reflects the content of \cite{Louis:2009xd,Louis:2010ui}.
% The content of \cite{Braun:2009bh} has not been included in the thesis.

\cleardoublepage
%%%%%%%%%%%%%%%%%%%%%%%%%%%%%%%%%%%%%%%%%%%%%%%
\chapter{$SU(n)\times SU(n)$-structure backgrounds} \label{section:Generalized}
%%%%%%%%%%%%%%%%%%%%%%%%%%%%%%%%%%%%%%%%%%%%%%%
In this chapter we will introduce the general setup of $SU(n)\times SU(n)$-structure backgrounds, $n=2,3$,
and gather basic information about such backgrounds by use of group-theoretical considerations. In particular we will see that unbroken supersymmetry forces the manifold to admit an $SU(n)\times SU(n)$ structure. We then will decompose the ten-dimensional field content in multiplets of the corresponding $d$-dimensional supersymmetry algebra. This corresponds to a rewriting of the ten-dimensional action such that it is only manifestly covariant with respect to this reduced supersymmetry algebra. Finally, we will describe geometric $SU(n)$ structures -- as a warm-up before introducing covariant formulations in Chapter~\ref{section:covariant}.

%%%%%%%%%%%%%%%%%%%%%%%%%%%%%%%%%%%%%%%%%%%%%%%
\section{Basics on $SU(n)\times SU(n)$ compactifications} \label{section:basics}
%%%%%%%%%%%%%%%%%%%%%%%%%%%%%%%%%%%%%%%%%%%%%%%

In backgrounds of the form \eqref{stringbackground} the ten-dimensional Lorentz group decomposes naturally into $SO(1,d-1)\times SO(10-d)$.
We are interested in backgrounds that are able to preserve at least some amount of supersymmetry in the low-energy effective theory. This naturally demands that the $d$-dimensional supersymmetry generator $\epsilon$ should lift to a ten-dimensional supersymmetry generator \cite{Grana:2004bg}
\begin{equation}
 \epsilon^{(10)}(x,y) = \epsilon(x) \otimes \eta(y) + \epsilon^c(x) \otimes \eta^c(y)
\end{equation}
in such a way that the corresponding supersymmetry tranformations relate the zero modes in the spectrum.
Here the superscript $c$ refers to the charge conjugate of the corresponding spinor. Since the zero modes of the Kaluza-Klein spectrum are usually nowhere-vanishing objects on $Y_{10-d}$, the same should hold for the internal spinor $\eta$.
The existence of such a spinor $\eta$ strongly restricts the internal manifold $Y$. If an object is globally-defined, it should not depend on the choice of charts on the manifold. More precisely, there should be charts on $Y$ such that $\eta$ does not depend on the chart used and therefore does not transform when one moves from one chart to another. This restricts the possible linear transformations that can be used for such chart transitions. The group of such linear transformations, the so-called \emph{structure group}, is therefore reduced in such a way that $\eta$ is in its singlet representation.
In the following we restrict to the case of even $d$.\footnote{The case of $d=5$ is discussed in \cite{Sim:2008,Triendl:2009ap}. The structure group in that case reduces to $SU(2)$.} Then, the structure group is reduced from the Lorentz group $SO(10-d)$ to $SU(n)$, $n=\frac{10-d}{2}$, which is the largest subgroup that admits a singlet spinor under the breaking.

In the following we concentrate on backgrounds for the type II string. This means we have two ten-dimensional supersymmetry generators $\epsilon^{(10)}_1$ and $\epsilon^{(10)}_2$, where each descends via some internal $SU(n)$ structure to a $d$-dimensional supersymmetry generator $\epsilon_1$ and $\epsilon_2$ where
\begin{equation}
  \epsilon^{(10)}_1 = \epsilon_{1} \otimes \eta_{1} + \epsilon^c_{1} \otimes \eta^c_{1} \ , \qquad
  \epsilon^{(10)}_2 = \epsilon_{2} \otimes \eta^c_{2} + \epsilon^c_{2} \otimes \eta_{2}
\end{equation}
holds for type IIA and
\begin{equation}
  \epsilon^{(10)}_i = \epsilon_{i} \otimes \eta_{i} + \epsilon^c_{i} \otimes \eta^c_{i}
\end{equation}
for type IIB, with $i=1,2$.
Each of the internal spinors $\eta_{1}$ and $\eta_{2}$ defines an $SU(n)$ structure, defining together a so-called $SU(n) \times SU(n)$ structure \cite{Jeschek:2004wy,Grana:2004bg}. There are two limiting cases for such $SU(n) \times SU(n)$ structures. If $\eta_{1}$ and $\eta_{2}$ are parallel at every point of $Y$, then this reduces to the case of a single $SU(n)$ structure. The other limiting case is when $\eta_{1}$ and $\eta_{2}$ can be chosen such that they are orthogonal at every point of $Y$. In this case, we have actually more supersymmetry generators in four dimensions since each of $\eta_{1}$ and $\eta_{2}$ give rise to $d$-dimensional supersymmetry generators for each ten-dimensional one.
The generic $SU(n) \times SU(n)$-structure case is more conveniently described by generalized geometry, which we introduce in Chapter~\ref{section:covariant}.

Let us be more concrete now. For a background \eqref{stringbackground} with $d=6$, $Y_4$ is a four-dimensional manifold and the Lorentz group is $SO(1,5)\times SO(4)$. The ten-dimensional Majorana-Weyl spinor representation ${\bf 16}$ decomposes accordingly as
\begin{equation}
  {\bf 16} \to ({\bf 4}, {\bf 2}) \oplus ({\bf \bar 4}, {\bf \bar 2})  \ .
\end{equation}
If we want a singlet on $Y_4$, we see that $SO(4)$ must be reduced to $SU(2)$ such that the spinor representations decompose as
\begin{equation} \label{SU2_spinor_doublet_decomposition}
 {\bf 2} \to {\bf 1} \oplus {\bf 1} \ , \qquad {\bf \bar 2} \to {\bf \bar 2} \ .
\end{equation}
The two singlets here are given by $\eta$ and $\eta^c$, which are both of the same chirality since charge conjugation in four dimensions preserves chirality. Since they are linearly independent, they together span the whole Weyl spinor space of given chirality.\footnote{Note that it is pure convention whether we denote the reduced representation in \eqref{SU2_spinor_doublet_decomposition} by ${\bf 2}$ or by ${\bf \bar 2}$, depending on the chirality of $\eta$.} This is the case of an $SU(2)$ structure, which preserves $16$ supercharges corresponding to $\cN=2$ in six dimensions.

For $d=4$, the Lorentz group decomposes into $SO(1,3)\times SO(6)$, and the ten-dimensional spinor representation accordingly as
\begin{equation}
  {\bf 16} \to ({\bf 2}, {\bf 4}) \oplus ({\bf \bar 2}, {\bf \bar 4})  \ .
\end{equation}
Under a breaking $SO(6) \to SU(3)$ the decomposition of the spinor bundles reads
\begin{equation}\label{4to31}
 {\bf 4} \to {\bf 3} \oplus {\bf 1} \ , \qquad {\bf \bar 4} \to {\bf \bar 3} \oplus {\bf 1} \ ,
\end{equation}
where the two singlets are again given by $\eta$ and $\eta^c$ but are of opposite chirality now. This is the case of an $SU(3)$ structure, which preserves $\cN=2$ supersymmetry in the four-dimensional theory.

If one imposes in this case the existence of a second, linearly independent spinor, the structure group is further broken to $SU(2)$, such that under $SO(6)\to SU(2)$ there is
\begin{equation}
 {\bf 4} \to {\bf 2} \oplus {\bf 1}\oplus {\bf 1}  \ , \qquad {\bf \bar 4} \to {\bf \bar 2} \oplus {\bf 1}\oplus {\bf 1} \ .
\end{equation}
Therefore, $SU(2)$-structure compactifications to four dimensions usually preserve $\cN=4$ supersymmetry because the number of four-dimensional supersymmetry generators is doubled.

One could further reduce the structure group by introducing additional nowhere-vanishing spinors on $Y$. This usually reduces the structure group from $SU(2)$ to the trivial subgroup. We will not discuss this case any further in this review and refer to \cite{Dabholkar:2002sy,Hull:2004in,Dabholkar:2005ve,Kachru:2002sk,Flournoy:2004vn,Dall'Agata:2005ff,Hull:2005hk,Dall'Agata:2005mj,Shelton:2005cf} for further details.

%%%%%%%%%%%%%%%%%%%%%%%%%%%%%%%%%%%%%%%%%%%%%%%
\section{Field decompositions} \label{section:spectrum}
%%%%%%%%%%%%%%%%%%%%%%%%%%%%%%%%%%%%%%%%%%%%%%%
Above we introduced the relevant structure groups for $\cN=2$ and $\cN=4$ compactifications of the type II string. The structure group times the $d$-dimensional Lorentz group emerges as a subgroup of the ten-dimensional Lorentz group. The massless spectrum of the ten-dimensional string, which is reviewed in Section \ref{section:flux_comp}, consists of ten-dimensional Lorentz representations and decomposes accordingly into representations of the $d$-dimensional Lorentz-group times the structure group. Furthermore, for massless fields we can use the light-cone gauge in order to reduce to the physical degrees of freedom that are then representations of the little group. More precisely, massless ten-dimensional fields come in representations of $SO(8)$ and decompose into representations of $SO(d-2)\times SU(n)$. Since string fields are combinations of left- and right-moving excitations, they are in representations of $SO(8)_L\times SO(8)_R$ and decompose into representations of $SO(d-2)\times SU(n)_L\times SU(n)_R$.

In this section we analyze how the ten-dimensional massless field content decomposes into representations of $SO(d-2)\times SU(n)_L\times SU(n)_R$. By identifying the emerging representations one is able to relate representations of the structure group with supersymmetry multiplets in $d$ dimensions. As we see below, in all cases the $d$-dimensional gravity multiplet sits in the singlet representation of the structure group while additional gravitino multiplets come in the $({\bf n},{\bf 1})$ and $({\bf 1},{\bf n})$ representations and the conjugates thereof, i.e.\ in the $n$-plet representations. Additional vector and possibly matter multiplets come as singlets or as higher representations of the structure group such as $({\bf n},{\bf n})$ and $({\bf \bar n},{\bf n})$ and conjugates thereof.

Note that the resulting fields still depend on all
coordinates of the ten-dimensional spacetime, i.e.\ we have not performed
any Kaluza-Klein truncation on the spectrum but really deal with
ten-dimensional backgrounds. This procedure just corresponds to a
rewriting of the ten-dimensional
supergravity in a form where instead of the
ten-dimensional Lorentz group only the $d$-dimensional Lorentz group times the structure group with eight or sixteen
supercharges is manifest.
This rewriting of the
ten-dimensional theory has been
pioneered in Ref.~\cite{deWit:1986mz} and applied to the case of $SU(3)\times
SU(3)$ structures in Refs.~\cite{Grana:2005ny,Koerber:2007xk}.

When we perform the Kaluza-Klein truncation in Section \ref{section:effective}, the above field content is reduced to the zero-modes of the fields. In this process, all additional gravitino multiplets should acquire a mass of order of the compactification scale in order to arrive at a low-energy action with the right amount of supersymmetry. The same happens to all other fields in the $n$-plet representation and truncating the theory below the compactification scale then removes all $n$-plets.
Therefore, whenever we discuss massless fields of the theory, including supersymmetric deformations of the background, we should project out the $n$-plet representation as they are removed in the compactification procedure.

Furthermore, the singlet representation of the structure group always corresponds to exactly one zero-mode, which is nowhere-vanishing on $Y$. Therefore, each $SU(n)\times SU(n)$ singlet descends to exactly one field in the low-energy effective action and their couplings are universal, i.e.\ independent of the precise form of the background. On the other hand, higher $SU(n)\times SU(n)$ representations may have an arbitrary number of massless fields in the effective action, whose number and couplings are determined by the background. Let us now discuss the field decompositions for $d=6$ and $d=4$.

%%%%%%%%%%%%%%%%%%%%%%%%%%%%%%%%%%%%%%%%%%%%%%%%%%%%%%%%%%%%%%%%%
\subsection{$SU(2)\times SU(2)$ field decompositions in $d=6$}\label{section:spectrumd6}
%%%%%%%%%%%%%%%%%%%%%%%%%%%%%%%%%%%%%%%%%%%%%%%%%%%%%%%%%%%%%%%%%
We start by discussing the case of an $SU(2)\times SU(2)$ structure in $d=6$, following \cite{Triendl:2009ap}.
This corresponds to decomposing the ten-dimensional massless field content for the breaking $SO(8)_L \times SO(8)_R \to SO(4)_{l} \times SU(2) \times SU(2)$, where for clarity we used the subscript $l$ to denote the ``little group'' of $SO(1,5)$ and to distinguish it from the Lorentz group on $Y_4$.
We gave the massless spectrum of the type II string in ten dimensions already in Section \ref{section:flux_comp}. It consists of the tensor product of the left-moving modes in the representation ${\bf 8}^v \oplus {\bf 8}^s$ with the right-moving representations ${\bf 8}^v \oplus {\bf 8}^c$ (${\bf 8}^v \oplus {\bf 8}^s$) for type IIA (IIB). Here, the vector representations give the NS sector while the spinor representations come from the R sector.
In order to understand the decomposition of the string modes, let us first recall the decomposition of the two Majorana-Weyl representations ${\bf 8}^s$ and ${\bf 8}^c$ and the vector representation ${\bf 8}^v$ under the breaking
\begin{equation}
SO(8) \to SO(4)_{l} \times SO(4) \to SO(4)_{l} \times SU(2) \ .
\end{equation}
We get
\begin{equation} \begin{aligned}
   {\bf 8}^s & \to {\bf 2}_{\bf 2} \oplus {\bf \bar{2}}_{\bf \bar{2}}
\to {\bf 2}_{\bf 2}  \oplus 2 \, {\bf 1}_{\bf \bar{2}} \ , \\
   {\bf 8}^c & \to {\bf 2}_{\bf \bar{2}} \oplus {\bf \bar{2}}_{\bf 2} \to {\bf 2}_{\bf \bar{2}}  \oplus 2\, {\bf 1}_{\bf 2} \ , \\
   {\bf 8}^v & \to {\bf 1}_{\bf 4} \oplus {\bf 4}_{\bf 1} \to {\bf
1}_{\bf 4}  \oplus 2 \, {\bf 2}_{\bf 1} \ ,
\end{aligned}\end{equation}
where the subscript denotes the representation under the group
$SO(4)_{l}$.

In type IIA string theory the massless fermionic degrees of freedom
originate from the $({\bf 8}^s,{\bf 8}^v)$ and $({\bf 8}^v,{\bf 8}^c)$
representation of $SO(8)_L \times SO(8)_R$, while in type IIB they
originate from the $({\bf 8}^s,{\bf 8}^v)$ and $({\bf 8}^v,{\bf 8}^s)$ representation.
Decomposing the fermions under $SO(8)_L \times SO(8)_R \to SO(4)_{l} \times
SU(2)_L \times SU(2)_R$ we find
\begin{equation}\label{fermiondec}\begin{aligned}
  ({\bf 8}^s,{\bf 8}^v) & \to 2 ({\bf 1},{\bf 1})_{\bf 6} \oplus 2 ({\bf 1},{\bf 1})_{\bf 2} \oplus 4 ({\bf 1},{\bf 2})_{\bf \bar{2}} \oplus ({\bf 2},{\bf 1})_{\bf \bar{6}} \oplus ({\bf 2},{\bf 1})_{\bf \bar{2}} \oplus 2 ({\bf 2},{\bf 2})_{\bf 2} \ , \\
  ({\bf 8}^v,{\bf 8}^s) & \to 2 ({\bf 1},{\bf 1})_{\bf 6} \oplus 2 ({\bf 1},{\bf 1})_{\bf 2} \oplus ({\bf 1},{\bf 2})_{\bf \bar{6}} \oplus ({\bf 1},{\bf 2})_{\bf \bar{2}} \oplus 4 ({\bf 2},{\bf 1})_{\bf \bar{2}} \oplus 2 ({\bf 2},{\bf 2})_{\bf 2} \ , \\
  ({\bf 8}^v,{\bf 8}^c) & \to 2 ({\bf 1},{\bf 1})_{\bf \bar{6}} \oplus 2 ({\bf 1},{\bf 1})_{\bf \bar{2}} \oplus ({\bf 1},{\bf 2})_{\bf 6} \oplus ({\bf 1},{\bf 2})_{\bf 2} \oplus 4 ({\bf 2},{\bf 1})_{\bf 2} \oplus 2 ({\bf 2},{\bf 2})_{\bf \bar{2}} \ .
\end{aligned}\end{equation}
We see that half of the gravitinos, denoted by the subscript $\bf 6$
and $\bf \bar{6}$, come in the $({\bf 1},{\bf 1})$ representation
while the other half is in the doublet representations $({\bf 1},{\bf
2})$ and $({\bf 2},{\bf 1})$ of $SU(2)_L \times SU(2)_R$. We will see
below that the $6d$ graviton is in the $({\bf 1},{\bf 1})$
representation and thus this representation labels the gravity
multiplet in six dimensions. Hence the $({\bf 1},{\bf 2})$ and $({\bf
2},{\bf 1})$ representations correspond to additional gravitino multiplets,
which acquire a mass at the Kaluza-Klein scale in the compactification process. We have to project out these representations to end up
with a standard $\cN =2$ supergravity in six
dimensions (as discussed above) \cite{Andrianopoli:2001zh,Andrianopoli:2001gm}.
After this projection, the fermionic components in the $({\bf 1},{\bf
1})$ representation become part of the gravity multiplet, while the $({\bf 2},{\bf
2})$ components correspond to the fermionic degrees of freedom in the
$\cN =2$ vector and tensor multiplets in type IIA and IIB,
respectively.\footnote{In type IIB, only the anti-self-dual part of the
antisymmetric two-tensor is part of the gravity multiplet \cite{Romans:1986er}. The
self-dual component forms a tensor multiplet together with scalars in
the R-R sector. This tensor multiplet is also in the $({\bf 1},{\bf 1})$ representation.}

The massless bosonic fields of type II supergravity can be decomposed
analogously. As stated in Section \ref{section:flux_comp}, the NS-NS sector consists of
a tensor $E_{MN}$ in the ${\bf 8}^v \otimes {\bf 8}^v$ representation, which gives the NS-NS fields
\begin{equation}
E_{MN} = g_{MN} + B_{MN} + \phi^{(10)} \, \eta_{MN} \ ,
\end{equation}
where $g_{MN}$ is symmetric and traceless and corresponds to metric degrees of freedom, $B_{MN}$ is an anti-symmetric tensor and $\eta_{MN}$ is the (fixed) ten-dimensional Minkowski metric so that the ten-dimensional dilaton $\phi^{(10)}$ corresponds to the trace of $E_{MN}$. $E_{MN}$ decomposes under the breaking $SO(8)_L \times SO(8)_R \to SO(4)_{l} \times SU(2) \times SU(2)$ as
\begin{equation}\begin{aligned}\label{NSdec}
 E_{\mu \nu} & : ({\bf 1},{\bf 1})_{{\bf 9}} \oplus ({\bf 1},{\bf 1})_{{\bf 1}} \oplus ({\bf 1},{\bf 1})_{{\bf 3} \oplus {\bf \bar{3}}} \ , \\
 E_{\mu m} & : 2 ({\bf 1},{\bf 2})_{\bf 4}  \ ,  \\
 E_{m \mu } & : 2 ({\bf 2},{\bf 1})_{\bf 4}  \ ,  \\
 E_{mn} & : 4 ({\bf 2},{\bf 2})_{\bf 1}  \ .
\end{aligned}\end{equation}
Projecting out the doublets eliminates the six-dimensional vectors
$E_{\mu m}$ and $E_{m \mu }$, and we are left with $E_{\mu \nu}$,
i.e.\ the metric, the six-dimensional dilaton and the antisymmetric two-tensor, which are part of the
gravity multiplet, and the scalars $E_{mn}$, which reside in vector
or tensor multiplets. Since the latter ones correspond to the
internal metric and $B$-field components, they can be associated with
deformations of the $SU(2) \times SU(2)$ background.

Finally, in the R-R sector we need to decompose the $({\bf 8}^s,{\bf
8}^c)$ representation in type IIA and the $({\bf 8}^s,{\bf 8}^s)$ in type IIB. One finds
\begin{equation}\begin{aligned}\label{RRdec}
  ({\bf 8}^s,{\bf 8}^c) & \to 4 ({\bf 1},{\bf 1})_{\bf 4} \oplus 2 ({\bf 1},{\bf 2})_{\bf \bar{3}} \oplus 2 ({\bf 1},{\bf 2})_{\bf 1} \oplus 2 ({\bf 2},{\bf 1})_{\bf 3} \oplus 2 ({\bf 2},{\bf 1})_{\bf 1} \oplus ({\bf 2},{\bf 2})_{\bf 4} \ , \\
  ({\bf 8}^s,{\bf 8}^s) & \to 4 ({\bf 1},{\bf 1})_{\bf \bar{3}} \oplus 4 ({\bf 1},{\bf 1})_{\bf 1} \oplus 2 ({\bf 1},{\bf 2})_{\bf 4} \oplus 2 ({\bf 2},{\bf 1})_{\bf 4} \oplus \phantom{2}({\bf 2},{\bf 2})_{\bf 3} \oplus ({\bf 2},{\bf 2})_{\bf 1} \ .
\end{aligned}\end{equation}
We see that in type IIA only six-dimensional vectors in the
R-R sector survive the projection. Those which are in the
$({\bf 1},{\bf 1})$ representation form the graviphotons in the
gravity multiplet, those in the $({\bf 2},{\bf 2})$ give the vectors
in the vector multiplets.

Projecting out all $SU(2) \times SU(2)$
doublets leaves a spectrum that  for type IIA combines into
a gravitational multiplets plus a vector multiplet of the non-chiral
$d=6, \cN =2$ supergravity. For type IIB we obtain instead
a gravitational multiplets and two tensor multiplet of the chiral
$\cN =2$ supergravity.
To be more precise, in type IIA the gravitational multiplet contains
the graviton, an antisymmetric tensor, two (non-chiral) gravitini, four vector
fields, four Weyl fermions and a real scalar. These degrees of freedom
precisely correspond to the $({\bf 1},{\bf 1})$ representation of the
decompositions given in \eqref{fermiondec}, \eqref{NSdec} and
\eqref{RRdec}.
The vector multiplet contains a vector field, four gaugini and four
real scalars. These arise in the $({\bf 2},{\bf 2})$ representation of
the above decompositions.
In type IIB the gravitational multiplet contains
the graviton, five self-dual antisymmetric tensor and two (chiral)
gravitini. These degree of freedom are found in the $({\bf 1},{\bf 1})$
representation of the above decompositions. In addition there are two
tensor multiplets each containing an anti-self-dual antisymmetric
tensor, two chiral fermions and five scalars. One of them also originates
from the $({\bf 1},{\bf 1})$
representation while the second one comes from the $({\bf 2},{\bf 2})$ representation of
the above decompositions.

%%%%%%%%%%%%%%%%%%%%%%%%%%%%%%%%%%%%%%%%%%%%%%%%%%%%%%%%%%%%%%%%%
\subsection{Field decompositions in $d=4$}\label{section:spectrumd4}
%%%%%%%%%%%%%%%%%%%%%%%%%%%%%%%%%%%%%%%%%%%%%%%%%%%%%%%%%%%%%%%%%
Let us now turn to the case $d=4$. One can decompose the spectrum for both the cases of an $SU(3)\times SU(3)$ or $SU(2)\times SU(2)$ structure.

The decomposition of the ten-dimensional spectrum with respect to $SO(2)\times SU(3)\times SU(3)$ is completely analogous to the discussion of last section and has been worked out in \cite{Grana:2005ny}. We only state the results here. Under $SU(3)\times SU(3)$, the $({\bf 1},{\bf 1})$ representation gives the gravity and a tensor multiplet. The triplet representations form massive gravitino multiplets, which are supposed to be projected out when performing the Kaluza-Klein reduction. In type IIA, the $({\bf 3},{\bf 3})$ and  $(\bar{\bf 3},\bar{\bf 3})$ form the hypermultiplet sector while the $(\bar{\bf 3},{\bf 3})$ and  $({\bf 3},\bar{\bf 3})$ form vector multiplets. In type IIB, the representations of vector and hypermultiplets are exchanged.

Let us now turn to field decompositions with respect to an $SU(2)\times SU(2)$ structure and find the corresponding $\cN =4$ multiplets. This can be easily achieved by using the results of the last section and reduce $SO(1,5)\to SO(1,3)$. Alternatively, one can use the above results for $SU(3)\times SU(3)$ structures and the reduction $SU(3)\times SU(3) \to SU(2)\times SU(2)$.
The resulting fields form $\cN=4$ multiplets. More precisely, we find the gravity multiplet
plus three $\cN=4$ vector multiplets. The gravity multiplet contains the
graviton, four gravitini, six vector fields, four Weyl fermions and
two scalars all
in the  ${({\bf 1},{\bf 1})}$ representation.
The vector multiplets each contain one vector, four gaugini and six scalars.
Two of them are also in the  ${({\bf 1},{\bf 1})}$ representation
while the third vector multiplet is in the ${({\bf 2},{\bf 2})}$
representation. We see that, in contrast to $d=6$, not all fields in
the ${({\bf 1},{\bf 1})}$
representation are part of the gravity multiplet but they also form
two vector multiplets. This corresponds to the fact that the
six-dimensional gravity multiplet reduces to a four-dimensional
gravity multiplet plus two vector multiplets.
The doublet representations again contain massive gravitino multiplets, which are projected out in the compactification procedure.

As we already discussed at the begin of Section \ref{section:spectrum}, these multiplets still consist of ten-dimensional fields that are reordered in such a way that they form $\cN =4$ multiplets. In the corresponding rewriting of the action only $SO(1,3)\times SO(6)$ symmetry and $\cN =4$ supersymmetry are manifest. Then we projected out the $SU(2)\times SU(2)$ doublets to obtain a theory that actually allows only for $\cN =4$ supersymmetry.

The couplings of the resulting theories are well-known and constrained by $\cN=2$ and $\cN=4$ supersymmetry, respectively. The fields in the $({\bf 1},{\bf 1})$ representation descend to the same number of four-dimensional fields in the compactification procedure.
The number of fields in higher representations can vary depending on the geometry of the internal manifold $Y$.
The aim of the first part of the review is to identify the geometry of the scalar field spaces.

%%%%%%%%%%%%%%%%%%%%%%%%%%%%%%%%%%%%%%%%%%%%%%%
\section{Geometric $SU(n)$ structures} \label{section:SUn}
%%%%%%%%%%%%%%%%%%%%%%%%%%%%%%%%%%%%%%%%%%%%%%%
Before studying the pure spinor formalism and its application to $SU(n)\times SU(n)$ structures, let us discuss the subclass of geometric $SU(n)$-structure compactifications. On a manifold $Y$ of dimension $2n$, an $SU(n)$ structure is characterized by a single nowhere-vanishing spinor $\eta$, which specifies the supersymmetry generator at low energies. By forming spinor bilinears, the parameter space of $\eta$ then can be mapped to the parameter space of two nowhere-vanishing differential forms. More precisely, an $SU(n)$ structure is specified in terms of a real two-form $J$ and a complex, locally decomposable $n$-form $\Omega$.
In contrast, $SU(2)$-structure manifolds of dimension six admit two nowhere-vanishing spinors $\eta_i$, $i=1,2$ that are linearly independent at any point of $Y$. As we show below, their parameter space can be equivalently described by a real two-form $J$, a complex, locally decomposable two-form $\Omega$ and a complex one-form $K$. The latter defines an almost product structure that decomposes the tangent space into a two-dimensional identity-structure part parameterized by $K$ and a four-dimensional $SU(2)$-structure piece parameterized by $J$ and $\Omega$.

A special case arises if one considers Calabi-Yau and K3 compactifications. These are compactifications on manifolds of $SU(n)$ holonomy, which are reviewed in \cite{Greene:1996cy} for $n=3$ and in \cite{Aspinwall:1996mn} for $n=2$. The holonomy group is defined as the group of transformations that appear in the Levi-Cevita connection and thereby is the group that acts on parallel transports along closed loops. $SU(n)$ holonomy means that the nowhere-vanishing spinor on $Y$ is moreover covariantly constant.
From the definitions \eqref{definition_two-forms} and \eqref{SU3_forms_definitions} we see that the forms $J$ and $\Omega$ (and $K$) are closed if $\eta$ is covariantly constant. However, for general $SU(n)$ structures these forms are generically not closed. Indeed, their exterior derivatives can be computed in terms of the intrinsic torsion of $Y$ \cite{Chiossi:2002tw}. The torsion classes classify the manifold $Y$ and determine its properties.

%%%%%%%%%%%%%%%%%%%%%%%%%%%%%%%%%%%%%%%%%%%%%%%%%%%%%%%%%%%%%%%%%
\subsection{$SU(2)$ structures on $Y_4$}\label{section:SU2d6}
%%%%%%%%%%%%%%%%%%%%%%%%%%%%%%%%%%%%%%%%%%%%%%%%%%%%%%%%%%%%%%%%%
Let us start with a four-dimensional compact manifold $Y_4$. A prominent example of such a
manifold is $K3$.
As discussed in Section \ref{section:basics}, the defining spinor $\eta$ and its charge conjugate $\eta^c$ are both
globally defined and nowhere-vanishing and therefore they are both
singlets under the structure group $SU(2)$. Moreover, they are linearly
independent and have the same chirality.

From the two singlets one can construct three distinct globally
defined real two-forms $J$, $\Re \Omega$ and $\Im \Omega$ by appropriately contracting with $SO(4)$ $\gamma$-matrices \cite{Gauntlett:2003cy}
\begin{equation}
\label{definition_two-forms}
 \bar{\eta} \gamma_{mn} \eta = - \iu J_{mn} \ , \quad \bar{\eta}^c \gamma_{mn} \eta = \iu \Omega_{mn} \ , \quad \bar{\eta} \gamma_{mn} \eta^c = \iu \bar{\Omega}_{mn} \ , \quad m,n=1,\dots,4 \ ,
\end{equation}
where the normalization $\bar{\eta}\eta = 1$ is chosen and $\gamma_{mn}$ denotes the anti-symmetric product of gamma matrices as defined in Appendix \ref{section:conventions}.
However, these two-forms are not independent but satisfy
\begin{equation}
\label{relations_forms}
 \Omega \wedge \bar{\Omega} = 2 J \wedge J \ne 0 \ , \qquad \Omega \wedge J = 0 \ , \qquad \Omega \wedge \Omega = 0 \ ,
\end{equation}
which follows from the Fierz identities given in \eqref{Fierz_identities_4d}.
Conversely, the Fierz identities also
  show that the choice of a real two-form $J$ and a complex two-form $\Omega$ determines $\eta$
  completely (up to normalization) if they satisfy the above
relations. Therefore, $J$ and $\Omega$
  equivalently define an $SU(2)$ structure on the
  manifold.

Alternatively one can also define an
$SU(2)$-structure in terms of stable forms~\cite{Hitchin:2001rw}. A
stable $p$-form $\omega \in \Lambda^p V^*$ on a vector space $V$ is
defined as a form whose orbit under the action of $Gl(V)$ is open in
$\Lambda^p V^*$, i.e.\ a $p$-form $\Psi$ is stable if any ``nearby'' $p$-form $\tilde \Psi$ can be reached by some linear transformation $G$ such that $\tilde \Psi = G \Psi$.
It can be shown that for a stable two-form $\omega$ on a $2m$-dimensional space this means that
$\omega^m \ne 0$. Thus, a stable two-form on an even-dimensional space
defines a symplectic structure on it.

On a four-dimensional manifold $Y_4$ the stable two-form $J$ satisfies $J \wedge J
\sim \operatorname{vol}_4$  and  locally defines a %(not necessary integrable)
 symplectic structure that reduces the structure group
from $Gl(4)$ to $Sp(4,\mathbb{R})$.\footnote{Note that this symplectic structure may be non-integrable in
  the sense that
  $\diff \omega \ne 0$. Therefore, our notion of a symplectic structure differs from the usual mathematical
  terminology.}
The existence of additional
stable forms can reduce the structure group even further. In this
case one has to ensure that these stable forms do not reduce the
structure group in the same way. For example, one can take two
linearly independent stable
 two-forms $J_i, i =1,2$ that satisfy
\begin{equation}\label{stable_forms}
J_i \wedge J_j = \delta_{ij} \operatorname{vol}_4 \  .
% J_1 \wedge J_2 = 0 \ ,\qquad J_i \wedge J_i \sim \operatorname{vol}_4\ .
\end{equation}
$J_1$ and $J_2$ then define a holomorphic two-form $\Omega =
J_1 + \iu J_2$, which globally defines a holomorphic subbundle in the
tangent space and therefore breaks the structure group to
$Sl(2,\mathbb{C}) \equiv Sp(2,\mathbb{C})$.

Analogously, in the case of three stable two-forms $J_i, i=1,2,3$,
which satisfy \eqref{stable_forms}
the structure group is reduced even further. Since $J_3$ is orthogonal
to $\Omega = J_1 + \iu J_2$ and  its complex conjugate, it defines a
product between the holomorphic and the anti-holomorphic tangent
bundle. Therefore, the $Sl(2,\mathbb{C})$ is further broken to the
$SU(2)$ subgroup which preserves this product.\footnote{Of course, this breaking is just the well-known
  relation $Sl(n,\mathbb{C}) \cap Sp(2n,\mathbb{R}) = SU(n)$.}
If one defines
\begin{equation}
\label{stable_forms_definition}
 J= J_3 \ , \quad \Omega = J_1+ \iu J_2 \ ,
\end{equation}
it is straightforward to check that~\eqref{stable_forms}
and~\eqref{relations_forms} are indeed equivalent.

In terms of stable forms it is easy to identify the parameter space of $SU(2)$ structures. A triple of stable forms $J_i$ has to fulfill \eqref{stable_forms} in order to define an $SU(2)$ structure on $Y_4$. Thus the $J_i$ span a three-dimensional subspace in the space of two-forms. By choosing some volume form $\operatorname{vol}_4$, i.e.\ some orientation on $Y_4$, we can interpret the wedge product as a scalar product of split signature $(3,3)$ on the space of two-forms.
With respect to this scalar product, the $J_i$ form an orthonormal basis for a space-like subspace.
The orbit of such a triple of $J_i$ under $SO(3,3)$, the group of linear transformations that preserve the scalar product and thereby lengths and angles in the space of two-forms, gives all possible configurations that respect the orthonormality condition \eqref{stable_forms}. Thus, the configuration space can be written as $SO(3,3)$ divided by the stabilizer of the $J_i$.
The stabilizer consists of $SO(3)$ rotations in the subspace orthogonal to the $J_i$, which leave the $SU(2)$ structure invariant. Therefore, the configuration space for the $J_i$ is $SO(3,3)/SO(3)$.
The $SO(3)$ rotations in the stabilizer correspond to the action of the $SU(2)$ structure group on the space of two-forms. The $J_i$ are singlets under the $SU(2)$ structure group while the space orthogonal to them forms an $SU(2)$ triplet.

One should note that there is some redundancy in  the descriptions of
  $SU(2)$ structures on a manifold.
Any rescaling of the $J_i$ does not change the unbroken $SU(2)$ and
therefore does not correspond to a degree of freedom for the $SU(2)$ structure. Hence we can fix
the normalization by~\eqref{stable_forms}. Furthermore,
there is a rotational $SO(3)$
symmetry between the three forms $J_i$. However, this symmetry is not
obvious from the definition \eqref{definition_two-forms}.  It
corresponds to $SU(2)$
rotations on the Weyl-spinor doublet $(\eta, \eta^c)$ which is a
symmetry because $\eta$ and $\eta^c$ have the same chirality on a
four-dimensional manifold.
One can check that the three two-forms $J_i$ indeed form a triplet under the action of this $SU(2)$.
By modding out this symmetry, we arrive at the parameter space of an $SU(2)$ structure over a point on the manifold $Y_4$, which is
\begin{equation} \label{moduli_space_geom_local}
 \mathcal{M}_{J_i} = \frac{SO(3,3)}{SO(3) \times SO(3)} \ .
\end{equation}

We can use the double cover of the groups appearing in \eqref{moduli_space_geom_local} to rewrite the result. We know that
\begin{equation}
SO(3,3) = Sl(4,\mathbb{R}) / \mathbb{Z}_2
\end{equation}
and
\begin{equation}
SO(3) \times SO(3) = (SU(2) \times SU(2)) /\mathbb{Z}_2^2 = SO(4)/\mathbb{Z}_2 \ .
\end{equation}
Therefore, we can express the result \eqref{moduli_space_geom_local} as
\begin{equation}
 \mathcal{M}_{J_i} = \frac{Sl(4,\mathbb{R})}{SO(4)} \ .
\end{equation}
If one compares this with the parameter space $Gl(4,\mathbb{R})/SO(4)$ of the metric over a point of $Y_4$, we see that the parameter space of $SU(2)$ structures incorporates all metric degrees of freedom except the volume factor. The missing degree of freedom corresponding to the volume factor can be associated with the normalization of the $J_i$ in \eqref{stable_forms}.\footnote{Note that the choice of a triple of normalized $J_i$ is just equivalent to the choice of a Hodge operator on the space of two-forms. This is reflected by the fact that the $J_i$ just span the positive eigenspace of a Hodge operator in the space of two-forms.}

%%%%%%%%%%%%%%%%%%%%%%%%%%%%%%%%%%%%%%%%%%%%%%%
\subsection{$SU(3)$ and $SU(2)$ structures on $Y_6$} \label{section:SU2d4}
%%%%%%%%%%%%%%%%%%%%%%%%%%%%%%%%%%%%%%%%%%%%%%%
For $SU(3)$ structures on $Y_6$ the same analysis as in Section \ref{section:SU2d6} can be done, see for reference \cite{Chiossi:2002tw,Gauntlett:2003cy}. However, there are various differences compared to the case of $Y_4$. First of all, charge conjugation changes the chirality of a spinor. Therefore, the $SU(3)$ singlets $\eta$ and $\eta^c$ have opposite chirality and out of them one can construct a real two-form $J$ and a complex three-form $\Omega$ given by
\begin{equation}\label{SU3_forms_definitions}
 J = - \iu \bar \eta \gamma_{mn} \eta\diff x^m \wedge \diff x^n \ , \quad \Omega = - \iu \bar \eta^c \gamma_{mnp} \eta \diff x^m \wedge \diff x^n \wedge \diff x^p \ ,
\end{equation}
where $\gamma_{m_1 \dots m_p}$ now are anti-symmetrized products of six-dimensional gamma matrices.
Using the Fierz identities \eqref{Fierz_identities_6d}, all spinor bilinears can be expressed in terms of $J$ and $\Omega$, thereby fixing the spinor $\eta$. Furthermore, the Fierz identities imply that
\begin{equation} \label{SU3_forms_conditions}
 J\wedge J \wedge J = \tfrac34 \iu \Omega \wedge \bar \Omega \ne 0 \ , \qquad J \wedge \Omega = 0 \ .
\end{equation}
Again, since $J$ and $\Omega$ can be used to define $\eta$, they actually define the $SU(3)$ structure. A non-degenerate real two-form $J$ is invariant under $Sp(6,\mathbb{R})$ transformations, while a complex three-form $\Omega$ with $\Omega \wedge \bar \Omega \ne 0$ is invariant under $Sl(3,\mathbb{C})$. Together this gives an $SU(3) \equiv Sp(6,\mathbb{R}) \cap Sl(3,\mathbb{C})$ structure group if \eqref{SU3_forms_conditions} is fulfilled.
Therefore, deformations of $J$ and $\Omega$ over a single point are parameterized by the coset space
\begin{equation}
 {\cal M}_{J,\,\Omega} = \frac{Sl(6,\mathbb{R})}{SU(3)} \ .
\end{equation}
Since $SU(3) \subset SO(6)$, this also defines a metric on $Y$. The additional degrees of freedom rotating the $SU(3)$ inside $SO(6)$ correspond to $SU(3)$ triplets, which are projected out in the compactification process, and one gauge degree of freedom multiplying $\Omega$ and thereby $\eta$ by a phase. By modding out these degrees of freedom, one arrives at the (physical) metric parameter space
\begin{equation}
 {\cal M}_{g} = \frac{Sl(6,\mathbb{R})}{SO(6)} \ .
\end{equation}

One can actually show that the three-form $\Omega$ is already determined by its real part $\Re \Omega$ \cite{Hitchin:2001rw}. $\Re \Omega$ is required to be stable, which means that it transforms in an open orbit under the action of $Gl(6,\mathbb{R})$, i.e.\ every nearby point in parameter space can be reached by some $Gl(6,\mathbb{R})$ transformation.

If we assume that there are two such spinors $\eta_{1}$ and $\eta_{2}$
that are orthogonal at each point,  the structure group is broken
further to $SU(2)$. Each spinor defines an $SU(3)$ structure on its own, parameterized by $(J^{(i)},\Omega^{(i)})$, $i=1,2$, as defined in \eqref{SU3_forms_definitions}, cf.\ \cite{Gauntlett:2003cy,Bovy:2005qq}.
With the use of the Fierz identities \eqref{Fierz_identities_6d} one can express them in
terms of an $SU(2)$ structure:
\begin{equation}\begin{aligned}
\label{SU(3)_structure_forms_SU2_splitting}
 J^{(1)} &= J + \tfrac{\iu}{2} K \wedge \bar{K} \ , \quad \Omega^{(1)} = \Omega \wedge K \ ,\\
 J^{(2)} &= J - \tfrac{\iu}{2} K \wedge \bar{K} \ , \quad \Omega^{(2)} = \Omega \wedge \bar{K} \ .
\end{aligned}\end{equation}
The $SU(2)$ structure is defined by the complex one-form
\cite{Gauntlett:2003cy,Bovy:2005qq}
\begin{equation} \label{definition_one-form_K}
 K_m := \bar{\eta}^c_{2} \gamma_m \eta_{1}
\end{equation}
and the two-forms $J$ and $\Omega$ given by
\begin{equation} \label{definition_two-forms_6d}
 J_{mn} = - \tfrac{1}{2} \iu \left(\bar{\eta}_{1} \gamma_{mn} \eta_{1}
  + \bar{\eta}_{2} \gamma_{mn} \eta_{2}\right) \ ,\qquad
 \Omega_{mn} = \iu \bar{\eta}_{2} \gamma_{mn} \eta_{1} \ .
\end{equation}
$J$ and $\Omega$ fulfill~\eqref{relations_forms}, while $K$ satisfies
\begin{equation}\label{K_compatible}
   K_m K^m =0 \ , \qquad \bar{K}_m K^m = 2 \ ,\qquad
 \iota_K J =0 \ , \qquad \iota_K \Omega = \iota_{\bar{K}} \Omega =0 \ .
\end{equation}
$K$ also specifies an almost product structure
\begin{equation}
\label{almost_product_structure}
 P_m^{\phantom{m}n} := K_m\bar{K}^n + \bar{K}_m K^n - \delta_m^{\phantom{m}n} \ ,
\end{equation}
i.e.\ it obeys
\begin{equation}
 P_m^{\phantom{m}n} P_n^{\phantom{n}p} = \delta_m^{\phantom{m}p} \ .
\end{equation}
As can be seen from \eqref{SU(3)_structure_forms_SU2_splitting}, this almost product structure is related to the almost complex structures $J^{(i)}$ of the two $SU(3)$ structures by
\begin{equation} \label{almost_product_structure_relation}
 P_m^{\phantom{m}n} = - J^{(1) \phantom{m}p}_{\phantom{(1)}m} \ J^{(2) \phantom{p}n}_{\phantom{(2)}p} \ .
\end{equation}
From \eqref{K_compatible} we can see that $K_m$ and $\bar{K}_m$ are
both eigenvectors of $P_m^{\phantom{m}n}$ with eigenvalue $+1$. The
vectors orthogonal to $K_m, \bar{K}_m$ have eigenvalue $-1$ as can be seen from \eqref{almost_product_structure}. Therefore, $K_m$
and $\bar{K}_m$ even span the $+1$ eigenspace.

In terms of stable forms, an $SU(2)$-structure on $Y_6$ can be defined
by a global complex one-form\footnote{Note that every one-form is stable by definition.} $K$ which breaks the structure group
$SO(6)$ to $SO(4)$  and -- as on $Y_4$ -- by three
stable two-forms $J_i$ that reduce this group further to
$SU(2)$. To assure this breaking of the structure group, all of
these forms have to be compatible with each other in that
they satisfy~\eqref{stable_forms}
and~\eqref{K_compatible}.

Actually, an almost product structure
$P_m^{\phantom{m}n}$ that has a positive eigenspace of dimension two
and a globally defined, nowhere-vanishing spinor
$\eta$ are enough to define an $SU(2)$ structure on a manifold of
dimension six. The reason is that $P_m^{\phantom{m}n}$ reduces the structure
group to the group of those $SO(6)$ transformations that leave its two- and four-dimensional eigenspaces intact, in other words $SO(2) \times SO(4)$. This group acts on $\eta$ via its double cover, which is $U(1) \times SU(2)\times SU(2)$, where we used $Spin(4)=SU(2)\times SU(2)$. The spinor bundle correspondingly decomposes under $SU(4)\to U(1) \times SU(2)\times SU(2)$ as ${\bf 4} \to ({\bf 2},{\bf 1})_{\bf +1} \oplus ({\bf 1},{\bf 2})_{\bf -1}$.
In order for $\eta$ to be a singlet of the structure group, the structure group must therefore reduced to $SU(2)$. Hence, $P_m^{\phantom{m}n}$ reduces the $SU(3)$ structure
defined by $\eta$ to an $SU(2)$ structure.
This fits nicely with the
fact that the existence of $P_m^{\phantom{m}n}$ is already enough to
assure that the forms given in \eqref{SU3_forms_definitions}, which parameterize an $SU(3) \times SU(3)$ structure, are of the
form~\eqref{SU(3)_structure_forms_SU2_splitting} and thereby indeed
define an $SU(2)$ structure on the manifold.
Correspondingly, the two globally defined spinors that reduce the structure group to $SU(2)$ are $\eta$ and $(v_{m} \gamma^{m} \eta^c)$ with $v_m$ is any (real) $+1$-eigenvector of $P$.

Now let us derive the parameter space of $SU(2)$ structures. As before, we have to ensure that we compactify to $\cN=4$, and therefore project out
all $SU(2)$ doublets, as explained in section \ref{section:spectrumd4}.
As shown in \cite{Triendl:2009ap} this projection forces the almost product structure $P$ to be rigid. Therefore, the parameter space splits into a part for the two-dimensional identity structure and one for deformations of the $SU(2)$ structure in the four-dimensional subspace. The former is parameterized by $K$, the latter one by $J$ and $\Omega$.
The local parameter space of the $SU(2)$ structure part was already derived in section \ref{section:SU2d6} and is given by \eqref{moduli_space_geom_local}. The identity structure is parameterized by the complex one-form $K$ in a two-dimensional space. Its length corresponds to $K \wedge \bar{K}$ and parameterizes the volume of the two-dimensional space. The group $SU(1,1) \equiv Sl(2,\mathbb{R})$ leaves $K \wedge \bar{K}$ invariant, while it acts freely on $K$. Therefore, its action parameterizes the remaining freedom in choosing $K$.
Since the phase of $K$ is of no relevance, we have to mod out this degree of freedom, and end up with the parameter space ${Sl(2,\mathbb{R})}/{SO(2)}$. Hence, after including the degree of freedom that correspond to the volume of the four-dimensional subspace, we end up with the parameter space
\begin{equation}
\mathcal{M}_{K, J_i} = \frac{SO(3,3)}{SO(3) \times SO(3)} \times \mathbb{R}_+ \times \frac{Sl(2,\mathbb{R})}{SO(2)} \times \mathbb{R}_+ \ ,
\end{equation}
where the two $\mathbb{R}_+$ factors are spanned by the ``volume'' factors of the two- and four-dimensional eigenspaces given by $K \wedge \bar{K}$ and $J\wedge J$.

We see that $SU(n)$ structures are conveniently described by nowhere-vanishing forms. Using this parametrization we could determine the parameter space of $SU(n)$ structures. This parameter space appears in the ten-dimensional theory rewritten in terms of a $d$-dimensional language as the target space of scalar fields corresponding to geometric deformations of the internal space $Y$. In the next chapter we want to improve on this result by including all other scalar fields in the parameter space. The main tool for this turns out to be generalized geometry.

\cleardoublepage
%%%%%%%%%%%%%%%%%%%%%%%%%%%%%%%%%%%%%%%%%%%%%%%
\chapter{Generalized geometry and $SU(n)\times SU(n)$ structures} \label{section:covariant}
%%%%%%%%%%%%%%%%%%%%%%%%%%%%%%%%%%%%%%%%%%%%%%%
We have seen in the last chapter that the geometry of the parameter space of $SU(n)$ structures can be determined by translating the parameter space of a spinor $\eta$ into the language of forms by use of spinor bilinears. In this chapter we shall use the same strategy to define formal sum of forms coming from spinor bilinears to describe $SU(n)\times SU(n)$ structures. One then realizes that these formal sums of forms describe pure spinors in generalized geometry and automatically include the NS-NS $B$ field.
It turns out that generalized geometry is manifestly covariant under string symmetries that transform the fields in the NS-NS sector, more precisely the metric and the $B$-field, into each other. Improving on this idea then leads to exceptional generalized geometry, which incorporates also the R-R sector.

%%%%%%%%%%%%%%%%%%%%%%%%%%%%%%%%%%%%%%%%%%%%%%%
\section{Pure spinors and $SU(n)\times SU(n)$ structures} \label{section:pure_spinors}
%%%%%%%%%%%%%%%%%%%%%%%%%%%%%%%%%%%%%%%%%%%%%%%
One can generalize the $SU(n)$ structures discussed in the previous
section by assuming that the manifold admits two globally defined,
nowhere-vanishing spinors $\eta_1$ and $\eta_2$. Each of them defines
an $SU(n)$ structure on its own and if they are identical everywhere
on the manifold, this reduces to the case discussed in the previous
section.
In the other limiting case where $\eta_1$ and $\eta_2$ are orthogonal
at each point, the two $SU(2)$ structures intersect in some
identity structure, which means that the spinor bundle is trivial and
compactification on this backgrounds preserves twice as many supercharges. However, in principle one can also have the intermediate case of two globally defined,
nowhere-vanishing spinors $\eta_1$ and $\eta_2$ that are linearly
independent at generic points but become parallel at some points on the
manifold.

Analogously to the last section, there is an equivalent formulation of
$SU(n) \times SU(n)$ structures in terms of globally defined stable
forms. This is elegantly captured by the notion of pure spinors and
generalized geometry~\cite{Gualtieri:2003dx,Hitchin:2004ut,Grana:2006kf}. The latter is covariant with respect to the T-duality group $SO(2n,2n)$, in which the structure group $SU(n)\times SU(n)$ can easily be embedded. Furthermore, the covariance with respect to the T-duality group enables one to glue together charts not only by geometric rotations in $Gl(n,\mathbb{R})$ but also by more general transformations. Due to this, flux for the two-form $B$ can also be described in terms of a twist of the generalized tangent bundle and even more general twists of this generalized tangent bundle, termed non-geometric fluxes, can be incorporated into the background.
Let us briefly review this concept for a $2n$-dimensional manifold $Y$
and then apply this formalism to the case of $SU(n)\times SU(n)$ structures
afterwards.

%%%%%%%%%%%%%%%%%%%%%%%%%%%%%%%%%%%%%%%%%%%%%%%
\subsection{Pure spinors in generalized geometry} \label{section:GG_pure_spinors}
%%%%%%%%%%%%%%%%%%%%%%%%%%%%%%%%%%%%%%%%%%%%%%%
In generalized geometry  one considers a generalized tangent bundle
$\mathcal{T}Y$ which locally looks like $TY \oplus T^*Y$ and
therefore admits a scalar product $\mathcal{I}$ of split signature
that is induced by the canonical pairing between tangent and cotangent
space, i.e.\
\begin{equation}
 \mathcal{I}(v + \xi, w + \zeta) = \xi(w) + \zeta(v) \ ,
\end{equation}
where in the considered local patch $v$ and $w$ are vectors and $\xi$ and $\zeta$ are one-forms.
On a $2n$-dimensional manifold $Y$ this bundle thus has a structure
group contained in $SO(2n,2n)$.
The elements of $\mathcal{T}Y$ obey the Clifford algebra
\begin{equation}
 \{v + \xi, w + \zeta\} = \xi(w) + \zeta(v)= \mathcal{I}(v + \xi, w + \zeta) \ ,
\end{equation}
and we can construct in the usual way the group $Spin(2n,2n)$ out of this Clifford algebra. In the following we can denote a basis of $\mathcal{T}Y$ by $\Gamma^\Pi$, $\Pi=1,\dots,12$. For the Lie algebra one finds locally
\begin{equation} \label{spin2n2n}
 so(2n,2n) \equiv \Lambda^2 \mathcal{T}Y = gl(2n,\mathbb{R})\oplus \Lambda^2 TY \oplus \Lambda^2 T^*Y \ .
\end{equation}
Therefore, one can understand the structure group $SO(2n,2n)$ to be generated by the algebra of geometric transformations together with bi-vectors and two-forms.

Similarly to our discussion in the last section, one can introduce
objects that break the structure group $SO(2n,2n)$. For example, an almost
complex structure $\mathcal{J}$, defined by its property
\begin{equation}
\mathcal{J}^2 = -1 \ ,
\end{equation}
 can be defined if (and only if)
the complexified generalized tangent bundle globally splits into its eigenspaces, i.e.\
\begin{equation}
\label{decomposition_generalized_acs}
\left( \mathcal{T}Y \right)_\mathbb{C} = L_+ \oplus L_- \ ,
\end{equation}
where $L_\pm$ are the eigenspaces of $\mathcal{J}$ with the eigenvalues
$\pm \iu$. If $\mathcal{J}$ is globally defined on $Y$, the structure group
of $\mathcal{T}Y$ is broken from $SO(2n,2n)$ to $U(n,n)$.

When two generalized almost complex structures $\mathcal{J}_{1,2}$ exist, the notion of
compatibility can be defined. More precisely,
$\mathcal{J}_1$ and
$\mathcal{J}_2$ are called compatible if
\begin{enumerate}
 \item $\mathcal{J}_1$ and $\mathcal{J}_2$ commute and
 \item $\mathcal{G} := \mathcal{I} \mathcal{J}_1 \mathcal{J}_2$ is a positive definite metric on $\mathcal{T}Y$, where $\mathcal{I}$ is the canonical scalar product on $\mathcal{T}Y$.
\end{enumerate}
The first condition ensures that the splittings~\eqref{decomposition_generalized_acs} can be done simultaneously, i.e.\ that
\begin{equation}
\label{decomposition_comp_generalized_acs}
\left( \mathcal{T}Y \right)_\mathbb{C} = L_{++} \oplus L_{-+} \oplus L_{+-} \oplus L_{--}\ ,
\end{equation}
where the indices correspond to the eigenvalues of $\mathcal{J}_{1,2}$.
The second condition ensures that each of the four components
in~\eqref{decomposition_comp_generalized_acs} is $n$-dimensional
such that the two compatible generalized almost complex structures
reduce the structure group to $U(n) \times U(n)$, where each $U(n)$
acts on two of the four components.

Let us now briefly review how to reformulate  generalized geometry
in terms of pure spinors $\Phi$ \cite{Gualtieri:2003dx,Hitchin:2004ut}. One first defines the annihilator
space  $L_\Phi$ of a complex $SO(2n,2n)$ Weyl spinor $\Phi$ as the
subspace of complexified gamma matrices $\Gamma$ which map
$\Phi$ to zero, i.e.\
\begin{equation}
 L_\Phi \equiv \left\lbrace \Gamma \in \left(\mathcal{T}Y \right)_\mathbb{C} | \Gamma \Phi = 0 \right\rbrace \ .
\end{equation}
Note that  $L_\Phi$ is always isotropic as for each element
$\Gamma$ of $L_\Phi$ we have
\begin{equation}
 0 = \Gamma^2 \Phi = \mathcal{I}(\Gamma,\Gamma)\, \Phi \ ,
\end{equation}
which implies  $\mathcal{I}(\Gamma,\Gamma)=0$ for all $\Gamma \in
L_\Phi$.

A complex Weyl spinor $\Phi$ of $SO(2n,2n)$ is called pure if its annihilator space
has maximal dimension, i.e.\ $\operatorname{dim} L_\Phi = 2n$.
$\Phi$ is called normalizable if
\begin{equation}
\langle \Phi, \bar{\Phi} \rangle > 0 \ ,
\end{equation}
where the brackets denote the usual spinor product.
As a consequence of Chevalley's theorem~\cite{Chevalley:1996}, which states
\begin{equation}
\label{Chevalley_theorem}
 \operatorname{dim} L_\Phi \cap L_\Psi = 0 \quad \Leftrightarrow \quad \langle \Phi , \Psi \rangle \ne 0 \ ,
\end{equation}
normalizable pure spinors define a splitting
\begin{equation}
 \left( \mathcal{T}Y \right)_\mathbb{C} = L_\Phi \oplus L_{\bar{\Phi}} \ .
\end{equation}
By matching the annihilator space $L_\Phi$ with the $+\iu$
eigenspace of a generalized almost complex structure, one can show
that both are equivalent up to the normalization factor of $\Phi$.
Thus, generalized almost complex structures are equivalent to
lines of pure $SO(2n,2n)$ spinors. A pure spinor breaks the
structure group of $\mathcal{T}Y$ further from $U(n,n)$ to $SU(n,n)$
in that fixing its
phase eliminates the $U(1)$ factor \cite{Hitchin:2004ut,Gualtieri:2003dx}.

The compatibility conditions for two generalized almost complex
structures translate into a compatibility condition
on the corresponding pure spinors.
Two normalizable pure $SO(2n,2n)$ spinors $\Phi_{1,2}$ are compatible if and only
if their annihilator spaces intersect in a space of dimension
$n$, i.e.\
\begin{equation}
\label{Compatibility_gacs}
\operatorname{dim}(L_{\Phi_1} \cap L_{\Phi_2}) =n \ .
\end{equation}
Thus the pair $\Phi_{1,2}$
%of pure normalizable compatible $SO(2n,2n)$ spinors
breaks the structure group to $SU(n) \times SU(n)$
(instead of $U(n) \times U(n)$).
Therefore, pure spinors of generalized geometry provide a convenient
framework to deal with $SU(n) \times SU(n)$ structures.
Whenever $\mathcal{T}Y=TY \oplus T^*Y$ globally,
both $SU(n)$ factors can be projected to the
tangent space $TY$. In this case the intersection of these
projections defines the
structure group of the tangent bundle.

The compatibility condition of two pure spinors also restricts their
chirality.  Since $SO(n,n)$ transformations do not mix
chiralities, one can always assume $\Phi_1$ and $\Phi_2$ to be
of definite chirality.
Furthermore, two pure spinors $\Phi_1$ and $\Phi_2$ have the same chirality if and only if \cite{Charlton:1996PhD}
\begin{equation}
\label{compatible_pure_spinors_chirality}
 \operatorname{dim}(L_{\Phi_1} \cap L_{\Phi_2}) = 2 k
\end{equation}
for $k \in \mathbb{N}$.
Therefore, two compatible pure spinors are of the same chirality if
$n$ is even and of opposite chirality for
$n$ being odd.

One can construct pure $SO(2n,2n)$ spinors out of the two globally
defined $SO(2n)$ spinors $\eta_1$ and $\eta_2$ discussed at the
beginning of this section, as follows
\begin{equation} \label{tensor_product_spinors}
\eta_1 \otimes \bar{\eta}_2 = \frac{1}{4}\sum_{k=0}^{2n} \frac{1}{k!} \left(
\bar{\eta}_2 \gamma_{m_1\dots m_k} \eta_1 \right) \gamma^{m_k\dots m_1}   \ ,
\end{equation}
where $\gamma^{m_1\dots m_k}$ is the totally antisymmetric product of
$SO(2n)$ $\gamma$-matrices.
One can act with $SO(2n)$ gamma matrices from the left or from the
right which in turn defines an $SO(2n,2n)$ action
on the bi-spinor $\eta_1 \otimes
\bar{\eta}_2$. By
extensive use of Fierz identities given in Appendix \ref{section:conventions} one can show that $\eta_1 \otimes \bar{\eta}_2$ is pure and normalizable. The same holds for
$\eta_1 \otimes \bar{\eta}_2^c$ and these two pure spinors are
moreover compatible. Thus they can be used to discuss $SU(n)\times
SU(n)$ structures.

The map
\begin{equation}\label{mapping_bispinors_forms}
\tau \ : \ \eta_1 \otimes \bar{\eta}_2 \longmapsto \tau(\eta_1 \otimes \bar{\eta}_2) \equiv \frac{1}{4}\sum_{k=0}^{2n} \frac{1}{k!} \left( \bar{\eta}_2 \gamma_{m_1\dots m_k} \eta_1 \right) e^{m_k} \wedge \dots \wedge e^{m_1} \ ,
\end{equation}
with $e^{m_k}$ being a local basis of one-forms, identifies
$SO(2n,2n)$ spinors with formal sums of differential forms.\footnote{This
isomorphism is canonical up to the choice of a volume form on the
manifold \cite{Grana:2005ny}.} The group $SO(2n,2n)$ acts on these formal sums of differential forms via the representation \eqref{spin2n2n}.
Note that \eqref{mapping_bispinors_forms} maps negative (positive)
chirality spinors to differential forms of odd (even) degree.
Moreover, it is an isometry with respect to the spinor
product and the ``Mukai pairing'' of differential forms, defined by\footnote{Here, $[ \cdot ]$ is the floor function which rounds down its argument to the next integer number.}
\begin{equation}\label{Mukai_pairing}
 \langle \Psi , \chi \rangle = \sum_p (-)^{[(p-1)/2]} \Psi_p \wedge
 \chi_{2n - p}\ ,
\end{equation}
which maps two formal sums of differential forms to a form of top degree. Like the spinor product, it is symmetric for $n$ even, and anti-symmetric for $n$ being odd.
Using the definition
\begin{equation} \label{lambda_definition}
 \lambda\, \alpha_{p} = (-1)^{[(p-1)/2]} \alpha_{p}
\end{equation}
for a $p$-form $\alpha_{p}$,
we can write the Mukai pairing also in the form
\begin{equation}\label{Mukai_pairing_2}
 \langle \Psi , \chi \rangle = \left[ \Psi \wedge \lambda\, \chi \right]_{\operatorname{deg}=2n} \ .
\end{equation}
In the following we frequently use the isomorphism \eqref{mapping_bispinors_forms}.

Before we go on, let us state two important facts. One can show that a pure spinor $\Phi$ is always of the form~\cite{Gualtieri:2003dx}
\begin{equation} \label{general_pure_spinor_form}
 \Phi = \e^{-B} \wedge \e^{- \iu J} \wedge \Omega \ ,
\end{equation}
where $B$ and $J$ are real two-forms and $\Omega$ is some complex
$k$-form, $k \leq 2n$, that is locally decomposable into complex
one-forms.
Furthermore, one can prove that two compatible
pure spinors are always of the form~\cite{Grana:2006kf}
\begin{equation} \label{general_comp_pure_spinor_form}
 \Phi_1 = \e^{-B} \wedge \tau( \eta_1 \otimes \bar{\eta}_2 )\ ,\qquad
 \Phi_2 = \e^{-B} \wedge \tau( \eta_1 \otimes \bar{\eta}^c_2)\ ,
\end{equation}
where the isomorphism $\tau$ is defined in \eqref{mapping_bispinors_forms}.

For later use let us define the
generalized Hodge operator~\cite{Jeschek:2004wy,Cassani:2007pq}
\begin{equation}
\label{Hodge_star_generalized}
\ast_B = \e^{-B} \ast \lambda\, \e^B\ ,
\end{equation}
which acts on the space of forms, with $\lambda$ defined by \eqref{lambda_definition}.
Under the isomorphism given in \eqref{mapping_bispinors_forms} the generalized Hodge operator is mapped to charge conjugation on the space of $SO(2n,2n)$ spinors.
Analogously to the conventional Hodge operator, the generalized version can define a positive definite metric $G(\cdot , \cdot ) \equiv \langle \cdot , \ast_B\, \cdot \rangle$ on the space of forms, which is just the composition of $\ast_B$ with the Mukai pairing.
From \eqref{Mukai_pairing_2} it is easy to see that $G$ acts
on the space of forms by
\begin{equation}
G( \e^{-B}\wedge \Psi , \e^{-B} \wedge \chi ) = \langle \e^{-B} \wedge \Psi , \ast_B\,  \e^{-B} \wedge  \chi \rangle =
  \left[ \Psi \wedge \ast\, \chi \right]_{\operatorname{deg}=2n} = \sum_{p=0}^{2n} \Psi_p \wedge \ast\, \chi_p \ ,
\end{equation}
which indeed is positive definite. Therefore, the Mukai pairing and the generalized Hodge operator have the same signature.

As stated at the beginning of this section the generalized tangent bundle
$\mathcal{T}Y$ locally has the structure $TY \oplus T^*Y$.
In the following, we first perform a
local analysis  and consider the
algebraic structure of the bundle over a point on the manifold.
Therefore, we abuse the notation $TY \oplus T^*Y$ to denote
$\mathcal{T}Y$.

%%%%%%%%%%%%%%%%%%%%%%%%%%%%%%%%%%%%%%%%%%%%%%%%%%%%%%%%%%%%%%%%%%%%%%%%%%%%
\subsection{$SU(2)\times SU(2)$ structures on $Y_4$}\label{section:SU2SU2d6}
%%%%%%%%%%%%%%%%%%%%%%%%%%%%%%%%%%%%%%%%%%%%%%%%%%%%%%%%%%%%%%%%%%%%%%%%%%%%
%%%%%%%%%%%%%%%%%%%%%%%%%%%%%%%%%%%%%%%%%%%%%%%%%%%%%%%%%%%%%%%%%%%%%%%%%%%%
\subsubsection{Pure spinors and $SU(2)\times SU(2)$ structures on $Y_4$}\label{section:SU2SU2d6pure}
%%%%%%%%%%%%%%%%%%%%%%%%%%%%%%%%%%%%%%%%%%%%%%%%%%%%%%%%%%%%%%%%%%%%%%%%%%%%

Let us now apply the previous discussion to the case of $SU(2)\times
SU(2)$ structures on a four-dimensional manifold $Y_4$. We
start with the simpler case of $SU(2)$ structures or in other words
with the case where a single $SO(4)$ spinor $\eta$ exists on $Y_4$, which defines two pure $SO(4,4)$ spinors $\eta \otimes
\bar{\eta}$ and $\eta \otimes \bar{\eta}^c$.
Using the definitions \eqref{definition_two-forms} and \eqref{mapping_bispinors_forms} we
identify
\begin{equation} \label{pure_SU2_spinors_4d}
\tau(\eta \otimes \bar{\eta}) = \tfrac{1}{4}\, \e^{- \iu J} \ , \qquad
\tau(\eta \otimes \bar{\eta}^c ) = \tfrac{1}{4}\, \iu \Omega \ .
\end{equation}
The pure spinors in \eqref{pure_SU2_spinors_4d} actually are not of the most general form. To cover all deformations of the pure spinors, we note that we can additionally shift these pure $SO(4,4)$ spinors by a $B$-field leaving all conditions unchanged. Thus, we arrive at
\begin{equation}
\label{pure_spinors_4d}
\Phi_1 =  \tfrac{1}{4}\, \e^{-B - \iu J} \ , \qquad
\Phi_2 = \tfrac{\iu}{4} \, \e^{-B} \wedge \Omega \ .
\end{equation}

Let us now turn to the case  of general $SU(2) \times SU(2)$ structures.
We first analyze the conditions for spinors to
be pure and compatible. For the case at hand this is simplified by the triality
property of $SO(4,4)$ which isomorphically
permutes the spinor representation, its
conjugate and the  vector representation among each other, cf.\ for example \cite{Adams:1980hp}.
In particular, the quadratic form $\langle \cdot
, \cdot \rangle$ on the spinor space is mapped to the usual scalar product on
the vector space. This fact is used in the following.

For a pure spinor $\Phi$ the annihilator space $L_\Phi$
has dimension four. In addition Chevalley's theorem
\eqref{Chevalley_theorem} implies
\begin{equation}
\label{pure_spinor_SO44}
 \langle \Phi , \Phi \rangle = 0 \ .
\end{equation}
As shown in~\cite{Charlton:1996PhD}, this condition is also sufficient for $\Phi$ to be pure. Since the spinor product is mapped to the standard vector product of $SO(4,4)$ under triality, the purity condition \eqref{pure_spinor_SO44} translates into the corresponding complex vector being light-like vectors under the triality map.
Furthermore,  for $\Phi$ to be normalizable we need
\begin{equation}
 \label{normalization_pure_spinor_SO44}
 \langle \Phi , \bar{\Phi} \rangle > 0 \ .
\end{equation}

Now let us consider two pure normalizable spinors $\Phi_i$, $i=1,2$
which by definition satisfy
\begin{equation} \label{conditions_two_pure_spinors}
 \langle \Phi_i,\Phi_i\rangle = 0 \ ,\qquad
 \langle \Phi_i,\bar{\Phi}_i\rangle > 0  \ .
\end{equation}
If they are compatible, they also satisfy
\eqref{Compatibility_gacs}, which on $Y_4$ reads
\begin{equation}\label{dim2}
 \operatorname{dim} (L_{\Phi_1} \cap L_{\Phi_2}) = 2 \ .
\end{equation}
{}From \eqref{compatible_pure_spinors_chirality} we conclude that both $\Phi_i$ have
the same chirality which, together with \eqref{Chevalley_theorem},
implies that \eqref{dim2} is equivalent to
\begin{equation} \label{comp_pure_spinor_conditions}
 \langle \Phi_1, \Phi_2\rangle  = 0  \ , \qquad
 \langle \Phi_1, \bar{\Phi}_2\rangle  = 0 \ .
\end{equation}
Finally, we can choose the normalization
\begin{equation}\label{normalization_pure_spinors}
 \langle \Phi_1 , \bar{\Phi}_1 \rangle = \langle \Phi_2 , \bar{\Phi}_2 \rangle \ne 0 \ .
\end{equation}

Let us now analyze which possible cases of $SU(2)\times
SU(2)$ structures can occur on four-dimensional manifolds.
We just argued that $\Phi_1$ and $\Phi_2$ have the same chirality
so that the corresponding forms are of odd or even degree.
We start with the case where both spinors $\Phi_1$ and $\Phi_2$ have
negative chirality.
From \eqref{general_pure_spinor_form} we see that both pure spinors
are of the form
\begin{equation} \label{4d_spinors_odd_degree}
 \Phi_i = U_i \wedge \e^{-\iu J_i} \ ,\quad i=1,2
\ ,
\end{equation}
where $U_i$ are two complex one-forms while $J_i$ are two
non-vanishing real two-forms.\footnote{For simplicity we ignore the $B$-field which
however can be easily included.}
In addition, the compatibility condition~\eqref{comp_pure_spinor_conditions} implies
\begin{equation} \label{4d_spinors_odd_degree_comp}
 U_1 \wedge U_2 \wedge (J_1 - J_2) =0 \ , \qquad U_1 \wedge \bar{U}_2 \wedge (J_1 + J_2) =0 \ ,
\end{equation}
while the normalization \eqref{normalization_pure_spinors} translates into
\begin{equation} \label{4d_spinors_odd_degree_norm}
 U_1 \wedge \bar{U}_1 \wedge J_1  = U_2 \wedge \bar{U}_2 \wedge J_2 \ne 0 \ .
\end{equation}
Since $U_2 = a U_1 + b \bar{U}_1$ does not solve \eqref{4d_spinors_odd_degree_comp} and
  \eqref{4d_spinors_odd_degree_norm} we conclude
that $U_2$ is linearly independent of $U_1$ and
$\bar{U}_1$ and therefore $U_1,\bar{U}_1,U_2,\bar{U}_2$ form a basis
of $T^*Y_4$.
Thus, we can find four one-forms that form a basis at every point
of $Y_4$ and hence the manifold $Y$ is parallelizable. This means
that the two factors of the $SU(2) \times SU(2)$ structure
just intersect in the identity. Thus, the structure group of the
manifold is trivial. This in turn implies that $Y_4$ admits four
globally defined
$SO(4)$ spinors corresponding to string backgrounds with $32$
supercharges. This fact can also be seen from
\eqref{general_comp_pure_spinor_form}. Since $\Phi_{1,2}$ are of odd
degree, $\eta_1$ and $\eta_2$ are of opposite chirality. Together with
their charge conjugated spinors they lead to four globally defined spinors.
Since in this paper we focus on backgrounds with $16$ supercharges, we
do not discuss this case any further.

Let us turn to the case where both spinors are of even degree. The
most general form for those two spinors is given in
\eqref{general_comp_pure_spinor_form}, where now $\eta_1$ and $\eta_2$
are of the same chirality to ensure that $\Phi_1$ and $\Phi_2$ are of
even degree. As we explained below Eq.~\eqref{SU2_spinor_doublet_decomposition} a spinor $\eta_1$ and its charge conjugate $\eta^c_1$ are linearly independent and thus span the whole space of Weyl spinors
of a given chirality. Therefore, $\eta_2$ has to be a linear
combination of $\eta_1$ and $\eta_1^c$. However, this  means that
we can rotate $\Phi_1$ and $\Phi_2$ in such a way that they are of the
form
\begin{equation}
 \Phi_1 = \e^{-B} \wedge \tau(\eta_1 \otimes \bar{\eta}_1) \ , \qquad  \Phi_2 = \e^{-B} \wedge \tau(\eta_1 \otimes \bar{\eta}_1^c) \ .
\end{equation}
Therefore, they give a single $SU(2)$
structure on the manifold, which takes the form
\eqref{pure_spinors_4d}.\footnote{Strictly speaking, we can only call this
a proper $SU(2)$ structure for geometric compactifications since for
non-geometric backgrounds there is globally no projection map
$\mathcal{T} Y \to T Y$ such that we can compare the two $SU(2)$
factors. However, we can do this projection locally, and thus may
compare both $SU(2)$ structures pointwise. In this sense, we can
define proper $SU(2)$ structures even for non-geometric backgrounds.
}

To summarize, due to the fact that the pure spinors have definite chirality
there is no case which interpolates between the trivial structure and
the $SU(2)$ structure case. This can also be understood from the fact
that a pair of nowhere-vanishing spinors $\eta, \eta^c$ spans
the space of given chirality. Therefore, all linearly independent spinors
have to be of opposite chirality and thus cannot be parallel to
$\eta$ at any point in $Y_4$.
Thus, generic $SU(2)\times SU(2)$ structures cannot exist but always
have to be $SU(2)$ or trivial structures.
Note that our conclusion crucially depends on the assumption that $\eta_1$ and $\eta_2$ are nowhere-vanishing. This is not necessarily true for warped compactifications. Therefore the case of a generic warp factor deserves a separate analysis
which, however, we do not go into here.

%%%%%%%%%%%%%%%%%%%%%%%%%%%%%%%%%%%%%%%%%%%%%%%%%%%%%%%%%%%%%%%%%%%%%%%%%%%%
\subsubsection{The deformation space of $SU(2)\times SU(2)$ structures on $Y_4$}\label{section:SU2SU2d6def}
%%%%%%%%%%%%%%%%%%%%%%%%%%%%%%%%%%%%%%%%%%%%%%%%%%%%%%%%%%%%%%%%%%%%%%%%%%%%

In Section \ref{section:spectrumd6} we decomposed the ten-dimensional fields in representations of the structure group. Let us do the same now for deformations of the pure spinors $\Phi_{1,2}$. This will enable us to derive the $\cN=4$ space of scalars in Section \ref{section:moduliSU2}.

Let us first observe that an eight-dimensional
 Weyl spinor of $SO(4,4)$ decomposes under  $SU(2) \times SU(2)$
as\footnote{We showed above that for
backgrounds with $16$ supercharges both $SU(2)$ factors must be the
same after projection to the tangent space. However, as long as we
stay in the framework of generalized geometry and consider pure $SO(4,4)$
spinors, these two factors are different. Therefore we do
the decomposition for $SU(2) \times SU(2)$.
}
\begin{equation}
  {\bf 8}^s \to ({\bf 2},{\bf 2}) \oplus 4 ({\bf 1},{\bf 1}) \ , \qquad {\bf 8}^c \to 2 ({\bf 2},{\bf 1}) \oplus 2 ({\bf 1},{\bf 2}) \ .
\end{equation}
Note that, exactly as in~\eqref{SU2_spinor_doublet_decomposition},
the two conjugate spinors decompose differently. Eq.~\eqref{choice_chirality} gives a canonical choice for the sign of the chirality operator. Hence,
the ${\bf 8}^s$ (${\bf 8}^c$) representation corresponds to forms of even
(odd) degree.
Let us denote the space of forms transforming in the $({\bf r}, {\bf
s})$ representation of $SU(2) \times SU(2)$
 by ${ U_{{\bf r},{\bf s}}}$. As done for $SU(3)\times SU(3)$ representations in \cite{Grana:2006hr}, they can be arranged in a diamond as
given in Table~\ref{diamond}, where the prime is used to distinguish the several singlets.
\begin{table}[htdp]
\begin{center}
\begin{tabular}{ccccc}
 &&  $U_{ {\bf 1},{\bf 1'}}$  && \\
& $U_{ {\bf 2},{\bf 1'}}$ && $U_{ {\bf 1},{\bf 2}}$ & \\
$U_{ {\bf 1'},{\bf 1'}}$ &&  $U_{ {\bf 2},{\bf 2}}$  && $U_{ {\bf 1},{\bf 1}}$ \\
& $U_{ {\bf 1'},{\bf 2}}$ && $U_{ {\bf 2},{\bf 1}}$ & \\
 &&  $U_{ {\bf 1'},{\bf 1}}$  &&
\end{tabular}
 \caption{\small
\textit{Generalized $SU(2) \times SU(2)$ diamond.}}\label{diamond}
\end{center}
\end{table}

In section~\ref{section:spectrumd6} we showed that
for a background to have 16 supercharges  it is necessary to
remove  all massive gravitino
multiplets which corresponds to projecting out all $SU(2)$
doublets. This
eliminates the entire ${\bf 8}^c$ representation (or equivalently all
odd forms in $U_{ {\bf 2},{\bf 1'}}, U_{ {\bf 1},{\bf 2}},
U_{ {\bf 1'},{\bf 2}}, U_{ {\bf 2},{\bf 1}}$) leaving only  the
${\bf 8}^s$ (i.e.\ the even forms in Table~\ref{diamond}).
This is consistent with the result of the previous section
that backgrounds with 16 supercharges require an $SU(2)$
  structure described by pure spinors of positive chirality.

Now we are able to derive the parameter space of $SU(2)\times SU(2)$ structures, which will be very helpful in order to identify the scalar field space of the low-energy effective action. For this, let us first discuss the parameter space of one single normalizable pure $SO(4,4)$ spinor.
The purity and normalization conditions \eqref{pure_spinor_SO44} and \eqref{normalization_pure_spinor_SO44} have a natural
interpretation in the isomorphic picture where $\Phi$ is a complex
vector. Equation~\eqref{pure_spinor_SO44}
and~\eqref{normalization_pure_spinor_SO44} ensure that the real
and imaginary part of $\Phi$ form a pair of space-like orthogonal
vectors. Therefore, $\Phi$ is left invariant by the group
$SO(2,4)$. From section~\ref{section:SU2SU2d6} we know that a pure normalizable $SO(4,4)$ spinor breaks the structure group to $SU(2,2)$. Both pictures are consistent with each other since $SU(2,2)$ is just the double cover of $SO(2,4)$.
The pure spinor $\Phi$ therefore parameterizes the space
$SO(4,4)/{SO(2,4)}$.
However, the phase of $\Phi$ does not affect the $SU(2,2)$
structure. Hence the actual parameter space of a single pure spinor is
\begin{equation}
{\cal M}_\Phi = \frac{SO(4,4)}{SO(2) \times SO(2,4)} \ .
\end{equation}

The parameter space of $SU(2)$ structures
 is more conveniently
discussed in terms of the real and imaginary
parts of the two spinors $\Phi_i$ or in other words in terms of four
real vectors $\Psi_a, a=1, \dots, 4$ in the space of even
forms. Then the compatibility conditions \eqref{comp_pure_spinor_conditions}
and \eqref{normalization_pure_spinors} just translate into the
conditions
\begin{equation}
\label{conditions_real_spinors}
 \langle \Psi_a, \Psi_b \rangle = c \, \delta_{ab}\, \operatorname{vol}_4 \ ,
\end{equation}
where $c$ parameterizes the scale of the $\Psi_a$.
The four $\Psi_a$ form the singlet corners in Table~\ref{diamond}
since they are (as the $\Phi_i$) globally defined and thus must be singlets of the
structure group.

In order to understand the signature of the $SU(2)\times
SU(2)$ diamond (Table~\ref{diamond}) we use the
generalized Hodge operator $\ast_B$ defined in \eqref{Hodge_star_generalized}.
Since there is $\ast_B^2 = 1$ on forms of even degree, we see that the generalized Hodge operator
corresponds to an almost
product structure on $\Lambda^\textrm{even}T^*Y$. For $B=0$, it coincides on two-forms with the conventional Hodge star operator, which is of split signature over each point. In this case, the forms $1\pm \operatorname{vol}$ are eigenvectors of $\ast_B$ with eigenvalue $\mp 1$ and we see that $\ast_B$ has split signature on $\Lambda^\textrm{even}T^*Y$ over each point of $Y$. Since $B$ refers to a continuous $SO(4,4)$ transformation, it can be continuously switched on. Therefore the signature is independent of $B$ and the eigenspaces of $\ast_B$ (with eigenvalue $\pm 1$) at a given point on the manifold have the same dimension.
As with the standard Hodge operator, the generalized Hodge operator $\ast_B$ can be globally defined on $Y$ and therefore must be invariant under the $SU(2) \times SU(2)$ structure group. Hence, $\ast_B$ leaves the $SU(2)\times SU(2)$ representations invariant and its eigenspaces coincide with these representations.

Using the form~\eqref{pure_spinors_4d} and the fact that the $J_i$,
$i=1,2,3$, defined in~\eqref{stable_forms_definition}, are self-dual
with respect to the standard Hodge operator one can show that the $\Psi_a$ are eigenvectors of $\ast_B$ with eigenvalue $+1$.
This implies that the eigenspace with eigenvalue $+1$ is spanned by the four
spinors $\Psi_a$, i.e.\ by the $SU(2) \times SU(2)$
singlets, which is consistent with~\eqref{conditions_real_spinors}.
Therefore, the orthogonal complement $U_{{\bf 2},{\bf 2}}$ is the
eigenspace with eigenvalue $-1$.
This shows that a choice of $\Psi_a$ already determines the eigenspaces of
the generalized Hodge operator and thus the operator itself.
Since the composition of the Mukai pairing with $\ast_B$ is positive definite, the eigenvalue corresponding to some eigenvector of $\ast_B$ gives also its signature under the Mukai
pairing. Therefore we conclude that the Mukai pairing is positive definite on the
$SU(2) \times SU(2)$ singlets and negative definite on $U_{{\bf 2},{\bf 2}}$.

Thus, we see that the $\Psi_a$ define a space-like
four-dimensional subspace in $\Lambda^\textrm{even}T^*Y_4$ in that they are due to \eqref{conditions_real_spinors} an orthonormal basis for this subspace, or in other words
they parameterize the space
${SO(4,4)}/{SO(4)}$, where $SO(4)$ denotes rotations inside the $-1$ eigenspace of $\ast_B$, which leave the $\Psi_a$ invariant.
However, we also need to divide out the rotational $SO(4)$ symmetry
among the $\Psi_a$ since it does
not change the $SU(2) \times SU(2)$ structure. This leaves as the
physical parameter space
\begin{equation}\label{local_moduli_space_4d}
{\cal M}_{\Psi_a} =  \frac{SO(4,4)}{SO(4)\times SO(4)} \ .
\end{equation}
This Grassmannian is the NS-NS parameter space of four-dimensional $SU(2)\times SU(2)$ structures. As we discuss in Section \ref{section:RRd6}, this already gives the complete parameter space of type IIA on $Y_4$, as there are no scalars coming from the R-R sector. For type IIB, we shall see in the same section that the space in \eqref{local_moduli_space_4d} is enlarged by the R-R scalars to another Grassmannian.

%%%%%%%%%%%%%%%%%%%%%%%%%%%%%%%%%%%%%%%%%%%%%%%
\subsection{$SU(n)\times SU(n)$ structures on $Y_6$} \label{section:SUnSUn_d4}
%%%%%%%%%%%%%%%%%%%%%%%%%%%%%%%%%%%%%%%%%%%%%%%
%%%%%%%%%%%%%%%%%%%%%%%%%%%%%%%%%%%%%%%%%%%%%%%
\subsubsection{$SU(3)\times SU(3)$ structures and pure spinors} \label{section:SU3SU3_pure_spinors}
%%%%%%%%%%%%%%%%%%%%%%%%%%%%%%%%%%%%%%%%%%%%%%%
Now let us turn to the pure spinor description for $SU(3)\times SU(3)$ structures on a six-dimensional manifold $Y_6$.

It has been shown in \cite{Hitchin:2004ut} that pure (complex) $SO(6,6)$ spinors are in one-to-one correspondence to stable (real) spinors.\footnote{Here, stable again means that any nearby point in spinor space can be reached by some $SO(6,6)$ transformation.}
More precisely, a pure spinor $\Phi$ is completely determined by its real part $\chi \equiv \Re \Phi$, which is a stable spinor. Indeed, a stable $SO(6,6)$ spinor $\chi$ defines an almost-complex structure ${\cal J}$ on $TY \oplus T^*Y$ by \cite{Hitchin:2004ut,Grana:2005ny}
\begin{equation}
 {\cal J} = - \frac{1}{\sqrt{3 \langle \chi, \Gamma_{\Lambda \Gamma} \chi \rangle \langle \chi, \Gamma^{\Lambda \Gamma} \chi \rangle}} \langle \chi, \Gamma^{\Pi\Sigma} \chi \rangle \Gamma_{\Pi\Sigma} \ .
\end{equation}
This generalized almost-complex structure then enables one to define the pure spinor
\begin{equation}
 \Phi = \chi - \iu {\cal J} \chi \ ,
\end{equation}
and induces also a complex structure on the parameter space of $\chi$.
Furthermore, one defines the Hitchin functional
\begin{equation} \label{Hitchin_functional}
 H(\chi) = \sqrt{\tfrac{1}{12} \langle \chi, \Gamma_{\Pi\Sigma} \chi \rangle \langle \chi, \Gamma^{\Pi\Sigma} \chi \rangle} = \iu \langle \bar  \Phi , \Phi \rangle \ ,
\end{equation}
which is homogeneous of degree two. One can then show that the parameter space of $\chi$ is special K\"ahler with a K\"ahler potential \cite{Hitchin:2004ut,Grana:2005sn}
\begin{equation} \label{Kahler_Hitchin}
 K = - \ln H(\chi) \ .
\end{equation}
We come back to this in more detail in Section~\ref{section:special_Kahler}.

The next step is to consider the case of two compatible pure spinors. Again, a group-theoretical decomposition of the pure $SO(6,6)$ spinor deformations in representations of $SU(3)\times SU(3)$ helps to understand the moduli space \cite{Gualtieri:2003dx,Grana:2005sn}.
Under $SO(6,6)\to SO(6) \times SO(6)$ the spinors decompose as
\begin{equation}
 {\bf 32}^+ \to ({\bf 4}, \bar {\bf 4}) \oplus (\bar {\bf 4}, {\bf 4}) \ , \qquad  {\bf 32}^- \to ({\bf 4}, {\bf 4}) \oplus (\bar {\bf 4}, \bar{\bf 4}) \ .
\end{equation}
Here, we conjugated the representations of the second $SU(3)$ (related to right-movers) to be consistent with the conventions of Section~\ref{section:spectrum}.
From \eqref{4to31} we see that under $SO(6,6)\to SU(3) \times SU(3)$ we get for each spinor eight representations that can all together be arranged in a diamond given in Table \ref{SU3diamond} \cite{Gualtieri:2003dx,Grana:2005sn}.
\begin{table}[htdp]
\begin{center}
\begin{tabular}{ccccccc}
 &&&  $U_{ {\bf 1},\bar{\bf 1}}$  &&& \\
&& $U_{ \bar{\bf 3},\bar{\bf 1}}$ && $U_{ {\bf 1},{\bf 3}}$ && \\
&$U_{ {\bf 3},\bar{\bf 1}}$ &&  $U_{ \bar{\bf 3},{\bf 3}}$  && $U_{ {\bf 1},\bar{\bf 3}}$ &\\
$U_{ \bar{\bf 1},\bar{\bf 1}}$ &&  $U_{ {\bf 3},{\bf 3}}$  && $U_{\bar {\bf 3},\bar{\bf 3}}$ && $U_{ {\bf 1},{\bf 1}}$ \\
&$U_{ \bar{\bf 1},{\bf 3}}$ &&  $U_{ {\bf 3},\bar{\bf 3}}$  && $U_{ \bar{\bf 3},{\bf 1}}$ &\\
&& $U_{ \bar{\bf 1},\bar{\bf 3}}$ && $U_{ {\bf 3},{\bf 1}}$ && \\
 &&&  $U_{ \bar{\bf 1},{\bf 1}}$  &&&
\end{tabular}
 \caption{\small
\textit{Generalized $SU(3) \times SU(3)$ diamond.}}\label{SU3diamond}
\end{center}
\end{table}
Here, $U_{ {\bf r},{\bf s}}$ denote the set of $({\bf r},{\bf s})$ forms (the singlets coming from the $\bar{\bf 4}$ are denoted by $\bar{\bf 1}$, abusing notation). The deformations of the pure spinor $\Phi^+$ are given in the odd rows while those of $\Phi^-$ are given in the even ones. Note that all representations $U_{ {\bf r},{\bf s}}$ may be differential forms of mixed degree. The singlets in the corners of Table \ref{SU3diamond} are the real and imaginary parts of the two pure spinors $\Phi^+$ and $\Phi^-$. The remaining representations on the outside of the diamond are $SU(3)\times SU(3)$ triplets, which are deformations that are projected out in the compactification process. The space of physical deformations consists of the representations in the interior of the diamond, as was already derived in Section~\ref{section:spectrum}. Here one sees that it indeed coincides with the deformations of the pair of pure spinors. In type IIA for instance, the deformations of $\Phi^+$ give the scalars in the vector multiplets while the deformations of $\Phi^-$ form part of the hypermultiplet sector.

Now let us turn to the compatibility condition. From \eqref{Compatibility_gacs} and \eqref{compatible_pure_spinors_chirality} one observes that the two pure spinors are of opposite chirality. In the following they are denoted by $\Phi^+$ and $\Phi^-$. Furthermore, the compatibility condition \eqref{Compatibility_gacs} can be rephrased as \cite{Grana:2005ny}
\begin{equation}\label{SU3_compatibility}
 \langle \Phi^+, \Gamma_\Pi \Phi^- \rangle =  \langle \Phi^+, \Gamma_\Pi \bar \Phi^- \rangle = 0 \ ,
\end{equation}
where $\Gamma_\Pi$ is a basis element of $TY \oplus T^*Y$. This is a condition on the space of deformations of the pure spinors. Therefore, it seems that the product structure of the parameter space of $\Phi^+$ and $\Phi^-$ is destroyed.
However, the elements of $TY \oplus T^*Y$ decompose under $SO(6,6)\to SO(6)\times SO(6) \to SU(3)\times SU(3)$ as
\begin{equation}
 {\bf 12} \to ({\bf 6},1) \oplus (1,{\bf 6}) \to ({\bf 3},{\bf 1}) \oplus (\bar{\bf 3},{\bf 1}) \oplus ({\bf 1},{\bf 3}) \oplus ({\bf 1},\bar {\bf 3})  \ .
\end{equation}
From this one observes that the set of $\Gamma_\Pi$ consists of $SU(3)\times SU(3)$ triplets only. Furthermore, the full expression in \eqref{SU3_compatibility} needs to be a singlet, and therefore, \eqref{SU3_compatibility} is only a condition on the triplet deformations of $\Phi^+$ and $\Phi^-$. Since these are projected out in $\cN=2$ compactifications, the compatibility condition \eqref{SU3_compatibility} is trivially satisfied. However, this was the only condition relating $\Phi^+$ and $\Phi^-$ and therefore the deformation space of $\Phi^+$ and $\Phi^-$ stays a product of two special K\"ahler spaces.

It is instructive to apply this analysis to the case of a single $SU(3)$ structure, i.e.\ to $\eta_2 = \eta_1$. It then follows from \eqref{mapping_bispinors_forms}, \eqref{general_comp_pure_spinor_form} and \eqref{SU3_forms_definitions} that
\begin{equation}
 \Phi^+ = \tfrac14 \e^{-B + \iu J} \ , \qquad \Phi^- = \tfrac14 \iu \Omega \ .
\end{equation}
Therefore, the formalism of generalized geometry naturally reproduces the parametrization of the geometric degrees of freedom in terms of forms as discussed in Section \ref{section:SU2d4} but complexifies the deformation space of $J$ by incorporating the $B$ field. The $B$ field here represents $SO(6,6)$ transformations which are generated by a two-form in the Lie algebra \eqref{spin2n2n}.

On a six-dimensional manifold the group $SO(6,6)$ is the T-duality group, which means it is the symmetry group of the NS-NS sector over each point of $Y$. Generalized geometry is constructed in such a way that it is covariant with respect to the T-duality group. Since the geometrical degrees of freedom of $Y$ and the scalars coming from the $B$-field transform into each other under the T-duality group, the $B$ field is naturally incorporated into the parametrization of generalized geometry.

%%%%%%%%%%%%%%%%%%%%%%%%%%%%%%%%%%%%%%%%%%%%%%%%%%%%%%%%%%%%%%%%%%%%%%%%
\subsubsection{$SU(2)\times SU(2)$ structures on $Y_6$}\label{section:SU2SU2d4}
%%%%%%%%%%%%%%%%%%%%%%%%%%%%%%%%%%%%%%%%%%%%%%%%%%%%%%%%%%%%%%%%%%%%%%%%
Now we want to describe $SU(2)\times SU(2)$ structures on a six-dimensional space $Y_6$ in terms of generalized geometry.
In the last section we reviewed the case of an $SU(3)\times SU(3)$ structure  on $Y_6$.
We already stated that a normalizable pure $SO(6,6)$ spinor $\Phi$ is in one-to-one
correspondence with a real stable $SO(6,6)$ spinor and hence looses half of its degrees of freedom.
Two normalizable pure  $SO(6,6)$ spinors $\Phi^+$ and $\Phi^-$ are
compatible if they are of opposite chirality and \eqref{SU3_compatibility} holds.
In addition, we can impose the normalization condition
\begin{equation}
\label{compatibility_conditions_normalization}
 \langle \Phi^+, \bar{\Phi}^+ \rangle = \langle \Phi^-, \bar{\Phi}^- \rangle  \ ,
\end{equation}
since the prefactor of each pure spinor is not physical.

The pure spinors $\Phi^+$ and $\Phi^-$ obeying \eqref{SU3_compatibility} and
\eqref{compatibility_conditions_normalization} only define
an $SU(3) \times SU(3)$ structure on $Y_6$. In order to construct an $SU(2) \times
SU(2)$ structure, one has to introduce further objects that are
globally defined and compatible with the spinors
introduced so far. One way to proceed is by mimicking the $SU(2)$
structure construction of Section \ref{section:SU2d4} and defining
two $SU(3) \times SU(3)$ structures with compatibility conditions
imposed such that they intersect in an $SU(2)\times SU(2)$
structure. Each $SU(3) \times SU(3)$ structure already defines a
generalized metric on $T Y_6 \oplus T^*Y_6$, and these two generalized
metrics must coincide for consistency. It turns out that
an $SU(2) \times SU(2)$ structure can alternatively be
defined by a pair of compatible pure spinors $\Phi^+$, $\Phi^-$
and a generalized almost-product structure $\mathcal{P}$ which has
the following properties \cite{Triendl:2009ap}:
\begin{enumerate}
 \item $\mathcal{P}^2 = 1$ .
 \item $\mathcal{P}$ is symmetric with respect to $\mathcal{I}$.
 \item $\mathcal{P}$ commutes with the generalized almost-complex structures $\mathcal{J}_{\Phi^\pm}$.
 \item The eigenspaces of $\mathcal{P}$ to the eigenvalues $-1$ and $+1$ are of dimension $8$ and $4$, respectively.
\end{enumerate}
Note that the second and third conditions ensure that $\mathcal{P}$ is also
symmetric with respect to the metric defined by
$\mathcal{J}_{\Phi^+}$ and $\mathcal{J}_{\Phi^-}$.
Furthermore, $\mathcal{P}$ is also symmetric with respect to the canonical $SO(6,6)$ scalar product by construction. This
implies that the canonical pairing is block-diagonal with respect to
the splitting of the bundle induced by $\mathcal{P}$. Therefore,
$\mathcal{P}$ reduces the structure group to $SO(4,4) \times
SO(2,2)$. Since it commutes with $\mathcal{J}_{\Phi^+}$ and
$\mathcal{J}_{\Phi^-}$, both generalized almost-complex structures
are similarly block-diagonal with respect to this splitting.

Thus, we conclude that reducing an $SU(3) \times SU(3)$ structure to
an $SU(2) \times SU(2)$ structure corresponds to the fact that one
is able to globally split  $TY_6 \oplus T^*Y_6$ into
\begin{equation}
\label{splitting_generalized_tangent_bundle}
 TY_6 \oplus T^*Y_6 = (T_2 Y_6 \oplus T_2^* Y_6) \oplus (T_4 Y_6 \oplus
 T_4^* Y_6) \ ,
\end{equation}
where $T_4 Y_6 \oplus T_4^* Y_6$ is the eight-dimensional vector bundle that is the $-1$ eigenspace of $\mathcal{P}$ at every point, and $T_2 Y_6 \oplus T_2^* Y_6$ is correspondingly the four-dimensional vector bundle that forms the $+1$ eigenspace of $\mathcal{P}$ at every point.\footnote{Properly written, \eqref{splitting_generalized_tangent_bundle}~reads
$\mathcal{T}Y_6 = \mathcal{T}_2 Y_6 \oplus \mathcal{T}_4 Y_6.$}
The pure spinor pair $\Phi^\pm$, corresponding to $\mathcal{J}_{\Phi^\pm}$, defines an $SU(2)\times SU(2)$ structure on $T_4 Y_6 \oplus T_4^* Y_6$ and an identity structure on $T_2 Y_6 \oplus T_2^* Y_6$, i.e.\ $T_2 Y_6 \oplus T_2^* Y_6$ is the trivial bundle.
On $T_4 Y_6 \oplus T_4^* Y_6$, we can redo the analysis of
Section~\ref{section:SU2SU2d6def} since the dimension of the bundle $T_4 Y_6 \oplus T_4^* Y_6$ is eight.

Let us make this structure slightly more explicit by considering
the pure spinors $\Phi^\pm$ that correspond to $\mathcal{J}_{\Phi^\pm}$.
First, let us fix the generalized almost-product structure
$\mathcal{P}$ and investigate the structure of $\Phi^+$ and
$\Phi^-$. Eq.~\eqref{splitting_generalized_tangent_bundle} induces a
splitting of the $SO(6,6)$ spinor space $\Lambda^\bullet T^* Y_6$, i.e.\
\begin{equation}
 \Lambda^\bullet T^* Y_6 = \Lambda^\bullet T^*_2 Y_6 \wedge \Lambda^\bullet T^*_4 Y_6 \ ,
\end{equation}
where $\Lambda^\bullet T^*_2 Y_6$ and $\Lambda^\bullet T^*_4 Y_6$ are the $SO(2,2)$ and the $SO(4,4)$ spinor bundles over $Y_6$, respectively.
This decomposition carries over to the chiral subbundles
\begin{equation}\begin{aligned}
\label{spinor_bundle_decomposition}
 \Lambda^{\textrm{even}} T^* Y_6\ &=\  \Lambda^{\textrm{even}} T^*_2 Y_6 \wedge  \Lambda^{\textrm{even}} T^*_4 Y_6\ \oplus\  \Lambda^{\textrm{odd}} T^*_2 Y_6 \wedge \Lambda^{\textrm{odd}} T^*_4 Y_6 \ , \\
 \Lambda^{\textrm{odd}} T^* Y_6\ &=\ \Lambda^{\textrm{even}} T^*_2 Y_6 \wedge  \Lambda^{\textrm{odd}} T^*_4 Y_6\ \oplus \ \Lambda^{\textrm{odd}} T^*_2 Y_6 \wedge \Lambda^{\textrm{even}} T^*_4 Y_6  \ .
\end{aligned}
\end{equation}
The direct sum on the right-hand side holds globally,
since, by use of $\mathcal{P}$, we can define chirality operators for
$\Lambda^\bullet T^*_2 Y_6$ and $\Lambda^\bullet T^*_4 Y_6$ independently. In other words, the structure group does not mix the spinor bundles $\Lambda^{\textrm{even}} T^*_4 Y_6$ and $\Lambda^{\textrm{odd}} T^*_4 Y_6$ and the spinor bundles $\Lambda^{\textrm{even}} T^*_2 Y_6 $ and $\Lambda^{\textrm{odd}} T^*_2 Y_6$.

Moreover, since the generalized almost-complex structures commute with
$\mathcal{P}$, they split
under~\eqref{splitting_generalized_tangent_bundle} into a generalized
almost-complex structure on each component. Correspondingly,
using~\eqref{spinor_bundle_decomposition}, the pure spinors $\Phi^+$
and $\Phi^-$ globally decompose into pure spinors on the spinor
sub-bundles.
As already argued above, the spinor bundles on the right-hand side of Eq.~\eqref{spinor_bundle_decomposition} do not mix under the action of the structure group, and therefore, the components of $\Phi^+$ and $\Phi^-$ on the sub-bundles can be analyzed separately. Their components on $\Lambda^\bullet T^*_4 Y_6$ must define an $SU(2) \times SU(2)$ structure.
However, we already discussed the case of an $SU(2) \times SU(2)$ structure group on a vector bundle of dimension eight in Section~\ref{section:SU2SU2d6def}.
We know from Section~\ref{section:SU2SU2d6} that an $SU(2) \times SU(2)$ structure group on $T_4 Y_6 \oplus T_4^* Y_6$ is defined by two pure spinors that must have the same chirality. Any additional nowhere-vanishing pure spinor would break the structure group further. Thus, we can distinguish two cases: Either both spinor components on $\Lambda^\bullet T^*_4 Y_6$ lie in $\Lambda^\textrm{odd} T^*_4 Y_6$ or in $\Lambda^\textrm{even} T^*_4 Y_6$.
Note that in both cases we are left with two pure spinors of opposite chirality in $\Lambda^{\bullet} T^*_2 Y_6$ which define a trivial structure on $T_2 Y_6 \oplus T_2^* Y_6$.

In the first case, both pure spinors
on $T_4 Y_6 \oplus T_4^* Y_6$ are of negative chirality. As we showed in
Section~\ref{section:SU2SU2d6}, these two pure spinors define an
$SU(2)\times SU(2)$ structure where the two $SU(2)$ factors have
trivial intersection.
Thus $Y_6$ admits a trivial structure, i.e.\ is parallelizable, and the background has 32 supercharges.
As in Section~\ref{section:SU2SU2d6}, we do not discuss this case any further.

The second possibility is that both spinor components are of positive
chirality and define -- analogously to section~\ref{section:SU2SU2d6} -- a proper $SU(2)$ structure on the
manifold. Thus, also on $Y_6$ the possibility of an intermediate $SU(2)\times SU(2)$
structures does not exist. Instead one can only have an $SU(2)$
structure or a trivial structure, as we already concluded in our analysis for $Y_4$ in
Section~\ref{section:SU2SU2d6}.
In the $SU(2)$-structure case we can write
\begin{equation}\label{SUtwoY6}
 \Phi^+ = \Theta_+ \wedge \Phi_1 \ , \qquad
 \Phi^- = \Theta_- \wedge \Phi_2 \ ,
\end{equation}
where $\Theta_\pm$ are $SO(2,2)$ spinors of opposite chirality and
therefore define a trivial structure on $T_2 Y_6 \oplus T_2^* Y_6$. The
$SO(4,4)$ spinors $\Phi_1$ and $\Phi_2$ are pure and of even
chirality and define the $SU(2)$ structure on $T_4 Y_6
\oplus T_4^* Y_6$. This is precisely the situation we already discussed
in Section~\ref{section:SU2d4}. There the $SU(2)$ structure was defined in terms of the two spinors $\eta_i$. The relation between the $\eta_i$ and the $\Phi_\pm$ is analogously to \eqref{general_comp_pure_spinor_form} described by
\begin{equation}\label{SUtwoY6_eta}
 \Phi^+ = \e^{-B} \wedge \tau(\eta_1 \otimes \bar{\eta}_2) \ , \qquad
 \Phi^- = \e^{-B} \wedge \tau(\eta_1 \otimes \bar{\eta}_2^c) \ ,
\end{equation}
where $B$ is the NS-NS two-form, which is not determined by the $\eta_i$. We can insert the definition of $\tau$ \eqref{mapping_bispinors_forms} and relate the components in \eqref{SUtwoY6} to the quantities $K, J, B$ and $\Omega$ via
\eqref{definition_one-form_K} and \eqref{definition_two-forms_6d}.
We end up with
\begin{equation}\label{thetaplus}
\Theta_+ = \e^{- B_{(2)} +\tfrac{1}{2} K \wedge \bar{K}} \ , \quad \Theta_- = K \ ,  \quad \Phi_1 = \tfrac{\iu}{4}\, \e^{-B_{(4)}} \wedge \Omega \ , \quad \Phi_2 = \tfrac{1}{4}\, \e^{-B_{(4)} - \iu J } \ ,
\end{equation}
and therefore
\begin{equation}
\label{pure_spinors_6d}
\Phi^+=
\tfrac{\iu}{4}\, \e^{-B_{(2)} + \tfrac{1}{2} K\wedge \bar{K}} \wedge \e^{-B_{(4)}} \wedge \Omega \ , \qquad
\Phi^-=
\tfrac{1}{4}\, K\wedge \e^{-B_{(4)} - \iu J} \ ,
\end{equation}
where we denoted the components of $B$ on $\Lambda^2 T^*_2 Y_6$ by $B_{(2)}$ and on $\Lambda^2 T^*_4 Y_6$ by $B_{(4)}$, respectively.
As mentioned earlier, there is some gauge freedom in choosing $\eta_1$ and $\eta_2$ out of the space of $SU(2)$ singlets, which translates into a rotational gauge freedom between $\Phi_1$ and $\Phi_2$. Therefore, it is more convenient in the following not to specify the $\Phi_i$ in terms of $J$ and $\Omega$.

Now we determine the parameter space of $\Phi^\pm$.
By the splitting we described above, we can do this independently for
the pure spinors on $T_2 Y_6 \oplus T_2^* Y_6$ and the ones on $T_4 Y_6
\oplus T_4^* Y_6$. On the eight-dimensional subspace $T_4 Y_6 \oplus
T_4^* Y_6$ the arguments are the same as on $Y_4$ and
thus $\Phi_1$ and $\Phi_2$ form the moduli
space $SO(4,4)/(SO(4)\times SO(4))$ given in \eqref{local_moduli_space_4d}.

Additionally, the $SO(2,2)$ spinors
$\Theta_+$ and $\Theta_-$ each parameterize a moduli space on their own. The reason is that the Lie algebra splits according to
\begin{equation} \label{SO22_splitting}
so(2,2) = sl(2,\mathbb{R})_T \oplus sl(2,\mathbb{R})_U \ .
\end{equation}
The first sub-algebra $sl(2,\mathbb{R})_T$ just acts on $\Theta_+$, while
the second $sl(2,\mathbb{R})_U$ acts on $\Theta_-$.
The degrees of freedom in
$\Theta_+$ correspond to a two-form acting on the
negative eigenspace of $\mathcal{P}$ and a form of
degree zero. Together they form an $Sl(2,\mathbb{R})_T$ doublet. Furthermore, we have to mod out
the gauge degree of freedom corresponding to the phase of $\Theta_+$.
From \eqref{general_pure_spinor_form} we learn that
the remaining complex
degree of freedom of $\Theta_+$ is given by the volume and the
$B$-field. It spans the parameter space
$Sl(2,\mathbb{R})_T/SO(2)$.
Similarly, $\Theta_-$ can be expanded in the basis of
one-forms on the negative eigenspace of $\mathcal{P}$, which is
two-dimensional, and therefore defines an $Sl(2,\mathbb{R})_U$ doublet analogously to $\Theta_+$, exhibiting the same normalization and gauge degree of freedom. Hence, $\Theta_-$ spans the moduli space ${Sl(2,\mathbb{R})_U}/{SO(2)}$. Note that
for $Y_6= K3\times T^2, \Theta_\pm$ parameterize the K\"ahler and
complex structure deformations of the $T^2$, respectively.

The formalism of generalized geometry automatically incorporates the $B$-field degrees of freedom. We can also incorporate the other string fields.
Additionally to the parameter space of the pure spinors, we have the dilaton field $\phi$ in the NS-NS sector, which is complexified by the dualized $B$ field in four dimensions, and forms the moduli space
${Sl(2,\mathbb{R})_S}/{SO(2)}$.
So altogether we have in the NS-NS sector the (local) scalar space
\begin{equation}
\label{local_moduli_space_6d}
   \mathcal{M}_{\Theta_\pm,\Phi_i} = \frac{SO(4,4)}{SO(4)\times SO(4)} \times \frac{Sl(2,\mathbb{R})_S}{SO(2)} \times \frac{Sl(2,\mathbb{R})_T}{SO(2)}  \times \frac{Sl(2,\mathbb{R})_U}{SO(2)}  \ .
\end{equation}
In order to incorporate the scalars coming from the R-R sector, it is more convenient to enlarge the manifestly covariant symmetry group from $SO(6,6)$ to the (local) symmetry group of all string scalars. We discuss this in the next section.

%%%%%%%%%%%%%%%%%%%%%%%%%%%%%%%%%%%%%%%%%%%%%%%
\section{Exceptional generalized geometry} \label{section:Exceptional}
%%%%%%%%%%%%%%%%%%%%%%%%%%%%%%%%%%%%%%%%%%%%%%%

Generalized geometry is a natural generalization of $G$-structures
since it covers the complete moduli space of the NS-NS sector of
string theory. This is due to the fact that the group acting on
the generalized tangent bundle coincides with the T-duality group $SO(m,m)$, which arises as symmetry group of the NS-NS sector in $m$-torus compactifications.
However, it is also possible to include the R-R sector of type II
string theories by extending the T-duality group $SO(m,m)$ to the larger
U-duality group $E_{m+1(m+1)}$ which also includes transformations between the NS-NS
and the R-R sector~\cite{Hull:1994ys} and gives the complete symmetry group for type II. To do so, one extends the
generalized tangent bundle $\mathcal{T}Y$ to the exceptional
generalized tangent bundle ${\cal E} Y$~\cite{Hull:2007zu,Pacheco:2008ps}. The spin
group over this bundle is then the U-duality group which coincides
with the (non-compact version of the) exceptional group $E_{m+1}$. It seems natural that the formalism of pure spinors should extend to the case of exceptional generalized
geometry \cite{Grana:2009im}.
In this section, following \cite{Grana:2009im} we discuss how we can parameterize the set of all scalar light fields in type II compactifications with the help of exceptional generalized geometry. By application to $SU(n)\times SU(n)$ structure backgrounds we derive then the parameter space of all $d$-dimensional scalars of the ten-dimensional theory. As we shall see, the resulting parameter spaces have the properties required by $\cN=2$ and $\cN=4$ supersymmetry.

As before, we first consider the simpler case of a four-dimensional $Y_4$ and turn to the case of a six-dimensional space afterwards. In both cases we first study the embedding of the T-duality group $SO(m,m)$ into the U-duality group $E_{m+1(m+1)}$ and find the geometric realizations of the $E_{m+1(m+1)}$ representations by comparison of the BPS spectrum of the related $m$-torus compactification, which fills out the fundamental representation of $E_{m+1(m+1)}$ and is realized in terms of a direct sum of geometric bundles over the $m$-torus~\cite{Hull:1994ys}. Since the identification of $E_{m+1(m+1)}$ representations with fibre bundles should be independent of the compactification manifold, it should hold for general compactifications on $Y_m$. In a second step we then embed the pure spinor spaces of generalized geometry in appropriate bundles and analyze the parameter space of the corresponding sections. Finally, we identify the scalar spaces of $\cN=2$ and $\cN=4$ compactifications by projecting out all representations containing massive gravitino multiplets.

%%%%%%%%%%%%%%%%%%%%%%%%%%%%%%%%%%%%%%%%%%%%%%%%%%%%%%%%%%%%%%%
\subsection{Exceptional generalized geometry on $Y_4$}\label{section:RRd6}
%%%%%%%%%%%%%%%%%%%%%%%%%%%%%%%%%%%%%%%%%%%%%%%%%%%%%%%%%%%%%%%

Let us examine the construction of exceptional generalized geometry for the case of $SU(2)$-structures on $Y_4$. In this case the U-duality group is $E_{5(5)} = SO(5,5)$ with the T-duality subgroup being $SO(4,4)$. Let us first look at the decomposition of the representations of $SO(5,5)$ in terms of its maximal subgroup $SO(4,4) \times \mathbb{R}_+$. The extra $\mathbb{R}_+$-factor corresponds to shifts of the dilaton.
The vector representation of $SO(5,5)$ decomposes as \cite{Slansky:1981yr}
\begin{equation}
 {\bf 10} \to {\bf 8}^v_{0} \oplus {\bf 1}_{+2} \oplus {\bf 1}_{-2}  \ ,
\label{splitting_vector_rep}
\end{equation}
while for the spinor representation we have
\begin{equation}
 {\bf 16} \to {\bf 8}^c_{+1} \oplus {\bf 8}^s_{-1}  \ .
\label{splitting_fund_rep}
\end{equation}
The subscript denotes the charge of the representation under shifts of
the dilaton.
Finally, the adjoint of $SO(5,5)$ decomposes as
\begin{equation} \label{splitting_adjoint_rep}
 {\bf 45} \to {\bf 28}_0 \oplus {\bf 8}^v_{+2} \oplus {\bf 8}^v_{-2} \oplus {\bf 1}_0  \ .
\end{equation}
Note that because of $SO(4,4)$ triality, the three ${\bf 8}$
representations can be interchanged pairwise, which, however,
has to be done in all three decompositions simultaneously.

Let us now determine the geometric realizations of these
representations. For the T-duality group $SO(4,4)$, the vector
representation ${\bf 8}^v$ is (locally) given in geometrical terms by
$T Y_4 \oplus T^* Y_4$ and analogously the spinor representations ${\bf
8}^s$ and ${\bf 8}^c$ by $\Lambda^\textrm{even}T^* Y_4$ and
$\Lambda^\textrm{odd}T^* Y_4$, respectively.
 However, $SO(4,4)$-triality can interchange the three
eight-dimensional bundles $T Y_4 \oplus T^* Y_4$, $\Lambda^\textrm{even}T^*
Y_4$ and $\Lambda^\textrm{odd}T^* Y_4$.
%One of these three bundles is the vector component in
%\eqref{splitting_vector_rep} and \eqref{splitting_adjoint_rep} while
%the other two form the components in \eqref{splitting_fund_rep}.
To assign them to the representations in the right way, we note -- as explained in~\cite{Hull:1994ys} -- that the NS-NS and R-R charges
together form the ${\bf 16}$ representation of
$E_{5(5)}$.\footnote{With charges we mean those solutions which are
  point-like particles in six dimensions that are charged under the
  NS-NS and R-R vectors.
%that are displayed in Table \ref{N=2d=6multipletsA} and
%\ref{N=2d=6multipletsB}.
In the NS-NS sector, the charges are formed by the momentum and winding modes of the fundamental string that are charged under $g_{m\mu}$ and $B_{m\mu}$, respectively, while the R-R charges descend from ten-dimensional D-brane solutions.}
Winding modes are winding one-cycles and therefore are represented by $T^* Y_4$, while
Kaluza-Klein modes are the momentum modes of the compactification, which correspond to the translation generators in $T Y_4$.
Hence, the NS-NS charges live in the
geometrical bundle $T Y_4 \oplus T^* Y_4$. The R-R charges arise from D-branes
wrapped on internal cycles. For type IIA (IIB), D-branes have an even (odd) number of space dimensions and therefore wrap even(odd)-dimensional cycles of $Y$ to give point particles in the $6$-dimensional theory. They therefore sit in
$\Lambda^\textrm{even}T^* Y_4$ for type IIA and $\Lambda^\textrm{odd}T^*Y_4$
for type IIB, respectively.
Hence the ${\bf 16}$ representation corresponds to $T Y_4 \oplus T^* Y_4 \oplus \Lambda^\textrm{even}T^* Y_4 $ in type IIA and to $T Y_4 \oplus T^* Y_4 \oplus \Lambda^\textrm{odd}T^* Y_4 $ in type IIB~\cite{Hull:2007zu}. Consequently, the representation ${\bf 8}^v$ is associated
with $\Lambda^\textrm{odd}T^* Y_4$  in type IIA and
$\Lambda^\textrm{even}T^* Y_4$ in type IIB, respectively.
Altogether we thus have
\begin{equation}\begin{aligned}
 \label{representation_IIA_geom}
 {\bf 10} &= (\Lambda^\textrm{odd}T^* Y_4 )_0 \oplus (\mathbb{R})_{+2}  \oplus (\mathbb{R})_{-2}  \ , \\
 {\bf 16} &= (T Y_4 \oplus T^* Y_4)_{+1}  \oplus (\Lambda^\textrm{even}T^* Y_4)_{-1} \ , \\
{\bf 45} &= (so(T Y_4 \oplus T^* Y_4))_0 \oplus (\Lambda^\textrm{odd}T^*
Y_4)_{+2} \oplus (\Lambda^\textrm{odd}T^* Y_4)_{-2} \oplus
(\mathbb{R})_0
\end{aligned}\end{equation}
for type IIA, while in type IIB we have
\begin{equation}\begin{aligned}
 \label{representation_IIB_geom}
 {\bf 10} &= (\Lambda^\textrm{even}T^* Y_4)_0 \oplus (\mathbb{R})_{+2}  \oplus (\mathbb{R})_{-2}  \ , \\
 %\label{representation_IIB_fundamental}
 {\bf 16} &= (T Y_4 \oplus T^*Y_4)_{+1} \oplus (\Lambda^\textrm{odd}T^*Y_4)_{-1} \ , \\
 %\label{representation_IIB_adjoint}
 {\bf 45} &= (so(T Y_4 \oplus T^* Y_4))_0 \oplus (\Lambda^\textrm{even}T^* Y_4)_{+2} \oplus (\Lambda^\textrm{even}T^* Y_4)_{-2} \oplus (\mathbb{R})_0 \ .
\end{aligned}\end{equation}
Here $so(T Y_4 \oplus T^* Y_4)$ denotes the Lie-Algebra of $SO(4,4)$
that acts on  $T Y_4 \oplus T^* Y_4$.
The subscripts give the charges under shifts of the dilaton, which do not have a geometric interpretation.
Note that the bundle $\Lambda^\textrm{even}T^* Y_4$ appears in different representations in \eqref{representation_IIA_geom} and in \eqref{representation_IIB_geom}. This shows that the embedding of the pure $SO(4,4)$ spinors
$\Phi_1, \Phi_2 \in \Lambda^\textrm{even}T^* Y_4$ has to be different for type IIA and type IIB.

In type IIA backgrounds with 16 supercharges the situation is
straightforward. We already argued that in this case we have to
project out all $SU(2)\times SU(2)$ doublets or correspondingly
$\Lambda^\textrm{odd}T^* Y_4$ together with $T Y_4 \oplus T^* Y_4$.
Eq.~\eqref{representation_IIA_geom} then implies
that $SO(5,5)$ is broken to $SO(4,4) \times \mathbb{R}_+$ by the
projection.
This in turn says that all scalar degrees of freedom
coming from the R-R sector are projected out together with
the massive gravitinos. Of course this conclusion is also reached by
direct inspection of the massless type IIA spectrum
discussed in section~\ref{section:spectrumd6}.
%given in Table~\ref{N=2d=6multipletsA}.
This observation also immediately says that the local moduli space is unchanged
and given by ${\cal M}_{\Psi_a}$ in \eqref{local_moduli_space_4d}.

The analogous discussion in  type IIB is slightly more involved. From
\eqref{representation_IIB_geom} we see that neither the additional
generators of $SO(5,5)$ are projected out nor can we embed the pure
spinors into the spinor representation of $SO(5,5)$. However,
from~\eqref{representation_IIB_geom} we see that we can embed the
$SO(4,4)$ spinors into the vector representation of
$SO(5,5)$.
More precisely,
we can either embed the complex  pure spinors $\Phi_1$ and
$\Phi_2$ into complex $SO(5,5)$ vectors or, alternatively, use their
real and imaginary parts denoted by  $\Psi_a$ in the previous section
and embed them into real $SO(5,5)$ vector representations.
We use \eqref{representation_IIB_geom} to decompose the $SO(5,5)$ vector into its components
\begin{equation} \label{decompose_SO55_vector}
 \zeta=(\zeta^+, \zeta^s, \tilde{\zeta}^s ) \ ,
\end{equation}
where $\zeta^+$ lives in $\Lambda^\textrm{even}T^* Y_4$ while $\zeta^s,
\tilde{\zeta}^s$ are the two singlets.
Then the embedding of the four $\Psi_a$ into $SO(5,5)$ vectors $\zeta_a$ is
given by
\begin{equation} \label{SO55_embedding_SO44_spinors}
 \zeta_a = (\Psi_a , 0 , 0) \ , \qquad a=1,\dots ,4 \ ,
\end{equation}
which are orthonormal due to \eqref{conditions_real_spinors}.
This results in a set of four orthonormal space-like $SO(5,5)$ vectors
$\zeta_a$ which -- after modding out the rotational symmetry between
them -- parameterize the space
\begin{equation}\label{IIBint}
\mathcal{M}_{\zeta_a} = \frac{SO(5,5)}{SO(4) \times SO(1,5)} \ .
\end{equation}

However, this cannot be the correct parameter space yet. As one can
read off from~\eqref{representation_IIB_geom}, the four vectors
$\zeta_a$ are not charged under the dilaton shift.
Thus, the dilaton is not  yet included in the parameter space \eqref{IIBint}.
Reconsidering the splitting of the fundamental representation \eqref{splitting_vector_rep} shows that the
two singlets are charged under
dilaton shifts, and together form a real $SO(1,1)$ vector. If we
impose a normalization condition on this vector, it parameterizes
$SO(1,1)$ and therefore the dilaton degree of freedom $\phi$.
We can embed this $SO(1,1)$ vector into an $SO(5,5)$ vector
$\zeta_5$ using \eqref{decompose_SO55_vector}, i.e.\
\begin{equation}
\label{SO55_embedding_SO11_vector}
 \zeta_5 = \tfrac{1}{\sqrt{2}} (0 , \e^{\phi} , \e^{-\phi} ) \ .
\end{equation}
We see that the $\zeta_I$, $I=1, \dots, 5$, are all space-like and
satisfy (due to \eqref{conditions_real_spinors})
\begin{equation}
\label{relations_SO55_vectors}
 \langle \zeta_I , \zeta_J \rangle_5 = \delta_{IJ} \ ,
\end{equation}
where we gauge-fixed the parameter $c$ in
\eqref{conditions_real_spinors} to be $1$.

The stabilizer of this set of vectors is naturally given by $SO(5)
\subset SO(5,5)$ which are the rotations in the space perpendicular to
all $\zeta_I$. Therefore, the $\zeta_I$, $I=1, \dots, 5$, obeying \eqref{relations_SO55_vectors}, span the space ${SO(5,5)}/{SO(5)}$.

The embedding of the NS-NS sector into $SO(5,5)$ given in
\eqref{SO55_embedding_SO44_spinors} and
\eqref{SO55_embedding_SO11_vector} is not yet generic, but can be rotated by some $SO(5,5)$ rotation.
Part of these rotations just rotate $\zeta_5$ and the $\zeta_a$, $a=1,\dots,4$, into each
other. The remaining parameters rotations genuinely modify the embedding and therefore correspond
to additional physical scalar degrees of freedom. They are precisely
the R-R scalars of type IIB coming from the real spinor
\begin{equation}
 C = C_0 + C_2 + C_4 \ \in \Lambda^\textrm{even} T^* Y_4  \ .
\end{equation}
Their embedding into the group $SO(5,5)$ is given in \cite{Triendl:2009ap}.
The previous discussion shows that in type IIB, apart from the pure
spinors $\Psi_a$, the dilaton and the R-R scalars $C$ are also part of
the moduli space and therefore the parameter space given in \eqref{IIBint} has to be modified.
We just argued that the basic objects are five $SO(5,5)$ vectors $\zeta_I$
that satisfy \eqref{relations_SO55_vectors} and which are stabilized
by $SO(5)$.
In addition there is an $SO(5)$ gauge symmetry rotating
the vectors into each other.
Therefore the physical parameter space is
\begin{equation}\label{moduli_space_IIB_d6}
{\cal M}_{\zeta_I}\ =\  \frac{SO(5,5)}{SO(5)\times SO(5)}  \ .
\end{equation}
This is of course still the parameter space over a point of the ten-dimensional spacetime. In Section~\ref{section:moduliSU2} we derive from \eqref{moduli_space_IIB_d6} the scalar field space of general compactifications on an $SU(2)$-structure manifold $Y_4$.

%%%%%%%%%%%%%%%%%%%%%%%%%%%%%%%%%%%%%%%%%%%%%%%
\subsection{Exceptional generalized geometry on $Y_6$}\label{section:EGG}
%%%%%%%%%%%%%%%%%%%%%%%%%%%%%%%%%%%%%%%%%%%%%%%
Now let us discuss the more involved case of a six-dimensional space $Y_6$.
The U-duality group for $d=4$ is $E_{7(7)}$ with the
T-duality subgroup being $SO(6,6)$. Let us first recall the
decomposition of the representations of $E_{7(7)}$ in terms of the
maximal subgroup $Sl(2,\mathbb{R})_S \times SO(6,6)$. The factor
$Sl(2,\mathbb{R})_S$ is the S-duality subgroup acting on the
four-dimensional dilaton $\phi$ complexified by the dualized $B$-field.
The fundamental representation of $E_{7(7)}$ decomposes as~\cite{Cremmer:1978ds}
\begin{equation}
\label{E7_decomposition_fundamental}
 {\bf 56} \to ({\bf 2},{\bf 12}) + ({\bf 1},{\bf 32})  \ ,
\end{equation}
while the adjoint of $E_{7(7)}$ decomposes as
\begin{equation}
\label{E7_decomposition_adjoint}
 {\bf 133} \to ({\bf 3},{\bf 1}) + ({\bf 1},{\bf 66}) + ({\bf 2}, {\bf \bar{32}}) \ .
\end{equation}

In full analogy to Section~\ref{section:GG_pure_spinors}, we define the bundle ${\cal E} Y$ which should form the fundamental representation of the U-duality group $E_{7(7)}$ and therefore should be of dimension $56$. To understand the relation to the generalized tangent bundle ${\cal T}Y$ we consider \eqref{E7_decomposition_fundamental} and associate geometrical bundles with the representations of $SO(6,6)$.
It was shown in Ref.~\cite{Hull:1994ys} that the electric and magnetic
charges form the $ {\bf 56}$ representation of $E_{7(7)}$.
The $({\bf 2},{\bf 12})$ part in \eqref{E7_decomposition_fundamental}
represents the NS-NS charges, i.e.\ winding and momentum modes as well as NS$5$-branes and KK-monopoles, and thus corresponds to a doublet in
$T Y_6 \oplus T^*Y_6$.\footnote{In contrast to~\cite{Hull:2007zu}, we do not distinguish the bundles $T Y_6 \oplus T^* Y_6$
  and $\Lambda^5 T Y_6 \oplus \Lambda^5 T^* Y_6$ because they are related by a volume form on $Y_6$. Such a volume form we
  already chose to identify the $SO(6,6)$ spinor bundles with $\Lambda^{\textrm{even}}T^* Y_6$ and $\Lambda^{\textrm{odd}}T^*
  Y_6$. Thus, we can identify the bundles $T Y_6 \oplus T^* Y_6$ and $\Lambda^5 T Y_6 \oplus \Lambda^5 T^* Y_6$, and write them as
  a doublet under the S-duality group.
}
The $({\bf 1},{\bf 32})$ represents
the R-R charges, which
correspond to ten-dimensional D-brane solutions.
In type IIA, they are elements of  $\Lambda^{\textrm{even}}T^* Y_6$, while in type IIB they live in the bundle $\Lambda^{\textrm{odd}}T^* Y_6$~\cite{Hull:1994ys}. Therefore, \eqref{E7_decomposition_fundamental} is realized geometrically by~\cite{Hull:2007zu}
\begin{equation}
\label{E7_decomposition_fundamental_geometrically_IIA}
{\bf 56}^{\textrm{IIA}} = (T Y_6 \oplus T^* Y_6)_{\bf 2} \oplus (\Lambda^{\textrm{even}}T^* Y_6)_{\bf 1}
\end{equation}
for type IIA and by
\begin{equation}
\label{E7_decomposition_fundamental_geometrically_IIB}
{\bf 56}^{\textrm{IIB}} = (T Y_6 \oplus T^* Y_6)_{\bf 2} \oplus (\Lambda^{\textrm{odd}}T^* Y_6)_{\bf 1}
\end{equation}
for type IIB. The subscript denotes the representation under the S-duality group $Sl(2,\mathbb{R})_S$, which has no geometric realization.
Correspondingly, the decomposition of the adjoint of the U-duality group is realized geometrically by
\begin{equation}
\label{E7_decomposition_geometrically_IIA}
 {\bf 133}^{\textrm{IIA}} = ( \mathbb{R} )_{\bf 3} \oplus (so(TY_6 \oplus T^*Y_6))_{\bf 1} \oplus (\Lambda^{\textrm{odd}}T^* Y_6 )_{\bf 2}
\end{equation}
for type IIA and
\begin{equation}
\label{E7_decomposition_geometrically_IIB}
 {\bf 133}^{\textrm{IIB}} = ( \mathbb{R} )_{\bf 3} \oplus (so(TY_6 \oplus T^*Y_6))_{\bf 1} \oplus (\Lambda^{\textrm{even}}T^* Y_6 )_{\bf 2}
\end{equation}
for type IIB.
The spinor representations of $SO(6,6)$ are actually related to the R-R fields $C$. In type IIA, the $C$-fields define an $SO(6,6)$ spinor of odd chirality via\footnote{Note that we use the ``democratic'' formulation for the R-R fields, and that we only consider scalar degrees of freedom. Therefore, all legs of the forms in \eqref{C_fields_IIA} and \eqref{C_fields_IIB} are internal.}
\begin{equation}
\label{C_fields_IIA}
 C^{IIA} = C_1 + C_3 + C_5 \ \in \Lambda^\textrm{odd}T^*Y_6 \ ,
\end{equation}
while in type IIB, the spinor is of even chirality and defined by
\begin{equation}
\label{C_fields_IIB}
 C^{IIB} = C_0 + C_2 + C_4 + C_6 \ \in \Lambda^\textrm{even}T^*Y_6 \ .
\end{equation}
This fits nicely with the $SO(6,6)$ spinors appearing
in~\eqref{E7_decomposition_geometrically_IIA}
and~\eqref{E7_decomposition_geometrically_IIB}. However, in both
\eqref{E7_decomposition_geometrically_IIA}
and \eqref{E7_decomposition_geometrically_IIB} there appears a
doublet of
$SO(6,6)$ spinors in the adjoint of $E_{7(7)}$. As we will see below, one linear combination
of these spinors is in the stabilizer of the $SU(3) \times SU(3)$
structure, while the remaining linearly independent linear combination
corresponds to the R-R scalar fields.

We conclude that the bundle ${\cal E} Y$ should locally look like \eqref{E7_decomposition_fundamental_geometrically_IIA} in type IIA and like \eqref{E7_decomposition_fundamental_geometrically_IIB} in type IIB.
Analogously to Section~\ref{section:GG_pure_spinors}, a metric on ${\cal E} Y$ incorporates all scalar degrees of freedom.

%%%%%%%%%%%%%%%%%%%%%%%%%%%%%%%%%%%%%%%%%%%%%%%%%%%%%%%%%%%%%%%%%%%%%%%%%%
\subsection{$SU(3)\times SU(3)$ in exceptional generalized geometry} \label{section:EGG_SU3SU3}
%%%%%%%%%%%%%%%%%%%%%%%%%%%%%%%%%%%%%%%%%%%%%%%%%%%%%%%%%%%%%%%%%%%%%%%%%%
We are now interested in the extension of the pure spinor formalism to representations of $E_{7(7)}$. We know from Section \ref{section:SUnSUn_d4} that for $d=4$ a pair of pure spinors of opposite chirality $\Phi^+$ and $\Phi^-$ parameterize the degrees of freedom of the NS-NS sector. Both of them should embed into representations of $E_{7(7)}$. Since they have opposite chirality, we see from \eqref{E7_decomposition_fundamental} and \eqref{E7_decomposition_adjoint} that one of them should embed into the fundamental and the other one into the adjoint representation. From \eqref{E7_decomposition_fundamental_geometrically_IIA} and \eqref{E7_decomposition_fundamental_geometrically_IIB} we see that $\Phi^+$ fits into the fundamental representation in type IIA, while in type IIB it is $\Phi^-$. Let us restrict in the following to type IIA and remark that the exchange of chiralities leads to the corresponding type IIB analysis.

Let us now review the embedding of the pure $SO(6,6)$ spinors into $E_{7(7)}$ representations as performed in \cite{Grana:2009im}. Using the decomposition~\eqref{E7_decomposition_fundamental}, the spinor $\Phi^+$ of positive chirality is embedded into the fundamental representation via
\begin{equation}
\label{E7_embedding_spinor_fund}
\lambda =  (\lambda_i^A, \lambda^+) = ( 0 , \Re ( \Phi^+ )) \
, \quad i=1,2 \ ,
\end{equation}
where $\lambda$ consists of a doublet of $SO(6,6)$ vectors $\lambda_i^A$ and an $SO(6,6)$ spinor $\lambda^+$ of even chirality and therefore is an $E_{7(7)}$ vector.
The stabilizer of $\lambda$ has been
determined in \cite{Grana:2009im} to be
$E_{6(2)}$.
Furthermore the phase of $\Phi^+$ is just a gauge degree of freedom. In the $E_{7(7)}$ covariant formalism this gauge freedom manifests itself in the fact that $\lambda$ and
\begin{equation} \label{E7_embedding_spinor_fund2}
\tilde{\lambda} = ( 0 , \Im (\Phi^+ ) )
\end{equation}
describe the same $SU(3) \times SU(3)$ structure. They are related by the generalized almost complex structure $\mathcal{J}_{\Phi^+}$ which embeds into the adjoint of $E_{7(7)}$.
Therefore, after modding out the transformations generated by $\mathcal{J}_{\Phi^+}$, the parameter space for $\lambda$ is
\begin{equation}\label{E7_SK}
 \mathcal{M}_{SK} = \frac{E_{7(7)}}{E_{6(2)} \times U(1)} \ ,
\end{equation}
which one can show is special K\"ahler \cite{Gunaydin:1983rk,Gunaydin:1983bi,Cecotti:1988ad,deWit:1992wf,deWit:1993rr} with a K\"ahler potential given by the Hitchin functional \cite{Grana:2009im}
\begin{equation}
 K_{SK} = - \ln H(\lambda) = - \ln \sqrt{Q(\lambda)} \ ,
\end{equation}
where $Q(\lambda)$ is the quartic invariant of $E_{7(7)}$ defined by
\begin{equation}\begin{aligned}
 Q(\lambda) = & \tfrac{1}{48} \langle \lambda^+, \Gamma_{AB}\lambda^+\rangle \langle \lambda^+, \Gamma^{AB}\lambda^+\rangle - \tfrac12 \epsilon^{ij} \lambda_i^A \lambda_j^B \langle \lambda^+, \Gamma_{AB}\lambda^+\rangle \\ &
 + \tfrac12 \epsilon^{ij}\epsilon^{kl} \lambda_i^A \lambda_{k\,A} \lambda_j^B \lambda_{l\,B} \ .
\end{aligned}\end{equation}
Note that this is the straight-forward generalization of the Hitchin functional defined in \eqref{Hitchin_functional}.

The pure $SO(6,6)$ spinor of negative chirality cannot be embedded
into the fundamental of $E_{7(7)}$ but only into its adjoint. However, from \eqref{E7_decomposition_adjoint} it can be seen that
it must be embedded as an $Sl(2,\mathbb{R})_S$ doublet. Therefore, a complex vector $u^i, {i=1,2},$ is introduced which is stable and normalized, i.e.\
\begin{equation} \label{E7_normalization_u}
|u|^2 = u^i \epsilon_{ij} \bar{u}^j = 1 \ .
\end{equation}
The $Sl(2,\mathbb{R})_S$ doublet $u^i$ describes the complexified dilaton degree of freedom. Note that the overall phase of $u^i$ is just a choice of gauge.
Then the pair $(u^i,\Phi^-)$ is embedded via
\begin{equation}
\label{E7_embedding_spinor_adjoint}
 \U_1 = (\admui^i_{\phantom{i}j} , \admuA^A_{\phantom{A}B} , \admuspin^{i-}) = ( 0 , 0 ,  \Re ( u^i \Phi^- ) ) \ .
\end{equation}
The calculation of the moduli space however is a bit more involved than expected. Naively one would think that analogously to the gauge freedom of $\lambda$ the gauge freedom in $\U_1$ is some phase rotation which relates $\U_1$ to
\begin{equation} \label{SU2embeddingU2}
\U_2 = ( 0 , 0 ,  \Im ( u^i \Phi^- ) ) \ .
\end{equation}
However, these two elements of the adjoint do not commute, and therefore determine a third one which reads
\begin{equation} \label{SU2embeddingU3}
 \U_3 = \tfrac{\iu}{4k} \langle \bar{\Phi}^- , \Phi^- \rangle ( u^i \bar{u}_j + \bar{u}^i u_j ,  \iu |u|^2  (\mathcal{J}_{\Phi^-})^A_{\phantom{A}B} , 0) \ ,
\end{equation}
where $\mathcal{J}_{\Phi^-}$ is the generalized almost complex structure corresponding to $\Phi^-$ and defined by \cite{Grana:2005ny}
\begin{equation}
(\mathcal{J}_{\Phi^-})_{AB} = \iu \frac{\langle \bar{\Phi}^- , \Gamma_{AB} \Phi^- \rangle}{\langle \bar{\Phi}^- , \Phi^- \rangle} \ .
\end{equation}
The normalization $k$ is defined as
\begin{equation}
k = \sqrt{\tfrac{1}{2} |u|^2 \langle \bar{\Phi}^- , \Phi^- \rangle} \ .
\end{equation}
As explained in Section \ref{section:GG_pure_spinors}, $\mathcal{J}_{\Phi^-}$ determines $\Phi^-$ up to a phase. As a consequence, $\U_3$ determines $\U_1$ and $\U_2$ up to a rotation between those two. Hence, each $\U_a$ determines the other two. It turns out that the $\U_a$ define a highest weight $SU(2)$ embedding of $u^i$ and the pure spinor $\Phi^-$ in $E_{7(7)}$ \cite{Swann:1990,Grana:2009im}. Indeed, the $\U_a$ fulfill the $su(2)$ algebra
\begin{equation}
[\U_a , \U_b] = 2 k \epsilon_{abc} \U_c \ .
\end{equation}
Purity of $\Phi^-$ together with \eqref{E7_normalization_u} is equivalent to the fact that the $\U_a$ indeed form an $su(2)$ algebra. Furthermore, the $\U_a$ share the same stabilizer and make the $SU(2)$ gauge freedom manifest.
One can compute the stabilizer to be the group $SO^*(12)$, which is a non-compact version of $SO(12)$. Therefore, the $\U_a$ locally span the space \cite{Grana:2009im}
\begin{equation}
 \mathcal{M}_{HK} = \frac{E_{7(7)}}{SO^*(12)} \times \mathbb{R}_* \ ,
\end{equation}
which can be shown to be hyper K\"ahler \cite{Wolf:1965zz,Alekseevskii:1975zz,Swann:1990} with a hyper-K\"ahler potential \cite{Kobak:2000mj}
\begin{equation}
 \chi = \sqrt{-\tfrac{1}{12} \operatorname{tr}(\U_a \U_a)} \ .
\end{equation}
By modding out the $SU(2)$ gauge freedom of the $\U_a$ one ends up with the parameter space
\begin{equation}\label{E7_QK}
 \mathcal{M}_{QK} = \frac{E_{7(7)}}{SU(2)\times SO^*(12)} \ .
\end{equation}

The compatibility condition \eqref{SU3_compatibility} can be rewritten in an $E_{7(7)}$-covariant way as
\begin{equation}\label{E7_compatibility_conditions}
 \U_a \cdot \lambda = 0 \ .
\end{equation}
This compatibility condition enforces the two stabilizer groups to intersect in
\begin{equation}
 SO^*(12) \cap E_{6(2)} = SU(6) \ .
\end{equation}
Therefore, the physical parameter space forms a fibre over both \eqref{E7_SK} and \eqref{E7_QK} and is
\begin{equation} \label{EGG_parameter_space_SU6}
  \mathcal{M} = \frac{E_{7(7)}}{U(2)\times SU(6)} \ ,
\end{equation}
where the $U(2)$ is the R-symmetry group of the theory.
This $SU(6)$ structure reduces to the usual $SU(3)\times SU(3)$ structure if one restricts $E_{7(7)}$ to $SO(6,6)$.

So far we reviewed how the $SU(3)\times SU(3)$ is embedded into $E_{7(7)}$ representations. In order to find an $\cN=2$ parameter space, one still has to project out all $SU(3)\times SU(3)$ triplets or, equivalently, all ${\bf 6}$ representations of $SU(6)$.
This eliminates all extra degrees of freedom for the fundamental representation \eqref{E7_decomposition_fundamental}, giving the special-K\"ahler space we already found in the last section. Comparing with the analysis in Section~\ref{section:spectrumd4}, one sees that this special-K\"ahler space is parameterized by the scalar degrees of freedom in the vector multiplet sector, consistent with $\cN=2$ supergravity.

In contrast, the projection in the adjoint representation \eqref{E7_decomposition_adjoint} leads to a space of larger dimension as compared to the special-K\"ahler manifold parameterized by the pure spinor $\Phi^-$. The inclusion of the R-R sector gives a fibration over this special-K\"ahler manifold.
Here, the fibre is parameterized by the deformations of the scalars coming from the R-R fields $C$ and by $u^i$, which parameterizes the complexified  dilaton. Since the classical type II supergravities in ten dimensions are symmetric under shifts in the $C$ fields and in the complexified dilaton, the fibre admits a transitively acting symmetry group of shift isometries.
As already anticipated, the resulting space is quaternionic-K\"ahler, parameterized by the scalars in the $\cN=2$ hypermultiplets. Quaternionic-K\"ahler spaces admitting the fibration structure are called special quaternionic-K\"ahler manifolds. The map from the special-K\"ahler base to the quaternionic-K\"ahler manifold is known as the c-map \cite{Cecotti:1988qn,Ferrara:1989ik}. We discuss them in more detail in Section \ref{section:special_quat}.

%%%%%%%%%%%%%%%%%%%%%%%%%%%%%%%%%%%%%%%%%%%%%%%%%%%%%%%%%%%%%%%%%%%%%%%%%%
\subsection{R-R scalars and $\cN=4$ compactifications on $Y_6$} \label{section:RRd4}
%%%%%%%%%%%%%%%%%%%%%%%%%%%%%%%%%%%%%%%%%%%%%%%%%%%%%%%%%%%%%%%%%%%%%%%%%%
Let us now discuss the $\cN=4$ case on $Y_6$. The main difference to the last section is the existence of a generalized almost-product structure $\mathcal{P}$, which has already been introduced in Section \ref{section:SUnSUn_d4}, and the projection to $\cN=4$ instead of $\cN=2$.
In particular, we can use the results of the discussion in the last section but do not project out the $SU(3)\times SU(3)$ triplet. Instead we reduce the $SU(3)\times SU(3)$ structure group to $SU(2)\times SU(2)$ and subsequently project out all $SU(2)\times SU(2)$ doublets.

Let us now discuss what happens when we project out all $SU(2)\times SU(2)$ doublets.
We already know from Section~\ref{section:SU2SU2d4} that  only the $SO(2,2) \times SO(4,4)$ subgroup of $SO(6,6)$ survives this projection. Therefore, the $({\bf 1},{\bf 66})$ component in \eqref{E7_decomposition_adjoint} is projected to the direct sum of the adjoints of $SO(2,2)$ and $SO(4,4)$.
Furthermore, the first component in \eqref{E7_decomposition_adjoint}, i.e.\ $({\bf 3}, {\bf 1})$, consists of $SO(6,6)$ singlets. Therefore, it is also a singlet under $SU(2) \times SU(2)$ and thus
invariant under the projection. Hence, we are left with the last component
in~\eqref{E7_decomposition_adjoint}, which is the $({\bf 2}, {\bf \bar{32}})$ representations, i.e.\ a doublet of $SO(6,6)$ spinors.

Due to the existence of the generalized almost-product structure
$\mathcal{P}$, we can decompose the $SO(6,6)$ spinor bundles as done
in~\eqref{spinor_bundle_decomposition}. Analogously to the discussion in
Section~\ref{section:SU2SU2d6def} one can argue that the space $\Lambda^\textrm{odd}_4 T^* Y_6$
consists of $SU(2)\times SU(2)$ doublets only and therefore is removed by the projection to $\cN
=4$. We are therefore left with only half of the degrees of freedom
\begin{equation}
 \begin{aligned}
 \Lambda^{\textrm{even}} T^* Y_6 &\longrightarrow \Lambda^{\textrm{even}} T^*_2 Y_6 \wedge  \Lambda^{\textrm{even}} T^*_4 Y_6 \ , \\
 \Lambda^{\textrm{odd}} T^* Y_6 &\longrightarrow \Lambda^{\textrm{odd}} T^*_2 Y_6 \wedge \Lambda^{\textrm{even}} T^*_4 Y_6  \ .
\end{aligned}
\end{equation}
Therefore, only part of the U-duality group $E_{7(7)}$ survives this
projection. In type IIA, we end up with the subgroup
$G^{\textrm{IIA}}$ whose adjoint is the subspace
of~\eqref{E7_decomposition_geometrically_IIA} given by
\begin{equation}
\label{SO66_subgroup_IIA}
 g^{\textrm{IIA}} =  ( \mathbb{R} )_{\bf 3} \oplus so(T_2 Y_6 \oplus T^*_2 Y_6)_{\bf 1} \oplus so(T_4 Y_6 \oplus T^*_4 Y_6 )_{\bf 1}
  \oplus  (\Lambda^{\textrm{odd}}T_2^* Y_6 \wedge \Lambda^{\textrm{even}}T_4^* Y_6 )_{\bf 2}  \ .
\end{equation}
In type IIB, we find the subgroup $G^{\textrm{IIB}}$ whose adjoint is
the subspace of~\eqref{E7_decomposition_geometrically_IIB} given by
\begin{equation}
\label{SO66_subgroup_IIB}
 g^{\textrm{IIB}} = ( \mathbb{R} )_{\bf 3} \oplus so(T_2 Y_6 \oplus T^*_2 Y_6)_{\bf 1} \oplus so(T_4 Y_6 \oplus T^*_4 Y_6)_{\bf 1}
  \oplus  (\Lambda^{\textrm{even}}T_2^* Y_6 \wedge \Lambda^{\textrm{even}}T_4^* Y_6)_{\bf 2} \ .
\end{equation}
Both $G^{\textrm{IIA}}$ and $G^{\textrm{IIB}}$ define $SO(6,6) \times Sl(2,\mathbb{R})_{T/U}$ subgroups of $E_{7(7)}$, as shown in \cite{Triendl:2009ap}. The $Sl(2,\mathbb{R})_{T/U}$ factor is generated by one of the two sub-algebras in \eqref{SO22_splitting}, depending on whether one considers type IIA (T) or type IIB (U). $Sl(2,\mathbb{R})_{T}$ acts on the K\"ahler part of the identity structure on $T_2 Y_6 \oplus T^*_2 Y_6$ and forms the extra factor in type IIA, while $Sl(2,\mathbb{R})_{U}$ acts on its complex structure part and forms the extra factor in type IIB.

Finally, we determine the parameter space of $SU(2)
\times SU(2)$ structure compactifications of type II theories in
$d=4$. We will mainly consider type IIA but the type
IIB results are easily obtained by swapping chiralities and an
exchange of $sl(2,\mathbb{R})_T$ with $sl(2,\mathbb{R})_U$ in the Lie
algebra $so(2,2)$.

For this we use the embedding of the pure $SO(6,6)$ spinors into $E_{7(7)}$ representations discussed in Section~\ref{section:EGG}. The spinor $\Phi^+$ of positive chirality is embedded into the fundamental representation as in \eqref{E7_embedding_spinor_fund}.
From \eqref{SUtwoY6} we see that $\Phi^+$ and thus $\lambda$ transform as a doublet under $Sl(2,\mathbb{R})_{T}$. Thus, $\lambda$ is mapped into the $({\bf 12}, {\bf 2})$ representation in \eqref{E7_decomposition_fundamental} under the projection $E_{7(7)}\to SO(6,6) \times Sl(2,\mathbb{R})_{T}$. Therefore, $\lambda$ descends to a doublet of $SO(6,6)$ vectors, which are stabilized by $SO(4,6) \times SO(2) \subset SO(6,6) \times Sl(2,\mathbb{R})_T$.
Furthermore, we have an $U(1)$ gauge freedom since $\lambda$ in \eqref{E7_embedding_spinor_fund} and $\tilde \lambda$ in \eqref{E7_embedding_spinor_fund2} describe the same $SU(2) \times SU(2)$ structure. They are related by the generalized almost-complex structure $\mathcal{J}_{\Phi^+}$ which embeds into the adjoint of $E_{7(7)}$.
Therefore, after modding out the transformations generated by $\mathcal{J}_{\Phi^+}$, the parameter space for $\lambda$ is
\begin{equation}\label{E7_local_moduli_space_fundamental}
 \mathcal{M}_\lambda = \frac{SO(6,6)}{SO(2) \times SO(4,6)} \times \frac{Sl(2,\mathbb{R})_T}{SO(2)} \ ,
\end{equation}
which is related to the unconstrained special-K\"ahler parameter space of $\lambda$, given in \eqref{E7_SK}, by projecting out all $SU(2)\times SU(2)$ doublets. Here, the first factor in \eqref{E7_local_moduli_space_fundamental} is spanned by the real and imaginary parts of $\Phi_1$, which are embedded as $SO(4,4)$ vectors into the space of $SO(6,6)$ vectors. The second factor is spanned accordingly by $\Theta_+$, as has been discussed below \eqref{SO22_splitting}.

In Section \ref{section:EGG} it already turned out that the pure $SO(6,6)$ spinor of negative chirality is tensored with an $Sl(2,\mathbb{R})_S$ doublet $u^i$ and can then be embedded into the adjoint of $E_{7(7)}$ as done in \eqref{E7_embedding_spinor_adjoint}. As discussed there, the embedding has some $SU(2)$ gauge choice as the embeddings \eqref{SU2embeddingU2} and \eqref{SU2embeddingU3} are completely equivalent and together with \eqref{E7_embedding_spinor_adjoint} actually form the $SU(2)$ generators.
According to \eqref{SUtwoY6}, $\mu_1$ is a singlet under $Sl(2,\mathbb{R})_{T}$ and therefore is mapped into the adjoint of $SO(6,6)$ in \eqref{E7_decomposition_adjoint} under the projection $E_{7(7)}\to SO(6,6) \times Sl(2,\mathbb{R})_{T}$. From \eqref{E7_embedding_spinor_adjoint} and \eqref{SUtwoY6} one can see that $\mu_1+\iu \mu_2$ is actually mapped to the antisymmetric tensor product $\mu_1+\iu \mu_2=(u^i\Theta_-) \wedge \Phi_2$. The stabilizer of this object consists of those elements of $SO(6,6)$ that either leave both $u^i\Theta_-$ and $\Phi_2$ invariant or rotate them into each other. The stabilizer therefore can be identified with $SU(2)\times SO(2,6)$, which together with the $SU(2)$ gauge freedom combines into $SO(4) \times SO(2,6)$. We end up with the parameter space
\begin{equation}\label{E7_local_moduli_space_adjoint}
 \mathcal{M}_\admu = \frac{SO(6,6)}{SO(4) \times SO(2,6)} \ ,
\end{equation}
which is related to the unconstrained quaternionic-K\"ahler parameter space of $\mu_a$, given in \eqref{E7_QK}, by projecting out all $SU(2)\times SU(2)$ doublets. As discussed above, the spacelike four-plane in \eqref{E7_local_moduli_space_adjoint} is spanned by the real and imaginary parts of $\Re (u^i\Theta_-)$ and $\Phi_2$, understood as $SO(2,2)$ and $SO(4,4)$ vectors embedded into $SO(6,6)$.

Finally, we consider the common parameter space of both objects, imposing the compatibility condition \eqref{E7_compatibility_conditions}.
The common stabilizer of both $\lambda$ and $\mu$ is the common subgroup of $SO(4,6)$ and $SO(2,6)$ and can be determined to be $SO(6)$. The gauge freedom on the other hand is enhanced from $SO(2)\times SO(4)$ to $SO(6)$ due to the fact that rotations between the $\Lambda^\textrm{even}_4 T^*Y$ components of $\Phi^+$ and $\Phi^-$ are pure gauge freedom, as already discussed in Section~\ref{section:RRd6}.
Thus, the parameter space is
\begin{equation} \label{E7_local_moduli_space}
 \mathcal{M}_{\lambda ,\admu} = \frac{SO(6,6)}{SO(6) \times SO(6)} \times \frac{Sl(2,\mathbb{R})_T}{SO(2)} \ ,
\end{equation}
which is the parameter space of the spacelike six-plane spanned by the real and imaginary parts of $\Re (u^i\Theta_-)$, $\Phi_1$ and $\Phi_2$.
The embeddings \eqref{E7_embedding_spinor_fund} and \eqref{E7_embedding_spinor_adjoint} can be deformed by some $E_{7(7)}$ transformation. This corresponds to the $SO(6,6)$ deformations which we discussed already in the last sections and to additional degrees of freedom that can be identified with the R-R scalars.

The spaces \eqref{local_moduli_space_4d}, \eqref{moduli_space_IIB_d6} and \eqref{E7_local_moduli_space} give the central result of this Chapter, reflecting the parameter spaces of $SU(2)\times SU(2)$ structures in $d=4$ and $d=6$. They are the target spaces of all $SO(1,d-1)$ scalars in the rewritten ten-dimensional theory. The next step is the dimensional reduction to the $d$-dimensional, effective theory, in which we truncate the theory in order to find a finite-dimensional field content.
In Section~\ref{section:moduliSU2} we shall use the results \eqref{local_moduli_space_4d}, \eqref{moduli_space_IIB_d6} and \eqref{E7_local_moduli_space} to identify the scalar field space of the $d$-dimensional effective theory in $d=4$ and $d=6$.
As we will see, its dimension may differ from the parameter space of the ten-dimensional theory but it inherits the geometry of the parameter space.

\cleardoublepage
%%%%%%%%%%%%%%%%%%%%%%%%%%%%%%%%%%%%%%%%%%%%%%%
\chapter{The effective action} \label{section:effective}
%%%%%%%%%%%%%%%%%%%%%%%%%%%%%%%%%%%%%%%%%%%%%%%
The aim of this chapter is to discuss consistent Kaluza-Klein truncations and to derive the $d$-dimensional low-energy effective field theory for compactifications to $\cN=2$ and $\cN=4$ supergravities. For simplicity, we shall focus on identifying the scalar field space of the low-energy effective action. A derivation of possible gaugings and the potentials will we presented in \cite{LST}.

So far we just parameterized the ten-dimensional field content over each point of $10d$ spacetime in a form where instead of the
ten-dimensional Lorentz symmetry group only the subgroup $SO(1,d-1)\times SO(10-d)$ (and consequently, only $8$ or $16$ of the $32$
supercharges) is manifest. In this formulation no Kaluza-Klein truncation
is performed but instead all fields still carry the full
ten-dimensional coordinate dependence
of the background \eqref{stringbackground}. This is the approach of Ref.~\cite{deWit:1986mz}, which
we reviewed in Section~\ref{section:spectrum}.
In the corresponding effective theories the spaces derived in \eqref{local_moduli_space_4d} and \eqref{moduli_space_IIB_d6} for $d=6$ and \eqref{EGG_parameter_space_SU6} and \eqref{E7_local_moduli_space} for $d=4$ appear as target space of
the Lorentz-scalar deformations, consistent with the
constraints of the corresponding $d$-dimensional supergravity.

Alternatively, one can perform a Kaluza-Klein truncation and only keep
the light modes of the background \eqref{stringbackground}. We shall do exactly this in the following in order to derive the scalar field spaces of the effective theories.

%%%%%%%%%%%%%%%%%%%%%%%%%%%%%%%%%%%%%%%%%%%%%%%
\section{Consistent truncations and fluxes} \label{section:truncation}
%%%%%%%%%%%%%%%%%%%%%%%%%%%%%%%%%%%%%%%%%%%%%%%
The Kaluza-Klein reduction on a compact space $Y_{10-d}$ in general leads to an infinite tower of massive states over a massless, $d$-dimensional spectrum. However, most of the massive states are negligible in the low-energy limit and therefore can be removed from the spectrum.
This corresponds to a truncation of the theory at a given energy scale such that only the light modes with masses below this scale appear in the resulting action.
There are several consistency conditions for such a truncation to work.

First of all, light modes must survive this truncation, as otherwise we do not describe the physics at this energy scale correctly. Furthermore, it is crucial that this truncation preserves supersymmetry.
Therefore, we assume that there is a finite-dimensional subspace $\Lambda^\bullet_\textrm{finite} T^*Y \subset \Lambda^\bullet T^*Y $ of forms on $Y$ that contains the light modes of the effective theory and that we can expand the pure spinors in this basis.
Secondly, the Mukai pairing $\langle \cdot , \cdot \rangle$ should be non-degenerate on $\Lambda^\textrm{even}_\textrm{finite} T^*Y$ and $\Lambda^\textrm{odd}_\textrm{finite} T^*Y$.
This ensures that the generalized Hodge-star operator $\ast_B$ is well-defined on $\Lambda^\textrm{even}_\textrm{finite} T^*Y$ and $\Lambda^\textrm{odd}_\textrm{finite} T^*Y$.
Furthermore, for $SU(3)\times SU(3)$-structure compactifications this assumption is enough to guarantee that the special-K\"ahler structure of the moduli spaces of $\Phi^+$ and $\Phi^-$ each survive the compactification.
For $SU(2)\times SU(2)$-structure compactifications, we find that this even ensures that the $d$-dimensional scalar field space takes the form of a symmetric space, more precisely, of a Grassmannian.

As already discussed in Section \ref{section:spectrum}, there should be no $SU(n)\times SU(n)$ $n$-plets in the spectrum $\Lambda^\bullet_\textrm{finite} T^*Y$, since we want to stay at energy scales below the mass of the massive gravitino multiplets. As a further assumption, one wants the exterior derivative $\diff$ to be well-defined on $\Lambda^\bullet_\textrm{finite} T^* Y$, in other words $\diff$ maps from $\Lambda_\textrm{finite}^p T^* Y$ to $\Lambda_\textrm{finite}^{p+1} T^* Y$ so that one can also expand field strengths and torsion in this finite set of forms.

On a four-dimensional manifold $Y_4$ with an $SU(2)\times SU(2)$ structure, the above assumptions are already very strong. As discussed in Section \ref{section:SU2SU2d6def}, projecting out the $SU(2)\times SU(2)$ doublets eliminates all elements in $\Lambda_\textrm{finite}^\textrm{odd} T^* Y$. As a consequence, the exterior derivative $\diff$ maps $\Lambda_\textrm{finite}^\textrm{even} T^* Y$ to zero so that all elements in $\Lambda_\textrm{finite}^\textrm{even} T^* Y$ are closed and the manifold has even $SU(2)$ holonomy, i.e.\ we deal with K3.

In order to find non-trivial examples in this case, we have to relax some of the above conditions. Indeed we can allow $SU(2)\times SU(2)$ doublets in $\Lambda^\bullet_\textrm{finite} T^* Y$ which do not correspond to light deformations but only to torsion classes. These torsion classes are related to a non-trivial warp factor in the compactification and we do not study them here any further. Let us instead turn to the case $d=6$.

On a six-dimensional manifold $Y_6$ the Mukai pairing $\langle \cdot , \cdot \rangle$ given in \eqref{Mukai_pairing_2} is anti-symmetric, giving a natural symplectic structure on $\Lambda^\textrm{even} T^* Y$ and $\Lambda^\textrm{odd} T^* Y$. Let us for later convenience denote the dimension of $\Lambda^\textrm{even} T^* Y$ and $\Lambda^\textrm{odd} T^* Y$ by $2(n_\textrm{v}+1)$ and $2n_\textrm{h}$, respectively. We can choose a symplectic basis
\begin{equation}\label{symplectic_basis}
 \Sigma^\Lambda = (\omega_I,\tilde \omega^I) \ , \quad I=1,\dots, n_\textrm{v}+1, \qquad \tilde \Sigma^{\tilde \Lambda}=(\alpha_A,\beta^A) \ , \quad A=1,\dots, n_\textrm{h},
\end{equation}
with respect to the Mukai pairing for both $\Lambda^\textrm{even}_\textrm{finite} T^*Y$ and $\Lambda^\textrm{odd}_\textrm{finite} T^*Y$ such that
\begin{equation}
 \langle \omega_I , \tilde \omega^J \rangle = \delta_I^J \ , \qquad \langle \alpha_A , \tilde \beta^B \rangle = \delta_A^B \ ,
\end{equation}
with all other Mukai pairings vanishing.
Then the exterior derivative $\diff$ induces a natural algebra of charges on $\Lambda^\bullet_\textrm{finite} T^*Y$, given by
\begin{equation}\label{SU3SU3_charges}\begin{aligned}
\diff \alpha_A \sim & p_A^I \omega_I + e_{AI} \tilde \omega^I \ , \qquad \diff \beta^A \sim q^{AI} \omega_I + m^A_{I} \tilde \omega^I \ , \\
\diff \omega_I \sim & m^A_I  \alpha_A - e_{AI} \beta^A  \ , \qquad \diff \tilde \omega^I \sim - q^{AI} \alpha_A + p_A^{I} \beta^A \ .
\end{aligned}\end{equation}
Here, the symbol $\sim$ denotes ``equal up to terms that vanish under the Mukai pairing''.
The charges $p_A^I, e_{AI}, q^{AI}, m^A_{I}$ parameterize the intrinsic torsion of ${\cal M}$ as well as background flux of the NS three-form $H$. The parameters $e_{AI}$ and $m^A_I$ already appear in $SU(3)$-structure compactification while $p_A^I$ and $q^{AI}$ only arise in genuine $SU(3)\times SU(3)$-structure compactifications and are often referred to as non-geometric fluxes. Together they form a doubly symplectic matrix \cite{Berglund:2005dm,Dall'Agata:2006nr,Grana:2006hr}
\begin{equation}\label{doubly_sympl_matrix}
 {\cal Q}^\Lambda_{\tilde \Lambda} = \left( \begin{array}{ccc}
 p_A^I && e_{AI}\\ q^{AI} && m^A_{I}
\end{array}
\right)
\end{equation}
such that we can rewrite \eqref{SU3SU3_charges} as
\begin{equation}\label{SU3SU3_charges_covariant}
 \diff \Sigma^\Lambda \sim {\cal Q}^\Lambda_{\tilde \Lambda} \tilde \Sigma^{\tilde \Lambda} \ , \qquad \diff \tilde \Sigma^{\tilde \Lambda} \sim {\cal Q}_\Lambda^{\tilde \Lambda} \Sigma^{\Lambda} \ .
\end{equation}
Note that the same matrix ${\cal Q}^\Lambda_{\tilde \Lambda}$ must appear in both equations of \eqref{SU3SU3_charges_covariant} in order to ensure the validity of partial integration, i.e.\ the property
\begin{equation}\label{partialint_basis}
\int_Y \langle \diff \Sigma^\Lambda, \tilde \Sigma^{\tilde \Lambda}\rangle = \int_Y \langle \Sigma^\Lambda, \diff  \tilde \Sigma^{\tilde \Lambda}\rangle \ .
\end{equation}
Furthermore, $\diff^2=0$ implies the quadratic constraints
\begin{equation} \label{doubly_sympl_quadratic_constr}
 {\cal Q}^\Lambda_{\tilde \Lambda} {\cal Q}_\Lambda^{\tilde \Sigma} = 0 \ , \qquad {\cal Q}^\Lambda_{\tilde \Lambda} {\cal Q}_\Sigma^{\tilde \Lambda} = 0 \ .
\end{equation}

An additional piece of information in the compactification is the amount of R-R flux on $Y$. This amounts to the fact that the field strengths of the R-R fields are only locally given by the exterior derivative of the R-R fields $C$ as $F^\pm=\diff C$ but globally admit a piece that is not exact, which we call the R-R form flux $F^\pm_\textrm{flux}$. Here, the superscript indicates whether we consider type IIA or IIB: In type IIA (IIB) the R-R field strengths are a formal sum of differential forms of even (odd) degree. In the remainder of this chapter, we focus on type IIA and discuss even-degree form flux $F^+_\textrm{flux}$. The formulas for type IIB compactifications are completely analogous and can be obtained by interchanging even and odd forms and thereby pure spinor chiralities.

In a consistent truncation, we can expand $F^+_\textrm{flux}$ in terms of $\Lambda^\textrm{even}_\textrm{finite} T^*Y$, leading to
\begin{equation}\label{SU3SU3_flux}
 F^+_\textrm{flux} = f^\Lambda \Sigma_{\Lambda} = f^I \omega_I + \tilde f_I \tilde \omega^I \ .
\end{equation}
It turns out that all entries of ${\cal Q}^\Lambda_{\tilde \Lambda}$ and $f^\Lambda$ are indeed charges in the effective theory \cite{D'Auria:2004tr,D'Auria:2004wd,Grana:2006hr}.

All charges appearing in \eqref{doubly_sympl_matrix} and \eqref{SU3SU3_flux} are quantized, i.e.\ integer-valued. More generally, charge quantization can be imposed for all charges, independent of their ten-dimensional origin (e.g.\ also for charges coming from non-geometric fluxes). The reason is that massive BPS states wrapping internal cycles of $Y$ descend to massive particles with quantized charges in the (full) $d$-dimensional theory. Even though all of these states are removed by the truncation, they give rise to Dirac charge quantization of all other charges in the theory, including those coming from ${\cal Q}^\Lambda_{\tilde \Lambda}$ and $f^\Lambda$. The responsible BPS spectra are generalizations of the spectrum discussed for instance in \cite{Hull:1994ys}.

If the structure group on $Y_6$ is $SU(2)\times SU(2)$, we can furthermore apply \eqref{spinor_bundle_decomposition} to $\Lambda^\bullet_\textrm{finite} T^* Y_6$ and use the fact that $\Lambda^{\textrm{odd}}_\textrm{finite} T^*_4 Y_6$ is projected out. Let us choose $v^i,i=1,2,$ as a basis of $\Lambda^{\textrm{odd}}_\textrm{finite} T^*_2 Y_6$ and $\Omega^I, I=1,\dots,n+6$ as a basis of $\Lambda^{\textrm{even}}_\textrm{finite} T^*_4 Y_6$. The basis of $\Lambda^{\textrm{even}}_\textrm{finite} T^*_2 Y_6$ then consists of $(1,v^1\wedge v^2)$.
As a consequence of this splitting, the charge algebra \eqref{SU3SU3_charges} simplifies to \cite{ReidEdwards:2008rd,Spanjaard:2008zz,Louis:2009dq}
\begin{equation}\label{SU2SU2_charges}
\begin{aligned}
\diff v^i &\sim t^i v^1\wedge v^2+ T^i_I \Omega^I\ , \\
\diff \Omega^I &\sim {\tilde T}_{iJ}^I v^i \wedge \Omega^J\ .
\end{aligned}
\end{equation}

As we already remarked, the above analysis crucially depends on the assumption of a constant warp factor. For non-trivial warping, both \eqref{SU3SU3_charges} and \eqref{SU2SU2_charges} can change and $\Lambda^\bullet_\textrm{finite} T^* Y_6$ may also contain $n$-plet representations \cite{Grana:2005sn}.

After this initial discussion, let us now discuss the low-energy effective action of $SU(2)\times SU(2)$- and $SU(3)\times SU(3)$-structure compactifications in more detail.
%%%%%%%%%%%%%%%%%%%%%%%%%%%%%%%%%%%%%%%%%%%%%%%%%%%%%%%%%%%%%%%%%%%%%%
\section{$\cN=4$ compactifications} \label{section:moduliSU2}
%%%%%%%%%%%%%%%%%%%%%%%%%%%%%%%%%%%%%%%%%%%%%%%%%%%%%%%%%%%%%%%%%%%%%%
The aim of this section is to determine the scalar field space of $SU(2)$
structure compactifications for $d=4$ and $d=6$.
The results should be consistent with $\cN=4$ four-dimensional supergravity and its six-dimensional counterparts, respectively. In a supergravity theory with $16$ supercharges, only discrete choices can be made. More precisely, the number of vector (or tensor) multiplets together with the choice of gauge group completely determine the supergravity theory (see for instance \cite{Schon:2006kz} and references therein). As a consequence, the scalar field space is highly constrained. For example, for type IIA the scalar field spaces are given by Grassmannian spaces of the form \eqref{N=4coset}.

In Chapter \ref{section:covariant} we rewrote the ten-dimensional theory in a way that is only manifestly covariant with respect to the $d$-dimensional Lorentz group and the structure group and gave the parameter space of the corresponding scalar fields. However, the result was an equivalent description of the ten-dimensional theory since no reduction had been performed.
In order to derive the scalar field space of the $d$-dimensional theory, we make use of a consistent truncation as discussed in the last section and reduce thereby the spectrum. This will lead us from the (rewritten) ten-dimensional theory to the truncated supergravity in $d$ dimensions.

Now we want to determine the scalar field space of $SU(2)$ structures.
Let us start with the case of type IIA for $d=6$, for which we determined the local parameter space to be given by \eqref{local_moduli_space_4d}, parameterized by four real spinors
$\Psi_a$ in $\Lambda^\textrm{even} T^* Y_4 $.
We use a consistent truncation as introduced in the last section and truncate the space of even forms to a finite subspace $\Lambda^\textrm{even}_\textrm{finite} T^* Y_4 $
with the assumption
that the Mukai pairing is non-degenerate on it.
Concretely this means that we can expand the four real spinors
$\Psi_a$ in a basis of
$\Lambda^\textrm{even}_\textrm{finite}T^* Y_4 $.
The  generalized Hodge-star operator $\ast_B$,
which, via the $\Psi_a$, is globally defined,
splits $\Lambda^\textrm{even}_\textrm{finite}T^* Y_4 $
into the eigenspaces of $\ast_B$, i.e.\
\begin{equation} \label{split_KKmodes}
\Lambda^\textrm{even}_\textrm{finite}T^* Y_4  = \Lambda^\textrm{even}_+ T^* Y_4 \oplus \Lambda^\textrm{even}_- T^* Y_4 \ ,
\end{equation}
where the subscripts $\pm$ denote the eigenvalues with respect to $\ast_B$. Furthermore,
these eigenspaces are orthogonal to each other with respect to the Mukai pairing.
$\Lambda^\textrm{even}_+ T^* Y_4 $  consists of $SU(2)\times SU(2)$ singlets
only, and thus over each point it is spanned by the
$\Psi_a$. Therefore, each element of $\Lambda^\textrm{even}_+ T^* Y_4 $ can be
written as a linear combination of the $\Psi_a$ where the
coefficients may depend on the base point on $Y_4$, i.e.\ are functions
on $Y_4$. However, only constant coefficients survive the Kaluza-Klein
truncation, and thus $\Lambda^\textrm{even}_+ T^* Y_4 $ is spanned by the
$\Psi_a$ only and has dimension four.

Now let us turn to the eigenspace $\Lambda^\textrm{even}_- T^* Y_4 $. It
consists of sections in $U_{ {\bf 2},{\bf 2}}$ and therefore we
cannot make the same argument as for $\Lambda^\textrm{even}_+ T^* Y_4 $. In
contrast to the bundle of $SU(2)\times SU(2)$ singlets, $U_{ {\bf
2},{\bf 2}}$ might be twisted over the manifold and the dimension of
$\Lambda^\textrm{even}_- T^* Y_4 $ may differ from four,
say $n+4$.
Thus, $\Lambda^\textrm{even}_\textrm{finite} T^* Y_4 $ is a vector space
of signature $(4,n+4)$. The four spinors $\Psi_a$
satisfy the orthonormality conditions \eqref{conditions_real_spinors}, and therefore span a
four-dimensional space-like subspace in
$\Lambda^\textrm{even}_\textrm{finite} T^* Y_4 $. The parameter space
describing these configurations is just $ \mathbb{R}_+ \times {SO(4,n+4)}/{SO(n+4)}$,
where the $\mathbb{R}_+$ factor corresponds to the gauge freedom contained in the choice of the parameter $c$ in the orthonormality condition \eqref{conditions_real_spinors}.

In order to find  the moduli space we still have to remove all gauge redundancies. This amounts to removing the scale factor $c$ and the $SO(4)$-symmetry between the four real
components of the compatible pure spinors.
Modding out both redundancies we finally arrive at the moduli space
\begin{equation}
\label{moduli_space_4d_NS}
  \mathcal{M}^\textrm{IIA}_{d=6}\ =\ \frac{SO(4,n+4)}{SO(4) \times SO(n+4)} \times \mathbb{R}_+  \ ,
\end{equation}
where we now also included the dilaton via the $\mathbb{R}_+$
factor consistently with (non-chiral) $\cN=4$ supergravity.

The derivation presented so far did not use the absence of $SU(2)$
doublets. We merely confined our attention to deformations of the
pure spinors, which are of even degree. However, let us note that
$\diff \Psi_a$ is an $SU(2)\times SU(2)$ doublet and therefore it
cannot correspond to a deformation
parameter after projecting out the doublets. At best it can be related to the
warp factor, as already discussed in the last section. Thus,
in the absence of a warp factor and without any doublets we have
$\diff \Psi_a=0$ and $Y_4$ has to be a $K3$ manifold. This is consistent
with the moduli space \eqref{moduli_space_4d_NS}, which for $n=16$
coincides with the moduli space of K3 manifolds (modulo the
$\mathbb{R}_+$ factor).

We want to stress that the arguments given below \eqref{split_KKmodes} can be made for any
vector bundle that consists of only singlets under the structure
group. Since the structure group does not act on the vector bundle, it
must be the trivial bundle and we can give a number of nowhere-vanishing sections of the vector bundle that form an orthonormal basis at every point. Since these sections are globally defined and nowhere-vanishing,
they can be associated with objects that define the structure group
and therefore they (or locally rescaled versions thereof) survive the
Kaluza-Klein truncation. Furthermore, we can conclude that any further mode in this bundle
that survives the
truncation can be expanded in this basis. The coefficients of the expansion are constrained by the truncation to be constant. Therefore any further mode can just be a linear combination of the sections in the orthonormal basis.
Therefore, the space of light modes resulting from a
Kaluza-Klein truncation on this vector bundle has the same dimension
as the bundle itself.

Let us apply this theorem to $SU(2)\times SU(2)$-structure compactifications of type IIB (in $d=6$), starting from the target space \eqref{moduli_space_IIB_d6}, which is parameterized by five vectors $\zeta_I$, $I=1, \dots, 5$, obeying \eqref{relations_SO55_vectors} in a ten-dimensional space $\mathbb{R}^{5,5}$ of split signature and therefore defining a spacelike five-dimensional subspace ${\cal Z}_+$. By use of the $SO(5,5)$ metric the orthogonal complement is defined as a five-dimensional timelike subspace ${\cal Z}_-$. Since ${\cal Z}_+$ consists of only structure-group singlets, we can apply the theorem, which tells that the Kaluza-Klein truncation leads to a five-dimensional spacelike space spanned by the $\zeta_I$. On the other hand, ${\cal Z}_-$ descends to a timelike space of arbitrary dimension, say $n+5$. Therefore, the $\zeta_I$ parameterize a spacelike five-plane in a space of signature $(5,n+5)$, which represents the moduli space
\begin{equation}
\label{moduli_space_IIB_d6_final}
  \mathcal{M}^\textrm{IIB}_{d=6}\ =\ \frac{SO(5,n+5)}{SO(5) \times SO(n+5)} \ .
\end{equation}

In the same way we can determine the $\cN=4$ moduli space of $SU(2)$ structure compactifications to four dimensions, starting from the parameter space \eqref{E7_local_moduli_space}.
For this we note that the subspace of positive signature in the first factor in \eqref{E7_local_moduli_space} is spanned by $SU(2) \times SU(2)$ singlets. By applying the above theorem, we know that after the Kaluza-Klein truncation this subspace is still of dimension six.
However, the space of negative signature is spanned by $({\bf 2}, {\bf
2})$ representations of $SU(2) \times SU(2)$. Therefore, its dimension
can be different globally, say $n+6$.
Note that also the second factor in \eqref{E7_local_moduli_space} consists of $SU(2) \times SU(2)$ singlets only and we can apply the theorem here, too. Thus, the scalar field space is given by
\begin{equation} \label{E7_moduli_space_IIA}
 \mathcal{M}^{\rm IIA}_{d=4}= \frac{SO(6,6+n)}{SO(6) \times SO(6+n)}  \times \frac{Sl(2,\mathbb{R})_T}{SO(2)} \ .
\end{equation}
Similarly, the type IIB scalar field space can be determined by exchanging ${Sl(2,\mathbb{R})_T}/{SO(2)}$ and ${Sl(2,\mathbb{R})_U}/{SO(2)}$ and therefore reads
\begin{equation} \label{E7_moduli_space_IIB}
 \mathcal{M}^{\rm IIB}_{d=4}= \frac{SO(6,6+n)}{SO(6) \times SO(6+n)}  \times \frac{Sl(2,\mathbb{R})_U}{SO(2)} \ .
\end{equation}

The scalar field spaces derived in this section are the final result of the first part of this review.
We see that by assuming the existence of an arbitrary consistent truncation, we could already derive the scalar field spaces of $SU(2)$-structure compactifications. It turns out that these scalar field spaces are consistent with the requirements of four-dimensional $\cN=4$ supergravity and its higher-dimensional analogues. Furthermore, the parametrization of the scalar field spaces in terms of pure spinors and its generalizations in exceptional generalized geometry can be shown to reflect the construction of the scalar field space from its superconformal cone, as we further discuss in \cite{Triendl:2009ap}.
Let us now review the analogous discussion in $\cN=2$ compactifications to set the stage for the discussion of $\cN=1$ vacua in gauged $\cN=2$ supergravities.

%%%%%%%%%%%%%%%%%%%%%%%%%%%%%%%%%%%%%%%%%%%%%%%
\section{$\cN=2$ effective actions} \label{section:EffActionN=2}
%%%%%%%%%%%%%%%%%%%%%%%%%%%%%%%%%%%%%%%%%%%%%%%
We just saw that in $SU(2)\times SU(2)$-structure compactifications the scalar field space is restricted to be a symmetric space of Grassmannian type, in accordance with $\cN=4$ supergravity. In $SU(3)\times SU(3)$-structure compactifications one would like to draw a similar conclusion. However, $\cN=2$ supergravity is less restrictive and therefore a full determination of the $\cN=2$ theory does crucially depend on the considered background $Y$. Still there are a number of features that are generically present in the effective theory of $SU(3)\times SU(3)$-structure compactifications.
The low-energy theory turns out to be a gauged $\cN=2$ supergravity theory. As we discuss in the following, the scalar fields parameterize a product of a special K\"ahler and a (special) quaternionic-K\"ahler manifold. Furthermore, the charges appearing in \eqref{SU3SU3_charges} and \eqref{SU3SU3_flux} correspond to gaugings in the low-energy effective action. Before we go into more detail, let us first review the basic facts of $\cN=2$ supergravity (for a comprehensive review, see \cite{Andrianopoli:1996cm}).

%%%%%%%%%%%%%%%%%%%%%%%%%%%%%%%%%%%%%%%%%%%%%%%
\subsection{$\cN=2$ supergravity}\label{section:N2SUGRA}
%%%%%%%%%%%%%%%%%%%%%%%%%%%%%%%%%%%%%%%%%%%%%%%
The possible field content and couplings of $\cN=2$ supergravity are highly restricted by its supersymmetric nature. In particular, each field must come in a representation of the $\cN=2$ supersymmetry group. The representations of supersymmetry algebras can be classified for any number of supersymmetries \cite{Strathdee:1986jr}. In four-dimensional $\cN=2$ supergravity, the standard field content consists of\footnote{Additionally, one could dualize scalars in the vector and hypermultiplets to build tensor multiplets. In non-Abelian theories, these multiplets can become inequivalent. We will not discuss such theories in the following.}
\begin{itemize}
 \item the gravitational multiplet, consisting of the metric $g_{\mu\nu}$, an $SU(2)$ doublet of spin-$3/2$ fields $\Psi^{\cal A}, {\cal A}=1,2,$ called gravitini, and a vector $A^0$, the graviphoton,
 \item $n_{\rm v}$ vector multiplets, which each contain a vector $A^i$, $i=1,\dots,n_{\rm v}$, a doublet of spinors $\lambda^{i\cal A}$, the gaugini, and a complex scalar field $t^i$.
 \item $n_{\rm h}$ hypermultiplets, which contain a doublet of spinors $\zeta^\alpha$, $\alpha=1,\dots,2n_{\rm h}$, the hyperini, and four real scalar fields $q^u$, $u=1,\dots,4n_{\rm h}$.
\end{itemize}
It turns out that the scalars $t^i$ and $q^u$ form a non-linear sigma model in which they can be understood as coordinates on a moduli space
\begin{equation}\label{scalarM_app}
{\cal M} = {\cal M}_{\rm v} \times {\cal M}_{\rm h} \ .
\end{equation}
Here, ${\cal M}_{\rm v}$ is a special-K\"ahler manifold spanned by the complex scalars $t^i$ and ${\cal M}_{\rm h}$ is a quaternionic-K\"ahler manifold of holonomy $SU(2) \times Sp(2n_{\rm h})$. The $SU(2)$ piece is the mentioned R-symmetry of the theory, so the scalars in the hypermultiplets are charged under the R-symmetry. We discuss the properties of the scalar field spaces in more detail in the following two sections.

The Lagrangian of ungauged $\cN=2$ supergravity is well-known, see for instance \cite{Andrianopoli:1996cm}. The bosonic part we are interested in is
\begin{equation}\begin{aligned}\label{sigmaint}
{\cal L}\ =\  - \mathrm{Im} \mathcal{N}_{IJ}\,F^{I}_{\mu\nu}F^{\mu\nu\, J}
- \mathrm{Re} \mathcal{N}_{IJ}\,
F^{I}_{\mu\nu} F_{\rho\sigma}^{J}\epsilon^{\mu\nu\rho\sigma}
+ g_{i\bar \jmath}(t,\bar t)\, \partial_\mu t^i \partial^\mu\bar t^{\bar \jmath}
+ h_{uv}(q)\, \partial_\mu q^u \partial^\mu q^v
%- V(t,\bar t,q) %+\ldots
\ ,
\end{aligned}\end{equation}
where the field strengths $F^{I}_{\mu\nu}, I=0,\ldots, n_{\rm v}$ include the
graviphoton and their kinetic matrix $\mathcal{N}_{IJ}$ is a function of the $n_{\rm v}$ scalars $t^i$, which we will define in \eqref{Ndef}.
In the kinetic terms of the scalars $g_{i\bar \jmath}$ and $h_{uv}$ denote the metrics on ${\cal M}_{\rm v}$ and ${\cal M}_{\rm h}$, respectively.

Let us now discuss the geometry of the two factors in \eqref{scalarM_app} and derive their origin in $SU(3)\times SU(3)$-structure compactifications. We start with the vector multiplet scalars on ${\cal M}_{\rm v}$ and afterwards turn to ${\cal M}_{\rm h}$.

%%%%%%%%%%%%%%%%%%%%%%%%%%%%%%%%%%%%%%%%%%%%%%%
\subsection{The vector multiplet sector and special-K\"ahler manifolds}\label{section:special_Kahler}
%%%%%%%%%%%%%%%%%%%%%%%%%%%%%%%%%%%%%%%%%%%%%%%
In the vector multiplet sector $g_{i\bar \jmath}$ denotes the metric
of the $2n_{\rm v}$-dimensional space ${\cal M}_{\rm v}$, which
${\cal N}=2$ supersymmetry constrains to be a  special-K\"ahler manifold \cite{deWit:1984pk,Craps:1997gp}.

Each special-K\"ahler manifold is in particular K\"ahler, i.e.\ the metric can locally be expressed as the second derivative of some function of the coordinates called the K\"ahler potential $K^{\rm v}$, i.e.\
\begin{equation}
 g_{i\bar \jmath} = \partial_i \partial_{\bar \jmath} K^{\rm v}\ .
\end{equation}
Equivalently, K\"ahler manifolds admit a K\"ahler two-form which is closed and non-de\-ge\-ne\-rate and locally reads
\begin{equation}
 {\cal K}^{\rm v} = -\frac{\iu}{2\pi} \partial \bar \partial K^{\rm v} \ .
\end{equation}
A special K\"ahler manifold of dimension $n_{\rm v}$ furthermore admits patchwise an embedding into a flat symplectic space of dimension $2n_{\rm v}+2$, which is parameterized by a symplectic vector of holomorphic functions $V^\Lambda = (X^I(t), {\cal F}_I(t))$. If we equip the symplectic space with the standard symplectic metric
\begin{equation}
 \Omega = \left( \begin{aligned}
                  0 && \mathbbm{1}  \\ -\mathbbm{1} &&  0
                \end{aligned} \right) \ ,
\end{equation}
the K\"ahler potential can be expressed in terms of $V^\Lambda$ via
\begin{equation}\label{Kdef}
K^{\rm v}= -\ln \iu \left(\bar {V}^\Lambda \Omega_{\Lambda\Sigma} {V}^\Sigma \right)= -\ln \iu \left( \bar X^I {\cal F}_I - X^I\bar {\cal F}_I \right)\ .
\end{equation}
Note that with help of the K\"ahler potential, we can define a K\"ahler-covariant derivative $\nabla_i = \partial_i +K^{\rm v}_i$, where $ K^{\rm v}_i = \partial_i K^{\rm v}$.

Furthermore, we see that the K\"ahler potential is invariant under symplectic rotations ${\cal S}$, where
${\cal S}$ is an $(2n_{\rm v}+2)\times(2n_{\rm v}+2)$ matrix % of $Sp(n_{\rm v}+1)$
that leaves the metric $\Omega$ of $Sp(n_{\rm v}+1)$ invariant, i.e.\ $S$
obeys ${\cal S}^T\Omega {\cal S}=\Omega$.
In terms of $(n_{\rm v}+1)\times(n_{\rm v}+1)$ matrices $S$ is given by
\begin{equation}
  \label{uvzwg}
  {\cal S}\ = \left(
    \begin{array}{cc}
      U & Z \\
      [1mm] W & V
    \end{array}
  \right) \ ,
\end{equation}
where $U$, $V$, $W$ and $Z$ %are constant, real,  which
obey
\begin{equation}\begin{aligned}
  \label{spc2}
  U^{\rm T} V- W^{\rm T} Z &= V^{\rm T}U - Z^{\rm T}W =
  {\bf 1}\, ,\\
  U^{\rm T}W = W^{\rm T}U\,, & \quad Z^{\rm T}V= V^{\rm T}Z\ .
\end{aligned}
\end{equation}

In $\cN=2$ supergravity, these symplectic rotations reflect electric-magnetic duality transformations of the vector multiplet sector of the Lagrangian.
They act on the
symplectic vector $H^\Lambda\equiv (F^I, G_I)$ according to
\begin{equation}\label{Sptrans}
H^\Lambda \to H^{\prime \Lambda} = {{\cal S}^\Lambda}_\Sigma H^\Sigma\ ,
\end{equation}
where $G_I\equiv \partial L/\partial F^I$ are the field
strengths of the dual magnetic gauge bosons, which only appear on-shell in that they are not part of the Lagrangian \eqref{sigmaint}.
The matrix of gauge couplings is also expressed in terms of the holomorphic vectors $(X^I(t), {\cal F}_I(t))$ and reads
\begin{equation}
  \label{Ndef}
  {\cal N}_{IJ} = \bar {\cal F}_{IJ} +2\iu\ \frac{\mbox{Im} {\cal F}_{IK}\mbox{Im}
    {\cal F}_{JL} X^K X^L}{\mbox{Im} {\cal F}_{LK}  X^K X^L} \ ,
\end{equation}
where ${\cal F}_{IJ}=\partial_I {\cal F}_J$, and transforms under a symplectic rotation non-linearly, according to
\begin{equation}
  \label{nchange}
  {\cal N} \to (V {\cal N}+ W) \,(U+ Z {\cal N})^{-1} \ .
\end{equation}
The matrix ${\cal N}$ has the following properties:
\begin{equation}
 \mathcal F_I = \mathcal N_{IJ} X^J \ , \quad \nabla_k \mathcal F_I = \bar{\mathcal N}_{IJ} \nabla_k X^J \ .
\end{equation}
The equations of motion derived from $\cal L$ are invariant under $Sp(n_{\rm v}+1)$ rotations acting via \eqref{Sptrans} and \eqref{nchange}.
Due to the symplectic invariance it is a matter of convention which vector fields are called electric and which are called magnetic. It is customary to denote the gauge fields which do appear in $\cal L$ as electric and their duals as magnetic.

%%%%%%%%%%%%%%%%%%%%%%%%%%%%%%%%%%%%%%%%%%%%%%%%%%%%%%%%%%%%%%%%%%%%%%%%%%%%%%%%%%%%%%%%%%%%%%%%%%%%%%%%%%%%%%%%%%%%%%%%%%%%
%%%%%%%%%%%%%%%%%%%%%%%%%%%%%%%%%%%%%%%%%%%%%%%%%%%%%%%%%%%%%%%%%%%%%%%%%%%%%%%%%%%%%%%%%%%%%%%%%%%%%%%%%%%%%%%%%%%%%%%%%%%%

Let us now make contact with the effective action of $SU(3)\times SU(3)$-structure compactifications.
In Section~\ref{section:truncation} we classified the spectrum of light modes in terms of the charges ${\cal Q}^\Lambda_{\tilde \Lambda}$ and $(f^I,f_I)$.
Now we expand the pure spinor $\Phi^+$ in terms of the basis vectors $\Sigma^{\Lambda}$ and $\tilde \Sigma^{\tilde \Lambda}$ introduced in \eqref{symplectic_basis}, respectively, leading to
\begin{equation} \label{sympl_hol_vectors}
 \Phi^+ = V^\Lambda \Sigma_{\Lambda} =X^I \omega_I + {\cal F}_I \tilde \omega^I \ .
\end{equation}
The coordinate vector $V^\Lambda=(X^I, {\cal F}_I)$ is a symplectic vector of complex dimension $2n_\textrm{v}+2$ and depends holomorphically on the $n_\textrm{v}$ complex coordinates $t^i$ spanning the K\"ahler manifold ${\cal M}_{\rm  v}$.
It turns out that one can always find a symplectic frame in which $X^I=(1,t^i)$ and furthermore ${\cal F}_I = \partial_I {\cal F}$, where ${\cal F}$ is the holomorphic prepotential, which is homogeneous of degree two,
such that the K\"ahler potential \eqref{Kahler_Hitchin} with \eqref{Hitchin_functional} leads to the standard form of special-K\"ahler geometry \eqref{Kdef}.
Note that the parameter space is projective because the complex prefactor of $\Phi^+$ is pure gauge and has to be modded out.

Before we turn to the hypermultiplet sector, we want to introduce some techniques from special geometry that will prove useful in the remainder of this review. On any special K\"ahler manifold of dimension $n_{\rm v}$ one can define the projection operator ${\Proj}_I^{\phantom{I}J}$ by~\cite{Ferrara:1989ik}
\begin{equation} \label{projection_v}
 {\Proj}_I^{\phantom{I}J} =
\tfrac12 \e^{-K^{\rm v}}{\Proj}_{I\, \underline{b}}
\bar{\Proj}_K^{\phantom{K}\underline{b}} (\Im  \mathcal G)^{-1\,KJ} =
\delta_I^J +
2 \e^{K^{\rm v}}{(\Im {\cal F})_{IK}\bar X^K} X^J= \delta_I^J + K^{\rm v}_I X^J\ ,
\end{equation}
where $K^{\rm v}_I$ denotes the holomorphic derivative of the K\"ahler potential $K^{\rm v}$ given in \eqref{Kdef} and ${\Proj}_I^{\phantom{I}\underline{j}}$ is given by
\begin{equation}\label{projector_vielbein_V}
{\Proj}_I^{\phantom{I}\underline{j}} =({\Proj}_0^{\phantom{0}\underline{j}}, {\Proj}_i^{\phantom{i}\underline{j}}) =(-e_i^{\phantom{i}\underline{j}}X^i, e_i^{\phantom{i}\underline{j}}) \ ,
\end{equation}
where $e_i^{\phantom{i}\underline{j}}$ denotes the vielbein of ${\cal M}_{\rm v} $.
From the definition follows
\begin{equation}\label{projection_der}
\nabla_i X^J = \Proj_i^{\phantom{i}J}\ , \qquad \textrm{and} \qquad \nabla_i {\cal F}_J = \Proj_i^{\phantom{i}K}{\cal F}_{KJ} \ .
\end{equation}
Furthermore, ${\Proj}_I^{\phantom{I}J}$ has the properties
\begin{equation} \label{properties_projection_v}
X^I {\Proj}_I^{\phantom{I}J} = 0 \ , \qquad {\Proj}_I^{\phantom{I}J} \Im {\cal F}_{JK} \bar X^K = 0 \ , \qquad
{\Proj}_I^{\phantom{I}J} {\Proj}_J^{\phantom{J}K} = {\Proj}_I^{\phantom{I}K} \ ,
\end{equation}
and  therefore is indeed a projection map which projects to the space orthogonal to $X^I$.
From the definition \eqref{projection_v} we we see that ${\Proj}_I^{\phantom{I}J}$ fulfils the reality condition
\begin{equation} \label{relation_projectorbar_v}
 (\Im {\cal F})^{-1 \, LI} \bar{\Proj}_I^{\phantom{I}J} (\Im {\cal F})_{JK} = {\Proj}_K^{\phantom{K}L} \ .
\end{equation}
From \eqref{period_matrix} and \eqref{projection_v} we see that
\begin{equation}\label{MG_subspaces}
X^I {\cal N}_{IJ}=X^I{\cal F}_{IJ} \ , \qquad {\Proj}_I^{\phantom{I}J} \bar{\cal N}_{JK} = {\Proj}_I^{\phantom{I}J} {\cal F}_{JK} \ .
\end{equation}
The projection ${\Proj}_I^{\phantom{I}J}$ canonically leads to the decompositions
\begin{equation} \label{decomposition}
\begin{aligned}
 C_I =& C^{(Z)}_I + C^{(P)}_I = - K^{\rm v}_I X^J C_J + {\Proj}_I^{\phantom{I}J} C_J \ , \\
 {\tilde C}^I =& {\tilde C}^{(Z)\,I} + {\tilde C}^{(P)\,I} = - {\tilde C}^J K^{\rm v}_J X^I  + {\tilde C}^J {\Proj}_J^{\phantom{J}I}
\end{aligned}
\end{equation}
for any vectors $C_I$ and ${\tilde C}^I$. Note that $C^{(Z)}_I$ and ${\tilde C}^{(Z)\,I}$ each live in a one-dimensional subspace, while
$C^{(P)}_I$ and ${\tilde C}^{(P)\,I}$ parameterize the remaining $n$ directions.
With \eqref{decomposition} one can easily show that $(\Im {\cal F})_{IJ}$ is of signature $(n_{\rm v},1)$ \cite{Ceresole:1995ca}:
Using \eqref{MG_subspaces} we find
\begin{equation} \label{signature_G}
\bar C^I (\Im {\cal F})_{IJ} C^J= \bar C^{(Z)\,I} (\Im{\cal N})_{IJ} C^{(Z)\,J} - \bar C^{(P)\,I} (\Im{\cal N})_{IJ} C^{(P)\,J} \ .
\end{equation}
Since $(\Im{\cal N})_{IJ}$ is negative definite~\cite{Cremmer:1984hj}, we conclude that $(\Im {\cal F})_{IJ}$ is of signature $(n_{\rm v},1)$.

%%%%%%%%%%%%%%%%%%%%%%%%%%%%%%%%%%%%%%%%%%%%%%%
\subsection{The hypermultiplet sector, quaternionic-K\"ahler manifolds and the c-map}\label{section:special_quat}
%%%%%%%%%%%%%%%%%%%%%%%%%%%%%%%%%%%%%%%%%%%%%%%
In the hypermultiplet sector $h_{uv}$ denotes the metric on the $4n_{\rm h}$-dimensional space  ${\cal M}_{\rm h}$, which ${\cal N}=2$ supersymmetry constrains to be a quaternionic-K\"ahler manifold \cite{Bagger:1983tt,deWit:1984px}.  Such manifolds have a holonomy group given by $SU(2)\times Sp(n_{\rm h})$. Equivalently, they admit a
triplet of complex structures $J^x, x=1,2,3$ which satisfy the quaternionic algebra \begin{equation}\label{jrel}
J^x J^y = -\delta^{xy}{\bf 1} + \epsilon^{xyz} J^z \ .
\end{equation}
The metric $h_{uv}$ is Hermitian with respect to all
three complex structures. Correspondingly, a  quaternionic-K\"ahler manifold admits a triplet of hyper-K\"ahler two-forms $K^x_{uv} = h_{uw} (J^x)^w_v$ which are covariantly closed with respect to the $Sp(1)$ connection $\omega^x$,
i.e.\
\begin{equation}\label{deriv_Sp(1)_curvature_app}
\nabla K^x \equiv dK^x + \epsilon^{xyz} \omega^y \wedge K^z=0 \ .
\end{equation}
In other words, $K^x$ is proportional to the $Sp(1)$ field strength of $\omega^x$, thus leading to
\begin{equation} \label{def_Sp(1)_curvature_app}
 K^x =\diff \omega^x + \tfrac12 \epsilon^{xyz} \omega^y\wedge \omega^z\ .
\end{equation}

One can introduce a set of vielbein one-forms on the quaternionic-K\"ahler manifold ${\cal M}_{\rm h}$ by ${\mathcal U}^{\mathcal A\alpha} = {\mathcal U}_u^{\mathcal A\alpha} \diff q^u$ that defines the metric $h_{uv}$ via
\begin{equation}\label{Udef}
h_{uv} \diff q^u \diff q^v = {\mathcal U}^{\mathcal A\alpha} \varepsilon_\mathcal{AB} \mathcal C_{\alpha \beta} \mathcal U^{\mathcal B\beta} \ ,
\end{equation}
where $\mathcal C_{\alpha \beta}$ is the $Sp(n_{\rm h})$ invariant metric. ${\mathcal U}^{\mathcal A\alpha}$ satisfies the reality condition
\begin{equation}\label{Ureal}
{\mathcal U}_{\mathcal A\alpha} = \varepsilon_\mathcal{AB} \mathcal C_{\alpha \beta} \mathcal U^{\mathcal B\beta} = ({\mathcal U}^{\mathcal A\alpha})^* \ .
\end{equation}
One can also express the triple of two-forms $K^x_{uv}$ in terms of the vielbein via
\begin{equation}
 K^x = {\mathcal U}_{\mathcal A\alpha} \wedge {\mathcal U}^{\mathcal B\alpha} (\sigma^x)^{\cal A}_{\cal B} \ .
\end{equation}

Let us now consider isometries on ${\cal M}_{\rm h}$, which are necessary in order to discuss gaugings in the theory, as we do in Section \ref{section:N=2}. Such isometries should respect the quaternionic-K\"ahler structure of the manifold. This means that the generating Killing vector $k_\lambda$ should be tri-holomorphic, i.e.\ it should preserve the quaternionic-K\"ahler two-forms up to $SU(2)$ rotations
\begin{equation}\label{SU2compensator}
 {\cal L}_{k_\lambda} K^x = \epsilon^{xyz} K^y W^z_{\lambda} \ , \qquad  {\cal L}_{k_\lambda} \omega^x = \nabla W^x_{\lambda} \ ,
\end{equation}
where $W^x_{\lambda}$ is an $SU(2)$ compensator associated with ${k_\lambda}$ and ${\cal L}_{k}$ denotes the Lie derivative with respect to the vector field $k$, defined on a differential form $\alpha$ by
\begin{equation}
 {\cal L}_{k} \alpha = i_k \diff \alpha + \diff (i_k \alpha) \ .
\end{equation}
Here, $i_k \alpha$ denotes the contraction of the vector $k$ with the differential form $\alpha$. Note that Eq.\ \eqref{SU2compensator} tells us that the $SU(2)$ compensator is zero if the Lie derivative of the $SU(2)$ connection $\omega^x$ vanishes.
With the help of the curvature two-forms $K^x$ one can map such tri-holomorphic Killing vectors to an $SU(2)$ triplet $P^x_\lambda$ of functions, defined by
\begin{equation}\label{Pdef}
2 k^u_\lambda\,K_{uv}^x \ = \  -\nabla_v P_\lambda^x\ .
\end{equation}
This equation is solved by using the $SU(2)$ connection and the $SU(2)$ compensator
\begin{equation}
 P^x_{\lambda} = \omega^x(k_\lambda) + W^x_{\lambda}\ .
\end{equation}
Note that for a vanishing Lie derivative of the connection one-forms $\omega^x$, cf.\ \eqref{SU2compensator}, the Killing prepotentials take a simple form in terms of the $SU(2)$ connection $\omega^x$ \cite{Michelson:1996pn}:
\begin{equation} \label{prepotential_no_compensator}
 P^x_\lambda = \omega^x_u k_\lambda^u \ .
\end{equation}

Now let us investigate how this quaternionic-K\"ahler space appears in the low-energy effective action of $SU(3)\times SU(3)$-structure compactifications. Analogously to ${\cal M}_{\rm v}$, the space parameterized by $\Phi^-$ is special K\"ahler.
One can expand $\Phi^-$ in terms of the basis vector $\tilde \Sigma_{\tilde \Lambda}$ given in \eqref{symplectic_basis} and finds
\begin{equation} \label{sympl_hol_vectors2}
 \Phi^- = U^{\tilde \Lambda} \tilde \Sigma_{\tilde \Lambda} =Z^A \alpha_A + {\cal G}_A \beta^A \ .
\end{equation}
 This leads to the symplectic vector $U^{\tilde \Lambda}=(Z^A,{\cal G}_A)$ of complex dimension $2n_{\rm h}$ that is holomorphic in the coordinates $z^a$ of this special-K\"ahler space, which we call ${\cal M}_{\rm sk}$ in the following. Analogously to ${\cal M}_{\rm v}$, the K\"ahler potential can be computed, using \eqref{Kahler_Hitchin} with \eqref{Hitchin_functional}, to be in the standard form
 \begin{equation}\label{Khdef}
K^{\rm h}= -\ln \iu \left(\bar {U}^{\tilde \Lambda} \Omega_{\tilde \Lambda \tilde \Sigma} {U}^{\tilde \Sigma} \right)= -\ln \iu \left( \bar Z^A {\cal G}_A - Z^A\bar {\cal G}_A \right)\ .
\end{equation}
Note that one can always find a symplectic frame where $Z^A=(1,z^a)$.
Analogously to \eqref{Ndef}, the period matrix is defined by
\begin{equation} \label{period_matrix}
\mathcal M_{AB} = \bar{\mathcal G}_{AB} + 2 \iu \frac{(\Im  \mathcal G)_{AC} Z^C (\Im  \mathcal G)_{BD} Z^D}{Z^E (\Im  \mathcal G)_{EF} Z^F} \ ,
\end{equation}
with ${\cal G}_{AB} = \partial_A\partial_B {\cal G}$. It satisfies
\begin{equation}
\mathcal G_A = \mathcal M_{AB} Z^B \ , \quad \nabla_c \mathcal G_A = \bar{\mathcal M}_{AB} \nabla_c Z^B \ .
\end{equation}
One can define the same projection operators and decompositions for ${\cal M}_{\rm sk}$ as done in the last section from \eqref{projection_v} on for ${\cal M}_{\rm v}$.

In Section \ref{section:EGG} we discussed that the R-R scalar fields together with the four-dimensional complexified dilaton form a fibration over the space parameterized by $\Phi^-$. After the truncation this corresponds to a fibration over the special-K\"ahler manifold ${\cal M}_\textrm{sk}$ of dimension $2n_{\rm h}-2$, forming the space ${\cal M}_{\rm  h}$ of dimension $4 n_{\rm h}$. Before the truncation we saw that the fibration of the ten-dimensional parameter space is described by the c-map \cite{Cecotti:1988qn,Ferrara:1989ik}. Due to the consistency of the Kaluza-Klein truncation, this structure carries over to the moduli space of the truncated theory. Therefore, ${\cal M}_{\rm  h}$ is the c-map image of ${\cal M}_\textrm{sk}$ and in particular it is special quaternionic-K\"ahler.
As discussed above, ${\cal M}_{\rm  sk}$ is spanned by the complex coordinates $z^a, a=1,\ldots,n_{\rm h}-1$. The other hypermultiplet scalars are the dilaton $\phi$, the axion $\tilde \phi$ which is dual to the four-dimensional tensor field $B_{\mu\nu}$ and  $2n_{\rm h}$ scalars $\xi_{\tilde \Lambda}= (\tilde\xi_A,\xi^A), A=1,\ldots,n_{\rm h},$ that arise by expanding the R-R form fields $C \in \Lambda^\textrm{odd} T^*Y$ in the basis $\tilde \Sigma^{\tilde \Lambda}=(\alpha_A,\beta^A)$ defined in \eqref{symplectic_basis} as
\begin{equation}\label{expansion_RRfields}
 C = \xi_{\tilde \Lambda} \tilde \Sigma^{\tilde \Lambda} + A_\Lambda \Sigma^{\Lambda}= \xi^A \alpha_A + \tilde\xi_A \beta^A + A^I \omega_I + B_I \tilde \omega^I \ ,
\end{equation}
where the coefficients $A_\Lambda = (A^I, B_I)$ are one-forms in four-dimensions and parameterize the gauge fields in the vector multiplets.
Due to their Ramond-Ramond origin the couplings of $\xi^A, \tilde\xi_A$ are strongly restricted. Furthermore, the dilaton $\phi$ and the axion $\ax$ have universal properties that are independent of the chosen compactification manifold. Together these scalars define a $G$-bundle over ${\cal M}_{\rm  sk}$, where $G$ is the semidirect product of a $(2n_{\rm h}+1)$-dimensional Heisenberg group with $\mathbb{R}$. As a consequence $(2n_{\rm h}+2)$ independent isometries exist, as we shall discuss further shortly.

In \cite{Ferrara:1989ik} it was observed that there is a specific parametrization of the quaternionic vielbein $\mathcal U^{\mathcal A\alpha}$ \eqref{Udef} which turns out to be useful on special quaternionic-K\"ahler manifolds. Specifically, one defines the quaternionic vielbein as\footnote{Our notation follows Ref.~\cite{Cassani:2007pq}.}
\begin{equation} \label{quat_vielbein}
\mathcal U^{\mathcal A\alpha}= \tfrac{1}{\sqrt{2}}
\left(\begin{aligned}
 \bar{u} && \bar{e} && -v && -E \\
\bar{v} && \bar{E} && u && e
\end{aligned}\right) \ ,
 \end{equation}
where the one-forms are defined as
\begin{equation} \label{one-forms_quat}
 \begin{aligned}
  u\ = &\ \iu \e^{K^{\rm h}/2+\phi}Z^A(\diff \tilde\xi_A - \mathcal M_{AB} \diff \xi^B) \ , \\
  v\ = &\ \tfrac{1}{2} \e^{2\phi}\big[ \diff \e^{-2\phi}-\iu (\diff \ax +\tilde\xi_A \diff \xi^A-\xi^A \diff \tilde \xi_A  ) \big] \ , \\
  E^{\,\underline{b}}\ = &\ -\tfrac{\iu}{2} \e^{\phi-K^{\rm h}/2} {\Proj}_A^{\phantom{A}\underline{b}} (\Im  \mathcal G)^{-1\,AB}(\diff \tilde\xi_B - \mathcal M_{BC} \diff \xi^C) \ , \\
  e^{\,\underline{b}} \ = &\ {\Proj}_A^{\phantom{A}\underline{b}} \diff Z^A \ .
 \end{aligned}
\end{equation}
In these expressions ${\Proj}_A^{\phantom{A}\underline{b}}$ is defined analogously to \eqref{projector_vielbein_V} by
\begin{equation}\label{projector_vielbein}
{\Proj}_A^{\phantom{A}\underline{b}} =({\Proj}_0^{\phantom{0}\underline{b}}, {\Proj}_a^{\phantom{a}\underline{b}}) =(-e_a^{\phantom{a}\underline{b}}Z^a, e_a^{\phantom{a}\underline{b}}) \ ,
\end{equation}
where $e_a^{\phantom{a}\underline{b}}$ is the vielbein of ${\cal M}_{\rm sk}$,
i.e.\ it satisfies
$g_{a\bar b} = e_a^{\phantom{a}\underline{b}}
\bar e_{\bar
  b}^{\phantom{a}\bar{\underline{c}}}\delta_{\underline{b}\bar{\underline{c}}},~(
\underline{a},\underline{b} = 1,\ldots,n_{\rm v}-1$ with $g_{a\bar b})$ being
the metric on ${\cal M}_{\rm sk}$. By definition, ${\Proj}_A^{\phantom{A}\underline{b}}$ has the property ${\Proj}_A^{\phantom{A}\underline{b}} Z^A =0$.
Note that the quaternionic-K\"ahler geometry is completely determined in terms of the holomorphic prepotential ${\cal G}$ of the special K\"ahler submanifold ${\cal M}_{\rm sk}$.
Furthermore, it is important to mention that the parametrization  of the vielbein specified by \eqref{quat_vielbein} and \eqref{one-forms_quat} singles out a particular $SU(2)$ frame on ${{\cal M}}_{\rm h}$. We come back to this issue in Section \ref{section:breaking_hyper}.

Due to its specific construction, ${{\cal M}}_{\rm h}$ has $(2n_{\rm h}+2)$ isometries which are generated by the following set of Killing vectors \cite{deWit:1990na,deWit:1992wf}
\begin{equation}\label{Killing}
\begin{aligned}
\Kdil  \ &=\ \tfrac{1}{2} \frac{\partial}{\partial \phi} -  \ax \frac{\partial}{\partial \ax} - \tfrac{1}{2} \xi^A \frac{\partial}{\partial \xi^A} - \tfrac{1}{2} \tilde \xi_A \frac{\partial}{\partial \tilde \xi_A} \ , \\
\Kax  \ &=\ - 2 \frac{\partial}{\partial \ax} \ , \\
 \Kxi_A \ &= \ \frac{\partial}{\partial \xi^A} + \tilde \xi_A \frac{\partial}{\partial \ax} \ , \\
\Ktxi^A \ &=\ \frac{\partial}{\partial \tilde \xi_A} - \xi^A \frac{\partial}{\partial \ax} \ .
\end{aligned}
\end{equation}
They act transitively on the $G$-fibre coordinates $(\phi, \ax, \xi^A,\tilde \xi_A)$ and the subset $\{\Kxi_A, \Ktxi^A,\Kax\} $ spans a Heisenberg algebra which is graded with respect to $k_\phi $. The corresponding commutation relations are given by
\begin{equation}\label{isometries_fibre}
\begin{aligned}
 && [\Kdil,\Kax]\  =& \Kax \ , \qquad \qquad
 && [\Kdil,\Kxi_A]\ \ = & \tfrac{1}{2} \Kxi_A \ ,  \\
 && [\Kdil,\Ktxi^A]\  =& \tfrac{1}{2} \Ktxi^A \ , \qquad \qquad
 && [\Kxi_A,\Ktxi^B]\  = & - \delta_A^B \Kax \ ,
\end{aligned}
\end{equation}
while all other commutators vanish. In the following, we collect the Killing vectors $\{\Kxi_A, \Ktxi^A\}$ in a symplectic vector $k_{\tilde \Lambda}=(\Kxi_A, \Ktxi^A)$.

In Section \ref{section:breaking_hyper} we shall also need the explicit form of the Killing prepotentials $P^x_\lambda$, which were defined in \eqref{Pdef}. For special quaternionic-K\"ahler manifolds it can be checked that Killing prepotentials take the form \eqref{prepotential_no_compensator}.
Finally, using the explicit form of the vielbein \eqref{quat_vielbein} given above, one can calculate $\omega^x$ in terms of the one-forms \eqref{one-forms_quat} \cite{Ferrara:1989ik}
\begin{equation}\label{quat_connection}
\begin{aligned}
\omega^1\ & =\  \iu (\bar u- u)  \ , \\
\omega^2\ & =\ u + \bar u \ , \\
\omega^3\ & =\ \tfrac{\iu}{2} (v-\bar v) - \iu \e^{K^{\rm h}} \left(Z^A (\Im  \mathcal G)_{AB}\diff \bar Z^B - \bar Z^A (\Im  \mathcal G)_{AB} \diff Z^B \right) \ .
 \end{aligned}
\end{equation}

It is rather surprising that all $\cN=2$ supergravities coming from $SU(3)\times SU(3)$-structure compactifications admit $(2n_{\rm h}+2)$ isometries in the hypermultiplet sector. As we discuss next, these isometries can also be gauged in the low-energy effective action.
%%%%%%%%%%%%%%%%%%%%%%%%%%%%%%%%%%%%%%%%%%%%%%%%%%%%%%%%%%%%%%%%%%%%%%%%%%%%%%%%%%%%%%%%%%%%%%%%%%%%%%%%%%%%%%%%%%%%%%%%%%%%%%%%
\subsection{Electric and magnetic gaugings in $\cN=2$ supergravity}
\label{section:N=2}
%%%%%%%%%%%%%%%%%%%%%%%%%%%%%%%%%%%%%%%%%%%%%%%%%%%%%%%%%%%%%%%%%%%%%%%%%%%%%%%%%%%%%%%%%%%%%%%%%%%%%%%%%%%%%%%%%%%%%%%%%%%%%%%%

In the last two sections we discussed the couplings of $\cN=2$ scalars in terms of geometry. Furthermore, scalar and vector fields can be coupled via the standard gauge principle. For this we promote a global symmetry of the scalar field space to a local symmetry by coupling this symmetry to a gauge vector of the theory via a covariant derivative. The scalar field corresponding to this symmetry in turn is charged under the gauge vector. The corresponding theory is called gauged supergravity \cite{Andrianopoli:1996cm}.

The symplectic invariance of the Lagrangian under electric-magnetic duality transformations, as discussed in Section \ref{section:special_Kahler}, is broken by the charges appearing in the gauging. The resulting theory crucially depends on which charges (electric or magnetic) the fermions and scalars carry. If all matter fields carry only electric charges, i.e.\ are charged with respect to the gauge fields which are declared electric in the ungauged case, then the Lagrangian is given by a standard $\cN=2$ gauged supergravity. However, it is possible that some fraction of the matter fields also carry magnetic charges, as frequently occurs in string compactifications. In this case it is still possible to symplectically rotate the vectors to the electric frame such that all the matter fields are electrically charged i.e.\ the initial electric and  magnetic charges are constrained to be mutually local. However, as the theory is no longer symplectically covariant the Lagrangian in the electric frame might not be of the standard supergravity form. In particular, ${\cal F}_I$ is no longer constrained to be the derivative of a prepotential, or in other words, a holomorphic $\cal F$ might not exist in the given symplectic frame  \cite{Ceresole:1995jg,deWit:2002vt,deWit:2005ub}.
Recently the formalism of the embedding tensor has been introduced in \cite{deWit:2002vt,deWit:2005ub} to discuss both electric and magnetic charges in the theory.
It treats the electric vectors $A_\mu^{~~I}$ and the magnetic vectors $B_{\mu I}$ on the same footing and naturally allows for arbitrary gaugings. Let us now briefly introduce the formalism, following \cite{deWit:2002vt,deWit:2005ub}.

The $\cN=2$ theory has a group $G_0$ of global isometries on ${\cal M}$ generated by the Killing vectors $k_{\hat\a}, \hat \a = 1,\ldots, {\rm dim}(G_0)$. The electric and magnetic charges of the $\cN=2$ theory are collected in the embedding tensor $ \Theta_\Lambda^{~~\hat\a}$, whose electric and magnetic components we denote by $ \Theta_\Lambda^{~~\hat\a} = (\Theta_\L^{~~\hat\a},-\Theta^{\L\hat\a})$, where $\Lambda$ labels the $(2n_{\rm v}+2)$ electric and magnetic gauge fields.\footnote{Here the minus sign in $\Theta_\Lambda^{~~\hat\a}$ is introduced such that $\Theta^{\Lambda\hat\a} = \Omega^{\Lambda \Sigma}  \Theta_{\Lambda}^{~~\hat\a}= (\Theta^{\L\hat\a}, \Theta_\L^{~~\hat\a})$ transforms covariantly under symplectic rotations, where $\Omega^{\Lambda\Sigma}$ is the inverse $Sp(n_{\rm v}+1)$ metric.}
Given the set of global generators $k_{\hat\a}$, the embedding tensor selects the gauged subset
\begin{equation}\label{generators}
T_\Lambda =
\Theta_\Lambda^{~~\hat\a} k_{\hat\a} = (\Theta_\L^{~~\hat\a}
k_{\hat\a},-\Theta^{\L\hat\a} k_{\hat\a})\ ,
\end{equation}
i.e.\ it selects the generators of the local gauge group $G$. The embedding tensor itself is a spurionic object, which means that it is formally considered to transform as defined by its index structure (adjoint $\times$ fundamental) such that  $Sp(n_{\rm v}+1)$-covariance is restored in the Lagrangian. In this way, electric and magnetic gaugings are treated on the same footing. Choosing a specific value for the embedding tensor then fixes the gauge group $G$ and breaks the global symmetry $G_0$ to $G\times H$, where $H$ is the maximal commutant of $G$. Consistency of the embedding tensor projection onto the local subset requires that the generators $T_\Lambda$ form a closed subalgebra $G$ of $G_0$. This is ensured by the quadratic constraint
\begin{equation}\label{embedding_constraint_closure}
f^{\hat\gamma}_{\hat{\alpha}\hat{\beta}} \Theta_{M}^{\hat{\alpha}} \Theta_{N}^{\hat{\beta}} + \Theta_{M}^{\hat{\alpha}} (k_{\hat{\alpha}})_N^{\phantom{N}P} \Theta_{P}^{\hat{\gamma}} = 0 \ ,
\end{equation}
where  $f^{\hat\gamma}_{\hat{\alpha}\hat{\beta}}$ are the structure constants of the global symmetry group $G_0$. In addition, supersymmetry imposes a linear constraint
\begin{equation}\label{embedding_constraint_linear}
 \Theta_{(\Lambda}^{\hat{\a}} (k_{\hat{\a}})_{\Sigma\Xi)} = 0 \ .
\end{equation}
It is also important to note that the requirement of mutually local charges is expressed as an additional constraint on the embedding tensor
\begin{equation}
 \label{embedding_constraint_q}
 \Omega^{\Lambda\Sigma} \Theta_{\Lambda}^{\hat{\alpha}}
 \Theta_{\Sigma}^{\hat{\beta}}\ =\ \Theta^{I[\hat{\alpha}}
 \Theta_{I}^{\phantom{I}\hat{\beta}]}\ =\ 0 \ ,
\end{equation}
where $\Omega^{\Lambda\Sigma}$ is the inverse $Sp(n_{\rm v}+1)$ metric.

In principal, the gauge group $G$ that is selected by the embedding tensor can be Abelian or non-Abelian. In $\cN=2$ supersymmetry any non-Abelian gauge group $G$ always has a Coulomb branch, where the scalars $t^i$ in the adjoint representation have a vacuum expectation value which breaks $G\to [U(1)]^{\rm rank(G)}$. Thus, in general we can go far out on the Coulomb branch and safely integrate out all massive vector multiplets, leaving an Abelian theory with charged hypermultiplets at low energies.
Furthermore, only Abelian gaugings appear in generic string compactifications since the special K\"ahler manifold ${\cal M}_{\rm v}$ in general does not admit any isometries.
As we argue in more detail in the next chapter, we only need to consider charged hypermultiplets there, but not non-Abelian vector multiplets. Therefore we restrict to Abelian gaugings in the following.

For an Abelian theory, no isometries on ${\cal M}_{\rm v}$ are gauged and the non-trivial Killing vectors -- denoted by $k_{\lambda}$ -- act only on ${\cal M}_{\rm h}$ (see, for instance, \cite{Galicki:1986ja,D'Auria:1990fj}). This immediately implies that the constraints \eqref{embedding_constraint_closure} and \eqref{embedding_constraint_linear} are  trivially satisfied and we only have to impose \eqref{embedding_constraint_q}.
The gauge transformation of the scalar fields $q^u$ in the hypermultiplets then takes the form:
\begin{equation}\begin{aligned}
\delta_\alpha q^u \ &= \ \alpha^\Lambda \Theta_\Lambda^{~~\lambda}
k_{\lambda}^u(q)  \ ,
\end{aligned}\end{equation}
where $k_{\lambda}^u(q)$ are the components of the Killing vectors $k_{\lambda}$, and $\alpha^\Lambda$ are the transformation parameters. In the kinetic terms for the scalars $q^u$ the ordinary derivative is replaced by the covariant derivative
\begin{equation}\begin{aligned}\label{d2}
\partial_\mu q^u\to D_{\mu} q^u &= \pt_{\mu} q^u - A^{~~\Lambda}_{\mu} T_\Lambda
q^u\\
&= \pt_{\mu}  q^u - A^{~~\L}_{\mu} \Theta_\L^{~~\lambda} k_{\lambda}^u + B_{\mu\L} \Theta^{\L~{\lambda}} k_{\lambda}^u \ ,
\end{aligned}\end{equation}
while the derivatives of the $t^i$ are unchanged. Inserting the replacement \eqref{d2} into the Lagrangian \eqref{sigmaint} introduces both electric and magnetic vector fields. This upsets the counting of degrees of freedom and leads to unwanted equations of motions. Therefore, the Lagrangian has to be carefully augmented by a set of two-form gauge potentials $B_{\mu\nu}^M$ with couplings that keep supersymmetry and gauge invariance intact. As we do not need these couplings in this review, we refer the interested reader to the literature for further details~\cite{deWit:2002vt,deWit:2005ub,deVroome:2007zd}.

An analysis of the symplectic extension of gauged $\cN=2$ supergravity Lagrangian in $d=4$ including electric and magnetic charges has been carried out in \cite{Dall'Agata:2003yr,Sommovigo:2004vj,D'Auria:2004yi}.\footnote{The case of global $\cN=2$ supersymmetry has been studied in the embedding tensor formalism in \cite{deVroome:2007zd}.} Due to the discussion in the next chapter, we are specifically interested in the scalar part of supersymmetry variations, i.e.\
\begin{eqnarray}\label{susytrans2}
\delta_\epsilon \Psi_{\mu {\cal A}} &=& D_\mu \epsilon_{\cal A} - S_{\cal AB} \gamma_\mu \epsilon^{\cal B} + \ldots \, ,\nonumber\\
\delta_\epsilon \lambda^{i {\cal A}} &=& W^{i{\cal AB}}\epsilon_{\cal B}+\ldots \, ,\\
\delta_\epsilon \zeta_{\alpha} &=& N_\alpha^{\cal A} \epsilon_{\cal A}+\ldots \, ,\nonumber
\end{eqnarray}
where the ellipses indicate further terms which vanish in a maximally symmetric ground state. $\gamma_\mu$ are Dirac matrices and $\epsilon^{\cal A}$ is the $SU(2)$ doublet of spinors parameterizing the $\cN=2$ supersymmetry transformations.\footnote{Note that the $SU(2)$ R-symmetry acts as the $Sp(1)$ introduced above on the quaternionic-K\"ahler manifold.}
$S_{\cal AB}$ is the mass matrix of the two gravitini, while $W^{i {\cal AB}}$ and $N_\alpha^{\cal A}$ are related to the mass matrices of the spin-$\tfrac12$ fermions, cf.\ for instance \cite{D'Auria:2001kv}. The symplectic extensions of these expressions in the embedding tensor formalism are given by are
\begin{eqnarray}\label{susytrans3}
S_{\cal AB} &=& \tfrac{1}{2} \e^{K^{\rm v}/2} {V}^\Lambda \Theta_\Lambda^{~~\lambda} P_{\lambda}^x
(\sigma^x)_{\cal AB} \ ,\nonumber\\
W^{i{\cal AB}}
&=& \mathrm{i} \e^{K^{\rm v}/2} g^{i\bar \jmath}\,
(\nabla_{\bar \jmath}\bar {V}^\Lambda) \Theta_\Lambda^{~~\lambda} P_{\lambda}^x (\sigma^x)^{\cal AB}
\ ,\\
N_\alpha^{\cal A}
&=& 2 \e^{K^{\rm v}/2} \bar {V}^\Lambda \Theta_\Lambda^{~~\lambda} U^{\cal A}_{\alpha u} k^u_{\lambda}
\ .\nonumber
\end{eqnarray}
Let us explain the notation used in equations.~\eqref{susytrans3}. The matrices $(\sigma^x)_{\cal AB}$ and $(\sigma^x)^{\cal AB}$ are found by applying the $SU(2)$ metric $\varepsilon_{\cal{AB}}$ (and its inverse) to the standard Pauli matrices $(\sigma^x)_{\cal A}^{~~\cal B}$, $x=1,2,3,$ and are given in Appendix~\ref{section:conventions}.

Given the supersymmetry variations \eqref{susytrans2} and \eqref{susytrans3}, a Ward identity leads to the general formula for the classical scalar potential $V$ \cite{Cecotti:1984wn,D'Auria:2001kv}:
 \begin{equation} \label{potential_identity}
 V \delta^{\cal A}_{\cal B} = -12 S_{\cal BC} \bar S^{\cal AC} + g_{i\bar \jmath} W^{i \cal {AC}} W^{\bar \jmath}_{\cal BC}
+ 2 N_\alpha^{\cal A} N_{\cal B}^\alpha \ ,
\end{equation}
and it has been argued that this expression holds true in the presence of magnetic charges \cite{Michelson:1996pn,deWit:2005ub}.

At the end of this section let us discuss how charges appear in gauged $\cN=2$ supergravities coming from $SU(3)\times SU(3)$-structure compactifications. In order to have gaugings in the scalar sector, we need isometries on ${\cal M}_\textrm{v} \times {\cal M}_\textrm{h}$. While there is no a-priori reason for ${\cal M}_\textrm{v}$ to admit any isometries, we know that ${\cal M}_\textrm{h}$ is a quaternionic-K\"ahler manifold ${\cal M}_\textrm{h}$ in the image of the c-map and therefore has $2 n_{h}+2$ isometries. It is well-known that the dilaton direction $k_\phi$ gets loop corrections that break the isometry, i.e.\ it is anomalous and therefore is not gauged in $SU(3)\times SU(3)$ compactifications (see for instance \cite{RoblesLlana:2006ez} for the computation of the one-loop correction).
All other Killing vectors can be gauged in the effective action of such compactifications. The electric and magnetic charges for the Killing vectors $\Kxi_A, \Ktxi^A$ and $\Kax$ in the effective low-energy effective action are induced by fluxes, intrinsic torsion classes and non-geometric fluxes, i.e.\ by the coefficients in \eqref{SU3SU3_charges} and \eqref{SU3SU3_flux} (see \cite{D'Auria:2004tr,D'Auria:2004wd,Grana:2006hr} and references therein).

By considering the supersymmetry transformation of the two massless gravitini one can then derive the Killing prepotentials and thereby the gravitino mass matrix \cite{House:2005yc,Grana:2005ny,Grana:2006hr}
\begin{equation}\label{SU3SU3_gravitinomatrix}
S_{\cal AB} = \left( \begin{array}{ccc}
 -2 \e^{\tfrac12 K_- +\phi} \int_Y \langle \Phi^+, \diff \bar \Phi^- \rangle && - \e^{2\phi} \int_Y \langle \Phi^+, F^+ \rangle \\ - \e^{2\phi} \int_Y \langle \Phi^+, F^+ \rangle && 2 \e^{\tfrac12 K_- +\phi} \int_Y \langle \Phi^+, \diff \Phi^- \rangle
\end{array}\right) \ .
\end{equation}
By comparing this result with \eqref{susytrans3} and the Killing prepotentials for the vectors in \eqref{Killing} we can read off the corresponding embedding tensor as
\begin{equation}\label{Etensorcharges}
 \Theta^{\ \tilde \Lambda}_\Lambda = \left( \begin{aligned}
                            e_{AI} && p_A^I \\ m^A_I && q^{AI}
                          \end{aligned} \right) \ , \qquad
\Theta^{\ \ax}_\Lambda = (f_{ I} \ , \ f^I ) \ .
\end{equation}
In particular, the doubly symplectic matrix ${\cal Q}_\Lambda^{\tilde \Lambda}=\Theta^{\ \tilde \Lambda}_\Lambda$ describes gaugings of the Killing vectors $k_{\tilde \Lambda} = (\Kxi_A,\Ktxi^A)$ and the R-R flux $f^\Lambda$ gaugings of the Killing vector $k_{\tilde \phi}$ under the gauge fields $A^{\Lambda} = (A^I, B_I)$ coming from the R-R sector (see \eqref{expansion_RRfields}).
Thus, we identify the embedding tensor $\Theta_\Lambda^{~~\lambda}$ with
\begin{equation} \label{charges_N=2comp}
\Theta_\Lambda^{~~\lambda} = ({\cal Q}^\Lambda_{\tilde \Lambda},f^\Lambda)
\end{equation}
for the $2n_{\rm h}+1$ Killing vectors $k_\lambda=(k_{\tilde \Lambda},k_{\tilde \phi})$.
Note that the quadratic constraints of ${\cal Q}$ given in \eqref{doubly_sympl_quadratic_constr} correspond to the locality constraint \eqref{embedding_constraint_q} and the requirement that the gauged Killing vectors with the algebra \eqref{isometries_fibre} commute \cite{D'Auria:2004wd,D'Auria:2007ay}.

We see that the gaugings in $\cN=2$ supergravity are completely determined by the chosen consistent truncation in the $SU(3)\times SU(3)$-structure compactification. Moreover, all Killing vectors in the Heisenberg algebra of special quaternionic-K\"ahler manifolds can be gauged. In Section \ref{section:breaking_hyper} we shall use these gaugings to construct $\cN=2$ supergravities with $\cN=1$ supersymmetric vacua.

\cleardoublepage
%%%%%%%%%%%%%%%%%%%%%%%%%%%%%%%%%%%%%%%%%%%%%%%
\chapter{$\cN=1$ vacua of $\cN=2$ supergravity and type II compactifications}\label{section:SPSB}
%%%%%%%%%%%%%%%%%%%%%%%%%%%%%%%%%%%%%%%%%%%%%%%
Now let us discuss the appearance and properties of $\cN=1$ vacua in gauged $\cN=2$ supergravities and apply this discussion to $SU(3)\times SU(3)$-structure compactifications, based on \cite{Louis:2009xd,Louis:2010ui}. In such $\cN=1$ vacua, $\cN=2$ supersymmetry is spontaneously broken to $\cN=1$ by a super-Higgs mechanism that gives mass to one of the gravitini. This leads effectively to an $\cN=1$ supergravity with an additional, massive $\cN=1$ gravitino multiplet, which subsequently can be integrated out. In order to show the existence of $\cN=1$ vacua, we first review the no-go theorem of \cite{Cecotti:1984rk,Cecotti:1984wn} and then show how to circumvent it by including magnetic charge. Afterwards, we determine the properties of the corresponding $\cN=1$ vacuum by integrating out the massive gravitino multiplet. This leads to an $\cN=1$ effective theory whose couplings we determine in terms of the parental $\cN=2$ theory.
Furthermore, we focus on the class of $\cN=2$ supergravities coming from $SU(3)\times SU(3)$-structure compactifications and construct the general solution for the appearance of $\cN=1$ vacua and determine their couplings. Finally, we qualitatively discuss the lift to flux configurations in ten-dimensional $SU(3)\times SU(3)$-structure backgrounds and comment on relevant quantum effects in the corresponding string theory.

\section{Spontaneous $\cN=2 \rightarrow \cN=1$ supersymmetry breaking}\label{section:vectors}

Spontaneous $\cN=2\to \cN=1$ supersymmetry breaking in a Minkowski or AdS ground state requires that for one linear combination
of the two spinors $\epsilon^{\cal A},{\cal A}=1,2,$ parameterizing the supersymmetry transformations, say $\epsilon^{\cal A}_1$, the variations of the fermions given in \eqref{susytrans2} vanish, i.e.\ $\delta_{\epsilon_1} \lambda^{i {\cal A}} = \delta_{\epsilon_1} \zeta_\alpha = \delta_{\epsilon_1} \Psi_{\mu {\cal A}} =0$ (see e.g.\ \cite{Louis:2002vy,Gunara:2003td} for a review). Furthermore, in a supersymmetric Minkowski or AdS background the supersymmetry parameter obeys the Killing spinor equation\footnote{Note that the index of $\epsilon^*_{1\,\cal A}$ is not lowered with $\varepsilon_{\cal AB}$ but $\epsilon^*_{1\,\cal A}$ is related to $\epsilon^A_1$ just by complex conjugation. $|\mu|$ is related to the cosmological constant via $\Lambda = - 3 |\mu|^2$, while the phase of $\mu$ is unphysical.}
\begin{equation}
 D_\mu \epsilon_{1\,\cal A} = \tfrac12 \mu \gamma_\mu
 \epsilon^*_{1\,\cal A} \ .
\end{equation}
The requirement of a maximally symmetric ground state ensures that the terms which are indicated by the ellipses in \eqref{susytrans2} automatically vanish, so that one is left with
\begin{equation} \label{N=1conditions}
W_{i\cal AB}\, \epsilon^{\cal B}_1\ =\ 0\ =\ N_{\alpha \cal A}\,
\epsilon^{\cal A}_1 \ ,\qquad \textrm{and}\qquad  S_\mathcal{AB}\,
\epsilon^{\cal B}_1\ =\  \tfrac12 \mu \epsilon^*_{1\,\cal A} \ .
\end{equation}
Here we have chosen to write the parameter of the unbroken $\cN=1$ supersymmetry $\epsilon_1$ as a vector $\epsilon^{\cal A}_1$ in the space of $\cN=2$ parameters. For the second, broken generator, which we denote by $\epsilon^{\cal A}_2$, we should have
\begin{equation} \label{N=1conditions2}
W_{i\cal AB}\, \epsilon^{\cal B}_2 \neq 0\qquad \textrm{or}\qquad
N_{\alpha \cal A}\, \epsilon^{\cal A}_2\neq 0\ ,\qquad \textrm{and}\qquad  S_\mathcal{AB}\,
\epsilon^{\cal B}_2\ \ne\  \tfrac12 \mu' \epsilon^*_{2\,\cal A}
\end{equation}
for any $\mu'$ that obeys $|\mu'|=|\mu|$, i.e.\ only differs from $\mu$ by an unphysical phase.

Before we attempt to solve \eqref{N=1conditions} and \eqref{N=1conditions2} let us assemble a few more facts. A necessary condition for the existence of an $\cN=1$ ground state is that the two eigenvalues $m_{\Psi_1}$ and $m_{\Psi_2}$ of the gravitino mass matrix $S_{\cal AB}$ are non-degenerate, i.e.\ $m_{\Psi_1}\neq m_{\Psi_2}$. In a Minkowski ground state one also needs $m_{\Psi_1}=0$ or, more generally, one of the two gravitini has to become massive, while the second one stays massless, cf.\ \eqref{N=1conditions} and \eqref{N=1conditions2}. Furthermore, the unbroken $\cN=1$ supersymmetry implies that the massive gravitino has to be a member of an entire $\cN=1$ massive spin-$3/2$ multiplet, which has the spin content $s=(3/2,1,1,1/2)$. This means that two vectors, say $A_\mu^0, A_\mu^1$ and a spin-$1/2$ fermion $\chi$ have to become massive, in addition to the gravitino. Therefore, the would-be Goldstone fermion (the Goldstino), which gets eaten by the gravitino, is accompanied by two would-be Goldstone bosons (the sGoldstinos) \cite{Ferrara:1983gn}. The field content of the massive spin-$3/2$ multiplet in terms of massless ${\cal N}=1$ multiplets is then one spin-$3/2$ multiplet, one vector multiplet and one chiral (Goldstino) multiplet. Naively, one might think that both the $\cN=1$ vector and chiral multiplet come from $\cN=2$ vector multiplets in a non-Abelian theory, without the need for additional charged hypermultiplet scalars. However, vector multiplet scalars are singlets under the $SU(2)$ R-symmetry of $\cN=2$ supergravity and therefore cannot give rise to a mass splitting between the gravitini \cite{Ferrara:1985gj}. In an Abelian theory, on the other hand, the sGoldstinos have to be `recruited' out of a charged $\cN=2$ hypermultiplet, while the need for two gauge bosons implies that at least one $\cN=2$ vector multiplet has to be part of the spectrum. Thus, the minimal $\cN=2$ spectrum which allows for the possibility of a spontaneous breaking to $\cN=1$ consists of the $\cN=2$ supergravity multiplet, one hypermultiplet and one vector multiplet.

In geometric terms, the presence of two sGoldstinos in the hypermultiplet sector means that ${\cal M}_{\rm h}$ has to admit two commuting isometries, say $k_1^u,k_2^u$, and that these isometries have to be gauged \cite{Fre:1996js}. The definition \eqref{Pdef} then implies that we need to have two non-zero Killing prepotentials $P_1^x, P_2^x$ in the ground state. Furthermore, these prepotentials must not be proportional to each other because otherwise we could take linear combinations of $k_1^u$ and $k_2^u$ such that one combination has vanishing prepotentials.
However, we can use the local $SU(2)$ invariance of the
hypermultiplet sector to rotate into a
convenient $SU(2)$-frame where
$P_{1,2}^x$
both lie entirely in
the $x=1,2$-plane. Thus, without loss of generality we can arrange
\begin{equation}\label{P3}
 P^3_1 \ =\ P^3_2 \ =\ 0 \ =\ \partial_u P^3_1\ =\ \partial_u P^3_2\ .
\end{equation}
From \eqref{susytrans3} we learn that in such a frame both
$S_{\cal  AB}$ and $W^{i{\cal AB}}$ are diagonal
in $SU(2)$ space and hence one can further choose  the parameter of
the unbroken $\cN=1$ generator to be $\epsilon_1 = {\epsilon\choose 0}$
or $\epsilon_1 = {0\choose\epsilon}$. This corresponds to the choice of
$\Psi_{\mu\, 1}$ or $\Psi_{\mu\, 2}$ as the massless
$\cN=1$ gravitino.\footnote{Note that all our expressions can also
be written in an $SU(2)$-covariant way by replacing the
``$3$''-direction with $\epsilon_1^A \sigma^x_{\cal AB} \epsilon_2^B$
and the direction spanned by $(P^1-\iu P^2)$ with $\epsilon_1^A
\sigma^x_{\cal AB} \epsilon_1^B$. So, for instance, \eqref{P3} then
reads
%\begin{equation}
$\epsilon_1^A \sigma^x_{\cal AB} \epsilon_2^B P^x_{1,2} = \epsilon_1^A \sigma^x_{\cal AB} \epsilon_2^B \diff P^x_{1,2}=0.
$%\end{equation}
}
In Section~\ref{section:magnetic_vectors} we verify that two gauged Killing vectors with non-aligned Killing prepotentials are necessary for partial supersymmetry breaking to appear and also discuss the case of more than two gauged Killing vectors.

%%%%%%%%%%%%%%%%%%%%%%%%%%%%%%%%%%%%%%%%%%%%%%%%%%%%%%%%%%
\subsection{The electric no-go theorem} \label{section:no-go}
%%%%%%%%%%%%%%%%%%%%%%%%%%%%%%%%%%%%%%%%%%%%%%%%%%%%%%%%%%
After this initial discussion of the necessary ingredients, let us now discuss the obstructions to spontaneous $\cN=2$ to $\cN=1$ supersymmetry breaking. Using the superconformal tensor calculus, Cecotti et al.\ showed that an $\cN=2$ gauged supergravity with only electric charges cannot have an $\cN=1$ Minkowski ground state \cite{Cecotti:1984rk,Cecotti:1984wn}. More precisely, it was shown that in this case the gravitino mass matrix is proportional to the unit matrix and hence the masses are degenerate. This implies that the ground state either has the full $\cN=2$ supersymmetry or none at all, ruling out the possibility of spontaneous partial supersymmetry breaking.
We shall now review this no-go theorem for purely electric gaugings with the help of the embedding tensor formalism, without using superconformal tensor calculus.
It turns out that the no-go theorem follows from the gravitino and gaugino variations alone. The hyperino
equation gives additional constraints on the hypermultiplet sector, and we postpone its discussion to Section~\ref{section:hypermultiplets}.

Assume that we are at a point $X^I_0$ in the vector multiplet moduli space and at a point $q^u_0$ in the quaternionic-K\"ahler manifold at which supersymmetry is broken to $\cN=1$ and the conditions \eqref{N=1conditions} hold. For simplicity, we shall drop the subscript and simply denote this point by $X^I$ and $q^u$. The gravitino equation in \eqref{N=1conditions} for electric gaugings is given by
\begin{equation}\label{S_N=1}
  S_\mathcal{AB}\, \epsilon_1^{\cal B} = \tfrac{1}{2} \e^{K^{\rm v}/2} X^I
  \Theta_I^{\phantom{I}\lambda} P_\lambda^x \sigma^x_\mathcal{AB}
  \epsilon_1^{\cal B} = \tfrac12 \mu \epsilon^*_{1\,\cal A} \ .
\end{equation}
The (complex conjugate) of the gaugino variation in \eqref{N=1conditions} leads to
\begin{equation}\begin{aligned}\label{W_N=1}
 W_{i\cal AB}\, \epsilon_1^{\cal B} &= \iu \e^{K^{\rm
 v}/2}(\nabla_i X^I)
 \Theta_I^{\phantom{I}\lambda} P_\lambda^x \sigma^x_\mathcal{AB}
 \epsilon_1^{\cal B} \\
&=\iu
 \e^{K^{\rm v}/2}  (\partial_i X^I) \Theta_I^{\phantom{\Lambda}\lambda}
 P_\lambda^x \sigma^x_\mathcal{AB} \epsilon_1^{\cal B}
%+ 2 \iu K^{\rm v}_i S_\mathcal{AB} \epsilon_1^{\cal B} = 0 \ ,
+  \iu K^{\rm v}_i \mu \epsilon^*_{1\,\cal A}  = 0 \ ,
\end{aligned}\end{equation}
where in the second line we have used $\nabla_i X^I = \partial_i X^I + K^{\rm v}_i X^I$ and inserted the gravitino equation \eqref{S_N=1}. Note that in total \eqref{S_N=1} and \eqref{W_N=1} give $2(n_{\rm v}+1)$ equations to solve. Let us now specialize to a frame where a prepotential exists. We can then express $X^I$ in terms of special coordinates as $X^I = (1, t^i)$ and we find that the gaugino equation \eqref{W_N=1} simplifies to
\begin{equation} \label{nogo_Theta_i}
 \Theta_i^{\phantom{i}\lambda} P_\lambda^x \sigma^x_\mathcal{AB}
 \epsilon_1^{\cal B} = - \e^{-K^{\rm v}/2} \mu K^{\rm v}_i \epsilon^*_{1\,\cal A}  \ .
\end{equation}
Inserting this back into the gravitino equation \eqref{S_N=1} yields
\begin{equation}\label{noint}
 \Theta_0^{\phantom{0}\lambda} P_\lambda^x \sigma^x_\mathcal{AB}
 \epsilon_1^{\cal B}\ =\ \e^{-K^{\rm v}/2} \mu\, (1+t^i K^{\rm v}_i)\, \epsilon^*_{1\,\cal A}  \ .
\end{equation}
From the definition of the K\"ahler potential \eqref{Kdef} one derives the identity $X^I K^{\rm v}_I = -1$, which in special coordinates $X^I = (1, t^i)$ reads $1+t^i K^{\rm v}_i= - K^{\rm v}_0 $. This further simplifies \eqref{noint} to give
\begin{equation}\label{noint2}
 \Theta_0^{\phantom{0}\lambda} P_\lambda^x \sigma^x_\mathcal{AB}
 \epsilon_1^{\cal B} = - \e^{-K^{\rm v}/2} \mu K^{\rm v}_0 \epsilon^*_{1\,\cal A}  \ ,
\end{equation}
which allows us to combine \eqref{nogo_Theta_i} and \eqref{noint2} into the $2(n_{\rm v}+1)$ equations
\begin{equation}\label{noend}
 \Theta_I^{\phantom{I}\lambda} P_\lambda^x \sigma^x_\mathcal{AB}\,
 \epsilon_1^{\cal B} = - \e^{-K^{\rm v}/2} \mu K^{\rm v}_I  \epsilon^*_{1\,\cal A}  \ .
\end{equation}
To summarize, by using the existence of the special coordinates  $X^I = (1, t^i)$ we have been able to rewrite the original $2(n_{\rm v}+1)$ equations arising from the gravitino \eqref{S_N=1} and gaugino \eqref{W_N=1} variations in a compact manner \eqref{noend}.

If we now consider a Minkowski vacuum, i.e.\ setting $\mu=0$, the expression $\sigma^x_\mathcal{AB} \epsilon_1^{\cal B}$ is the only complex quantity appearing in \eqref{noend} and the left-hand side of \eqref{noend} simply describes an $su(2)$ variation of $\epsilon_1^{\cal A}$. For $\mu=0$, the $su(2)$ variation of the doublet has to vanish, which only happens for the trivial variation, i.e.\
\begin{equation}\label{Theta1}
  \Theta_I^{\phantom{I}\lambda} P_\lambda^x = 0 \ .
\end{equation}
If we then insert \eqref{Theta1} back into the matrices appearing in the supersymmetry transformations \eqref{susytrans3}, we see that $S_{\cal AB}$ (and $W_{i{\cal AB}}$) identically vanish, and thus partial supersymmetry breaking is not possible, i.e.\ we have recovered the original no-go theorem \cite{Cecotti:1984rk}. The important step in this derivation was using the existence of a prepotential and the special coordinates $X^I = (1, t^i)$ to find $n_{\rm v}$ independent equations in \eqref{nogo_Theta_i}. Therefore, this no-go theorem might be circumvented by using a symplectic frame in which no prepotential exists at the $\cN=1$ point. It turns out that this is possible, and the first examples of spontaneous partial supersymmetry breaking used precisely such frames where the prepotential does not exist \cite{Ferrara:1995gu,Ferrara:1995xi,Fre:1996js}. On the other hand, such symplectic frames are related to the standard one by a symplectic transformation which just rotates electric and magnetic charges into each other. Therefore, in the following  we still assume the existence of a prepotential but generalize our discussion by allowing for both electric and magnetic charges. This covers all possible gauged supergravities and in particular the examples mentioned above. In the next section, we show that this generalization indeed gives rise to the possibility of spontaneous partial supersymmetry breaking.

%%%%%%%%%%%%%%%%%%%%%%%%%%%%%%%%%%%%%%%%%%%%%%%%%%%%%%%%%%
\subsection{A way out - magnetic charges} \label{section:magnetic_vectors}
%%%%%%%%%%%%%%%%%%%%%%%%%%%%%%%%%%%%%%%%%%%%%%%%%%%%%%%%%%
We shall now repeat the discussion of Section~\ref{section:no-go} with magnetic gaugings included. We will also discuss partial supersymmetry breaking to both Minkowski and AdS vacua, i.e.\ we keep $\mu$ nonzero in \eqref{N=1conditions}. First, we note that the condition which comes from the vanishing of the gaugino variation \eqref{W_N=1}, now with electric and magnetic gaugings, gives rise to
\begin{equation}\label{N=1_condition_vectorsI}
\e^{K^{\rm v}/2}(\partial_i X^I \Theta_I^{\phantom{I}\lambda} - \partial_i {\cal F}_I \Theta^{I\lambda}) P_\lambda^x \sigma^x_\mathcal{AB} \epsilon_1^{\cal B} +  K^{\rm v}_i \mu \epsilon^*_{1\,\cal A}  = 0 \ ,
\end{equation}
where the second term in the brackets is due to the presence of magnetic charges $\Theta^{I\lambda}$. Contracting \eqref{N=1_condition_vectorsI} with $t^i$ and adding it to  $2 S_\mathcal{AB} \epsilon_1^{\cal B}= \mu \epsilon_{1\,\cal A}^*$ we arrive at
\begin{equation}\begin{aligned}\label{int}
\e^{-K^{\rm v}/2} \mu (1+t^i K^{\rm v}_i) \epsilon_{1\,\cal A}^* &= (X^I \Theta_I^{\phantom{I}\lambda} - {\cal F}_I \Theta^{I\lambda}) P_\lambda^x \sigma^x_\mathcal{AB} \epsilon_1^{\cal B} - t^i (\Theta_i^{\phantom{i}\lambda} - {\cal F}_{iJ} \Theta^{J\lambda}) P_\lambda^x \sigma^x_\mathcal{AB} \epsilon_1^{\cal B}\\
&= (\Theta_0^{\phantom{0}\lambda} - {\cal F}_{0J} \Theta^{J\lambda}) P_\lambda^x \sigma^x_\mathcal{AB} \epsilon_1^{\cal B} \ .
\end{aligned}\end{equation}
Using again $1+t^i K^{\rm v}_i= - K^{\rm v}_0 $ in \eqref{int} and combining it with \eqref{N=1_condition_vectorsI} yields $2(n_{\rm v}+1)$ equations, replacing the conditions \eqref{noend} of the previous section:
\begin{equation} \label{condition_vectors}
 (\Theta_I^{\phantom{I}\lambda} - {\cal F}_{IJ} \Theta^{J\lambda}) P_\lambda^x \sigma^x_\mathcal{AB} \epsilon_1^{\cal B} = - \e^{-K^{\rm v}/2} \mu K^{\rm v}_I \epsilon_{1\,\cal A}^*  \ .
\end{equation}
These equations give conditions on the embedding tensor and on the prepotential. However, in order to ensure that the second supersymmetry is broken the conditions \eqref{condition_vectors} should not simultaneously hold for the second supersymmetry generator
\begin{equation} \label{susy_generator_2}
 \epsilon_2^{\cal A} = (\varepsilon_{\cal AB} \epsilon_1^{\cal B})^* \ .
\end{equation}
Inserting \eqref{susy_generator_2} into \eqref{condition_vectors}, we arrive at the additional condition
\begin{equation} \label{condition_vectors_2}
 (\Theta_I^{\phantom{I}\lambda} - \bar{\cal F}_{IJ} \Theta^{J\lambda})
 P_\lambda^x \sigma^x_\mathcal{AB} \epsilon_1^{\cal B} \ne \e^{-K^{\rm
 v}/2} \bar \mu' \bar K^{\rm v}_I \epsilon_{1\,\cal A}^*  \quad
 \textrm{for some} \ I \ ,
\end{equation}
for any $\mu'$ that obeys $|\mu'|=|\mu|$.

\subsubsection{Minkowski vacua}

Let us proceed by first analyzing Minkowski vacua ($\mu=0$). For this case \eqref{condition_vectors} and \eqref{condition_vectors_2} simplify to
\begin{subequations}\label{condition_vectors_M}
 \begin{eqnarray}
 (\Theta_I^{\phantom{I}\lambda} - {\cal F}_{IJ} \Theta^{J\lambda})\,
 P_\lambda^x \sigma^x_\mathcal{AB} \epsilon_1^{\cal B}\ &=\ 0  \quad
 \textrm{for all} \ I \ ,\quad \label{condition_vectors_M1} \\
 (\Theta_I^{\phantom{I}\lambda} -  \bar{\cal F}_{IJ} \Theta^{J\lambda})\, P_\lambda^x
 \sigma^x_\mathcal{AB} \epsilon_1^{\cal B}\ &\ne\ 0  \quad \textrm{for some} \ I \ . \label{condition_vectors_M2}
\end{eqnarray}
\end{subequations}
The crucial point is that the existence of an ${\cal N}=1$ vacuum requires that there is a set of charges for which \eqref{condition_vectors_M1} vanishes while \eqref{condition_vectors_M2} does not.
If \eqref{condition_vectors_M2} were also to vanish for all $I$, then the vacuum would preserve the full ${\cal N}=2$ supersymmetry.\footnote{For the subset of $I$ for which \eqref{condition_vectors_M2} does also vanish, we can add and subtract the equations \eqref{condition_vectors_M1} and \eqref{condition_vectors_M2} such that $\sigma^x_\mathcal{AB} \epsilon_1^{\cal B}$ is the only complex quantity in the resulting equations. Analogously to the discussion above \eqref{Theta1}, this then leads to $(\Theta_I^{\phantom{I}\lambda} - {\cal F}_{IJ} \Theta^{J\lambda}) P^x_\lambda = 0$. If this is the case for all $I$, we have $S_{\cal AB}=0$ and thus an $\cN=2$ vacuum.}
On the other hand, for an $\cN=1$ vacuum it is sufficient to find that for some $I$ \eqref{condition_vectors_M2} does not vanish.
Let us also reiterate that \eqref{condition_vectors_M1} and \eqref{condition_vectors_M2} do not have to hold over all of field space but only at the $\cN=1$ point. As ${\cal N}=1$ supersymmetry is preserved, one can show that this point is a minimum of the potential, see \cite{Louis:2009xd}.

Before solving \eqref{condition_vectors_M} let us first recall that we must have two commuting isometries $k_1$ and $k_2$ on ${\cal M}_{\rm h}$, as discussed at the beginning of Section~\ref{section:vectors}, and that at the $\cN=1$ point the corresponding Killing prepotentials $P^x_1$ and $P^x_2$ are both non-vanishing and not proportional to each other.
%%%%%%%%%%%%%%%%%%%%%%%%%%%%%%%%%%%%%%%%%%%%%%%%%%%%%%%%%%%%%%%%%%%%%%%%%%%%%%%%%%%%%%%%%%%%%%%%%%%%%%%%%%%%%%%%%%%
Consider \eqref{condition_vectors_M} with just one gauged isometry, say $k_1$. In this case \eqref{condition_vectors_M1} factorizes into two parts, i.e. either $(\Theta_I^{\phantom{I}1} - {\cal F}_{IJ} \Theta^{J1})$ or $P_1^x \sigma^x_\mathcal{AB} \epsilon_1^{\cal B}$ must vanish. However, from \eqref{condition_vectors_M2} we see that both of these expressions have to be non-zero. Therefore, for one gauged isometry we can only have ${\cal N}=2$ or ${\cal N}=0$. We shall first study the case with two gauged isometries and discuss the case with more gauged isometries later.
%%%%%%%%%%%%%%%%%%%%%%%%%%%%%%%%%%%%%%%%%%%%%%%%%%%%%%%%%%%%%%%%%%%%%%%%%%%%%%%%%%%%%%%%%%%%%%%%%%%%%%%%%%%%%%%%%%%
Recall that we choose the $SU(2)$ frame where \eqref{P3} holds.
Furthermore, we can make use of the complex combination
\begin{equation}\label{Pc}
 P_{1,2}^\pm = P_{1,2}^1 \pm \iu P_{1,2}^2 \ .
\end{equation}

We will now construct an embedding tensor $\Theta_\Lambda^{1,2}$ such that in this $SU(2)$ frame the supersymmetry generated
by $\epsilon_1^{\cal A} = (\epsilon_1^1,0)$ is unbroken. Using \eqref{Pc}, \eqref{condition_vectors_M} becomes
\begin{subequations}
\begin{eqnarray}
 P^-_1(\Theta_I^{\phantom{I}1} - {\cal F}_{IJ} \Theta^{J1}) + P^-_2
 (\Theta_I^{\phantom{I}2} - {\cal F}_{IJ} \Theta^{J2})
%- \iu P^2_2(\Theta_I^{\phantom{I}2} - {\cal F}_{IJ} \Theta^{J2})
&=& 0 \quad \textrm{for all} \ I \ , \label{condition_vectors_Msolve1} \\
 P^-_1(\Theta_I^{\phantom{I}1} - \bar{\cal F}_{IJ}
 \Theta^{J1})+P^-_2 (\Theta_I^{\phantom{I}2} - \bar{\cal F}_{IJ}
 \Theta^{J2})
%- \iu P^2_2 (\Theta_I^{\phantom{I}2} - \bar{\cal F}_{IJ} \Theta^{J2})
& \ne & 0 \quad \textrm{for some} \ I  \ . \label{condition_vectors_Msolve2}
\end{eqnarray}
\end{subequations}
Applying the elementary identity
\begin{equation} \label{identity_v}
 \Im({\cal F}_{IJ}L^J) - {\cal F}_{IJ} \Im L^J = (\Im{\cal F})_{IJ} \bar L^J \ ,
\end{equation}
which holds for any complex vector $L^I$,
we can solve \eqref{condition_vectors_Msolve1} in
terms of an arbitrary
complex vector $\C^I$ by choosing
\begin{equation}\label{solution_embedding_tensor}
\begin{aligned}
\Theta_I^{\phantom{I}1} = & -\Im(P^+_2 {\cal F}_{IJ}\C^J ) \ , \qquad  \Theta^{I1} = & -\Im (P^+_2\C^I) \ , \\
\Theta_I^{\phantom{I}2} = & \quad \Im(P^+_1 {\cal F}_{IJ}\C^J ) \ , \qquad  \Theta^{I2} = & \quad \Im (P^+_1 \C^I) \ ,
\end{aligned}
\end{equation}
where the Killing prepotentials and ${\cal F}_{IJ}$ are evaluated at the local $\cN=1$ minimum.
Note that since $P^x_1$ and $P^x_2$ are not aligned, the expression \eqref{condition_vectors_Msolve2} cannot vanish for any non-zero $\C^I$.

We also need to enforce the mutual locality constraint \eqref{embedding_constraint_q}, which for the case at hand reads
\begin{equation}\label{locality_two_isometries}
 \Theta^{I1}\Theta_I^{\phantom{I}2} - \Theta^{I2} \Theta_I^{\phantom{I}1} = 0 \ .
\end{equation}
If we now insert the solutions \eqref{solution_embedding_tensor} into this constraint, we find a condition on the coefficients  $\C^{I}$:
\begin{equation} \label{constraint_embedding_tensor}
  \bar{\C}^{I} (\Im {\cal F})_{IJ} \C^{J}  = 0 \ .
\end{equation}
In deriving this we have used that $\Im(P^-_1 P^+_2)\ne0$, which holds because the prepotentials $P^x_1$ and $P^x_2$ are not aligned. Therefore, we have found that $\C^{I}$ has to be null with respect to $(\Im {\cal F})_{IJ}$. Since $(\Im {\cal F})_{IJ}$ is of signature $(n_{\rm v},1)$, as we have shown in \eqref{signature_G}, this constraint can be easily satisfied.
Therefore, we have found that breaking $\cN=2$ to $\cN=1$ supersymmetry is possible.

We can perform a symplectic rotation $\cal S$ given by \eqref{uvzwg}
to transform the charge vector $\Theta^{\Lambda1,2} = (\Theta^{I1,2}, \Theta^{\ 1,2}_I)$ such that we only have electric charges, in other words $\Theta^{I\,1,2}$ vanishes in the rotated frame. We see from \eqref{solution_embedding_tensor} that then also the symplectic vector
\begin{equation}
 \left( \begin{array}{cc}
  \C^I \\ {\cal F}_{IJ} C^J
 \end{array} \right)
 = \tfrac{P^-_1}{\Im(P^+_1 P^-_2)}
 \left( \begin{array}{cc}
  \Theta^{I1} \\ \Theta_I^{\ 1}
 \end{array} \right) + \tfrac{P^-_2}{\Im(P^+_1 P^-_2)}  \left( \begin{array}{cc}
  \Theta^{I2} \\ \Theta_I^{\ 2}
 \end{array} \right)
\end{equation}
has to become purely electric under ${\cal S}$, i.e.\
\begin{equation}
(U^I_{~J} + Z^{IK} {\cal F}_{KJ}) \C^J = 0 \ ,
\end{equation}
and thus the matrix $U^I_{~J} + Z^{IK} {\cal F}_{KJ}$ is not invertible.
As discussed in \cite{Ceresole:1995jg}, this precisely means that we transform into a symplectic frame where no prepotential exists at the $\cN=1$ point, as demanded by the no-go theorem we reviewed in Section~\ref{section:no-go}.

%%%%%%%%%%%%%%%%%%%%%%%%%%%%%%%%%%%%%%%%%%%%%%%%%%%%%%%%%%%%%%%%%%%%%%%%%%%%%%%%%%%%%%%%%%%%%%%%%%%%%%%%%%%%%%%%%%%
Let us now consider the case with $n$ gauged commuting isometries. We can always go to a new basis of Killing vectors $k_\lambda$ where there are only three Killing vectors that have $P_{\lambda}^x\ne 0$ at the $\cN=1$ point. Imposing \eqref{condition_vectors_M1} then tells us that at least one combination of the $P^x$ has to vanish and, therefore, there are effectively only two Killing vectors with non-vanishing $P_{\lambda}^x$ at the $\cN=1$ point. We can identify these two Killing vectors with those used above to construct the $\cN=1$ solution. The other Killing vectors do not play a role in the supersymmetry breaking, but could give rise to additional masses as the derivatives of their $P_{\lambda}^x$'s could be non-zero, as we discuss in Section \eqref{section:LEEA}.

%%%%%%%%%%%%%%%%%%%%%%%%%%%%%%%%%%%%%%%%%%%%%%%%%%%%%%%%%%%%%%%%%%%%%%%%%%%%%%%%%%%%%%%%%%%%%%%%%%%%%%%%%%%%%%%%%%%

The above result is quite surprising. By appropriately choosing the embedding tensor, the conditions for partial $\cN=1$ supersymmetry breaking arising from the gravitino and the gaugino variations can be fulfilled for \emph{any} point on \emph{any} special K\"ahler
manifold ${\cal M}_{\rm v}$ and for \emph{any} quaternionic-K\"ahler manifold ${\cal M}_{\rm h}$ that admits two commuting isometries with Killing prepotential $P^x_1$ and $P^x_2$ that are not proportional to each other at the $\cN=1$ point. Of course, we still have to satisfy the non-trivial condition $N_{\alpha \cal A} \epsilon_1^{\cal A} = 0$ of \eqref{N=1conditions}. We shall turn to this issue in Section~\ref{section:hypermultiplets} and finally show in Section \eqref{section:breaking_hyper} that it can be solved for any special quaternionic-K\"ahler manifold.

Before we consider the analysis of AdS vacua, let us discuss a simple example given by the four-dimensional quaternionic-K\"ahler manifold ${{\cal M}}_{\rm h}=SO(1,4)/SO(4)$ with arbitrary ${{\cal M}}_{\rm v}$.  ${{\cal M}}_{\rm h}$ is parameterized by the quaternionic coordinates $(q^0, q^1,q^2, q^3)$ and admits the commuting Killing vectors $k_\lambda=\tfrac{\partial}{\partial q^\lambda}$ for $\lambda=1,2,3$. The Killing prepotentials are given by \cite{Ferrara:1995gu,Ferrara:1995xi}
\begin{equation}\label{prepotentials_ex}
P_\lambda^x = \tfrac{1}{q^0} \delta_\lambda^x \ ,
\end{equation}
which, when inserted into our solution for the embedding tensor components \eqref{solution_embedding_tensor}, yield
\begin{equation}\label{solution_embedding_tensor_ex}
\begin{aligned}
\Theta_I^{\phantom{I}1} = & -\Re( {\cal F}_{IJ}\C^J ) \ , \qquad \Theta^{I1} = & -\Re \C^I \ , \\
\Theta_I^{\phantom{I}2} = &  \quad \Im( {\cal F}_{IJ} \C^J ) \ , \qquad \Theta^{I2} = & \quad \Im \C^I \ .
\end{aligned}
\end{equation}
In this case, it can easily be shown that the hyperino variation $N_{\alpha \cal A}  \epsilon_1^{\cal A} = 0$ is automatically satisfied and we recover the $\cN=1$ vacuum given in \cite{Ferrara:1995gu}. However, the example in \cite{Ferrara:1995gu} was for a specific choice of ${{\cal M}}_{\rm v}$, whereas we have just shown that partial supersymmetry breaking is possible for arbitrary ${{\cal M}}_{\rm v}$.

\subsubsection{AdS vacua}

Let us now consider partial supersymmetry breaking in an AdS vacuum, i.e.\ for
$\mu \ne 0$. We again require that there are two commuting Killing vectors with non-aligned
Killing prepotentials and choose an $SU(2)$ frame where $P^x_1$ and $P^x_2$ are in the $x=1,2$ plane. We shall also make use of the identity
\begin{equation}
K^{\rm v}_I = 2\e^{K^{\rm v}} (\Im {\cal F})_{IJ} \bar X^J \ ,
\end{equation}
which follows from the definition of the K\"ahler potential \eqref{Kdef}. We then find that the gaugino conditions \eqref{condition_vectors} simplify and, as a consequence, the first condition for partial supersymmetry breaking is\footnote{The second condition similarly follows from \eqref{condition_vectors_2}.}
\begin{equation}\label{condition_vectors_AdS}
 P^-_1(\Theta_I^{\phantom{I}1} - {\cal F}_{IJ} \Theta^{J1}) + P^-_2
 (\Theta_I^{\phantom{I}2} - {\cal F}_{IJ} \Theta^{J2}) = - 2 \e^{K^{\rm v}/2}\mu (\Im {\cal F})_{IJ} \bar X^J  \ .
\end{equation}
This is just the Minkowski condition \eqref{condition_vectors_Msolve1} with an additional inhomogeneity proportional to $\mu$. If we now again make use of the identity \eqref{identity_v}, the solution to \eqref{condition_vectors_AdS} can be obtained analogously to the Minkowski case \eqref{solution_embedding_tensor}
\begin{equation}\label{solution_embedding_tensor_AdS}
\begin{aligned}
\Theta_I^{\phantom{I}1} = & -\Im({\cal F}_{IJ}(P^+_2 \C_{\rm AdS}^J + \e^{K^{\rm v}/2} \tfrac{\bar \mu}{P^+_1} X^J) ) \ , \\
\Theta^{I1} = & -\Im (P^+_2\C_{\rm AdS}^I + \e^{K^{\rm v}/2} \tfrac{\bar \mu}{P^+_1} X^I)) \ , \\
\Theta_I^{\phantom{I}2} = & \quad \Im({\cal F}_{IJ}(P^+_1 \C_{\rm AdS}^J - \e^{K^{\rm v}/2} \tfrac{\bar \mu}{P^+_2} X^J) ) \ , \\
\Theta^{I2} = & \quad \Im (P^+_1 \C_{\rm AdS}^I- \e^{K^{\rm v}/2} \tfrac{\bar \mu}{P^+_2} X^I) \ ,
\end{aligned}
\end{equation}
where again $\C_{\rm AdS}^I$ is an arbitrary vector. The mutual locality constraint \eqref{constraint_embedding_tensor} now reads
\begin{equation}\label{constraint_embedding_tensor_AdS_g}
\begin{aligned}
 \bar{\C}_{\rm AdS}^{I} (\Im {\cal F})_{IJ} \C_{\rm AdS}^{J} + \tfrac{|\mu|^2}{2 |P_1|^2 |P_2|^2} = - 2 \tfrac{\Re(P^-_1 P^+_2 )}{\Im(P^-_1 P^+_2 )} \e^{K^{\rm v}/2} \Im \left(\tfrac{\bar \mu}{P^+_1 P^+_2} \bar{\C}_{\rm AdS}^{I} (\Im {\cal F})_{IJ} X^J\right)  \ .
\end{aligned}
\end{equation}
For instance, if we choose the phase of $\C_{\rm AdS}^{I}$ appropriately, the right-hand side of this constraint vanishes and we end up with
\begin{equation}\label{constraint_embedding_tensor_AdS}
 \bar{\C}_{\rm AdS}^{I} (\Im {\cal F})_{IJ} \C_{\rm AdS}^{J} = - \tfrac{|\mu|^2}{2 |P_1|^2 |P_2|^2} \ ,
\end{equation}
which tells us that $\bar{\C}_{\rm AdS}^{I}$ is timelike with respect to $(\Im {\cal F})_{IJ}$. Once again, as $(\Im {\cal F})_{IJ}$ is of signature $(n_{\rm v},1)$, cf.\ discussion in \eqref{signature_G}, this condition is easily satisfied. It is straightforward to check that the second condition \eqref{condition_vectors_2} is automatically satisfied and we find that the breaking from $\cN=2$ to $\cN=1$ is possible for any solution in \eqref{solution_embedding_tensor_AdS} with non-zero $\C_{\rm AdS}^I$ obeying \eqref{constraint_embedding_tensor_AdS}.
%%%%%%%%%%%%%%%%%%%%%%%%%%%%%%%%%%%%%%%%%%%%%%%%%%%%%%%%%%%%%%%%%%%%%%%%%%%%%%%%%%%%%%%%%%%%%%%%%%%%%%%%%%%%%%%%%%%
Similarly to the Minkowski case, the discussion for $n$ gauged commuting isometries always reduces to the above, i.e.\ to just two gauged isometries with non-vanishing prepotentials, while the other gauged isometries can only induce mass terms at the $\cN=1$ point.
%%%%%%%%%%%%%%%%%%%%%%%%%%%%%%%%%%%%%%%%%%%%%%%%%%%%%%%%%%%%%%%%%%%%%%%%%%%%%%%%%%%%%%%%%%%%%%%%%%%%%%%%%%%%%%%%%%%

This concludes our analysis of the gravitino and gaugino variations. We found that in both Minkowski and AdS spacetimes partial supersymmetry breaking does not constrain the special-K\"ahler geometry, but essentially only imposes a condition on  the structure of the embedding tensor. In other words, this is a constraint on the choice of gauge vectors. In addition, two commuting isometries have to exist on the scalar field space ${\cal M}_{\rm  h}$. This imposes additional constraints in the hypermultiplet sector, to which we now turn.

 %in particular, we have to ensure that $N^\alpha_{\cal A} \epsilon^{\cal A}_1=0$ holds so that $N=1$ supersymmetry is preserved. In the next we will extend our analysis of partial supersymmetry breaking to the hyperino variation.

\subsection{The hyperino equation} \label{section:hypermultiplets}
The solution of the hyperino equation is more model
dependent. We already stated that the  quaternionic-K\"ahler manifold
${\cal M}_{\rm h}$ has to  admit two commuting isometries with Killing
prepotential $P^x_1$ and $P^x_2$ that are not proportional to each
other at the $\cN=1$ point. In addition, the $\cN=1$ hyperino supersymmetry conditions
\begin{equation} \label{Ncond}
N_{\alpha \cal A}\, \epsilon_1^{\cal A}\ =\ N_{\alpha 1}\ =\ 0\ ,\qquad
N_{\alpha \cal A}\, \epsilon^{\cal A}_2\ =\ N_{\alpha  2}\ \neq\ 0\
\end{equation}
have to be satisfied. Before we continue, let us rewrite \eqref{Ncond} in a more convenient form.
The insertion of \eqref{susytrans3} into \eqref{Ncond} and subsequent complex conjugation implies
\begin{equation}\label{kcond}
 k^u\, {\mathcal U}_{\alpha u}^{2}\ =\ 0\ , \qquad
 k^u\, {\mathcal U}_{\alpha u}^{1}\ \neq\ 0\ ,
\end{equation}
where we have defined
\begin{equation}\label{knew}
 k^u =  {V}^\Lambda \big(\Theta_\Lambda^{~~1} \kk_{1}^u +\Theta_\Lambda^{~~2} \kk_{2}^u\big)\ .
\end{equation}
By contracting the decomposition \cite{D'Auria:2001kv}
\begin{equation}\label{Udecomp}
{\mathcal U}_{\alpha u}^{\mathcal A}{\mathcal U}_v^{\mathcal B\alpha}
= - \tfrac{\iu}{2} K^x_{uv}\sigma^{x {\mathcal A}{\mathcal B}} - \tfrac12 h_{uv}
\epsilon^{{\mathcal A}{\mathcal B}}\ ,
\end{equation}
with $k^v$ and using the explicit form of the Pauli matrices \eqref{sigma_conv}
we see that \eqref{kcond} implies
\begin{equation}\label{jhol}
k^u \left(J^{1~v}_{~~u} - \iu J^{2~v}_{~~u}\right) = 0 \ , \qquad
k^u J^{3~v}_{~~u} = \iu k^{v}\ .
\end{equation}
Note that these two conditions automatically satisfy both conditions in \eqref{kcond}. The second condition of \eqref{jhol} simply states that
$k$ is holomorphic with respect to the complex structure $J^3$. Furthermore, using the relation between the three $J$'s given in \eqref{jrel}, the first equation in \eqref{jhol} follows from the second one.
For our subsequent analysis it is convenient to
define a new pair of Killing vectors $k^u_{1,2}$
by using the real and imaginary parts of the $k^u$ defined in \eqref{knew}, such that the following holds\footnote{In order to keep the notation simple we shall use the same letter $k$ to denote the original Killing vectors, as well as the redefined ones. The same holds for the respective Killing prepotentials $P^x$.}
\begin{equation} \label{Jk}
J^{3~v}_{~~u} k_1^u = -k_2^v \ , \qquad J^{3~v}_{~~u} k_2^u = k_1^v \ .
\end{equation}
Note that this is nothing more than a change of basis in the space spanned by the two Killing vectors. The coefficients in this change of basis do not depend on the coordinates of ${\cal M}_{\rm h}$, as the embedding tensor components are constant, but only on the value of $V^\Lambda$. As the related Killing prepotentials $P^x_{1,2}$ will also not be proportional to each other, we can equally use the new Killing vectors to construct a partial supersymmetry breaking solution, instead of the original Killing vectors
$\kk_{1,2}$ appearing in \eqref{d2}.

The conditions \eqref{jhol}, or equivalently \eqref{Jk}, also constrain the Killing prepotentials. Written in terms of the associated K\"ahler two-forms the first condition of \eqref{jhol} reads
\begin{equation}\label{relationkK}
 k_1^uK^1_{uv}=-k_2^uK^2_{uv} \ , \qquad k_1^uK^2_{uv}=k_2^uK^1_{uv} \ ,
\end{equation}
which, together with the definition of the prepotentials \eqref{Pdef}, implies
\begin{equation}
 P_1^1=-P_2^2 \ , \qquad P_1^2=P_2^1 \ .
\end{equation}
This in turn simplifies our embedding tensor solutions \eqref{solution_embedding_tensor}, which after a redefinition of $C^I$ read
\begin{equation}\label{solution_embedding_tensor2}
\begin{aligned}
\Theta_I^{\phantom{I}1} = & \Re\big( {\cal F}_{IJ}\,\C^J \big) \ , \qquad  \Theta^{I1} = & \Re \C^I \ , \\
\Theta_I^{\phantom{I}2} = & \Im\big( {\cal F}_{IJ}\,\C^J \big) \ , \qquad  \Theta^{I2} = & \Im  \C^I \ ,
\end{aligned}
\end{equation}
and, similarly, the AdS solutions \eqref{solution_embedding_tensor_AdS} become
\begin{equation}\label{solution_embedding_tensor_AdS2}
\begin{aligned}
\Theta_I^{\phantom{I}1} = & \Re\big({\cal F}_{IJ}\,( \C_{\rm AdS}^J - \iu \e^{K^{\rm v}/2} \tfrac{\bar \mu}{P^+_1} X^J) \big) \ , \\
\Theta^{I1} = & \Re \big(\C_{\rm AdS}^I - \iu  \e^{K^{\rm v}/2} \tfrac{\bar \mu}{P^+_1} X^I\big) \ , \\
\Theta_I^{\phantom{I}2} = &  \Im\big({\cal F}_{IJ}\,(\C_{\rm AdS}^J +\iu \e^{K^{\rm v}/2} \tfrac{\bar \mu}{P^+_1} X^J) \big) \ , \\
\Theta^{I2} = &  \Im \big(\C_{\rm AdS}^I + \iu \e^{K^{\rm v}/2} \tfrac{\bar \mu}{P^+_1} X^I\big) \ .
\end{aligned}
\end{equation}

The hyperino conditions \eqref{Jk}, or equivalently \eqref{Ncond}, are difficult to solve in general. In Section \ref{section:breaking_hyper} we will show that for special quaternionic-K\"ahler manifolds the condition \eqref{Ncond} together with all other constraints can be fulfilled. In the following, however, we do not confine our analysis to this class of manifolds but instead only assume that an $\cN=1$ solution exists, i.e.\ we assume that equations \eqref{Jk}, \eqref{solution_embedding_tensor2} and \eqref{solution_embedding_tensor_AdS2} are satisfied without specifying a particular explicit solution.

\subsection{Massive, light and massless scalars}\label{section:scales}

The Minkowski and AdS ground states described above are local $\cN=1$
minimum in $\cN=2$ field space i.e.\ the $\cN=2$ supersymmetry
variations were solved for an $\cN=1$ vacuum which can be a point in
each of ${\cal M}_{\rm  h}$ and ${\cal M}_{\rm  v}$ or a higher-dimensional
vacuum manifold. In the latter case there are
exactly flat directions (moduli) of the minimum along which $\cN=1$
supersymmetry is preserved.
In addition, there can be light scalars in the spectrum
(i.e.\ with masses much smaller than $\mino$)
with couplings that either preserve $\cN=1$ supersymmetry or spontaneously break it
at a scale beneath $\mino$. This breaking is negligible at the scale $\mino$ and therefore we also include \emph{all} light scalar fields
in the definition of the $\cN=1$ field space.
%The couplings
%of $\hat{t}$ and $\hat{q}$ could induce further supersymmetry breaking
%at a scale beneath $\mino$, which will be captured by the
%superpotential and D-terms in the effective action.
As we will see
in Sections \ref{section:NoneW} and
\ref{section:NoneD} the light fields contribute to the
superpotential and D-terms in the effective action  and any spontaneous
$\cN=1$ supersymmetry breaking will be captured by these couplings.
In the following we
denote the scalars of the $\cN=1$ field space by $\hat{t}$ and
$\hat{q}$, where there is natural split into fields descending from
the $\cN=2$ vector and hypermultiplets, respectively.

Let us now give a more specific description of the distinction between
scalars with masses of ${\cal O}(\mino)$ and massless (or light)
scalar fields. The latter are the deformations which preserve the
$\cN=1$ supersymmetry conditions \eqref{N=1conditions}. Equivalently, we can say that \eqref{jhol} and
\eqref{solution_embedding_tensor2} or
\eqref{solution_embedding_tensor_AdS2} are preserved i.e.\  the
(constant) embedding tensor is still given by \eqref{solution_embedding_tensor2} or
\eqref{solution_embedding_tensor_AdS2} across the
$\cN=1$ field space. On the other hand, any deformation that violates
the $\cN=1$ supersymmetry conditions \eqref{N=1conditions}
should have
a mass of ${\cal O} (\mino)$. Consistency of the low-energy effective
theory implies that all fields with mass of ${\cal O} (\mino)$ should
be integrated-out along with the massive gravitino.

As an example, let us consider the Minkowski solution
\eqref{solution_embedding_tensor2} at a point $t=t_0$ and determine
the deformations $t=t_0+\delta t$ which preserve
\eqref{solution_embedding_tensor2}.
This implies
\begin{equation}\label{deformV}
{\cal F}_{IJk} C^J \delta t^k = 0 \ ,
\end{equation}
which, for generic prepotential ${\cal F}$, gives
$n^\textrm{v}+1$ homogeneous equations for $n^\textrm{v}$ deformation
parameters. Therefore all $n^\textrm{v}$ scalars in the vector
multiplets are generically stabilized with masses of ${\cal O}(\mino)$
and only for specific
prepotentials an $\cN=1$ moduli space can occur for the vector scalars.
A corresponding condition
arises for the scalars of ${\cal M}_{\rm h}$ from \eqref{Ncond} or
\eqref{jhol} which, however, cannot be stated as succinctly as
\eqref{deformV}.

%%%%%%%%%%%%%%%%%%%%%%%%%%%%%%%%%%%%%%%%%%%%%%%%%%%%%%%%%%%%%%
\section{The low energy effective $\cN=1$ theory}\label{section:LEEA}
%%%%%%%%%%%%%%%%%%%%%%%%%%%%%%%%%%%%%%%%%%%%%%%%%%%%%%%%%%%%%%%%
Let us now derive the low-energy effective $\cN=1$ theory that is
valid below the scale of supersymmetry breaking set by $\mino$. This derivation is based on \cite{Louis:2010ui}.
We will begin by outlining the procedure employed and
briefly summarizing the results which we obtain.

In the previous section we discussed the properties of an $\cN=2$
supergravity that admits $\cN=1$ Minkowski or AdS backgrounds. We expect the following features should appear in general. An $\cN=1$ massive spin-3/2 multiplet with spins $s=(3/2,1,1,1/2)$ and mass $\mino$ is generated along with a set of massive $\cN=1$ chiral- and vector multiplets, whose masses are also of ${\cal O}(\mino)$. All of these multiplets  have to be integrated out to obtain the $\cN=1$ low-energy effective action.\footnote{If the $\cN=2$ theory has a supersymmetric mass scale above $\mino$ then all multiplets at that scale are also integrated out.}  At the two-derivative level this is achieved by using the equations of motion of the massive fields to first non-trivial order in $p/\mino$, where $p\ll\mino$ is the characteristic momentum.  The low-energy effective theory should then contain the leftover light $\cN=1$ multiplets, i.e.\ the gravity multiplet, $n'_{\rm v}$ vector multiplets and $n_{\rm c}$ chiral multiplets. These multiplets either have a mass  below $\mino$ or are exactly massless. The case when all the multiplets are massless arises when the $\cN=2$ supergravity is gauged with respect to just the two Killing vectors that are responsible for the partial supersymmetry breaking.  If, on the other hand, the $\cN=2$ supergravity is gauged with respect to additional Killing vectors, then some of the $\cN=1$ multiplets can have a light mass or, more generally, contribute to the $\cN=1$ effective potential. However, the derivation of the low-energy effective action is insensitive to which additional gaugings appear. Whether or not such gaugings preserve the $\cN=1$ supersymmetry or spontaneously break it only becomes clear on examining the ground states of the effective potential.

Integrating out all massive fields of order of $\mino$ in the $\cN=2$
gauged supergravity should naturally lead to an $\cN=1$ effective
Lagrangian. Its bosonic matter Lagrangian therefore has a standard form, given by
\cite{Wess:1992cp,Gates:1983nr}
\begin{eqnarray}\label{N=1Lagrangian}
  \hat{\cal L}\ =\ - \ K_{\hat A  \hat{\bar B} } D_\mu M^{\hat A} D^{\mu} \bar M^{\hat{\bar B} } - \tfrac{1}{2} f_{\hat I \hat J}\ F^{\hat I -}_{\mu\nu}F^{\mu\nu\, \hat J -} - \tfrac{1}{2}\bar{f}_{\hat I \hat J}\ F^{\hat I +}_{\mu\nu} F_{\rho\sigma}^{\hat J + } - V \ ,
\end{eqnarray}
where
\begin{eqnarray}\label{N=1pot}
  V\ =\ V_F + V_{\cal D}\ =\
  e^K \big( K^{\hat A \hat{\bar B}} D_{\hat A} {\cal W} {D_{\hat{\bar B}} \bar {\cal W}}-3|{\cal W}|^2 \big)
  +\tfrac{1}{2}\,
  (\text{Re}\; f)_{\hat I \hat J} {\cal D}^{\hat I} {\cal D}^{\hat J}
  \ .
\end{eqnarray}
We will use hatted indices to label the fields of the $\cN=1$
effective theory. $M^{\hat A} = M^{\hat A} (\hat{t},\hat{q}) $ collectively denote
all complex scalars in the theory, i.e.\ those descending from both the
vector and hypermultiplet sectors in the original ${\cal N}=2$ theory.
$K_{\hat A \hat{\bar B} } $ is a K\"ahler metric satisfying $ K_{\hat
  A \hat{\bar B} } = \partial_{\hat A} \bar\partial_{\hat{\bar B}}
K(M,\bar M)$. $F^{\hat I \pm}_{\mu\nu}$ denote the self-dual and
anti-self-dual $\cN=1$ gauge field strengths and $f_{\hat I \hat J}$
is the holomorphic gauge kinetic function. The scalar potential $V$ is determined in terms of the holomorphic
superpotential ${\cal W}$, its K\"ahler-covariant derivative $D_{\hat A}
{\cal W}=
\partial_{\hat A} {\cal W} + (\partial_{\hat A} K)\, {\cal W}$ and the
D-terms ${\cal D}^{\hat I}$, defined by
\begin{equation}\label{NoneDterm}
  {\cal D}^{\hat I}\ =\ -2\, (\text{Re}\; f)^{-1 \hat I \hat J}\, {\cal P}_{\hat J}~,
\end{equation}
where ${\cal P}_{\hat J}$ is the $\cN=1$ Killing prepotential.

The objective of this section is to compute the coupling
functions $K,{\cal W},f$ and ${\cal P}$ of the effective $\cN=1$
theory in terms of $\cN=2$ `input data'. $\cN=1$ supersymmetry
constrains $W$ and $f$ to be holomorphic while the metric $K_{\hat A
  \hat{\bar B} }$ has to be K\"ahler. Showing that the low-energy
effective theory has these properties serves as an important
consistency check of our results.

Before we turn to the derivation of these couplings let us briefly anticipate the results. One interesting aspect relates to the $\cN=1$ scalar manifold that descends from the $\cN=2$ product space
${\cal M} = {\cal M}_{\rm h}\times {\cal M}_{\rm v}$, where ${\cal M}_{\rm v}$ is
already a K\"ahler manifold but ${\cal M}_{\rm h}$ is not. In
Section~\ref{section:NoneK} we will show that integrating out the two
heavy gauge bosons in the gravitino multiplet amounts to taking a
quotient of ${\cal M}_{\rm h}$ with respect to the two gauged isometries
$k_1,k_2$ discussed in the previous section. This quotient
\begin{equation}\label{quotient}
\hat{\cal M}_{\rm h}= {\cal M}_{\rm h}/\mathbb{R}^2
\end{equation}
has co-dimension two, corresponding to the fact that the two Goldstone
bosons giving mass to the two gauge bosons have been removed. We
shall see that the quotient $\hat{\cal M}_{\rm h}$ is indeed
K\"ahler, which establishes the consistency with $\cN=1$ supersymmetry.
In order to obtain the final $\cN=1$ scalar field space, we also have to
integrate out all additional scalars that gained a mass of ${\cal O}
(\mino)$. However, these scalars are not Goldstone bosons and thus
integrating them out corresponds to simply projecting ${\cal M}_{\rm
  v}\times \hat{\cal M}_{\rm h}$ to a K\"ahler subspace ${\cal M}^{\cN=1} = \hat{\cal M}_{\rm  v} \times \hat{\cal M}_{\rm  h}\ ,$
where $\hat{\cal M}_{\rm v}$ coincides with ${\cal M}_{\rm v}$ or is a
submanifold thereof.\footnote{Note that the $\hat{\cal M}_{\rm h}$ in
  \eqref{N=1product} can also be a subspace of the $\hat{\cal M}_{\rm h}$ given
  in \eqref{quotient}, but for notational simplicity we did not
  introduce a separate symbol for this situation.}

Integrating out the two massive gauge bosons also projects the
$\cN=2$ gauge kinetic function to a submatrix. In
Section~\ref{section:Nonef} we will show that one of the two massive gauge
bosons is always given by the graviphoton.\footnote{This is related to
the fact that solutions of gravitino and gaugino conditions \eqref{constraint_embedding_tensor}
only exist because the matrix $(\Im{\cal F})_{IJ}$ has signature
$(1,n_{\rm v})$, where the positive direction precisely corresponds to
the graviphoton.
}
Integrating out this vector
leads to a holomorphic gauge kinetic function $f$ that is the second
derivative of the holomorphic prepotential on $\hat{\cal M}_{\rm v}$, similarly to the case of $\cN=1$ truncations \cite{Andrianopoli:2001zh,Andrianopoli:2001gm}.

Finally, as our $\cN=1$ effective theory descends from an $\cN=2$ supergravity, its superpotential ${\cal W}$ and the D-terms can only be non-trivial if there are additional charged scalars present, i.e.\ if there are further gaugings at a scale beneath $\mino$. As discussed above, this precisely occurs when isometries other than $k_1$ and $k_2$ are gauged in the original $\cN=2$ theory.
Since both ${\cal W}$ and ${\cal D}$ appear in the $\cN=1$ supersymmetry
transformation of the gravitino and gaugini, we can consider the corresponding $\cN=2$
supersymmetry transformations restricted to $\cN=1$ fields and then read off the appropriate terms. We will carry this out in Sections
\ref{section:NoneW} and \ref{section:NoneD}. Using the complex structure of ${\cal M}^{\cN=1}$, we will then also check the holomorphicity of ${\cal W}$ in Section~\ref{section:NoneW}.

Let us now turn to the detailed derivation of the $\cN=1$ couplings,
starting with the metric on the quotient
$\hat{\cal M}_{\rm h}$.

%%%%%%%%%%%%%%%%%%%%%%%%%%%%%%%%%%%%%%%%%%%%%%%%%%%%%%%%%%%%%%%%%%%%%%%
\subsection{The K\"ahler metric on the quotient $\hat{\cal M}_{\rm h}$}
\label{section:NoneK}

The first step in determining the sigma-model metric on the quotient
$\hat{\cal M}_{\rm h}$ is to eliminate the two massive gauge bosons via their
field equations, which are algebraic in the limit $p\ll\mino$.  In
order to be able to use the constraints \eqref{jhol} and \eqref{Jk} derived from the hyperino conditions, we
first have to rewrite the combination $\Theta_\Lambda^\lambda\kk_\lambda,~ \lambda=1,2,$
that appears in \eqref{d2} in terms of the new
Killing vectors defined in \eqref{knew}.  This change of basis can be
compensated by an appropriate change of $\Theta_\Lambda^\lambda$, such that the
covariant derivatives given in \eqref{d2} continue to have the same
form, albeit with rotated $\kk_\lambda$ and $\Theta_\Lambda^\lambda$.
{}From \eqref{sigmaint} we then obtain
\begin{equation}\label{eofA}
  \frac{\partial \cal L}{\partial A_\mu^{\lambda}} =
  -2 k^v_{\lambda} h_{uv} \partial_{\mu} q^u
  + m_{\lambda\rho}^2 A_\mu^{\rho} = 0\ ,\qquad \lambda, \rho =1,2\ ,
\end{equation}
where we have defined
\begin{equation}\label{Acomb}
  A_\mu^\lambda \equiv A_\mu^\Lambda\Theta^\lambda_\Lambda =
  A_\mu^I \Theta^\lambda_I -B_{\mu I}\Theta^{I\lambda}\ ,
\end{equation}
and its mass matrix
\begin{equation}\label{mdef}
  m_{\lambda\rho}^2 = 2 k^u_{\lambda} h_{uv} k^v_{\rho}\ .
\end{equation}
Using the quaternionic algebra \eqref{jrel} and the hyperino conditions \eqref{jhol} written in terms of the associated
K\"ahler forms $K^x$, we see that this mass matrix is diagonal
\begin{equation}
  m_{\lambda\rho}^2 = m^2 \, \delta_{\lambda\rho}  \ ,
\end{equation}
where
\begin{equation}\label{massrel}
  m^2 = 2 |k_{1}|^2 = 2 |k_{2}|^2 \ .
\end{equation}
Inserting the algebraic field equations \eqref{eofA} back into the
Lagrangian yields a modified kinetic term for the hypermultiplet
scalars, which reads
\begin{equation}
  \hat{\cal L}\ =\ \hat h_{uv} \partial_{\mu} q^u \partial^{\mu} q^v\ .
\end{equation}
$\hat h_{uv}$ is the metric on the quotient $\hat{\cal M}_{\rm h}$
and is given by
\begin{equation}\label{hq}
  \hat h_{uv} =
  h_{uv} - \frac{2k_{1 u} k_{1 v} + 2k_{2 u} k_{2 v}}{m^{2}} = \PQ^w_u h_{wv}   \ ,
\end{equation}
where
$k_{\lambda u} = k^w_{\lambda} h_{wu}$ and
\begin{equation}\label{mproj}
  \PQ^u_v=\delta^u_v - \frac{2k_{1}^{u} k_{1 v} + 2k_{2}^{u} k_{2 v}}{m^{2}}~.
\end{equation}
From \eqref{hq} it is easy to see that $\hat h_{uv}$ satisfies
\begin{equation}
  \hat h_{uv}k^v_\lambda\ =\ 0\ ,\qquad
  \hat h_{uv}h^{vw}\hat h_{wr}\ =\ \hat h_{ur}\ , \label{hid}
\end{equation}
where $h^{vw}$ is the inverse metric of the original quaternionic
manifold ${\cal M}_{\rm h}$, i.e.\ $h^{vw} h_{wu} = \delta_u^v$. We can then
use \eqref{mproj} to define the inverse metric on the quotient as
$\hat h^{uv} = \PQ^u_w h^{wv}$. The first equation in \eqref{hid} states
that the rank of $\hat h_{uv}$ is reduced by two relative to $h_{uv}$, which precisely
corresponds to the two Goldstone bosons that have been integrated
out. The second equation in \eqref{hid} tells us that the inverse
metric on the quotient $\hat h^{uv} = \PQ^u_w h^{wv}$ actually coincides
with the inverse of the original metric $h^{vw}$.

Consistency with $\cN=1$ supersymmetry requires that $\hat h_{uv}$ is a
K\"ahler metric.  In order to show this we first need to find the
integrable complex structure on the K\"ahler manifold. It seems likely
that one of the three almost complex structure of the quaternionic
manifold descends to the complex structure on the quotient. Indeed,
$J^3$ plays a preferred role in that it points in the direction (in
$SU(2)$-space) normal to the plane spanned by $P_1^x, P_2^x$ and is
left invariant by the $U(1)$ rotation in that plane. One way to
calculate $J^3$ on the quotient is to employ the same method that we
just used for the metric and apply it to the two-form $K^3_{uv}$. This
is possible in an (auxiliary) two-dimensional $\sigma$-model of the
form\footnote{This Lagrangian has nothing to do with the theory
  considered so far and is only used to derive the form of the complex
  structure -- or rather its associated fundamental two-form -- on the
  quotient.}
\begin{equation}\label{Laux} {\cal L}_{K^3} =
  K^3_{uv}  D_{\alpha} q^u D_{\beta} q^v\epsilon^{\alpha\beta}\ ,\quad
  \alpha, \beta = 1,2\ ,
  % = h_{uv} \partial_{\mu} q^u \partial^{\mu} q^v + 2h_{uv}
  % \partial_{\mu} q^u k^v_{\lambda}A_\mu^{\lambda} + \frac12
  % m_{\lambda\rho}^2 A_\mu^{\lambda}A_\mu^{\rho}\ ,
\end{equation}
where the covariant derivatives are again given by \eqref{d2}.  As
above, we derive the algebraic equation of motion for
$A_\alpha^{\lambda}$ and insert it back into \eqref{Laux} to arrive at
\begin{equation} {\cal L}_{K^3} = \hat K_{uv} \epsilon^{\alpha\beta}
  \partial_{\alpha} q^u
  \partial_{\beta} q^v\ ,
\end{equation}
where
\begin{equation}\begin{aligned}\label{Khatdef}
    \hat K_{uv}\
    =\ K^3_{uv} - \frac{2k_{2 u} k_{1 v}-2k_{1 u} k_{2 v}}{m^2} =\
    \PQ^w_u K^3_{wv} \ .
  \end{aligned}\end{equation}
Here we have used the relations \eqref{Jk} to conclude that
$k^u_\lambda K^3_{uv} k^v_\rho = m^2 \epsilon_{\lambda\rho}$, where
$\epsilon_{21}=1$.
We find that the rank of $\hat K_{uv}$ is reduced by two due to
$k^u_{\lambda} \hat K_{uv} = 0$, analogous to the result for the metric
$h_{\mu\nu}$.
%%%%%%%%%%%%%%%%%%%%%%%%%%%%%%%%%%%%%%%%%%%%%%%%%%%%%%%%%%%%%%

For two commuting isometries $k_1$ and $k_2$ we have the identity
\cite{Andrianopoli:1996cm}
\begin{equation}
  2 k_1^u k_2^v K^x_{uv} + \epsilon^{xyz} P^y_1 P^z_2 = 0  \ ,
\end{equation}
which, together with \eqref {jhol}, allows us to simplify the
expression for the mass:
\begin{equation}\label{mass_prepot}
  m^2= P^1_1 P^2_2 - P^1_2 P^2_1 \ .
\end{equation}
On the other hand, from the definition of the prepotentials
\eqref{Pdef} we find
\begin{equation} \label{Killing_oneform}
  \begin{aligned}
    & k_{2v}\ =\  k^u_1 \,K_{uv}^3 \ = \  \omega^2_v P^1_1 - \omega^1_v P^2_1 \ , \\
    & k_{1v}\ =\ k^u_2 \,K_{uv}^3 \ = \ \omega^1_v P^2_2 - \omega^2_v
    P^1_2 \ ,
  \end{aligned}
\end{equation}
where we have used \eqref{P3} and \eqref{Jk}. Inserting \eqref{Killing_oneform} and
\eqref{mass_prepot} into \eqref{Khatdef} we arrive at
\begin{equation}\label{Kform}
  \hat K_{uv} = \partial_u \omega_v^3 - \partial_v \omega_u^3\ .
\end{equation}
Thus, on $\hat{\cal M}_{\rm h}$ there exists a fundamental two-form $\hat K$
which is indeed closed
\begin{equation}\label{juhu}
  \diff\hat K = 0 \ .
\end{equation}
Furthermore, we find that $\hat J$ defined via $\hat K_{uv} = \hat h_{uw} \hat J^w_v$
is the projected complex structure $J^3$, i.e.
\begin{equation}
  \hat J^u_v = \PQ^u_w J^{3w}_{v}~,
\end{equation}
and since $\PQ$ commutes with $J^3$, due to \eqref{jhol}, $\hat J$ is the
associated complex structure, i.e.\ it satisfies $\hat J^u_v \hat J^v_w = - \PQ^u_w$, which on the subspace reads $\hat J^2 =-{\bf  1}$.\footnote{This together with \eqref{juhu} implies that the
  Nijenhuis-tensor $N(\hat J)$ vanishes.} This completes the proof that $\hat{\cal M}_{\rm h}$ is a K\"ahler manifold,
with K\"ahler form $\hat K$ and complex structure $\hat J$.

In order to display the K\"ahler potential let us explicitly introduce
complex coordinates. Since $\hat J$ is an honest complex structure, we can
group the $4n_{\rm h}-2$ coordinates $q^u$ into two sets of
coordinates $q^{2a-1}$ and $q^{2b}, a,b=1,\ldots,2n_{\rm h}-1$ such
that $\hat J$ is constant and `block-diagonal' in this basis, taking the
form
\begin{equation}
  \hat J_u^v =  \left(\begin{array}{ccccc} 0 & -1 && \\ 1 &0&&& \\ && \ddots && \\ &&&0 & -1  \\ &&&1 &0 \end{array}\right)\ .
\end{equation}
We can then define the following complex coordinates
\begin{equation}\label{zdef}
  z^a := q^{2a-1} +\iu q^{2a}\ , \quad \bar z^{\bar a} := q^{2a-1} - \iu q^{2a}
\end{equation}
and the associated derivatives
\begin{equation}\label{dzdef}
  \partial_{a} =
  \tfrac12\big(\partial_{q^{2a-1}} - \iu \partial_{q^{2a}}\big)\ ,\qquad
  \bar\partial_{\bar a} =
  \tfrac12\big(\partial_{q^{2a-1}} + \iu \partial_{q^{2a}}\big)\ .
\end{equation}
From $\hat J^w_u \hat J^t_v \hat K_{wt} = \hat K_{uv}$ we see that, in terms of
complex coordinates, the two-form $\hat K_{uv}$ given in \eqref{Kform} has
no $(2,0)$ and $(0,2)$ parts. In other words, $\hat K_{ab}=
\partial_a\omega_b^3-\partial_b\omega_a^3=0$ and $\hat K_{\bar a\bar b}=
\bar \partial_{\bar a} \bar \omega_{\bar b}^3-\bar \partial_{\bar b}
\bar \omega_{\bar a}^3=0$ . This in turn implies\footnote{Note that
  one could add a further term in \eqref{Kahlerconnection} that does
  not contribute in \eqref{Kform} and corresponds to a K\"ahler
  transformation.}
\begin{equation}\label{Kahlerconnection}
  \omega_a^3 = \tfrac\iu2 \partial_a \hat K\ ,\qquad
  \bar\omega_{\bar a}^3 = -\tfrac\iu2 \bar\partial_{\bar a} \hat K\ ,
\end{equation}
where $\hat K$ is the (real) ${\cN=1}$ K\"ahler potential.  Inserting these expressions into
\eqref{Kform} one obtains the K\"ahler-form
\begin{equation}\label{Kdef_N=1}
  \hat K_{a\bar b} =  \partial_a\bar\omega_{\bar b}^3-\bar\partial_{\bar b}\omega_a^3
  = -\iu\partial_a\bar\partial_b \hat K\ .
\end{equation}

So far, we have only integrated out the two vector bosons of the
massive gravitino multiplet including their Goldstone degrees of
freedom.  As we have just shown, the removal of the two Goldstone bosons
amounts to taking the quotient of the original quaternionic-K\"ahler
manifold ${\cal M}_{\rm h}$ with respect to the two gauged isometries
$k_{1,2}$. This quotient $\hat{\cal M}_{\rm h}= {\cal M}_{\rm h}/\langle k_1,k_2 \rangle$ has
co-dimension two and is indeed a K\"ahler manifold consistent with the
unbroken $\cN=1$ supersymmetry.  However, additional scalars, both from vector and/or hypermultiplets, can acquire a mass of ${\cal O}(\mino)$ due to the partial supersymmetry breaking. Integrating out these scalar fields results in a submanifold $\hat{\cal M}_{\rm v}$ of the original $\cN=2$ special K\"ahler manifold ${\cal M}_{\rm v}$ and a submanifold of $\hat{\cal M}_{\rm h}$. Thus, the final $\cN=1$ field space is the K\"ahler manifold
\begin{equation}\label{NoneMod}
 {\cal M}^{\cN=1} = \hat{\cal M}_{\rm  v} \times \hat{\cal M}_{\rm h}
\end{equation}
with K\"ahler potential
\begin{equation}\label{Kone}
  K^{\cN=1} = \hat K^{\rm v} +\hat K \ .
\end{equation}

%%%%%%%%%%%%%%%%%%%%%%%%%%%%%%%%%%%%%%%%%%%%%%%%%%%%%%%%
\subsection{The gauge couplings}
\label{section:Nonef}

Let us now check the holomorphicity of the gauge couplings.  In
section~\ref{section:NoneK} we integrated out the two heavy gauge
bosons in the low-energy limit by neglecting their kinetic terms and
using their algebraic equations of motion. In order to compute the
gauge couplings of the remaining light gauge fields that descend to
the $\cN=1$ theory we have to explicitly project out the heavy gauge
bosons in the coupled kinetic terms in \eqref{sigmaint}. From
\eqref{Acomb} we see that the projection is determined by the
embedding tensor solutions given in \eqref{solution_embedding_tensor2}. In other
words, we should impose the projection
\begin{equation}\label{Fcond}
  \Theta^{\lambda I} G\,_I^\pm +  \Theta^{\lambda}_I\, F^{I \pm}\ =\ 0\ ,\qquad \lambda=1,2 \ ,
\end{equation}
and then compute the gauge couplings of the remaining gauge fields.
Taking complex combinations and inserting the embedding tensor solutions
\eqref{solution_embedding_tensor2} yields\footnote{We only discuss the
  Minkowski case here. The AdS case is completely equivalent, in that
  \eqref{solution_embedding_tensor_AdS2} only leads to a different
  prefactor (i.e.\ not $C^I$) but the conclusion remains the same.}
\begin{equation}\label{Fint}
  C^I ({\cal F}_{IJ}(\hat{t}) -  \mathcal{N}_{IJ}(\hat{t})) F^{J +}\ =\ 0 \ =\ \bar{C}^I
  (\bar{\cal F}_{IJ}(\hat{t}) -  \mathcal{N}_{IJ}(\hat{t})) F^{J +}
\end{equation}
and a similar set of equations for $F^{J -}$, where we have restricted $\mathcal{N}_{IJ}$ to $\cN=1$ fields. Note that
${\cal F}_{IJ}$ is evaluated in the $\cN=1$ background, which means that
scalar fields not obeying \eqref{deformV} are fixed at their
background values, while the scalars of the $\cN=1$ theory that do
obey \eqref{deformV}, which we denote by $\hat{t}$, can vary
arbitrarily, see Section \ref{section:scales}.

Using the definition of $\mathcal{N}_{IJ}$ \eqref{Ndef} we find that \eqref{Fint} implies
\begin{equation}\label{Fproj}
  X^I (\Im{\cal F})_{IJ} (\hat{t}) F^{J +}\ %\Big|_{\cN=1}
  =\ 0 \ ,
\end{equation}
where we have dropped a non-vanishing prefactor.
This condition projects out one linear combination of the $F^{I}$
that is heavy. Note that from now on we shall not explicitly write the $\hat{t}$-dependence. For the following analysis it will be useful to define
the related projection operator
\begin{equation}\label{Pidef}
  \bar \Pi^I_J \equiv \delta^I_J + 2 \e^{K^{\rm v}} \bar X^I X^K (\Im{\cal F})_{KJ} \ ,
\end{equation}
such that $(1-\bar\Pi)$ projects onto the heavy gauge boson while
$\bar\Pi$ projects onto the orthogonal gauge bosons.

Before we identify the second heavy gauge boson let us check which
physical field is projected out by \eqref{Fproj}. Looking at the full
$\cN=2$ gravitino variation \cite{Andrianopoli:1996cm}, we see that it
contains the `dressed' graviphoton term
\begin{equation}
  \tilde T_{\mu\nu}^{+} = -2 \iu X^{I} (\Im\mathcal{\bar N})_{IJ} F^{J + }_{\mu\nu} + \ldots~.
\end{equation}
It is straightforward to check that the projection $X^{I}
(\Im\mathcal{\bar N})_{IJ}$ appearing here coincides with
\eqref{Fproj}. Therefore, \eqref{Fproj} can be understood as
projecting out the graviphoton.

The second projection condition implied by \eqref{Fint} reads
\begin{equation} \label{FprojC}
C^{(P)\,I} (\Im{\cal F})_{IJ} F^{J+}\
  =\ 0 \ ,
\end{equation}
where we have defined $C^{(P)\,I}= \Pi^I_JC^J$.  Expressing this in terms of
the projection operator
\begin{equation}\label{Gammadef}
  \bar \Gamma_{J}^I \equiv \delta^I_J - \frac{\bar  C^{(P)\,I}   C^{(P)\,K}  (\Im{\cal F})_{KJ}}{ C^{(P)\,M}  (\Im{\cal F})_{MN} \bar  C^{(P)\,N}} \ ,
\end{equation}
we see that $(1-\bar\Gamma)$ projects onto the second heavy gauge
boson while $\bar\Gamma$ projects to the orthogonal gauge bosons.
With the help of the two projection operators, which one can show commute, we are now in the
position to define the light vector fields which remain in the $\cN=1$
theory by
\begin{equation} \label{FN=1} F^{\hat I +} \equiv F^{I +}\Big|_{\cN=1}
  = \bar \Pi^I_J \bar \Gamma^{J}_K F^{K +} \ ,
\end{equation}
where \ $\hat I = 1,\dots n'_v = (n_{\mathrm v}-1)$, i.e.\ we have
projected out two of the $\cN=2$ vectors.

Let us now return to our original task and compute the gauge coupling
functions of the $\cN=1$ action. This can be done by imposing the two
projections \eqref{Fproj} and \eqref{FprojC} in the gauge kinetic term
$\mathcal{N}_{IJ}\,F^{I +}_{\mu\nu}F^{\mu\nu\, J+}$ of
\eqref{sigmaint}. In other words, we should compute
$\mathcal{N}_{\hat I\hat J}\,F^{\hat I +}_{\mu\nu}F^{\mu\nu\, \hat
  J+}$ with $F^{\hat I +}$ given by \eqref{FN=1}. Inserting the
definition of $\mathcal{N}_{I J}$ \eqref{Ndef} we find that the
$\cN=1$ gauge coupling functions appearing in \eqref{N=1Lagrangian}
are giving by
\begin{equation}\label{fN=1}
  \bar f_{\hat I\hat J}(\hat{t})\ =\ -\iu \bar {\cal F}_{\hat I\hat J}\  \ ,
\end{equation}
where the second term of $\mathcal{N}_{IJ}$ drops out due to the
identity
\begin{equation}\label{Fproj2}
  X^I (\Im{\cal F})_{I\hat J} F^{\hat J +}\ =\ 0 \ ,
\end{equation}
which can be shown by inserting \eqref{FN=1} and using $e^{-K_{\rm v}}
= -2 \bar X^I {\rm Im}({\cal F})_{IJ} X^J$.

As promised, we see that the gauge couplings are manifestly
holomorphic. Furthermore, $f_{\hat I\hat J}(t)$ can only depend on the
scalar fields that descend from $\cN=2$ vector multiplets, but not on
those descending from hypermultiplets. In fact, this is analogous to
the situation in $\cN=2\to \cN=1$ truncations, where the graviphoton
also has to be projected out and, as a consequence, the gauge couplings
are holomorphic and only depend on the scalars of the vector
multiplets \cite{Andrianopoli:2001zh,Andrianopoli:2001gm}.

%%%%%%%%%%%%%%%%%%%%%%%%%%%%%%%%%%%%%%%%%%%%%%%%%%%%%%%%%%%%%%%%%%%%%%%
\subsection{The superpotential}
\label{section:NoneW}

Our next task is to determine the $\cN=1$ superpotential ${\cal W}$.
This is most easily done by comparing the supersymmetry transformation of the
$\cN=1$ gravitino $\Psi_{\mu\, 1}$ \eqref{susytrans2} with
the conventional $\cN=1$ transformation given, for example, in
\cite{Wess:1992cp}. (An analogous computation
for $\cN=1$ truncations of $\cN=2$ theories can be found in
\cite{Grana:2005ny,Andrianopoli:2001zh,Andrianopoli:2001gm}).  Focusing
on the scalar contribution one has
\begin{equation}\label{Nonegravitino}
  \delta_\epsilon \Psi_{\mu\,1}\ =\ D_\mu \epsilon - S_{11}
  \gamma_\mu \bar \epsilon  + \ldots \,\ =\ D_\mu \epsilon - \e^{\tfrac12K^{\cN=1}} {\cal W}
  \gamma_\mu \bar \epsilon  + \ldots \ ,
\end{equation}
where we have already inserted our choice $\epsilon_1
= {\epsilon\choose 0}$.

Using the definition of the gravitino mass matrices \eqref{susytrans3}
we find that the $\cN=1$ superpotential is given by
\begin{equation}\label{Wone}
  {\cal W}\ =\ \e^{-\tfrac12K^{\cN=1}} S_{11}\ =\ \tfrac{1}{2} \e^{-\hat K/2} {V}^\Lambda
  \Theta_\Lambda^{~~\lambda} P_{\lambda}^- \ .
\end{equation}
In this expression we have to appropriately project out all scalars
with masses of ${\cal O}(\mino)$. In other words, ${\cal W}$ should be
expressed in terms of $\cN=2$ input couplings restricted to the light
$\cN=1$ modes. As we discussed at the end of section \ref{section:NoneK},
this projection preserves the K\"ahler and complex structure
of ${\cal M}_{\rm v} \times \hat{\cal M}_{\rm h}$. Therefore, we should be able to
check the holomorphicity of ${\cal W}$ without knowing the precise
$\cN=1$ spectrum.

Before we continue,  let us
discuss the situation where the original $\cN=2$
supergravity is only gauged with respect to the two Killing vectors
$k_1,k_2$ that induce the partial supersymmetry breaking.
In this case the index $\lambda$
in \eqref{Wone} only takes the values $\lambda=1,2$ and
all  fields in the $\cN=1$ effective theory are exactly massless, i.e.\ they are $\cN=1$ moduli. Their vacuum expectation
values are not fixed, or, in other words, they parameterize the entire
$\cN=1$ background. As a consequence the superpotential has to be proportional
to the cosmological constant.  This can be seen explicitly by inserting the gravitino mass matrix
\eqref{N=1conditions} into  \eqref{Wone} which gives
\begin{equation}
|{\cal  W}|^2= \e^{-K^{\cN=1}} |S_{11}|^2 = -3 \e^{-K^{\cN=1}} |\mu|^2 \ ,
\end{equation}
in agreement with the standard $\cN=1$ relation \cite{Wess:1992cp}.

If additional Killing vectors are gauged, then their corresponding Killing prepotentials appear in \eqref{Wone}  and the index $\lambda$ runs over all non-trivial Killing directions. For this case we will now show that ${\cal W}$ is holomorphic with respect to the $\cN=1$ complex structure determined in the previous section.

Inspecting the superpotential ${\cal W}$ \eqref{Wone} we see that the scalars of ${\cal M}_{\rm v}$ already appear holomorphically via $V^\Lambda$. Therefore, we are left to show that the anti-holomorphic derivative of ${\cal W}$ with respect to the scalars of $\hat{\cal M}_{\rm h}$ vanishes, i.e.\
\begin{equation}\label{der_W}
  \bar\partial_{\bar a} {\cal W} = \tfrac12 \e^{-\hat K/2} {V}^\Lambda
  \Theta_\Lambda^{~~\lambda} (\bar \partial_{\bar a} P_\lambda^- - \tfrac12 (\bar \partial_{\bar a} \hat K)P_\lambda^-)  = 0\ .
\end{equation}
Let us first note that using \eqref{Kahlerconnection} we can
express $\bar \partial_{\bar a} \hat K$ in terms of $\omega^3_{\bar a}$.
Furthermore, from the definition of Killing prepotentials \eqref{Pdef}
we see that
\begin{equation}
  -2 K^-_{uv} k^v_\lambda = \partial_u P^-_\lambda + \iu \omega^-_u P^3_\lambda - \iu \omega^3_u P^-_\lambda \ ,
\end{equation}
which implies
\begin{equation}\label{der_W2}
  \bar\partial_{\bar a} {\cal W} = - \tfrac12 \e^{-\hat K/2} {V}^\Lambda
  \Theta_\Lambda^{~~\lambda} (2 K^-_{\bar a v} k^v_\lambda + \iu \bar \omega^-_{\bar a} P_\lambda^3) \ .
\end{equation}
From the quaternionic algebra \eqref{jrel} and $K^x_{uv} = h_{uw} (J^x)^w_v$ it is easy to see that $K^-$ is actually a $(2,0)$-form and thus only has  holomorphic indices. This immediately implies that the first term in the bracket vanishes. From \eqref{Killing_oneform} we can infer that both $\omega^1$ and $\omega^2$ live entirely in the space spanned by $k_{1v}$ and $k_{2v}$, which in fact is divided out. This implies that $\omega^-_{\bar a}$ is zero on $\hat{\cal M}_{\rm h}$ and therefore the second term in \eqref{der_W2} also vanishes. Thus, the superpotential ${\cal W}$ is holomorphic, consistent with $\cN=1$ supersymmetry.

%%%%%%%%%%%%%%%%%%%%%%%%%%%%%%%%%%%%%%%%%%%%%%%%%%%%%%%%%%%%%%%%%%%%%%%%%%%%%%%

\subsection{The D-terms}
\label{section:NoneD}

Our final task is to explicitly compute the $\cN=1$ D-terms appearing in the effective potential \eqref{N=1pot}. This proceeds analogously to the calculation of the superpotential in Section \ref{section:NoneW}, but by comparing the $\cN=2$ and $\cN=1$ gaugino variations instead of the gravitino variations. Once again, this procedure is similar to the one used in  $\cN=1$ truncations \cite{Andrianopoli:2001zh,Andrianopoli:2001gm}, but here we shall more closely follow the review given in \cite{Cassani:2007pq}.

The $\cN=2$ gaugino variation is given by \cite{Andrianopoli:1996cm}
\begin{equation}\label{gauginivar}
  \delta_\epsilon \lambda^{i {\cal A}} = \gamma^\mu\partial_\mu t^i \epsilon^{\cal A} - \tilde G_{\mu\nu}^{i -} \gamma^{\mu\nu} \varepsilon^{\cal AB}\epsilon_{\cal B} + W^{i{\cal AB}}\epsilon_{\cal B}+\ldots~,
\end{equation}
where $W^{i {\cal AB}}$ was defined in \eqref{susytrans3} and $\tilde
G_{\mu\nu}^{i -} = -g^{i\bar j} \nabla_{\bar j}\bar X^{I}
(\Im\mathcal{N})_{IJ} F^{J - }_{\mu\nu} + \ldots$ are the
`dressed' anti-self-dual field strengths, with the ellipses denoting higher-order fermionic contributions.

In order to identify the gaugini of the effective $\cN=1$ theory we evaluate \eqref{gauginivar} for our choice of the preserved supersymmetry parameter $\epsilon_1 = {\epsilon\choose 0}$ and obtain
\begin{eqnarray}
  \delta_\epsilon \lambda^{i 1} &=& \gamma^\mu\partial_\mu t^i
\bar\epsilon +  W^{i{11}} \epsilon~ +\ldots~, \label{trunc-fermion} \\
  \delta_\epsilon \lambda^{i 2} &=& - \tilde G_{\mu\nu}^{i -}
\gamma^{\mu\nu} \epsilon +  W^{i{ 21}}  \epsilon~ +\ldots~. \label{trunc-gaugino}
\end{eqnarray}
Comparing with the standard $\cN=1$ gaugino variation \cite{Wess:1992cp,Gates:1983nr}
\begin{equation}
  \delta_\epsilon \lambda^{\hat{I}} =  F_{\mu\nu}^{\hat{I} -}
\gamma^{\mu\nu} \epsilon + \mathrm{i} {\cal D}^{\hat I} \epsilon~ +\ldots \ , \label{NoneGauginoVar}
\end{equation}
we conclude that the $\lambda^{i 2}$ are candidates for $\cN=1$ gaugini.  However, not all $\lambda^{i 2}$ descend to the effective $\cN=1$ theory as some of them are massive and have to be integrated out. The $\cN=1$ gaugini should be defined as those with the light $\cN=1$ gauge fields \eqref{FN=1} appearing in their supersymmetry variations. Using the projection operators $\Pi$ \eqref{Pidef} and $\Gamma$ \eqref{Gammadef} and the definition \eqref{FN=1}, we can restrict the gauge fields appearing in $\cN=2$ gaugino variation \eqref{trunc-gaugino} to the light $\cN=1$ gauge fields. By comparing the resulting expression with the $\cN=1$ gaugino variation \eqref{NoneGauginoVar}, we can identify the $\cN=1$ gaugini as
\begin{equation}\label{NoneGaugino}
  \lambda^{\hat{I}}\ =\ -2e^{K^{\rm v}/2} \nabla_{i} X^{\hat I} \lambda^{i 2}  \ ,
\end{equation}
where we have used the same projector \eqref{FN=1} to define $\nabla_{i}X^{\hat I} = \Pi^I_J  \Gamma^{J}_K \nabla_{i}X^K$. In order to reach this result, we have made use of the special geometry
relation \cite{Andrianopoli:1996cm}
\begin{equation}\label{special}
  \nabla_{i} X^{\hat I}  g^{i\bar \jmath}\, \nabla_{\bar \jmath}\bar {X}^{\hat J} = -\tfrac{1}{2} e^{-K^{\rm v}}(\mathrm{Im} \cN)^{-1 ~\hat I \hat J} - X^{\hat I}  \bar {X}^{\hat J}
\end{equation}
and also that the projector \eqref{Pidef} is defined such that the following property holds
\begin{equation}\label{Xcond}
  X^{\hat I}= \Pi^I_J  \Gamma^{J}_K X^K =0 \ .
\end{equation}

We can now take the $\cN=1$ supersymmetry variation of \eqref{NoneGaugino}
(to lowest fermionic order), use \eqref{trunc-gaugino}, insert the
definition of $W^{i{ 21}}$ \eqref{susytrans3} and compare the result
with the standard $\cN=1$ expression \eqref{NoneGauginoVar} to read
off the D-term:
\begin{eqnarray}
  {\cal D}^{\hat I} &=& 2\mathrm{i}e^{K^{\rm v}/2} \nabla_{i} X^{\hat I} W^{i{ 21}} \nonumber \\
  &=&-2 e^{K^{\rm v}} \nabla_{i} X^{\hat I} g^{i\bar \jmath}\,
  \nabla_{\bar \jmath}\bar {X}^{\hat J}\left( \Theta_{\hat
      J}^{~~\lambda} - {\cal N}_{\hat J \hat K} \Theta^{\hat
      K\lambda}\right) P_{\lambda}^3 ~, \label{Dterm}
\end{eqnarray}
where we have used $\nabla_{i} F_{\hat J} = {\cal F}_{\hat J \hat K}\nabla_{i} X^{\hat K}$ in the second line. In order to see that this expression agrees with the standard $\cN=1$ D-term \eqref{NoneDterm}, we again make use of \eqref{special} and \eqref{Xcond} to see that it can be written as
\begin{equation}\label{D_terms}
  {\cal D}^{\hat I} = - (\mathrm{Re} f)^{-1 ~\hat I \hat J} \left( \Theta_{\hat J}^{~~\lambda} - f_{\hat J \hat K} \Theta^{\hat K\lambda}\right) P_{\lambda}^3 \ .
\end{equation}
This result agrees with the standard $\cN=1$ supergravity expression \eqref{NoneDterm} if we identify the $\cN=1$ Killing prepotential as follows
\begin{equation}\label{N=1prepotential} {\cal P}_{\hat J} =
  \tfrac{1}{2} \left( \Theta_{\hat J}^{~~\lambda} - f_{\hat J \hat K}
    \Theta^{\hat K\lambda}\right) P_{\lambda}^3 \ .
\end{equation}
If we now consider gaugings with respect to just the Killing vectors $k_1$ and $k_2$ responsible for partial supersymmetry breaking, we see that the D-term vanishes by our $\cN=1$ supersymmetry condition \eqref{P3}, as expected for a supersymmetric vacuum.

Note that both the D-terms \eqref{D_terms} and the Killing prepotentials \eqref{N=1prepotential} are complex,
in agreement with the analogous results from $\cN=1$ truncations \cite{D'Auria:2005yg,Cassani:2007pq}. The reason is that these quantities appear in the supersymmetry variations of the gaugini in \eqref{trunc-gaugino} which are paired with the (complexified) anti-self-dual field strengths $\tilde G_{\mu\nu}^{i -}$. Therefore, \eqref{D_terms} describes a complex linear combination of the electric and the magnetic D-terms. More precisely, from \eqref{N=1prepotential} we see that the electric and magnetic Killing prepotentials of the $\cN=1$ theory are given by $\frac{1}{2} \Theta_{\hat J}^{~~\lambda} P_{\lambda}^3$ and $\frac{1}{2} \Theta^{\hat K\lambda} P_{\lambda}^3$.\footnote{We thank G.\ Dall'Agata and D.\ Cassani for discussions on this points.}

Before we close this section let us
note that one can also check that the supersymmetry
transformation of the $\cN=1$ fermions in chiral multiplets that descend
from the $\cN=2$ gaugini $\lambda^{i1}$ (cf.~\eqref{trunc-gaugino})
correctly reproduce the F-terms. Furthermore,
one might expect that it is necessary to take field redefinitions of
the gaugini and the hyperini with respect to the Goldstino, such that
we can rewrite the fermionic Lagrangian in terms of physical fermions,
i.e.\ fermions that cannot be gauged away by further field
redefinitions of the massive gravitino $\Psi_{\mu 2}$ \cite{Gunara:2003td}.
However, it is
straightforward to check that any such field redefinitions are
projected out when one identifies the $\cN=1$ fields as in
\eqref{NoneGaugino}. In other words, the $\cN=1$ fermionic field space
is defined by quotienting the $\cN=2$ counterpart by the Goldstino
direction.

Let us draw attention to an important difference with $\cN=2 \rightarrow \cN=1$ supergravity truncations\cite{D'Auria:2005yg,Andrianopoli:2001gm}. For the case of partial supersymmetry breaking considered here, the condition \eqref{Xcond} does not fix any scalars, as the projection operators $\Pi^I_J$ and $\Gamma^{J}_K$ are field-dependent quantities which vary over the $\cN=1$ moduli space in such a way that \eqref{Xcond} is automatically fulfilled. In the case of $\cN=1$ truncations \cite{Andrianopoli:2001zh,Andrianopoli:2001gm},  the equivalent projection operators are constant and therefore some scalars are projected out by the condition $\Pi^I_J  X^J =0$.

This completes our analysis of the low-energy effective theory in the
$\cN=1$ vacua of $\cN=2$ gauged supergravity with electric and magnetic charges. We have proven that this theory enjoys $\cN=1$ supersymmetry, as is required for the consistency of
the partial supersymmetry breaking mechanism. We shall now focus on a
specific class of c-map examples, where the hypermultiplet scalars
parameterize a special quaternionic-K\"ahler manifold.

%%%%%%%%%%%%%%%%%%%%%%%%%%%%%%%%%%%%%%%%%%%%%%%%%%%%%%%%%%
\section[$\cN=2\to \cN=1$ breaking on special quaternionic manifolds]{Partial supersymmetry breaking on special qua\-ter\-nionic-K\"ahler manifolds}
\label{section:breaking_hyper}
%%%%%%%%%%%%%%%%%%%%%%%%%%%%%%%%%%%%%%%%%%%%%%%%%%%%%%%%%%
In Section \ref{section:hypermultiplets} we found that in order to realize $\cN=1$ vacua we need to have two commuting isometries on ${\cal M}_{\rm  h}$ which are furthermore holomorphic with respect to $J^3$. This is certainly not satisfied on a generic quaternionic-K\"ahler manifold and so ${\cal M}_{\rm  h}$ is constrained from the outset by this requirement. It is difficult to analyze this condition on an arbitrary ${\cal M}_{\rm  h}$ which admits two isometries. To proceed, we shall focus our attention on the subclass of special quaternionic-K\"ahler manifolds \cite{Cecotti:1988qn,Ferrara:1989ik}, which we already discussed in Section \ref{section:special_quat}. These manifolds arise at string tree-level in $SU(3)\times SU(3)$-structure compactifications of type II string theories. Beyond their interest in string compactifications, we have chosen to concentrate on this specific subclass as they have a large number of isometries.

Let us now return to the conditions for partial supersymmetry breaking arising from the hypermultiplet sector. The initial analysis in this section follows \cite{Cassani:2007pq}. It will be useful in the following to express the parameter of the unbroken $\cN=1$ supersymmetry in terms of a vector of complex coefficients
\begin{equation} \label{SUSYgenerator}
\epsilon_1^{\cal A}   = \left( \begin{aligned} n^1 \\ n^2 \end{aligned} \right) \epsilon_1 \ ,
\end{equation}
where the Killing spinor $\epsilon_1$ is the generator of the unbroken supersymmetry in $\cN=1$ notation. Inserting \eqref{SUSYgenerator} and \eqref{prepotential_no_compensator} into the gravitino equation \eqref{N=1conditions}, we obtain
\begin{equation}\label{qS}
 \begin{aligned}
 n^1 u(\kk) + \tfrac{1}{4} n^2 (v-\bar v)(\kk)= \tfrac \iu2
(n^1)^*\,  \e^{-K^{\rm v}/2} \mu \ , \\
 \tfrac{1}{4}n^1 (v-\bar v)(\kk) + n^2 \bar u(\kk) = \tfrac
\iu2 (n^2)^*\,  \e^{-K^{\rm v}/2} \mu  \ ,
 \end{aligned}
\end{equation}
where we have used the following abbreviations for the Killing vectors $k=k^u \partial_u$:
\begin{equation}\label{kVdef}
\kk \equiv {V}^\Lambda \Theta^{\ \lambda}_\Lambda
k_\lambda\ , \qquad \textrm{and} \qquad u(\kk)\equiv \kk^v u_v\ .
\end{equation}
In deriving \eqref{qS}, we also used the fact that the Killing vectors $k_\lambda$, defined in \eqref{Killing}, do not have a component in the base directions, i.e.\ $\diff Z^I (k_\lambda) = 0$ holds.

Turning to the hyperino equation \eqref{N=1conditions}, and making use of \eqref{susytrans3}, \eqref{quat_vielbein},
\eqref{one-forms_quat} and \eqref{SUSYgenerator}, we find
\begin{equation}\label{qH}
 \begin{aligned}
n^1 u(\kk) +n^2 v(\kk)  &= 0 \ , \\
  -n^1 \bar v(\kk)+n^2 \bar u(\kk)  &= 0
 \end{aligned}
\end{equation}
and
\begin{equation}\label{qE}
\begin{aligned}
  n^2 E^{\underline b}(\kk)  &= 0 \ , \\
  n^1 \bar E^{\underline b}(\kk) &= 0 \ .
\end{aligned}
\end{equation}
In \eqref{qE} we have used that all Killing vectors \eqref{Killing} are in the fibre directions and therefore $e(\kk)=\bar e(\kk)=0$. If we now take the difference of the gravitino \eqref{qS} and hyperino \eqref{qH} conditions, we arrive at
\begin{equation} \label{cmap_vbarv}
\begin{aligned}
n^2 (3v+\bar v) (\kk)  &= - 2 \iu (n^1)^* \e^{-K^{\rm v}/2} \mu  \ , \\
n^1 (v+3\bar v) (\kk)  &= 2 \iu (n^2)^* \e^{-K^{\rm v}/2} \mu  \ .
 \end{aligned}
\end{equation}
Here we see that possible solutions for Minkowski and AdS vacua preserving $\cN=1$  supersymmetry differ significantly due to the $\mu$-term on the right-hand side of \eqref{cmap_vbarv}. By comparing \eqref{cmap_vbarv} with the original hyperino constraint \eqref{qH}, we see that the only way to solve the conditions for a Minkowski vacuum with both $n^1$ and $n^2$ non-zero is to set $v(\kk)=\bar v(\kk)= 0$. As we shall describe further in the next section, one can then easily check that such a vacuum preserves $\cN=2$ supersymmetry \cite{Cassani:2007pq}. Therefore, in order to find an honest $\cN=1$ vacuum we are forced to set $n^1$ or $n^2$ to zero. On the other hand, for AdS vacua  a similar check shows that $n^1$, $n^2$ and $v(\kk)$ must all be non-zero in order to solve \eqref{cmap_vbarv}. Due to the different nature of these possible solutions, we analyze the Minkowski and AdS cases separately in the following.

%%%%%%%%%%%%%%%%%%%%%%%%%%%%%%%%%%%%%%%%%%%%%%%%%%%%%%%%%%%%%%%%%%%%%%%%%%
\subsection{Minkowski vacua}\label{section:Minkowski_vacua}
We will first consider the case of a Minkowski vacuum, setting $\mu=0$ in all the expressions above. As we have just discussed, there are two cases to consider, depending on whether both $n^1$ and $n^2$ are non-zero or not \cite{Frey:2003sd}. If both $n^1$ and $n^2$ are non-zero, one sees from \eqref{cmap_vbarv} that $(v-\bar v)(\kk)=0$ and then the original hyperino conditions \eqref{qH} implies that $u(\kk)=\bar u(\kk)=0$. Inserting this into  \eqref{prepotential_no_compensator} and \eqref{quat_connection} we see that all three prepotentials $P^x$ vanish separately and the vacuum actually has $\cN=2$ supersymmetry \cite{Cassani:2007pq}.\footnote{It is important
to keep in mind that this conclusion crucially depends on the fact that we confine our analysis to the Killing vectors \eqref{Killing}
which correspond to translations in the fibre. If on the other hand isometries in the special K\"ahler base exist, partial supersymmetry might be possible for this case.} If we consider instead the case where one of the components of $n^{\mathcal A}$ is zero we can evade this conclusion. In the remainder of this section we will show that such a solution does exist, and that the conditions for preserved $\cN=1$ supersymmetry \eqref{N=1conditions} can be solved for two commuting isometries.

To proceed, we will set one of the complex coefficients in \eqref{SUSYgenerator} to zero
\begin{equation}\label{nchoice}
n^2=0\ , \qquad n^1\neq 0\ .
\end{equation}
This leads to a simplified set of gravitino \eqref{qS} and hyperino \eqref{qH}, \eqref{qE} equations to solve:
\begin{equation} \label{condition_hypers_M}
v(\kk)=\bar v(\kk)=u(\kk)=\bar E^{\underline b} (\kk) = 0 \ ,
\end{equation}
with $\bar u(\kk)$ and $E^{\underline b}(\kk)$ undetermined. In order to avoid an $\cN=2$ vacuum we must ensure that $\bar u(\kk) \ne 0$, such that $P^x$ does not vanish and we can have the possibility of partial supersymmetry breaking. As we will see, this implies $E^{\underline b}(\kk) \ne 0$.

Our first task is to construct two commuting Killing vectors $k_1$ and $k_2$ out of the set provided by the c-map construction \eqref{isometries_fibre}. By considering the inner product of the quaternionic one-forms \eqref{one-forms_quat} with the Killing vectors \eqref{isometries_fibre}, we see that $\Kdil$ is not a good choice for our purposes as $(v+\bar v) (\Kdil) \ne 0$. Therefore, if we were to use this Killing vector we would not be able to satisfy the $\cN=1$ vacuum conditions \eqref{condition_hypers_M}. This leads us to make the following general ansatz in terms of the remaining Killing vectors
\begin{equation}\label{ex_Killing_vectors}
 \begin{aligned}
  k_1 =  \P_1^B \Kxi_B + \Q_{1\, A} \Ktxi^A + \A_1 \Kax \ , \\
  k_2 =  \P_2^B \Kxi_B + \Q_{2\, A} \Ktxi^A + \A_2 \Kax \ ,
 \end{aligned}
\end{equation}
where for the moment $\P_{1,2}^B, \Q_{1,2\, A}, \A_{1,2}$ are arbitrary real coefficients. By demanding that $k_1$ and $k_2$ commute, we then find a constraint on the coefficients\footnote{At this point, we can already see that we cannot have partial supersymmetry breaking in Minkowski space with just the universal hypermultiplet as the condition \eqref{comm_isometries} reads
\begin{displaymath}
\operatorname{det} \left( \begin{aligned}\P_1 && \P_2 \\ \Q_1 && \Q_2  \end{aligned} \right) = 0 \ .
\end{displaymath}
This in turn means that $k_1$ and $k_2$ are actually linearly dependent, i.e.\ only one linear combination of $\Kxi_A$ and $\Ktxi^A$ is gauged, the prepotentials $P^x_1$ and $P^x_2$ are aligned and no $\cN=1$ solution can be constructed, cf.\ Section~\ref{section:magnetic_vectors}.
}
\begin{equation} \label{comm_isometries}
  \P_1^A \Q_{2\, A} - \P_2^A \Q_{1\, A} = 0 \ .
\end{equation}
If we consider the inner product of the quaternionic one-forms \eqref{one-forms_quat} with our ansatz for the Killing vector combinations \eqref{ex_Killing_vectors}, we immediately observe that both $k_1$ and $k_2$ automatically satisfy the conditions $(v+\bar v) (k_{1,2}) = 0$ , while $(v-\bar v)(\kk) = 0$ imposes
\begin{equation}
{V}^\Lambda \Theta_\Lambda^{\ 1} ( \Q_{1\, A} \xi^A - \P_1^A \tilde\xi_A + \A_1 )  + {V}^\Lambda \Theta_\Lambda^{ \ 2} (\Q_{2\, A} \xi^A - \P_2^A \tilde\xi_A + \A_2) = 0 \ .
\end{equation}
The solution of this condition then fixes the two coefficients $\A_{1}$ and $\A_{2}$
\begin{equation} \label{hypermultiplet_restriction}
 \A_{1,2} = \P_{1,2}^A \tilde\xi_A -\Q_{1,2\, A} \xi^A \ ,
\end{equation}
where $\tilde\xi_A$ and $\xi^A$ are the Ramond-Ramond scalars evaluated at the $N=1$ vacuum. We can now make use of the solution for the embedding tensor components \eqref{solution_embedding_tensor} found from the gravity plus vector multiplet sector, which by construction fulfil \eqref{condition_vectors_Msolve1} and \eqref{condition_vectors_Msolve2}. We already solved the first two equations in \eqref{condition_hypers_M}. Since \eqref{condition_vectors_Msolve1} implies the gravitino and gaugino equation, we find that also $u(\kk)=0$, such that in \eqref{condition_hypers_M} it only remains to solve $\bar E^{\underline b} (\kk) = 0 $, which comes from the hyperino equation and gives further constraints on $\P_{1,2}^A$ and $\Q_{1,2\, A}$. We shall now rewrite the solution for the embedding tensor components \eqref{solution_embedding_tensor} in the notation of this section and then turn to solving the remaining equation $\bar E^{\underline b} (\kk) = 0 $.

Using \eqref{prepotential_no_compensator} and \eqref{quat_connection}, we see that the Killing prepotentials are given by
\begin{equation}
P^+_{1,2} = 2 \iu \bar u(k_{1,2}) \ ,
\end{equation}
where we have used the complex notation introduced in \eqref{Pc}. If we now insert the definition of the one-form $\bar u$ \eqref{one-forms_quat} and make use of \eqref{MG_subspaces}, we find that the solution for the embedding tensor components \eqref{solution_embedding_tensor} can be expressed as
\begin{equation}\label{solution_embedding_tensor_Min}
\begin{aligned}
\Theta_I^{\phantom{I}1} = & -\Im(\bar Z^A (\Q_{2\, A} - \bar {\cal G}_{AB}\P_{2}^B) {\cal F}_{IJ}\C^J ) \ , \\  \Theta^{I1} = & -\Im (\bar Z^A (\Q_{2\, A} - \bar {\cal G}_{AB}\P_{2}^B) \C^I) \ , \\
\Theta_I^{\phantom{I}2} = & \quad \Im(\bar Z^A (\Q_{1\, A} - \bar {\cal G}_{AB}\P_{1}^B) {\cal F}_{IJ} \C^J ) \ , \\  \Theta^{I2} = & \quad \Im (\bar Z^A (\Q_{1\, A} - \bar {\cal G}_{AB}\P_{1}^B)  \C^I) \ ,
\end{aligned}
\end{equation}
where we have absorbed the prefactor $2\e^{K^h/2+\phi}$ into $\C^I$.

Let us now solve $\bar E^{\underline b} (\kk) = 0 $. Inserting \eqref{solution_embedding_tensor_Min} into \eqref{one-forms_quat} we find
\begin{equation}\label{Ebar}
\begin{aligned}
 X^I (\Im {\cal F})_{IJ}  \bar{\C}^J {\Proj}_A^{\phantom{A}B} Z^C \big( & (\Q_{2\, B} - {\cal G}_{BD}\P_{2}^D) (\Q_{1\, C} - {\cal G}_{CE}\P_{1}^E) \\ & - (\Q_{1\, B} - {\cal G}_{BD}\P_{1}^D) (\Q_{2\, C} - {\cal G}_{CE}\P_{2}^E) \big) = 0 \ ,
 \end{aligned}
\end{equation}
where, for convenience, we have contracted the expression with ${\Proj}_{A\, \underline{b}}$ in order to introduce the projection operator $ {\Proj}_A^{\phantom{A}B}$, cf.\ \eqref{projection_v}.
Furthermore, we have used the identity \eqref{identity_v} to pull out the prefactor $ X^I (\Im {\cal F})_{IJ} \bar{\C}^J $. This prefactor is non-zero for all $\C^I$ fulfilling \eqref{constraint_embedding_tensor}, see \eqref{signature_G}, and can be neglected. We can parameterize the Killing vector coefficients $\P_{1,2}^A$ and $\Q_{1,2\, A}$ by
\begin{equation}\label{solution_beta}
 \P_{1,2}^A= \Im(\D^A_{1,2}) \ , \qquad \Q_{1,2\, A} = \Im({\cal G}_{AB}\D^B_{1,2}) \ ,
\end{equation}
where $\D^A_{1,2}$ are two complex vectors. We can then decompose $\D^A_{1,2}$ into the components canonically defined by the projection $ {\Proj}_A^{\phantom{A}B}$ as done in \eqref{decomposition}.
% \begin{equation}
%  \D^A_{1,2} = \D^{(Z)\,B}_{1,2} + \D^{(P)\,A}_{1,2} \equiv - \D^B_{1,2} K^{\rm h}_B Z^A  + \D^B_{1,2} {\Proj}_B^{\phantom{B}A} \ .
% \end{equation}
Using this, the condition \eqref{Ebar} simplifies to
\begin{equation}
 \D^{(P)\,A}_{1} \D^{(Z)\,B}_{2} = \D^{(P)\,A}_{2} \D^{(Z)\,B}_{1} \ .
\end{equation}
The only solution to this equation is $\D^A_{2} =a \D^A_{1}$ with a complex factor $a$, and in the following we will just write $\D^A$. Note that for $a$ real, the two Killing vectors are the same and the embedding tensor components \eqref{solution_embedding_tensor_Min} just cancel against each other, giving an ungauged supergravity with an $\cN=2$ vacuum. Furthermore, for any complex $a$, its real part drops out due to this cancellation. Thus, we can choose $a=\iu$, since any additional real prefactor can be absorbed into the embedding tensor.
After absorbing a prefactor $-\iu \bar Z^A (\Im {\cal G})_{AB} \D^B$ into the definition of $\C^I$, the embedding tensor \eqref{solution_embedding_tensor_Min} similarly to \eqref{solution_embedding_tensor_ex} simply reads
\begin{equation}\label{solution_embedding_tensor_Min2}
\begin{aligned}
\Theta_I^{\phantom{I}1} = & \Im({\cal F}_{IJ}\C^J ) \ , \qquad  \Theta^{I1} = & \Im \C^I \ , \\
\Theta_I^{\phantom{I}2} = & \Re({\cal F}_{IJ} \C^J ) \ , \qquad  \Theta^{I2} = & \Re \C^I \ .
\end{aligned}
\end{equation}
It remains to check that the two Killing vectors commute when the coefficients are parameterized by \eqref{solution_beta}. To do so, we insert \eqref{solution_beta} together with $\D^A= \D^A_{1} = - \iu \D^A_{2}$ into the commutation condition \eqref{comm_isometries} and find
\begin{equation} \label{null_alpha}
 0= \bar \D^A (\Im{\cal G})_{AB} \D^B \ .
\end{equation}
Thus, the complex vector $\D^A$ must be null with respect to the matrix $(\Im{\cal G})_{AB}$, which is of signature $(n_{\rm h}-1,1)$, cf.\ \eqref{signature_G}.

In order to make contact with the literature, we can rewrite the embedding tensor components in a more convenient basis. Instead of expressing $\Theta_\Lambda^{\ \tilde \Lambda}$ in the basis of $k_{1,2}$ plus the other (ungauged) isometries, we can make a change of basis and go back to the standard basis of c-map Killing vectors \eqref{Killing}. To do this, we collect, as in Section \ref{section:special_quat}, the Killing vectors $\Kxi_A$ and $\Ktxi^A$, as well as the fibre coordinates $\xi^A$ and $\tilde \xi_A$, in the $Sp(n_{\rm h})$ vectors
\begin{equation}\label{standard_k}
 k_{\tilde \Lambda} = \left( \begin{aligned}
                           \Ktxi^A \\ \Kxi_A
                          \end{aligned} \right)
\end{equation}
and
\begin{equation} \label{standard_xi}
 \xi_{\tilde \Lambda} = \left( \begin{aligned}
                           \xi^A \\ \tilde \xi_A
                          \end{aligned} \right) \ .
\end{equation}
The embedding tensor then reads
\begin{equation} \label{complete_solution_M}
\begin{aligned}
\Theta_\Lambda^{\ \tilde \Lambda} &= \Re \left(  \bar{\C}^J \D^B \left( \begin{aligned} \bar {\cal F}_{JI} {\cal G}_{BA} && \bar {\cal F}_{JI} \delta^A_B \\  \delta^I_J {\cal G}_{BA} && \delta^I_J \delta^A_B \end{aligned} \right) \right) \ , \\
\Theta_\Lambda^{\ \ax} &= - \Theta_\Lambda^{\ \tilde \Lambda} \xi_{\tilde \Lambda} = \Re \left(  \D^A (\tilde\xi_A - {\cal G}_{AB} \xi^B ) \bar{\C}^J \left( \begin{aligned} \bar {\cal F}_{JI} \\  \delta^I_J \end{aligned} \right) \right)  \ ,
\end{aligned}
\end{equation}
where $\D^A$ and $\C^I$ have to satisfy commutation \eqref{null_alpha} and mutual locality \eqref{constraint_embedding_tensor} conditions respectively.

Now let us give the explicit form of tensors $S_{\cal AB}$, $W^{i{\cal AB}}$ and $N^\alpha_{\cal A}$ for the embedding tensor solution \eqref{complete_solution_M}:
\begin{subequations}
\begin{eqnarray}
 S_{\cal AB} &=& 2 \e^{K^{\rm v}/2+K^{\rm h}/2+\phi} [X^I (\Im {\cal F})_{IJ} \bar{\C}^J] [\bar Z^A \Im {\cal G}_{AB} \D^B] \left( \begin{aligned}
                    0 && 0 \\ 0 && 1                                                                                  \end{aligned} \right) \ , \\
W_{i \cal AB} &=& 4 \iu \e^{K^{\rm v}/2+K^{\rm h}/2+\phi} [{\Proj}_i^{\phantom{i}J} (\Im {\cal F})_{JK} \bar{\C}^K] [\bar Z^A \Im {\cal G}_{AB} \D^B] \left( \begin{aligned}
                    0 && 0 \\ 0 && 1                                                                                  \end{aligned} \right) \ , \\
N_{\alpha \cal A} &= & 2 \sqrt{2} \iu \e^{K^{\rm v}/2+K^{\rm h}/2+\phi} [X^I (\Im {\cal F})_{IJ} \bar{\C}^J] \hskip4cm \nonumber \\ && \cdot \D^B\left( \begin{aligned}
                    0 && 0 \qquad && 0 \qquad && 0\\ 0 && [\tfrac12 \e^{-K^{\rm h}} {\Proj}_B^{\phantom{B}\underline{a}}] && [(\Im {\cal G})_{BA} \bar Z^A] && 0   \end{aligned} \right) \ ,
\end{eqnarray}
\end{subequations}
where we used the relations between the projector $\Proj_i^{\phantom{i}J}$ and the K\"ahler covariant derivatives of $X^J$ and ${\cal F}_J$ \eqref{projection_der}.

Note that the solution \eqref{complete_solution_M} can be constructed for \emph{any} point of the moduli space ${\cal M}_{\rm v}\times {\cal M}_{\rm h}$ and does only depend on the second derivatives of the prepotentials $\cal F$ and $\cal G$ at the $\cN=1$ point.
Furthermore, the solution is completely covariant under Mirror symmetry, which essentially exchanges the two special K\"ahler manifolds.

Now we want to compute the K\"ahler potential following \eqref{Kdef_N=1} and \eqref{Kahlerconnection}.
Note that the Killing vectors given in \eqref{ex_Killing_vectors} with \eqref{hypermultiplet_restriction} and \eqref{solution_beta} fulfill $P^3_{1,2} = 0$ in the $\cN=1$ locus but there is $\diff P^3_{1,2} \ne 0$, in constrast to \eqref{P3}.\footnote{This is due to the
  fact that some scalar directions can get a mass in the partial
  super-Higgs mechanism, see Section \ref{section:scales}.  These directions can be identified with the
  deformations of the expressions appearing in the solution
  \eqref{complete_solution_M} for the embedding tensor, i.e.\
  with the deformations of $D^A{\cal G}_{AB}(z)$ and $D^A (\tilde \xi_A -
  {\cal G}_{AB}(z) \xi^B )$. Therefore, in order to find the correct
  K\"ahler potential given by \eqref{Kdef_N=1}, we should integrate out
  such massive scalars. One of the consequences would be that we set $\Re(D^A (\diff \tilde \xi_A - {\cal G}_{AB}(z) \diff \xi^B )) = \Im(D^A (\diff \tilde \xi_A - {\cal G}_{AB}(z) \diff \xi^B ))= 0$ in the $\cN=1$ theory, which implies $\diff P^3_{1,2} = 0$.}
Therefore, we need to perform a local $SU(2)$ R-symmetry rotation to ensure $\diff P^3_{1,2} = 0$.
This is achieved by the transformation
\begin{equation}
 \Lambda = \e^{f^x \sigma^x} \ ,
\end{equation}
such that
\begin{equation} \label{f_def}
 f^x \Big|_{{\cal N}=1} = 0 \ , \quad \diff f^x \Big|_{{\cal N}=1} = (P^1_1 P^2_2 - P^1_2 P^2_1 )^{-1} (P^x_1 \diff P^3_2 - P^x_2 \diff P^3_1) \ .
\end{equation}
Then, \eqref{Kdef_N=1} is modified to
\begin{equation}\label{Kmodified}
 \hat K = \diff \omega^3 + \diff f^1\wedge \omega^2 - \diff f^2 \wedge \omega^1 \ .
\end{equation}
Differentiation of $\omega^3$ given in \eqref{quat_connection} results in
\begin{equation} \label{special_complexstructure}
 \diff \omega^3 = \iu (v \wedge \bar v + u \wedge \bar u + E \wedge \bar E  + e \wedge \bar e) \ .
\end{equation}
The additional terms in \eqref{Kmodified} are computed from \eqref{f_def} to be
\begin{equation}
\diff f^1\wedge \omega^2 - \diff f^2 \wedge \omega^1 \Big|_{{\cal N}=1} = \Im \Big[ \frac{D^A (\diff \tilde \xi_A - {\cal G}_{AB} \diff \xi^B) \wedge (\bar Z^A \diff \tilde \xi_A - \bar{\cal G}_{A} \diff \xi^A)}{D^A \Im({\cal G})_{AB}\bar Z^B}  \Big] \ .
\end{equation}
This determines the K\"ahler two-form in the $\cN=1$ locus to be
\begin{equation} \label{specialKahlertwoform}\begin{aligned}
  \hat K \Big|_{{\cal N}=1} =& \iu \big(v \wedge \bar v + u \wedge \bar u + E \wedge \bar E  + e \wedge \bar e \big) \\ & + \Im \Big[ \frac{D^A (\diff \tilde \xi_A - {\cal G}_{AB} \diff \xi^B) \wedge (\bar Z^A \diff \tilde \xi_A - \bar{\cal G}_{A} \diff \xi^A)}{D^A \Im({\cal G})_{AB}\bar Z^B}  \Big] \ .
\end{aligned} \end{equation}
By use of the complex structure related to this K\"ahler two-form we can identify the holomorphic component of $\omega^3$ in \eqref{quat_connection} to be $\omega^3_a=\tfrac{\iu}{2} (v_a - \partial_a K^{\rm h}) $. Plugging this into \eqref{Kahlerconnection} and integrating we then find
\begin{equation}\label{special_Kpot}
 \hat K = K^{\rm h} + 2 \phi \ .
\end{equation}

Before we proceed, let note that $\K$ given in \eqref{special_Kpot} is still expressed in terms of the original $\cN=2$ field variables. Using the $\cN=1$ complex structure $\hat J$ it is possible to express $\K$ in terms of proper holomorphic $\cN=1$ field variables. However, in general this computation is rather involved and we leave it for future investigation.

Inserting the Killing prepotentials \eqref{prepotential_no_compensator} into the general expression for the superpotential \eqref{Wone} we find
\begin{equation} \label{Wone_M}
 W = V^\Lambda \Theta^{\phantom{\Lambda}\tilde \Lambda}_\Lambda U_{\tilde \Lambda}  \ ,
\end{equation}
where the symplectic vector $U_{\tilde \Lambda}$ on ${\cal M}_{\rm  sk}$ was defined in \eqref{sympl_hol_vectors}.
Note that the superpotential in \eqref{Wone_M} is indeed holomorphic.

\subsection{AdS vacua}\label{section:AdS_vacua}

Let us now consider the case of an AdS vacuum preserving $\cN=1$ supersymmetry. For $\mu \ne 0$, we see from combined gravitino and hyperino condition \eqref{cmap_vbarv} that both $n^1$ and $n^2$ must be non-zero. By manipulating \eqref{cmap_vbarv}, we are led to the following conditions
\begin{subequations}\label{AdS_vbarv}
 \begin{eqnarray}
  n^1 n^2 (v+\bar v) (\kk) &=& - \tfrac12 \iu \e^{-K^{\rm v}/2}\mu (|n^1|^2 - |n^2|^2) \ , \label{AdS_vbarv1} \\
  n^1 n^2 (v-\bar v) (\kk) &=& - \iu \e^{-K^{\rm v}/2}\mu (|n^1|^2 + |n^2|^2) = - \iu \e^{-K^{\rm v}/2}\mu |\epsilon_1|^2\ . \label{AdS_vbarv2}
 \end{eqnarray}
\end{subequations}
If the $k_\phi$ direction is not gauged, then we have that $(v+\bar v) (k(L))=0$ and we can conclude that the complex coefficients of the preserved supersymmetry generator \eqref{SUSYgenerator} must be equal $|n^1|= |n^2|$ \cite{Cassani:2007pq}.\footnote{The dilaton isometry is spoilt by quantum corrections in $\cN=2$ supergravity (see \cite{RoblesLlana:2006ez} for the one-loop result). Therefore we do not consider gaugings with respect to this isometry.} This agrees with the result using a different approach in type II supergravity in ten dimensions \cite{Grana:2006kf}. In the following we shall restrict to $|n^1|= |n^2|$ and parameterize the coefficients as
\begin{equation}  \label{unbroken_SUSY_AdS}
n^1 =\e^{\iu \varphi/2}n \ , \quad \textrm{and} \quad n^2=\e^{-\iu \varphi/2}n \ ,
\end{equation}
where $\varphi$ is a phase.

Before we proceed to analyze the supersymmetry variations in detail, we shall make a remark about the amount of unbroken supersymmetry. For AdS vacua, we take the general ansatz for the Killing vectors $k_1$ and $k_2$  used in the Minkowski case \eqref{ex_Killing_vectors}, and demand that they commute i.e. that \eqref{comm_isometries} is satisfied. The embedding tensor components which solve the gravitino and gaugino equations are then given by \eqref{solution_embedding_tensor_AdS}, but as we now break to a different $\cN=1$ vacuum with a different preserved Killing spinor \eqref{SUSYgenerator} we must perform an $SU(2)$-rotation. By comparing \eqref{unbroken_SUSY_AdS} with the spinor used in Section \eqref{section:magnetic_vectors}, which has $n^1\ne 0$ and $n^2=0$, we see that the appropriate $SU(2)$-rotation is given by
\begin{equation}
 M^{\cal A}_{\phantom{\cal A} \cal B}= \tfrac{1}{\sqrt{2}}\left( \begin{aligned}
           \e^{\iu\varphi/2} && - \e^{\iu\varphi/2} \\ \e^{-\iu\varphi/2} &&\e^{-\iu\varphi/2}
           \end{aligned}\right) \ .
\end{equation}
The only term in the embedding tensor components \eqref{solution_embedding_tensor_AdS} that transforms non-trivially under this rotation is $P^+_{1,2}$:
\begin{equation} \label{twisted_prepotentials}
 P^-_{1,2} \longrightarrow \tilde P^-_{1,2} = \iu \Im(\e^{\iu \varphi} P^-_{1,2}) - P^3_{1,2} \ .
\end{equation}
In order to find the embedding tensor components which solve the gravitino and gaugino conditions \eqref{solution_embedding_tensor_AdS} we assumed that $P^3_{1,2}=0$. In the new $SU(2)$-frame we have to adjust $k_1$ and $k_2$ such that
\begin{equation} \label{vanishingP3}
 \tilde P^3_{1,2} = \Re(\e^{\iu \varphi} P^-_{1,2}) = 0 \ .
\end{equation}
Analogously to the Minkowski case \eqref{solution_beta}, we make the following ansatz for the Killing vector coefficients
\begin{equation}\label{solution_beta_AdS}
 \P_{1,2}^A= \Im(\D^A_{{\rm AdS}\,1,2}) \ , \qquad \Q_{1,2\, A} = \Im({\cal G}_{AB}\D^B_{{\rm AdS}\,1,2}) \ ,
\end{equation}
where we have used the decomposition \eqref{decomposition} with respect to the projector ${\Proj}_A^{\phantom{A}B}$ to express $\D^A_{{\rm AdS}\,1,2}$ as
\begin{equation} \label{decompose_AdS}
 \D^A_{{\rm AdS}\,1,2} =  \D^{(Z)\,A}_{{\rm AdS\,}1,2} + \D^{(P)\,A}_{{\rm AdS}\,1,2} \ .
\end{equation}
Inserting this ansatz into \eqref{vanishingP3} and using the expressions \eqref{prepotential_no_compensator}, \eqref{quat_connection} and \eqref{one-forms_quat} we find
\begin{equation}
 \Re(\e^{\iu \varphi} Z^A (\Im {\cal G})_{AB} \bar{\D}^{(Z)\,B}_{{\rm AdS}\,1,2}) = 0 \ ,
\end{equation}
which is solved by
\begin{equation} \label{TZ}
 \D^{(Z)\,A}_{{\rm AdS\,}1,2} = \iu \e^{\iu \varphi} \R_{1,2} Z^A\ ,
\end{equation}
where $\R_{1,2}$ are real numbers.
Inserting the above expressions into the transformation of the Killing prepotential \eqref{twisted_prepotentials} then leads to
\begin{equation} \label{prepotentials_AdS}
 \tilde P^-_{1,2} = \e^{2\phi} (\A_{1,2} -  \Im((\iu \R_{1,2} \e^{\iu\varphi}Z^A +\bar{\D}^{(P)\,A}_{{\rm AdS}\,1,2}) (\tilde \xi_A - {\cal G}_{AB} \xi^B)) ) + \iu \e^{-K^{\rm h}/2 + \phi} \R_{1,2} \ .
\end{equation}
We remind the reader that the prepotentials $\tilde P^x_1$ and $\tilde P^x_2$ should not be aligned for a proper $\cN=1$ vacuum.

We still have to solve the equations coming from the hyperino variation. In the Minkowski case we only had to solve the condition $\bar E(\kk)=0$, whereas we now see from \eqref{qE} that we that we have an addition condition $E(\kk)=0$ in the AdS case. Furthermore, \eqref{qH} also now gives an additional non-trivial condition, which is rephrased as \eqref{AdS_vbarv2}. Considering again the projector decomposition \eqref{decomposition} for $\D_{{\rm AdS}\,1,2}^A$,  we see that \eqref{AdS_vbarv2} gives a condition on $\C_{1,2}$, while \eqref{qE} restricts $\D^{(P)\,A}_{{\rm AdS}\,1,2}$ in \eqref{decompose_AdS}. Let us start with \eqref{qE}. By insertion of \eqref{ex_Killing_vectors} into \eqref{solution_beta_AdS} and using the definition \eqref{projection_v} and the relations \eqref{MG_subspaces}, we can write \eqref{qE} as
\begin{equation}\label{AdS_qE}
 \begin{aligned}
 (\tilde P^2_2 + \tfrac\iu2 \tilde P^1_2) \D^{(P)\,A}_{{\rm AdS}\,1}  - (\tilde P^2_1 + \tfrac\iu2 \tilde P^1_1)\D^{(P)\,A}_{{\rm AdS}\,2} = 0     \ , \\
 (\tilde P^2_2 - \tfrac\iu2 \tilde P^1_2) \D^{(P)\,A}_{{\rm AdS}\,1}-  (\tilde P^2_1 - \tfrac\iu2 \tilde P^1_1)\D^{(P)\,A}_{{\rm AdS}\,2} = 0 \ ,
 \end{aligned}
\end{equation}
where for simplicity we took  the complex conjugate in the first equation. As the prepotentials of $k_1$ and $k_2$ must not coincide in an $\cN=1$ vacuum, \eqref{AdS_qE} implies that both $\D^{(P)\,A}_{{\rm AdS}\,1}$ and $\D^{(P)\,A}_{{\rm AdS}\,2}$ must vanish. Then from the commutation relation \eqref{comm_isometries}, together with \eqref{solution_beta_AdS}, \eqref{decompose_AdS} and \eqref{TZ}, it follows that $\R_1$ or $\R_2$ is zero. We can choose $\R_2=0$ and note that by taking linear combinations of $k_1$ and $k_2$ we can always set $\A_1=0$. Furthermore, the resulting Killing vectors can be rescaled such that $\R_1=\A_2=1$.

Let us now solve \eqref{AdS_vbarv2}. Inserting the embedding tensor \eqref{solution_embedding_tensor_AdS} with \eqref{prepotentials_AdS} and \eqref{solution_beta_AdS}, we find
\begin{equation}\label{AdS_fixing}
 X^I (\Im {\cal F})_{IJ} \bar{\C}_{\rm AdS}^J = \iu \frac{3+4 \iu \E}{2+2 \iu \E} \e^{K^{\rm h}/2-K^{\rm v}/2-3\phi} \mu  \ ,
\end{equation}
where we abbreviated
\begin{equation}
  \E = \e^{K^{\rm h}/2 +\phi} \Re(\e^{\iu \varphi} (Z^A \tilde \xi_A - {\cal G}_A\xi^A)) \ .
\end{equation}
Using again the decomposition \eqref{decomposition} we can insert \eqref{AdS_fixing} into \eqref{solution_embedding_tensor_AdS}.
If we now go back to the standard basis of \eqref{standard_k} and \eqref{standard_xi}, the embedding tensor reads
\begin{equation}\label{complete_solution_AdS}
\begin{aligned}
\Theta_\Lambda^{\ \tilde \Lambda} &= - \Re \left( \left( \begin{aligned} {\cal F}_{IJ} \\  \delta^I_J \end{aligned} \right) (4\e^{K^{\rm h}/2 + K^{\rm v}/2-\phi} \bar \mu X^J + \C^{(P)\,J}_{\rm AdS})  \right) \cdot \Re( \e^{\iu \varphi}(\ {\cal G}_{A}  \ ,\  Z^A \ ))  \ , \\
\Theta_\Lambda^{\ \ax} &= \e^{-K^{\rm h}/2 -\phi} \Im \left( \left( \begin{aligned} {\cal F}_{IJ} \\  \delta^I_J \end{aligned} \right) (4 \e^{K^{\rm h}/2 + K^{\rm v}/2-\phi} (\tfrac12- \iu \E) \bar \mu X^J + (1-\iu \E) \C^{(P)\,J}_{\rm AdS})  \right)  \ ,
\end{aligned}
\end{equation}
where we have rescaled $\C^{(P)\,I}_{\rm AdS}$ by the factor $\iu\e^{2\phi}$. If we plug our result \eqref{complete_solution_AdS} into the constraint \eqref{constraint_embedding_tensor_AdS_g}, we find
\begin{equation}\label{constraint_AdS}
 \bar \C^{(P)\,J}_{\rm AdS} (\Im {\cal F})_{JI} \C^{(P)\,I}_{\rm AdS} = \e^{K^{\rm h} - 6\phi} \frac{|\mu|^2}{1+\rho^2} \ .
\end{equation}
This can be easily solved, since the left-hand side is naturally greater than zero (see the discussion in \eqref{signature_G}).

Finally, for the embedding tensor solution \eqref{complete_solution_AdS} the tensors appearing in the supersymmetry transformations $S_{\cal AB}$, $W^{i{\cal AB}}$ and $N^\alpha_{\cal A}$ are given by
\begin{subequations}
\begin{eqnarray}
 S_{\cal AB} &=& \mu  \left( \begin{aligned}
                    \e^{-\iu \varphi} && -\tfrac12 \\ -\tfrac12 && \e^{\iu \varphi}                                                                                  \end{aligned} \right) \ , \hskip4.5cm \\
W_{i \cal AB} &=& -\tfrac12 \e^{K^{\rm v}/2-K^{\rm h}/2+\phi} (\Im {\cal F})_{iJ} \bar \C^{(P)\,J}_{\rm AdS}
               \left( \begin{aligned}
                    \e^{-\iu \varphi} && -1 \\ -1 && \e^{\iu \varphi}                                                \end{aligned} \right) \ , \\
N_{\alpha \cal A} &= & \tfrac{1}{\sqrt{2}} \iu \mu \left( \begin{aligned}
                    \e^{-\iu \varphi} && 0 && -1 && 0\\ -1 && 0 && \e^{\iu \varphi} && 0   \end{aligned} \right) \ ,\hskip3.5cm
\end{eqnarray}
\end{subequations}
where we have again used \eqref{projection_der}.

The embedding tensor given by \eqref{complete_solution_AdS} can be defined at any point on ${{\cal M}}_{\rm v}\times {{\cal M}}_{\rm h}$. Furthermore, for any choice of the moduli spaces ${{\cal M}}_{\rm v}$ and ${{\cal M}}_{\rm h}$ -- as long as ${{\cal M}}_{\rm h}$ is in the image of the c-map -- we have found a construction for the gaugings that lead to $\cN=1$ AdS vacua. The only constraints on the solution \eqref{complete_solution_AdS} is \eqref{constraint_AdS}, which can easily be fulfilled. In this way, the results of this section are completely analogous to those of Section~\ref{section:Minkowski_vacua}.

We want now to determine the K\"ahler potential and the superpotential of the corresponding $\cN=1$ vacuum. In contrast to the Minkowski case, there are both $\tilde P^3_{1,2}=0$ and $\diff \tilde P^3_{1,2}=0$, cf.\ \eqref{vanishingP3}, and we can use \eqref{Kdef_N=1} and \eqref{vanishingP3} to compute the K\"ahler two-form $\hat K$ on $\hat {\cal M}_{\rm h}$
\begin{equation} \begin{aligned}
  \hat K =& \ 2\Im\big(\e^{\iu \varphi} u \big) \wedge \Re v - 2 \Im\Big(\e^{\iu \varphi}\bar E \wedge e \Big) \\
  &+ 2 \iu \e^{K^{\rm h}} \Re \big(\e^{\iu \varphi} u\big) \wedge \Big(Z^A (\Im  \mathcal G_{AB})\diff \bar Z^B - \bar Z^A (\Im  \mathcal G_{AB}) \diff Z^B \Big)  \ .
\end{aligned}\end{equation}
From this we can identify the holomorphic part of $\tilde \omega^3$ to be $\tilde \omega^3_a= 2 (\Im(\e^{\iu \varphi} u_a) - \iu (v+\bar v)_a)$. Inserting this into \eqref{Kahlerconnection} leads to the K\"ahler potential
\begin{equation}\label{special_Kpot_AdS}
 \hat K = 4 \phi \ .
\end{equation}
Finally, inserting the Killing prepotentials \eqref{twisted_prepotentials} and \eqref{prepotential_no_compensator} into the general expression for the superpotential \eqref{Wone}, we find
\begin{equation}
 W = \tfrac12  V^\Lambda (\Theta_\Lambda^{\phantom{\Lambda}\ax} + \Theta^{\tilde \Lambda}_\Lambda(\xi_{\tilde \Lambda} + 2 \iu \e^{K^{\rm h}/2-\phi}  \Im(\e^{\iu \varphi} U_{\tilde\lambda}) )  )  \ .
\end{equation}

The K\"ahler potential $ \hat K$ in \eqref{special_Kpot_AdS} also coincides with the expression obtained in orientifold truncations of the type II compactifications considered for instance in \cite{Benmachiche:2006df,Cassani:2007pq}.

%%%%%%%%%%%%%%%%%%%%%%%%%%%%%%%%%%%%
\section{Realization in string theory} \label{section:strings}
%%%%%%%%%%%%%%%%%%%%%%%%%%%%%%%%%%%%

Let us now comment on how the solutions of Section~\ref{section:breaking_hyper} can be realized in string theory. We shall only consider smooth $\cN=2$ compactifications of the type II string on here, but similar realizations should be possible for the heterotic string and for type II orientifolds. For further discussion of four-dimensional $\cN=1$ Minkowski and AdS vacua from string theory see \cite{Lust:2004ig,Behrndt:2005bv,House:2005yc,Grana:2005sn,Micu:2006ey,Grana:2006kf,Micu:2007rd,KashaniPoor:2007tr,Andriot:2008va,Anguelova:2008fm,Cassani:2009ck,Lust:2009zb,Cassani:2009na}.

Here, we use the results of Chapter~\ref{section:effective} where we determined the embedding tensor to be of the form \eqref{charges_N=2comp}, where its components are defined in \eqref{SU3SU3_charges_covariant} and \eqref{SU3SU3_flux}.
For the $\cN=1$ Minkowski solution \eqref{complete_solution_M} we can then identify the charges appearing in \eqref{Etensorcharges} as follows
\begin{subequations}\label{Msolution}
\begin{eqnarray}
e_{AI} &= & \Re(\bar {\cal F}_{IJ} \bar{\C}^J {\cal G}_{AB} \D^B)  \ , \\
p_A^I &= & \Re(\bar{\C}^I {\cal G}_{AB} \D^B) \ , \\
m^A_I  &= & \Re(\bar {\cal F}_{IJ} \bar{\C}^J \D^A) \ ,\\
q^{AI}  &= & \Re(\bar{\C}^I \D^A) \ ,\\
f_{I}  &= & \Re(\bar {\cal F}_{IJ} \bar{\C}^J (\xi^A {\cal G}_{AB}- \tilde \xi_B )\D^B) \ , \\
f^I &= & \Re(\bar{\C}^I (\xi^A {\cal G}_{AB}- \tilde \xi_B )\D^B) \ .
\end{eqnarray}
\end{subequations}
Let us recall that charges are quantized in string theory and therefore all entries of the embedding tensor are integral, as we already discussed in the paragraph after \eqref{SU3SU3_flux}.
This implies that partial supersymmetry breaking may only be possible at discrete points on ${\cal M}_{\rm v}$ and ${\cal M}_{\rm h}$, where the expressions in \eqref{Msolution} are integer-valued. This condition might restrict the form of the prepotential and therefore the allowed moduli spaces ${\cal M}_{\rm v} \times {\cal M}_{\rm h}$.

The issue of mirror symmetry in $SU(3)\times SU(3)$-structure compactifications has been discussed at length in Ref.\ \cite{Grana:2006hr}, where it was found that, apart from an exchange of the prepotentials ${\cal F} \leftrightarrow {\cal G}$, the charges are exchanged as follows \begin{equation}
m^A_I \leftrightarrow -p^A_I\ , \qquad e_{AI}\leftrightarrow
e_{IA}\ ,\qquad q^{AI}\leftrightarrow q^{IA}\ .
\end{equation}
An inspection of \eqref{Msolution} shows that the solutions indeed obey this symmetry if we also simultaneously exchange $\C^I\leftrightarrow  \D^A$.

If we set $p_A^I$ and $q^{AI}$ to zero in \eqref{Msolution}, the product $\bar \C^I \D^B$ must vanish and we end up with the trivial solution. Therefore, an $\cN=1$ Minkowski vacuum can only occur when non-geometric fluxes are turned on. This is in agreement with the compactification no-go-theorem \cite{Gibbons:1984kp,deWit:1986xg,Maldacena:2000mw}, which states that there can be no stable Minkowski vacuum with only fluxes turned on. This statement is believed to also be true for backgrounds with torsion. Here we explicitly see that non-geometric fluxes can compensate for the form field fluxes and torsion, leading to a vanishing energy density i.e.\ to vanishing $\mu$. In this way, the solution of Section~\ref{section:Minkowski_vacua} evades the no-go theorem.\footnote{A related result on the necessity of non-geometric fluxes for Minkowski vacua in orientifold compactifications has recently been found \cite{deCarlos:2009qm}.}

Before we turn to the AdS case, let us also note that the $\cN=1$ solutions given in \eqref{Msolution} are not within the class of solutions considered in \cite{Grana:2005sn} as one of the complex parameters $n^1$ or $n^2$ introduced in \eqref{SUSYgenerator} has to vanish. Rather, they  correspond to the class of solutions denoted Type A in \cite{Frey:2003sd}, which have been much less investigated. It would be interesting to further investigate this class of models.

We shall now consider the solution for $\cN=1$ AdS vacua. Comparing \eqref{complete_solution_AdS} with \eqref{Etensorcharges} we can read off
\begin{subequations}\label{Asolution}
\begin{eqnarray}
e_{AI} &= & -\Re( {\cal F}_{IJ} (4\e^{K^{\rm h}/2 + K^{\rm v}/2-\phi} \bar \mu X^J + \C^{(P)\,J}_{\rm AdS}) ) \quad \Re( \e^{\iu \varphi}  {\cal G}_{A} ) \ , \\
p_A^I &= & -\Re( (4\e^{K^{\rm h}/2 + K^{\rm v}/2-\phi} \bar \mu X^I + \C^{(P)\,I}_{\rm AdS}) \quad \Re( \e^{\iu \varphi}  {\cal G}_{A} ) \ , \\
m^A_I  &= & -\Re( {\cal F}_{IJ} (4\e^{K^{\rm h}/2 + K^{\rm v}/2-\phi} \bar \mu X^J + \C^{(P)\,J}_{\rm AdS}) ) \quad \Re( \e^{\iu \varphi} Z^A) \ ,\\
q^{AI}  &= & -\Re( (4\e^{K^{\rm h}/2 + K^{\rm v}/2-\phi} \bar \mu X^I + \C^{(P)\,I}_{\rm AdS}) \quad \Re( \e^{\iu \varphi} Z^A) \ ,\\
f_{I}  &= & \e^{-K^{\rm h}/2 -\phi}\Im( {\cal F}_{IJ} (4 \e^{K^{\rm h}/2 + K^{\rm v}/2-\phi} (\tfrac12- \iu \E) \bar \mu X^J + (1-\iu \E) \C^{(P)\,J}_{\rm AdS}) ) \ , \\
f^I & = & \e^{-K^{\rm h}/2 -\phi} \Im( 4 \e^{K^{\rm h}/2 + K^{\rm v}/2-\phi} (\tfrac12- \iu \E) \bar \mu X^I + (1-\iu \E) \C^{(P)\,I}_{\rm AdS} ) \ .
\end{eqnarray}
\end{subequations}
If we turn off non-geometric fluxes ($p_A^I=q^{AI} =0$), we see that non-trivial solutions do exist but must obey
\begin{equation}
\Re (X^I \bar \mu) = 0 \ .
\end{equation}
It would be interesting to further investigate the ten-dimensional origin of this condition.

Let us close this section by discussing possible quantum corrections in string theory. First of all, worldsheet instantons correct the K\"ahler potentials $K^{\rm v}$ in type IIA and $K^{\rm h}$ in type IIB. However, since we never used their explicit forms, all our results are unchanged and hold for any instanton-corrected K\"ahler potential. What we did use explicitly were the isometries resulting from the special fibration structure of ${\cal M}_{\rm h}$. Spacetime instanton effects generated from wrapped Euclidean branes generically break all of the isometries of ${\cal M}_{\rm h}$. However, it has been argued that the isometries which are gauged due to fluxes are precisely those protected (by the flux itself) from spacetime instanton effects \cite{KashaniPoor:2005si}.
It would be very interesting to identify \eqref{Msolution} and \eqref{Asolution} as solutions of the ten-dimensional supergravity equations of motion.

\cleardoublepage
%%%%%%%%%%%%%%%%%%%%%%%%%%%%%%%%%%%%%%%%%%%%%%%
\chapter{Conclusions}\label{section:conclusion}
%%%%%%%%%%%%%%%%%%%%%%%%%%%%%%%%%%%%%%%%%%%%%%%

In the first part of this review we showed that $SU(2) \times SU(2)$ structures always reduce to either $SU(2)$ or identity structures and we derived the general form of the scalar field space for all $SU(2)$ backgrounds in four and six spacetime dimensions.

We defined $SU(2)\times SU(2)$ structures in the pure spinor formalism and embedded it into the paradigm of exceptional generalized geometry. Thereby we derived the parameter space of $SO(1,d-1)$ scalars for the ten-dimensional theory. Furthermore, we showed how to derive the scalar field space of the $d$-dimensional effective theory for the class of consistent truncations.

In particular, for type IIA compactifications to six dimensions,
we derived the space of scalar degrees of freedom to be
\begin{equation}\label{conclusions_NSd6}
  \mathcal{M}^\textrm{IIA}_{d=6}\ =\ \frac{SO(4,n+4)}{SO(4) \times SO(n+4)} \times \mathbb{R}_+  \ ,
\end{equation}
where $n$ is some integer number and the $\mathbb{R}_+ $ corresponds to the six-dimensional dilaton.
The space $\mathcal{M}^\textrm{IIA}_{d=6}$ consists of only scalars coming from the NS-NS sector, as the corresponding type IIA compactification has no R-R scalar degrees of freedom.

In contrast, in the analogous type IIB setting it is necessary to embed both pure spinors into representations of the U-duality group $SO(5,5)$ before truncating the theory.
As we showed, R-R scalars enlarge the moduli space to
\begin{equation}
 \mathcal{M}_{d=6}^{\textrm{IIB}}\ =\ \frac{SO(5,n+5)}{SO(5) \times SO(n+5)}  \ .
\end{equation}

We used the same strategy to determine the scalar field spaces for $SU(2)$
structure compactifications to $d=4$. Additionally, we had to
introduce a generalized almost product structure to
force the structure group to be $SU(2)$, which divided the tangent bundle and its generalizations into a four-dimensional $SU(2)$-structure and a two-dimensional identity-structure piece.
By using the same techniques as in the case $d=6$, we derived the scalar field space to be of the form
\begin{equation}
 \mathcal{M}_{d=4}^\textrm{IIA/IIB}=\frac{SO(6,n+6)}{SO(6) \times SO(n+6)} \times \frac{Sl(2,\mathbb{R})_{T/U}}{SO(2)}  \ ,
\end{equation}
where the extra factor is either ${Sl(2,\mathbb{R})_T}/{SO(2)} $ acting on the complexified K\"ahler structure scalar of the identity structure
or ${Sl(2,\mathbb{R})_U}/{SO(2)}$ acting on the complex structure scalar of it depending on whether we consider type IIA or type IIB.
We also showed that we can interpret the flat superconformal cone over this space in terms of pure spinors and their embeddings into $E_{7(7)}$ representations.

In the derivation we mainly used algebraic properties of the pure
spinors but did not impose explicitly any differential constraint.
The reason being that the metric on scalar field space
is determined by the algebraic properties while differential
constraints affect the potential of the effective action.
However, by analyzing the light spectrum of the effective supergravity
we argued that we have to project out all $SU(2)$ doublet degrees of
freedom in order to remove the massive gravitino multiplets. Their
presence would alter the standard supergravity with $16$ supercharges
and in particular change the scalar geometry. Since the exterior
derivative of the pure spinors $\diff\Phi$ is an $SU(2)$ doublet this
effectively also constrains the class of compactification manifolds.
In the absence of a warp factor it implies that $K3$ is the
four-dimensional compactification manifold $Y_4$, while for
the higher-dimensional $Y_{5,6}$ a component of the almost
product structure appears locally as $K3$.

For all spaces the number of light modes is determined by the integer
$n$ with $n=16$ for $K3$. Generically, this number is related to the global twisting of the bundle of forms that are in the $({\bf 2},{\bf 2})$ representation of $SU(2) \times SU(2)$.
All other details of the dimensional reduction are encoded in the
possible gauging of the supergravity action and in the warp factor.
The moduli spaces which we derived here could already have been predicted from the general form of supergravity theories with $16$ supercharges. However, here we showed explicitly how these moduli spaces arise in the compactification procedure. More precisely, we gave an example how the U-duality covariant formalism can be used to determine the moduli space for backgrounds that break part of the supersymmetry.

%%%%%%%%%%%%%%%%%%%%%%%%%%%%%%%%%%%%%%%%%%%%%%%

In the second part of the review, we carried out a systematic analysis of when spontaneous $\cN=2 \rightarrow \cN=1$ supersymmetry breaking can take place in gauged supergravities with general vector multiplet couplings and special hypermultiplet couplings. Our results provide a new perspective on the circumvention of well-known no-go theorems which forbid partial supersymmetry breaking in a Minkowski vacuum for a class of supergravity theories \cite{Cecotti:1984rk,Cecotti:1984wn,Mayr:2000hh,Maldacena:2000mw}. In particular, we have found the general solution to the conditions for spontaneous $\cN=2 \rightarrow \cN=1$ supersymmetry breaking in Minkowski and AdS space.

Allowing for mutually local electric and magnetic charges, we evaded the absence of a holomorphic prepotential and translated the gravitino and gaugino equations into a set of conditions for spontaneous partial supersymmetry breaking in terms of the charges, encoded in the embedding tensor, which left the special-K\"ahler manifold without any constraint. In contrast, we showed the quaternionic-K\"ahler manifold must allow for a pair of Killing vectors which are constrained by the hyperino equation in that they must build a holomorphic vector with respect to a specific almost complex structure.

In a next step, we derived the $\cN=1$ low-energy effective action of a partially
broken $\cN=2$ gauged supergravity by integrating out all modes with a mass of the order of the massive $\cN=1$ gravitino multiplet that results from the super-Higgs mechanism. In particular, the two vectors gauging the pair of Killing vectors are part of the massive gravitino multiplet. Their removal corresponds to taking the quotient of the quaternionic-K\"ahler manifold with respect to the pair of isometries. We showed that the resulting space is K\"ahler, consistent with $\cN=1$ supersymmetry, and derived its K\"ahler potential.
The removal of the pair of gauge vectors furthermore lead to a projection on the gauge kinetic function. Since one of the two gauge vectors is given by precisely the graviphoton, one can show that the projected gauge kinetic function is holomorphic and thereby fulfills the $\cN=1$ constraints. Finally, we identified the superpotential and the D-terms that can be generated by additional gaugings at a scale below the $\cN=2 \rightarrow \cN=1$ breaking scale. Their form is determined by the $\cN=2$ data. Furthermore, we checked the holomorphicity of the superpotential.

We then focussed on the case of special quaternionic-K\"ahler manifolds, which appear in general $SU(3)\times SU(3)$-structure compactifications, and constructed for this class of manifolds a pair of Killing vectors out of the Heisenberg algebra of Killing vectors that arises in the c-map construction such that they solve the additional necessary conditions coming from the hyperino variation.
The resulting solutions for the embedding tensor components could be rephrased in terms of the second derivatives of the prepotentials. For the Minkowski case, we found that the set of conditions for partial supersymmetry breaking are mirror symmetric under the exchange of the prepotentials of the special K\"ahler (${\cal F}$) and special quaternionic-K\"ahler (${\cal G}$) geometry.
Our final conclusion is that spontaneous $\cN=2 \rightarrow \cN=1$ supersymmetry breaking is possible at any point on the special K\"ahler manifold and at any point on the special quaternionic-K\"ahler manifold in gauged supergravity.

It is natural to ask about the stringy realization of this mechanism for partial supersymmetry breaking. By comparing our solution for the embedding tensor components with the charges appearing in flux compactifications, we found that the charges needed to solve the $\cN=1$ Minkowski vacuum conditions include non-geometric fluxes. This explains how we have evaded the no-go theorem forbidding the compactification of supergravity to Minkowski space in four dimensions \cite{Gibbons:1984kp,deWit:1986xg,Maldacena:2000mw}, which applies only to geometric fluxes. For an $\cN=1$ AdS vacuum, we found that geometric fluxes alone are sufficient to solve the supersymmetry conditions. For both cases, a possible direction for future work would be to understand the lift of the general $\cN=1$ solutions.

Finally, we should note that the fluxes appearing in a supergravity derived from string theory are quantized, and therefore partial supersymmetry breaking may only be possible at discrete points on ${\cal M}_{\rm v} \times {\cal M}_{\rm h}$, where the second derivatives of the prepotentials obey an integer condition. Furthermore, flux quantization may put some constraints on the allowed moduli spaces. We shall leave a more thorough analysis of this point for future work.

\appendix

\cleardoublepage
%%%%%%%%%%%%%%%%%%%%%%%%%%%%%%%%%%%%%%%%%%%%%%%%%%%%%%
\chapter{Conventions and technical details}\label{section:conventions}
%In this appendix we supplement our discussion of $\cN=2$ gauged supergravity in $D=4$ in the main text with further details \cite{Andrianopoli:1996cm,xxx}.

In this appendix we collect our conventions used throughout the
review.

The $SU(2)$ matrices $(\sigma^x)_{\cal AB}$ which appear in the $\cN=2$ supersymmetry variations are given by
\begin{equation} \label{sigma_conv} \begin{aligned}
(\sigma^1)_{\cal AB} = & \left(\begin{array}{cc}1 &0\\ 0& -1 \end{array}\right)~, \qquad (\sigma^2)_{\cal AB} =  \left(\begin{array}{cc} -\iu &0\\ 0& -\iu \end{array}\right)~ , \qquad (\sigma^3)_{\cal AB} =  \left(\begin{array}{cc}0 &-1\\ -1& 0 \end{array}\right) \ , \\
(\sigma^1)^{\cal AB} = & \left(\begin{array}{cc}-1 &0\\ 0& 1 \end{array}\right)~, \ (\sigma^2)^{\cal AB} =  \left(\begin{array}{cc} -\mathrm{i} &0\\ 0& -\mathrm{i} \end{array}\right)~ , \ (\sigma^3)^{\cal AB} =  \left(\begin{array}{cc}0 &1\\ 1& 0 \end{array}\right) \ .
\end{aligned}\end{equation}
These can be found from the usual Pauli matrices by applying the antisymmetric $SU(2)$ metric $\epsilon_{\cal AB}$, which in our conventions has the properties
\begin{equation}
\epsilon^{\cal AB} \epsilon_{\cal BC}  = - \delta^{\cal A}_{\cal C}~, \qquad \epsilon^{12}  = \epsilon_{12} =+1 \ .
\end{equation}

For $SO(N)$ the gamma matrices $\gamma_m$ satisfy
\begin{equation} \label{appendix_Clifford_algebra}
 \{ \gamma_m , \gamma_n \} = 2 g_{mn}\ , \quad  m,n=1,\dots,N\ ,
\end{equation}
where $g_{mn}$ is the $SO(N)$ metric, which can be used to raise and
lower the index of the gamma matrices.
% The standard operations on the
% $\gamma$-matrices are defined as
% \begin{equation}
% \label{appendix_operations_gamma_matrices}
%  \gamma_m^\dagger = A \gamma_m A^{-1} \ , \quad - \gamma^T_m = C^{-1} \gamma_m C \ , \quad - \gamma^*_m = D^{-1} \gamma_m D  \ .
% \end{equation}
For $N$ even the chirality operator is given by $\gamma_0 =
\iu^{N/2} \frac{1}{N!} \epsilon^{m_1 \dots m_N} \gamma_{m_1 \dots
m_N}$, where $\epsilon$ specifies the orientation of the
manifold.
For antisymmetric products of gamma matrices we abbreviate
\begin{equation}
 \gamma_{m_1 \dots m_k} = \gamma_{[m_1} \dots \gamma_{m_k]} \ .
\end{equation}
The antisymmetric products of two gamma matrices $\gamma_{mn}$ fulfill the $SO(N)$ commutation relations and generate the action of $SO(N)$ on spinors $\eta$.
% With the help of $A, C$ and $D$ one defines
% \begin{equation}
% \bar{\eta} = \eta^\dagger A \ , \quad \eta^t = \eta^T C^{-1} \ , \quad \eta^c = D \eta^* \ ,
% \end{equation}
% where $\eta$ is a spinor of $SO(N)$ and $A,C$ and $D$ are regular matrices.
%Note that we can always choose a representation where $A=C=D=1$ is satisfied.

As explained e.g.\ in \cite{VanProeyen:1999ni}, for any $N$ one can define the charge conjugation matrix, which maps a spinor $\eta$ to its charge conjugate $\eta^c$.
For $N = 4k, k\in \mathbb{N}_0$, the charge conjugation matrix commutes with the chirality operator and therefore charge conjugation preserves the chirality of a spinor. For $N= 4k+ 2$, charge conjugation anti-commutes with the chirality operator and thus exchanges the chirality of spinors.
% Note that $\eta^t$ and $\bar{\eta}$ are in the dual space of $\eta$.

With~\eqref{definition_two-forms} and \eqref{tensor_product_spinors} we can
compute the $SO(4)$ Fierz identities to be
\begin{equation} \begin{aligned}\label{Fierz_identities_4d}
 \eta_\alpha \bar{\eta}^\beta & = \tfrac{1}{2}(P_-)_{\alpha}^{\phantom{\alpha} \beta} - \tfrac{1}{8} \iu J_{mn}\left( \gamma^{mn} P_- \right)_{\alpha}^{\phantom{\alpha} \beta} \ , \qquad \textrm{for} \ \alpha , \beta =1, \dots , 4 \ , \\
 (\eta^c)_\alpha \bar{\eta}^\beta & = \tfrac{1}{8} \iu \bar{\Omega}_{mn} \left( \gamma^{mn} P_- \right)_{\alpha}^{\phantom{\alpha} \beta} \ , \\
  \eta_\alpha (\bar{\eta}^c)^\beta & = \tfrac{1}{8} \iu \Omega_{mn} \left( \gamma^{mn} P_- \right)_{\alpha}^{\phantom{\alpha} \beta}   \ ,
\end{aligned} \end{equation}
where $P_\pm = \tfrac{1}{2} \left( 1 \pm \gamma_0 \right)$ are the
chiral
projection operators.

Analogously, the $SO(6)$ Fierz identities for two spinors $\eta_1$ and
$\eta_2$ can be derived  by using \eqref{tensor_product_spinors} together with the definitions \eqref{SU3_forms_definitions}, \eqref{definition_one-form_K} and \eqref{definition_two-forms_6d} to be
\begin{equation} \begin{aligned}\label{Fierz_identities_6d}
(\eta_i)_\alpha (\bar{\eta}_i)^\beta & = \frac{1}{2}(P_-)_{\alpha}^{\phantom{\alpha} \beta} - \frac{1}{4} \iu J^{(i)}_{mn}\left( \gamma^{mn} P_- \right)_{\alpha}^{\phantom{\alpha} \beta} \qquad \textrm{for} \ i=1,2, \ \textrm{and} \ \alpha , \beta =1, \dots , 8 \ ,  \\
(\eta_i^c)_\alpha (\bar{\eta}_i)^\beta & = \frac{1}{24} \iu \bar{\Omega}^{(i)}_{mnp} \left( \gamma^{mnp} P_- \right)_{\alpha}^{\phantom{\alpha} \beta} \ ,  \\
(\eta_i)_\alpha (\bar{\eta}^c_i)^\beta & = \frac{1}{24} \iu \Omega^{(i)}_{mnp} \left( \gamma^{mnp} P_+ \right)_{\alpha}^{\phantom{\alpha} \beta} \ ,  \\
(\eta_1)_\alpha (\bar{\eta}_2^c)^\beta & =  \frac{1}{2} K_m \left( \gamma^{m} P_+ \right)_{\alpha}^{\phantom{\alpha} \beta}
- \frac{1}{8} \iu K_m J_{np} \left( \gamma^{mnp} P_+ \right)_{\alpha}^{\phantom{\alpha} \beta} \ ,  \\
(\eta_2^c)_\alpha (\bar{\eta}_1)^\beta & =  \frac{1}{2} \bar{K}_m \left( \gamma^{m} P_- \right)_{\alpha}^{\phantom{\alpha} \beta} - \frac{1}{8} \iu \bar{K}_m J_{np} \left( \gamma^{mnp} P_- \right)_{\alpha}^{\phantom{\alpha} \beta} \ ,  \\
(\eta_2)_\alpha (\bar{\eta}_1)^\beta & =  \frac{1}{4} \iu \bar{\Omega}_{mn}\left( \gamma^{mn} P_- \right)_{\alpha}^{\phantom{\alpha} \beta} \ , \\
(\eta_1)_\alpha (\bar{\eta}_2)^\beta & =  \frac{1}{4} \iu \Omega_{mn}\left( \gamma^{mn} P_- \right)_{\alpha}^{\phantom{\alpha} \beta} \ .
\end{aligned} \end{equation}
With the help of
\begin{equation}
(\eta_1)_\alpha (\bar{\eta}_1)^\beta = (\eta_1)_\alpha (\bar{\eta}_2^c)^\delta (\eta_2^c)_\delta (\bar{\eta}_1)^\beta \ , \qquad
(\eta_1)_\alpha (\bar{\eta}_1^c)^\beta = (\eta_1)_\alpha (\bar{\eta}_2)^\delta (\eta_2)_\delta (\bar{\eta}^c_1)^\beta \ ,
\end{equation}
etc., we can derive the relations \eqref{K_compatible} and \eqref{SU(3)_structure_forms_SU2_splitting} for the forms involved.

For $SO(N,N)$ spinors, the gamma matrices $\Gamma_A$ are defined by
\begin{equation}
  \{ \Gamma_A , \Gamma_B \} = 2 \mathcal{I}_{AB}\ , \quad  A,B=1,\dots,2N \ ,
\end{equation}
where $\mathcal{I}$ is the $SO(N,N)$ metric.
We can  % define all
% operations of~\eqref{appendix_operations_gamma_matrices} analogously,
% and additionally
also write the gamma matrices in terms of raising and
lowering operators $\Gamma_{m^+}$ and $\Gamma_{m^-}$ such that
\begin{equation} \label{appendix_Clifford_split}
\begin{aligned}
\{ \Gamma_{m^+} , \Gamma_{n^+} \} & = 0 \ , \\
\{ \Gamma_{m^-} , \Gamma_{n^-} \} & = 0 \ , \\
\{ \Gamma_{m^+} , \Gamma_{n^-} \} & = 2 g_{mn} \quad \textrm{for all}
\ m,n=1,\dots, N \ ,
\end{aligned} \end{equation}
where $g_{mn}$ is the $SO(N)$ metric.
As for $SO(N)$ gamma matrices, we abbreviate the antisymmetric product of $SO(N,N)$ gamma matrices by
\begin{equation}
 \Gamma_{A_1 \dots A_k} = \Gamma_{[A_1} \dots \Gamma_{A_k]} \ .
\end{equation}
The antisymmetric products of two gamma matrices $\Gamma_{AB}$ fulfill the $SO(N,N)$ commutation relations, and generate the action of $SO(N,N)$ on spinors $\Phi$.
The chirality operator is given by $\Gamma_0 =
\frac{1}{(2N)!} \epsilon^{A_1 \dots A_d} \Gamma_{A_1 \dots
A_d}$, where $\epsilon$ is naturally normalized by
\begin{equation} \label{choice_chirality}
\epsilon^{m_1+m_1- \dots m_N+ m_N-} = 1 \ ,
\end{equation}
if $N$ is even. In this case it defines a canonical choice of positive chirality.

Over a point on a $k$-dimensional manifold $Y_k$ we can define $SO(k,k)$ gamma matrices via the operators
\begin{equation} \label{Gamma_matrices_geometrical}
\Gamma_{m^+} \equiv \diff x^m \wedge \ , \quad \Gamma_{m^-} \equiv \iota_{x^m} \ ,
\end{equation}
which act on forms and where $\iota_{x_m}$ denotes the insertion of the tangent vector $x_m$. They naturally
fulfill the Clifford algebra \eqref{appendix_Clifford_split} since
\begin{equation} \label{appendix_geometric_clifford_algebra}
[\diff x^m \wedge \ , \ \iota_{x^n}] \omega_p = \delta^m_{\phantom{m}n} \omega_p
\end{equation}
for any $p$-form $\omega_p$.
Therefore, we can canonically define an $SO(k,k)$ action on the space of forms $\Lambda^\bullet T^* Y_k$. The chirality operator $\Gamma_0$ acts on a $p$-form $\omega_p$ by
\begin{equation}
 \Gamma_0\ \omega_p = (-1)^{p}\ \omega_p \ ,
\end{equation}
hence the Weyl spinor bundle of positive (negative) chirality is given by the bundle of even (odd) forms.
The generators of this $SO(k,k)$ action naturally split into three types according to the number of raising and lowering operators. Transformations of the type $\Gamma_{m+n-}$ preserve the degree of a form and span the algebra of the geometrical group $Gl(k, \mathbb{R})$. The generators $\Gamma_{m+n+}$ and $\Gamma_{m-n-}$ correspond to two-forms and bi-vectors. Hence we conclude
\begin{equation} \label{appendix_decomposition_so_algebra}
 so(k,k) = gl(k, \mathbb{R}) \oplus \Lambda^2 T^* Y \oplus \Lambda^2 T Y \ .
\end{equation}

%%%%%%%%%%%%%%%%% Bibliography %%%%%%%%%%%%%%%%%%%%%%%%%%%%%%%
%\bibliography{Doktorarbeit}

\begin{thebibliography}{10%
0}

\bibitem{Green:1987sp}
M.~B. Green, J.~H. Schwarz and E.~Witten, {\it {Superstring theory}}, . (2
  vols), Cambridge, Uk: Univ. Pr. ( 1987) ( Cambridge Monographs On
  Mathematical Physics).

\bibitem{Lust:1989tj}
D.~Lust and S.~Theisen, {\it {Lectures on string theory}},  {\em Lect. Notes
  Phys.} {\bf 346} (1989) 1--346.
%%CITATION = LNPHA,346,1;%%

\bibitem{Polchinski:1998rq}
J.~Polchinski, {\it {String theory}}, . (2 vols), Cambridge, UK: Univ. Pr.
  (1998).

\bibitem{Becker:2007zj}
K.~Becker, M.~Becker and J.~H. Schwarz, {\it {String theory and M-theory: A
  modern introduction}}, . Cambridge, UK: Cambridge Univ. Pr. (2007) 739 p.

\bibitem{Kaluza:1921tu}
T.~Kaluza, {\it {On the problem of unity in physics}},  {\em Sitzungsber.
  Preuss. Akad. Wiss. Berlin (Math. Phys. )} {\bf 1921} (1921) 966--972.
%%CITATION = SPWPA,1921,966;%%

\bibitem{Klein:1926tv}
O.~Klein, {\it {Quantum theory and five-dimensional theory of relativity}},
  {\em Z. Phys.} {\bf 37} (1926) 895--906.
%%CITATION = ZEPYA,37,895;%%

\bibitem{Klebanov:2000hb}
I.~R. Klebanov and M.~J. Strassler, {\it {Supergravity and a confining gauge
  theory: Duality cascades and chiSB-resolution of naked singularities}},  {\em
  JHEP} {\bf 08} (2000) 052 [\href{http://arXiv.org/abs/hep-th/0007191}{{\tt
  hep-th/0007191}}].
%%CITATION = HEP-TH/0007191;%%

\bibitem{Koerber:2008sx}
P.~Koerber and L.~Martucci, {\it {Warped generalized geometry
  compactifications, effective theories and non-perturbative effects}},  {\em
  Fortsch. Phys.} {\bf 56} (2008) 862--868
  [\href{http://arXiv.org/abs/0803.3149}{{\tt 0803.3149}}].
%%CITATION = 0803.3149;%%

\bibitem{Martucci:2009sf}
L.~Martucci, {\it {On moduli and effective theory of N=1 warped flux
  compactifications}},  {\em JHEP} {\bf 05} (2009) 027
  [\href{http://arXiv.org/abs/0902.4031}{{\tt 0902.4031}}].
%%CITATION = 0902.4031;%%

\bibitem{Louis:1998rx}
J.~Louis, I.~Brunner and S.~J. Huber, {\it {The supersymmetric standard
  model}},  \href{http://arXiv.org/abs/hep-ph/9811341}{{\tt hep-ph/9811341}}.
%%CITATION = HEP-PH/9811341;%%

\bibitem{Bustamante:2009us}
M.~Bustamante, L.~Cieri and J.~Ellis, {\it {Beyond the Standard Model for
  Montaneros}},  \href{http://arXiv.org/abs/0911.4409}{{\tt 0911.4409}}.
%%CITATION = 0911.4409;%%

\bibitem{Gates:1984nk}
S.~J. Gates, Jr., C.~M. Hull and M.~Rocek, {\it {Twisted multiplets and new
  supersymmetric nonlinear sigma models}},  {\em Nucl. Phys.} {\bf B248} (1984)
  157.
%%CITATION = NUPHA,B248,157;%%

\bibitem{Strominger:1986uh}
A.~Strominger, {\it {Superstrings with torsion}},  {\em Nucl. Phys.} {\bf B274}
  (1986) 253.
%%CITATION = NUPHA,B274,253;%%

\bibitem{Hull:1986kz}
C.~M. Hull, {\it {Compactifications of the heterotic superstring}},  {\em Phys.
  Lett.} {\bf B178} (1986) 357.
%%CITATION = PHLTA,B178,357;%%

\bibitem{Hitchin:2000jd}
N.~J. Hitchin, {\it {The geometry of three-forms in six and seven dimensions}},
   \href{http://arXiv.org/abs/math/0010054}{{\tt math/0010054}}.
%%CITATION = MATH/0010054;%%

\bibitem{Hitchin:2001rw}
N.~J. Hitchin, {\it {Stable forms and special metrics}},
  \href{http://arXiv.org/abs/math/0107101}{{\tt math/0107101}}.
%%CITATION = MATH/0107101;%%

\bibitem{Gauntlett:2001ur}
J.~P. Gauntlett, N.~Kim, D.~Martelli and D.~Waldram, {\it {Fivebranes wrapped
  on SLAG three-cycles and related geometry}},  {\em JHEP} {\bf 11} (2001) 018
  [\href{http://arXiv.org/abs/hep-th/0110034}{{\tt hep-th/0110034}}].
%%CITATION = HEP-TH/0110034;%%

\bibitem{Gauntlett:2002sc}
J.~P. Gauntlett, D.~Martelli, S.~Pakis and D.~Waldram, {\it {G structures and
  wrapped NS5-branes}},  {\em Commun. Math. Phys.} {\bf 247} (2004) 421--445
  [\href{http://arXiv.org/abs/hep-th/0205050}{{\tt hep-th/0205050}}].
%%CITATION = HEP-TH/0205050;%%

\bibitem{Chiossi:2002tw}
S.~Chiossi and S.~Salamon, {\it {The intrinsic torsion of SU(3) and $G_2$
  structures}},  \href{http://arXiv.org/abs/math/0202282}{{\tt math/0202282}}.
%%CITATION = MATH/0202282;%%

\bibitem{Gauntlett:2003cy}
J.~P. Gauntlett, D.~Martelli and D.~Waldram, {\it {Superstrings with intrinsic
  torsion}},  {\em Phys. Rev.} {\bf D69} (2004) 086002
  [\href{http://arXiv.org/abs/hep-th/0302158}{{\tt hep-th/0302158}}].
%%CITATION = HEP-TH/0302158;%%

\bibitem{Aspinwall:1996mn}
P.~S. Aspinwall, {\it {K3 surfaces and string duality}},
  \href{http://arXiv.org/abs/hep-th/9611137}{{\tt hep-th/9611137}}.
%%CITATION = HEP-TH/9611137;%%

\bibitem{Greene:1996cy}
B.~R. Greene, {\it {String theory on Calabi-Yau manifolds}},
  \href{http://arXiv.org/abs/hep-th/9702155}{{\tt hep-th/9702155}}.
%%CITATION = HEP-TH/9702155;%%

\bibitem{Grana:2005jc}
M.~Grana, {\it {Flux compactifications in string theory: A comprehensive
  review}},  {\em Phys. Rept.} {\bf 423} (2006) 91--158
  [\href{http://arXiv.org/abs/hep-th/0509003}{{\tt hep-th/0509003}}].
%%CITATION = HEP-TH/0509003;%%

\bibitem{Douglas:2006es}
M.~R. Douglas and S.~Kachru, {\it {Flux compactification}},  {\em Rev. Mod.
  Phys.} {\bf 79} (2007) 733--796
  [\href{http://arXiv.org/abs/hep-th/0610102}{{\tt hep-th/0610102}}].
%%CITATION = HEP-TH/0610102;%%

\bibitem{Blumenhagen:2006ci}
R.~Blumenhagen, B.~Kors, D.~Lust and S.~Stieberger, {\it {Four-dimensional
  string compactifications with D-branes, orientifolds and fluxes}},  {\em
  Phys. Rept.} {\bf 445} (2007) 1--193
  [\href{http://arXiv.org/abs/hep-th/0610327}{{\tt hep-th/0610327}}].
%%CITATION = HEP-TH/0610327;%%

\bibitem{Wecht:2007wu}
B.~Wecht, {\it {Lectures on nongeometric flux compactifications}},  {\em Class.
  Quant. Grav.} {\bf 24} (2007) S773--S794
  [\href{http://arXiv.org/abs/0708.3984}{{\tt 0708.3984}}].
%%CITATION = 0708.3984;%%

\bibitem{Samtleben:2008pe}
H.~Samtleben, {\it {Lectures on gauged supergravity and flux
  compactifications}},  {\em Class. Quant. Grav.} {\bf 25} (2008) 214002
  [\href{http://arXiv.org/abs/0808.4076}{{\tt 0808.4076}}].
%%CITATION = 0808.4076;%%

\bibitem{Joyce:2000}
D.~Joyce, {\it {Compact manifolds with special holonomy}}, . Oxford, UK: Univ.
  Pr. (2000).

\bibitem{Gurrieri:2002wz}
S.~Gurrieri, J.~Louis, A.~Micu and D.~Waldram, {\it {Mirror symmetry in
  generalized Calabi-Yau compactifications}},  {\em Nucl. Phys.} {\bf B654}
  (2003) 61--113 [\href{http://arXiv.org/abs/hep-th/0211102}{{\tt
  hep-th/0211102}}].
%%CITATION = HEP-TH/0211102;%%

\bibitem{Jeschek:2004wy}
C.~Jeschek and F.~Witt, {\it {Generalised $G_2$ structures and type IIB
  superstrings}},  {\em JHEP} {\bf 03} (2005) 053
  [\href{http://arXiv.org/abs/hep-th/0412280}{{\tt hep-th/0412280}}].
%%CITATION = HEP-TH/0412280;%%

\bibitem{Grana:2004bg}
M.~Grana, R.~Minasian, M.~Petrini and A.~Tomasiello, {\it {Supersymmetric
  backgrounds from generalized Calabi-Yau manifolds}},  {\em JHEP} {\bf 08}
  (2004) 046 [\href{http://arXiv.org/abs/hep-th/0406137}{{\tt
  hep-th/0406137}}].
%%CITATION = HEP-TH/0406137;%%

\bibitem{Grana:2005sn}
M.~Grana, R.~Minasian, M.~Petrini and A.~Tomasiello, {\it {Generalized
  structures of N=1 vacua}},  {\em JHEP} {\bf 11} (2005) 020
  [\href{http://arXiv.org/abs/hep-th/0505212}{{\tt hep-th/0505212}}].
%%CITATION = HEP-TH/0505212;%%

\bibitem{Grana:2005ny}
M.~Grana, J.~Louis and D.~Waldram, {\it {Hitchin functionals in N = 2
  supergravity}},  {\em JHEP} {\bf 01} (2006) 008
  [\href{http://arXiv.org/abs/hep-th/0505264}{{\tt hep-th/0505264}}].
%%CITATION = HEP-TH/0505264;%%

\bibitem{Grana:2006hr}
M.~Grana, J.~Louis and D.~Waldram, {\it {SU(3) $\times$ SU(3) compactification
  and mirror duals of magnetic fluxes}},  {\em JHEP} {\bf 04} (2007) 101
  [\href{http://arXiv.org/abs/hep-th/0612237}{{\tt hep-th/0612237}}].
%%CITATION = HEP-TH/0612237;%%

\bibitem{Bovy:2005qq}
J.~Bovy, D.~Lust and D.~Tsimpis, {\it {N = 1,2 supersymmetric vacua of IIA
  supergravity and SU(2) structures}},  {\em JHEP} {\bf 08} (2005) 056
  [\href{http://arXiv.org/abs/hep-th/0506160}{{\tt hep-th/0506160}}].
%%CITATION = HEP-TH/0506160;%%

\bibitem{ReidEdwards:2008rd}
R.~A. Reid-Edwards and B.~Spanjaard, {\it {N=4 gauged supergravity from
  duality-twist compactifications of string theory}},  {\em JHEP} {\bf 12}
  (2008) 052 [\href{http://arXiv.org/abs/0810.4699}{{\tt 0810.4699}}].
%%CITATION = 0810.4699;%%

\bibitem{Spanjaard:2008zz}
B.~Spanjaard, {\it {Compactifications of IIA supergravity on SU(2)-structure
  manifolds}}, . DESY-THESIS-2008-016.

\bibitem{Lust:2009zb}
D.~Lust and D.~Tsimpis, {\it {Classes of AdS4 type IIA/IIB compactifications
  with SU(3)$\times$SU(3) structure}},  {\em JHEP} {\bf 04} (2009) 111
  [\href{http://arXiv.org/abs/0901.4474}{{\tt 0901.4474}}].
%%CITATION = 0901.4474;%%

\bibitem{Louis:2009dq}
J.~Louis, D.~Martinez-Pedrera and A.~Micu, {\it {Heterotic compactifications on
  SU(2)-structure backgrounds}},  {\em JHEP} {\bf 09} (2009) 012
  [\href{http://arXiv.org/abs/0907.3799}{{\tt 0907.3799}}].
%%CITATION = 0907.3799;%%

\bibitem{Hitchin:2004ut}
N.~Hitchin, {\it {Generalized Calabi-Yau manifolds}},  {\em Quart. J. Math.
  Oxford Ser.} {\bf 54} (2003) 281--308
  [\href{http://arXiv.org/abs/math/0209099}{{\tt math/0209099}}].
%%CITATION = MATH/0209099;%%

\bibitem{Gualtieri:2003dx}
M.~Gualtieri, {\it {Generalized complex geometry}},
  \href{http://arXiv.org/abs/math/0401221}{{\tt math/0401221}}.
%%CITATION = MATH/0401221;%%

\bibitem{Witt:2004vr}
F.~Witt, {\it {Generalised $G_2$ manifolds}},  {\em Commun. Math. Phys.} {\bf
  265} (2006) 275--303 [\href{http://arXiv.org/abs/math/0411642}{{\tt
  math/0411642}}].
%%CITATION = MATH/0411642;%%

\bibitem{Witt:2005sk}
F.~Witt, {\it {Special metric structures and closed forms}},
  \href{http://arXiv.org/abs/math/0502443}{{\tt math/0502443}}.
%%CITATION = MATH/0502443;%%

\bibitem{Giveon:1994fu}
A.~Giveon, M.~Porrati and E.~Rabinovici, {\it {Target space duality in string
  theory}},  {\em Phys. Rept.} {\bf 244} (1994) 77--202
  [\href{http://arXiv.org/abs/hep-th/9401139}{{\tt hep-th/9401139}}].
%%CITATION = HEP-TH/9401139;%%

\bibitem{Dabholkar:2002sy}
A.~Dabholkar and C.~Hull, {\it {Duality twists, orbifolds, and fluxes}},  {\em
  JHEP} {\bf 09} (2003) 054 [\href{http://arXiv.org/abs/hep-th/0210209}{{\tt
  hep-th/0210209}}].
%%CITATION = HEP-TH/0210209;%%

\bibitem{Hull:2004in}
C.~M. Hull, {\it {A geometry for non-geometric string backgrounds}},  {\em
  JHEP} {\bf 10} (2005) 065 [\href{http://arXiv.org/abs/hep-th/0406102}{{\tt
  hep-th/0406102}}].
%%CITATION = HEP-TH/0406102;%%

\bibitem{Dabholkar:2005ve}
A.~Dabholkar and C.~Hull, {\it {Generalised T-duality and non-geometric
  backgrounds}},  {\em JHEP} {\bf 05} (2006) 009
  [\href{http://arXiv.org/abs/hep-th/0512005}{{\tt hep-th/0512005}}].
%%CITATION = HEP-TH/0512005;%%

\bibitem{Kachru:2002sk}
S.~Kachru, M.~B. Schulz, P.~K. Tripathy and S.~P. Trivedi, {\it {New
  supersymmetric string compactifications}},  {\em JHEP} {\bf 03} (2003) 061
  [\href{http://arXiv.org/abs/hep-th/0211182}{{\tt hep-th/0211182}}].
%%CITATION = HEP-TH/0211182;%%

\bibitem{Flournoy:2004vn}
A.~Flournoy, B.~Wecht and B.~Williams, {\it {Constructing nongeometric vacua in
  string theory}},  {\em Nucl. Phys.} {\bf B706} (2005) 127--149
  [\href{http://arXiv.org/abs/hep-th/0404217}{{\tt hep-th/0404217}}].
%%CITATION = HEP-TH/0404217;%%

\bibitem{Dall'Agata:2005ff}
G.~Dall'Agata and S.~Ferrara, {\it {Gauged supergravity algebras from twisted
  tori compactifications with fluxes}},  {\em Nucl. Phys.} {\bf B717} (2005)
  223--245 [\href{http://arXiv.org/abs/hep-th/0502066}{{\tt hep-th/0502066}}].
%%CITATION = HEP-TH/0502066;%%

\bibitem{Hull:2005hk}
C.~M. Hull and R.~A. Reid-Edwards, {\it {Flux compactifications of string
  theory on twisted tori}},  {\em Fortsch. Phys.} {\bf 57} (2009) 862--894
  [\href{http://arXiv.org/abs/hep-th/0503114}{{\tt hep-th/0503114}}].
%%CITATION = HEP-TH/0503114;%%

\bibitem{Dall'Agata:2005mj}
G.~Dall'Agata, R.~D'Auria and S.~Ferrara, {\it {Compactifications on twisted
  tori with fluxes and free differential algebras}},  {\em Phys. Lett.} {\bf
  B619} (2005) 149--154 [\href{http://arXiv.org/abs/hep-th/0503122}{{\tt
  hep-th/0503122}}].
%%CITATION = HEP-TH/0503122;%%

\bibitem{Shelton:2005cf}
J.~Shelton, W.~Taylor and B.~Wecht, {\it {Nongeometric flux
  compactifications}},  {\em JHEP} {\bf 10} (2005) 085
  [\href{http://arXiv.org/abs/hep-th/0508133}{{\tt hep-th/0508133}}].
%%CITATION = HEP-TH/0508133;%%

\bibitem{Hull:1994ys}
C.~M. Hull and P.~K. Townsend, {\it {Unity of superstring dualities}},  {\em
  Nucl. Phys.} {\bf B438} (1995) 109--137
  [\href{http://arXiv.org/abs/hep-th/9410167}{{\tt hep-th/9410167}}].
%%CITATION = HEP-TH/9410167;%%

\bibitem{Cavalcanti:2005hq}
G.~R. Cavalcanti, {\it {New aspects of the ddc-lemma}},
  \href{http://arXiv.org/abs/math/0501406}{{\tt math/0501406}}.
%%CITATION = MATH/0501406;%%

\bibitem{Hitchin:2005cv}
N.~Hitchin, {\it {Instantons, Poisson structures and generalized K\"ahler
  geometry}},  {\em Commun. Math. Phys.} {\bf 265} (2006) 131--164
  [\href{http://arXiv.org/abs/math/0503432}{{\tt math/0503432}}].
%%CITATION = MATH/0503432;%%

\bibitem{Triendl:2009ap}
H.~Triendl and J.~Louis, {\it {Type II compactifications on manifolds with
  SU(2) $\times$ SU(2) structure}},  {\em JHEP} {\bf 07} (2009) 080
  [\href{http://arXiv.org/abs/0904.2993}{{\tt 0904.2993}}].
%%CITATION = 0904.2993;%%

\bibitem{Hull:2007zu}
C.~M. Hull, {\it {Generalised geometry for M-theory}},  {\em JHEP} {\bf 07}
  (2007) 079 [\href{http://arXiv.org/abs/hep-th/0701203}{{\tt
  hep-th/0701203}}].
%%CITATION = HEP-TH/0701203;%%

\bibitem{Pacheco:2008ps}
P.~P. Pacheco and D.~Waldram, {\it {M-theory, exceptional generalised geometry
  and superpotentials}},  {\em JHEP} {\bf 09} (2008) 123
  [\href{http://arXiv.org/abs/0804.1362}{{\tt 0804.1362}}].
%%CITATION = 0804.1362;%%

\bibitem{Sim:2008}
A.~Sim, {\it {Exceptionally generalised geometry and supergravity}},  {\em
  Imperial College London PhD Thesis} (October 2008).

\bibitem{Grana:2009im}
M.~Grana, J.~Louis, A.~Sim and D.~Waldram, {\it {$E_{7(7)}$ formulation of N=2
  backgrounds}},  {\em JHEP} {\bf 07} (2009) 104
  [\href{http://arXiv.org/abs/0904.2333}{{\tt 0904.2333}}].
%%CITATION = 0904.2333;%%

\bibitem{deWit:1986mz}
B.~de~Wit and H.~Nicolai, {\it {d = 11 supergravity with local SU(8)
  invariance}},  {\em Nucl. Phys.} {\bf B274} (1986) 363.
%%CITATION = NUPHA,B274,363;%%

\bibitem{Cassani:2007pq}
D.~Cassani and A.~Bilal, {\it {Effective actions and N=1 vacuum conditions from
  SU(3) $\times$ SU(3) compactifications}},  {\em JHEP} {\bf 09} (2007) 076
  [\href{http://arXiv.org/abs/0707.3125}{{\tt 0707.3125}}].
%%CITATION = 0707.3125;%%

\bibitem{Cassani:2008rb}
D.~Cassani, {\it {Reducing democratic type II supergravity on SU(3) $\times$
  SU(3) structures}},  {\em JHEP} {\bf 06} (2008) 027
  [\href{http://arXiv.org/abs/0804.0595}{{\tt 0804.0595}}].
%%CITATION = 0804.0595;%%

\bibitem{Cecotti:1988qn}
S.~Cecotti, S.~Ferrara and L.~Girardello, {\it {Geometry of type II
  superstrings and the moduli of superconformal field theories}},  {\em Int. J.
  Mod. Phys.} {\bf A4} (1989) 2475.
%%CITATION = IMPAE,A4,2475;%%

\bibitem{Ferrara:1989ik}
S.~Ferrara and S.~Sabharwal, {\it {Quaternionic manifolds for Type II
  superstring vacua of Calabi-Yau spaces}},  {\em Nucl. Phys.} {\bf B332}
  (1990) 317.
%%CITATION = NUPHA,B332,317;%%

\bibitem{deWit:1990na}
B.~de~Wit and A.~Van~Proeyen, {\it {Symmetries of dual quaternionic
  manifolds}},  {\em Phys. Lett.} {\bf B252} (1990) 221--229.
%%CITATION = PHLTA,B252,221;%%

\bibitem{deWit:1992wf}
B.~de~Wit, F.~Vanderseypen and A.~Van~Proeyen, {\it {Symmetry structure of
  special geometries}},  {\em Nucl. Phys.} {\bf B400} (1993) 463--524
  [\href{http://arXiv.org/abs/hep-th/9210068}{{\tt hep-th/9210068}}].
%%CITATION = HEP-TH/9210068;%%

\bibitem{Polchinski:1995sm}
J.~Polchinski and A.~Strominger, {\it {New vacua for type II string theory}},
  {\em Phys. Lett.} {\bf B388} (1996) 736--742
  [\href{http://arXiv.org/abs/hep-th/9510227}{{\tt hep-th/9510227}}].
%%CITATION = HEP-TH/9510227;%%

\bibitem{Michelson:1996pn}
J.~Michelson, {\it {Compactifications of type IIB strings to four dimensions
  with non-trivial classical potential}},  {\em Nucl. Phys.} {\bf B495} (1997)
  127--148 [\href{http://arXiv.org/abs/hep-th/9610151}{{\tt hep-th/9610151}}].
%%CITATION = HEP-TH/9610151;%%

\bibitem{D'Auria:2004tr}
R.~D'Auria, S.~Ferrara, M.~Trigiante and S.~Vaula, {\it {Gauging the Heisenberg
  algebra of special quaternionic manifolds}},  {\em Phys. Lett.} {\bf B610}
  (2005) 147--151 [\href{http://arXiv.org/abs/hep-th/0410290}{{\tt
  hep-th/0410290}}].
%%CITATION = HEP-TH/0410290;%%

\bibitem{D'Auria:2004wd}
R.~D'Auria, S.~Ferrara, M.~Trigiante and S.~Vaula, {\it {Scalar potential for
  the gauged Heisenberg algebra and a non-polynomial antisymmetric tensor
  theory}},  {\em Phys. Lett.} {\bf B610} (2005) 270--276
  [\href{http://arXiv.org/abs/hep-th/0412063}{{\tt hep-th/0412063}}].
%%CITATION = HEP-TH/0412063;%%

\bibitem{Louis:2002ny}
J.~Louis and A.~Micu, {\it {Type II theories compactified on Calabi-Yau
  threefolds in the presence of background fluxes}},  {\em Nucl. Phys.} {\bf
  B635} (2002) 395--431 [\href{http://arXiv.org/abs/hep-th/0202168}{{\tt
  hep-th/0202168}}].
%%CITATION = HEP-TH/0202168;%%

\bibitem{Dall'Agata:2003yr}
G.~Dall'Agata, R.~D'Auria, L.~Sommovigo and S.~Vaula, {\it {D = 4, N = 2 gauged
  supergravity in the presence of tensor multiplets}},  {\em Nucl. Phys.} {\bf
  B682} (2004) 243--264 [\href{http://arXiv.org/abs/hep-th/0312210}{{\tt
  hep-th/0312210}}].
%%CITATION = HEP-TH/0312210;%%

\bibitem{Sommovigo:2004vj}
L.~Sommovigo and S.~Vaula, {\it {D = 4, N = 2 supergravity with Abelian
  electric and magnetic charge}},  {\em Phys. Lett.} {\bf B602} (2004) 130--136
  [\href{http://arXiv.org/abs/hep-th/0407205}{{\tt hep-th/0407205}}].
%%CITATION = HEP-TH/0407205;%%

\bibitem{D'Auria:2004yi}
R.~D'Auria, L.~Sommovigo and S.~Vaula, {\it {N = 2 supergravity Lagrangian
  coupled to tensor multiplets with electric and magnetic fluxes}},  {\em JHEP}
  {\bf 11} (2004) 028 [\href{http://arXiv.org/abs/hep-th/0409097}{{\tt
  hep-th/0409097}}].
%%CITATION = HEP-TH/0409097;%%

\bibitem{Haack:2001iz}
M.~Haack, J.~Louis and H.~Singh, {\it {Massive type IIA theory on K3}},  {\em
  JHEP} {\bf 04} (2001) 040 [\href{http://arXiv.org/abs/hep-th/0102110}{{\tt
  hep-th/0102110}}].
%%CITATION = HEP-TH/0102110;%%

\bibitem{D'Auria:2002tc}
R.~D'Auria, S.~Ferrara and S.~Vaula, {\it {N = 4 gauged supergravity and a IIB
  orientifold with fluxes}},  {\em New J. Phys.} {\bf 4} (2002) 71
  [\href{http://arXiv.org/abs/hep-th/0206241}{{\tt hep-th/0206241}}].
%%CITATION = HEP-TH/0206241;%%

\bibitem{D'Auria:2003jk}
R.~D'Auria, S.~Ferrara, F.~Gargiulo, M.~Trigiante and S.~Vaula, {\it {N = 4
  supergravity Lagrangian for type IIB on $T^6/\mathbb{Z}_2$ in presence of
  fluxes and D3-branes}},  {\em JHEP} {\bf 06} (2003) 045
  [\href{http://arXiv.org/abs/hep-th/0303049}{{\tt hep-th/0303049}}].
%%CITATION = HEP-TH/0303049;%%

\bibitem{Derendinger:2004jn}
J.-P. Derendinger, C.~Kounnas, P.~M. Petropoulos and F.~Zwirner, {\it
  {Superpotentials in IIA compactifications with general fluxes}},  {\em Nucl.
  Phys.} {\bf B715} (2005) 211--233
  [\href{http://arXiv.org/abs/hep-th/0411276}{{\tt hep-th/0411276}}].
%%CITATION = HEP-TH/0411276;%%

\bibitem{Derendinger:2005ph}
J.~P. Derendinger, C.~Kounnas, P.~M. Petropoulos and F.~Zwirner, {\it {Fluxes
  and gaugings: N = 1 effective superpotentials}},  {\em Fortsch. Phys.} {\bf
  53} (2005) 926--935 [\href{http://arXiv.org/abs/hep-th/0503229}{{\tt
  hep-th/0503229}}].
%%CITATION = HEP-TH/0503229;%%

\bibitem{Schon:2006kz}
J.~Schon and M.~Weidner, {\it {Gauged N = 4 supergravities}},  {\em JHEP} {\bf
  05} (2006) 034 [\href{http://arXiv.org/abs/hep-th/0602024}{{\tt
  hep-th/0602024}}].
%%CITATION = HEP-TH/0602024;%%

\bibitem{deRoo:1984gd}
M.~de~Roo, {\it {Matter coupling in N=4 supergravity}},  {\em Nucl. Phys.} {\bf
  B255} (1985) 515.
%%CITATION = NUPHA,B255,515;%%

\bibitem{LST}
T.~Danckaert, J.~Louis, D.~Martinez-Pedrera, B.~Spanjaard and H.~Triendl, {\it
  {Reducing IIA on SU(2)-structure manifolds}},  {\em in preparation} (2010).

\bibitem{Andrianopoli:2001zh}
L.~Andrianopoli, R.~D'Auria and S.~Ferrara, {\it {Supersymmetry reduction of
  N-extended supergravities in four dimensions}},  {\em JHEP} {\bf 03} (2002)
  025 [\href{http://arXiv.org/abs/hep-th/0110277}{{\tt hep-th/0110277}}].
%%CITATION = HEP-TH/0110277;%%

\bibitem{Andrianopoli:2001gm}
L.~Andrianopoli, R.~D'Auria and S.~Ferrara, {\it {Consistent reduction of N = 2
  $\rightarrow$ N = 1 four-dimensional supergravity coupled to matter}},  {\em
  Nucl. Phys.} {\bf B628} (2002) 387--403
  [\href{http://arXiv.org/abs/hep-th/0112192}{{\tt hep-th/0112192}}].
%%CITATION = HEP-TH/0112192;%%

\bibitem{Andrianopoli:2002rm}
L.~Andrianopoli, R.~D'Auria, S.~Ferrara and M.~A. Lledo, {\it {Super-Higgs
  effect in extended supergravity}},  {\em Nucl. Phys.} {\bf B640} (2002)
  46--62 [\href{http://arXiv.org/abs/hep-th/0202116}{{\tt hep-th/0202116}}].
%%CITATION = HEP-TH/0202116;%%

\bibitem{Andrianopoli:2002vq}
L.~Andrianopoli, R.~D'Auria, S.~Ferrara and M.~A. Lledo, {\it {N = 2 super
  Higgs, N = 1 Poincar\'e vacua and quaternionic geometry}},  {\em JHEP} {\bf
  01} (2003) 045 [\href{http://arXiv.org/abs/hep-th/0212236}{{\tt
  hep-th/0212236}}].
%%CITATION = HEP-TH/0212236;%%

\bibitem{Grimm:2004uq}
T.~W. Grimm and J.~Louis, {\it {The effective action of N = 1 Calabi-Yau
  orientifolds}},  {\em Nucl. Phys.} {\bf B699} (2004) 387--426
  [\href{http://arXiv.org/abs/hep-th/0403067}{{\tt hep-th/0403067}}].
%%CITATION = HEP-TH/0403067;%%

\bibitem{Grimm:2004ua}
T.~W. Grimm and J.~Louis, {\it {The effective action of type IIA Calabi-Yau
  orientifolds}},  {\em Nucl. Phys.} {\bf B718} (2005) 153--202
  [\href{http://arXiv.org/abs/hep-th/0412277}{{\tt hep-th/0412277}}].
%%CITATION = HEP-TH/0412277;%%

\bibitem{Benmachiche:2006df}
I.~Benmachiche and T.~W. Grimm, {\it {Generalized N = 1 orientifold
  compactifications and the Hitchin functionals}},  {\em Nucl. Phys.} {\bf
  B748} (2006) 200--252 [\href{http://arXiv.org/abs/hep-th/0602241}{{\tt
  hep-th/0602241}}].
%%CITATION = HEP-TH/0602241;%%

\bibitem{Andrianopoli:1996cm}
L.~Andrianopoli {\em et.~al.}, {\it {N = 2 supergravity and N = 2 super
  Yang-Mills theory on general scalar manifolds: Symplectic covariance,
  gaugings and the momentum map}},  {\em J. Geom. Phys.} {\bf 23} (1997)
  111--189 [\href{http://arXiv.org/abs/hep-th/9605032}{{\tt hep-th/9605032}}].
%%CITATION = HEP-TH/9605032;%%

\bibitem{Cecotti:1984rk}
S.~Cecotti, L.~Girardello and M.~Porrati, {\it {Two into one won't go}},  {\em
  Phys. Lett.} {\bf B145} (1984) 61.
%%CITATION = PHLTA,B145,61;%%

\bibitem{Cecotti:1984wn}
S.~Cecotti, L.~Girardello and M.~Porrati, {\it {Constraints on partial super
  Higgs}},  {\em Nucl. Phys.} {\bf B268} (1986) 295--316.
%%CITATION = NUPHA,B268,295;%%

\bibitem{Bagger:1994vj}
J.~Bagger and A.~Galperin, {\it {Matter couplings in partially broken extended
  supersymmetry}},  {\em Phys. Lett.} {\bf B336} (1994) 25--31
  [\href{http://arXiv.org/abs/hep-th/9406217}{{\tt hep-th/9406217}}].
%%CITATION = HEP-TH/9406217;%%

\bibitem{Antoniadis:1995vb}
I.~Antoniadis, H.~Partouche and T.~R. Taylor, {\it {Spontaneous breaking of N=2
  global supersymmetry}},  {\em Phys. Lett.} {\bf B372} (1996) 83--87
  [\href{http://arXiv.org/abs/hep-th/9512006}{{\tt hep-th/9512006}}].
%%CITATION = HEP-TH/9512006;%%

\bibitem{Ferrara:1995gu}
S.~Ferrara, L.~Girardello and M.~Porrati, {\it {Minimal Higgs branch for the
  breaking of half of the supersymmetries in N=2 supergravity}},  {\em Phys.
  Lett.} {\bf B366} (1996) 155--159
  [\href{http://arXiv.org/abs/hep-th/9510074}{{\tt hep-th/9510074}}].
%%CITATION = HEP-TH/9510074;%%

\bibitem{Ferrara:1995xi}
S.~Ferrara, L.~Girardello and M.~Porrati, {\it {Spontaneous breaking of N=2 to
  N=1 in rigid and local supersymmetric theories}},  {\em Phys. Lett.} {\bf
  B376} (1996) 275--281 [\href{http://arXiv.org/abs/hep-th/9512180}{{\tt
  hep-th/9512180}}].
%%CITATION = HEP-TH/9512180;%%

\bibitem{Fre:1996js}
P.~Fre, L.~Girardello, I.~Pesando and M.~Trigiante, {\it {Spontaneous N = 2
  $\rightarrow$ N = 1 local supersymmetry breaking with surviving compact gauge
  groups}},  {\em Nucl. Phys.} {\bf B493} (1997) 231--248
  [\href{http://arXiv.org/abs/hep-th/9607032}{{\tt hep-th/9607032}}].
%%CITATION = HEP-TH/9607032;%%

\bibitem{Hohm:2004rc}
O.~Hohm and J.~Louis, {\it {Spontaneous N = 2 $\rightarrow$ N = 1 supergravity
  breaking in three dimensions}},  {\em Class. Quant. Grav.} {\bf 21} (2004)
  4607--4624 [\href{http://arXiv.org/abs/hep-th/0403128}{{\tt
  hep-th/0403128}}].
%%CITATION = HEP-TH/0403128;%%

\bibitem{Gibbons:1984kp}
G.~W. Gibbons, {\it {Aspects of supergravity theories}},  {\em GIFT Seminar}
  {\bf 0123} (1984). Three lectures given at GIFT Seminar on Theoretical
  Physics, San Feliu de Guixols, Spain, Jun 4-11, 1984.

\bibitem{deWit:1986xg}
B.~de~Wit, D.~J. Smit and N.~D. Hari~Dass, {\it {Residual supersymmetry of
  compactified D=10 supergravity}},  {\em Nucl. Phys.} {\bf B283} (1987) 165.
%%CITATION = NUPHA,B283,165;%%

\bibitem{Maldacena:2000mw}
J.~M. Maldacena and C.~Nunez, {\it {Supergravity description of field theories
  on curved manifolds and a no-go theorem}},  {\em Int. J. Mod. Phys.} {\bf
  A16} (2001) 822--855 [\href{http://arXiv.org/abs/hep-th/0007018}{{\tt
  hep-th/0007018}}].
%%CITATION = HEP-TH/0007018;%%

\bibitem{Mayr:2000hh}
P.~Mayr, {\it {On supersymmetry breaking in string theory and its realization
  in brane worlds}},  {\em Nucl. Phys.} {\bf B593} (2001) 99--126
  [\href{http://arXiv.org/abs/hep-th/0003198}{{\tt hep-th/0003198}}].
%%CITATION = HEP-TH/0003198;%%

\bibitem{Ceresole:1995jg}
A.~Ceresole, R.~D'Auria, S.~Ferrara and A.~Van~Proeyen, {\it {Duality
  transformations in supersymmetric Yang-Mills theories coupled to
  supergravity}},  {\em Nucl. Phys.} {\bf B444} (1995) 92--124
  [\href{http://arXiv.org/abs/hep-th/9502072}{{\tt hep-th/9502072}}].
%%CITATION = HEP-TH/9502072;%%

\bibitem{deWit:2002vt}
B.~de~Wit, H.~Samtleben and M.~Trigiante, {\it {On Lagrangians and gaugings of
  maximal supergravities}},  {\em Nucl. Phys.} {\bf B655} (2003) 93--126
  [\href{http://arXiv.org/abs/hep-th/0212239}{{\tt hep-th/0212239}}].
%%CITATION = HEP-TH/0212239;%%

\bibitem{deWit:2005ub}
B.~de~Wit, H.~Samtleben and M.~Trigiante, {\it {Magnetic charges in local field
  theory}},  {\em JHEP} {\bf 09} (2005) 016
  [\href{http://arXiv.org/abs/hep-th/0507289}{{\tt hep-th/0507289}}].
%%CITATION = HEP-TH/0507289;%%

\bibitem{Louis:2009xd}
J.~Louis, P.~Smyth and H.~Triendl, {\it {Spontaneous N=2 to N=1 supersymmetry
  breaking in supergravity and type II string theory}},  {\em JHEP} {\bf 02}
  (2010) 103 [\href{http://arXiv.org/abs/0911.5077}{{\tt 0911.5077}}].
%%CITATION = 0911.5077;%%

\bibitem{Louis:2010ui}
  J.~Louis, P.~Smyth and H.~Triendl,
  {\it {The N=1 Low-Energy Effective Action of Spontaneously Broken N=2
  Supergravities}}, \href{http://arXiv.org/abs/1008.1214}{{\tt 1008.1214}}.

\bibitem{Louis:2002vy}
J.~Louis, {\it {Aspects of spontaneous N = 2 $\rightarrow$ N = 1 breaking in
  supergravity}},  \href{http://arXiv.org/abs/hep-th/0203138}{{\tt
  hep-th/0203138}}.
%%CITATION = HEP-TH/0203138;%%

\bibitem{Gunara:2003td}
B.~E. Gunara, {\it {Spontaneous N=2 $\rightarrow$ N=1 supersymmetry breaking
  and the super-Higgs effect in supergravity}},  {\em Cuvillier} (2003). Ph.D.
  Thesis, Goettingen, Germany.

\bibitem{Cortes}
V.~Cortes, J.~Louis, P.~Smyth and H.~Triendl, {\it {K\"ahler quotients of
  quaternionic-K\"ahler manifolds}},  {\em in preparation} (2010).

\bibitem{Koerber:2007xk}
P.~Koerber and L.~Martucci, {\it {From ten to four and back again: How to
  generalize the geometry}},  {\em JHEP} {\bf 08} (2007) 059
  [\href{http://arXiv.org/abs/0707.1038}{{\tt 0707.1038}}].
%%CITATION = 0707.1038;%%

\bibitem{Romans:1986er}
L.~J. Romans, {\it {Selfduality for interacting fields: Covariant field
  equations for six-dimensional chiral supergravities}},  {\em Nucl. Phys.}
  {\bf B276} (1986) 71.
%%CITATION = NUPHA,B276,71;%%

\bibitem{Grana:2006kf}
M.~Grana, R.~Minasian, M.~Petrini and A.~Tomasiello, {\it {A scan for new N=1
  vacua on twisted tori}},  {\em JHEP} {\bf 05} (2007) 031
  [\href{http://arXiv.org/abs/hep-th/0609124}{{\tt hep-th/0609124}}].
%%CITATION = HEP-TH/0609124;%%

\bibitem{Chevalley:1996}
C.~Chevalley, {\it {The algebraic theory of spinors and Clifford algebras.
  Collected works. Vol. 2}},  {\em Berlin, Germany: Springer Verlag} (1996)
  227.

\bibitem{Charlton:1996PhD}
P.~Charlton, {\it {The geometry of pure spinors, with applications}},  {\em
  Ph.D. thesis} (1996)
  [\href{http://arXiv.org/abs/http://csusap.csu.edu.au/~pcharlto/charlton\_the%
sis.pdf}{{\tt http://csusap.csu.edu.au/~pcharlto/charlton\_thesis.pdf}}].

\bibitem{Adams:1980hp}
J.~F. Adams, {\it {Spin(8), triality, f(4) and all that. (talk)}}, . In
  *Cambridge 1980, Proceedings, Superspace and Supergravity*, 435-445.

\bibitem{Slansky:1981yr}
R.~Slansky, {\it {Group theory for unified model building}},  {\em Phys. Rept.}
  {\bf 79} (1981) 1--128.
%%CITATION = PRPLC,79,1;%%

\bibitem{Cremmer:1978ds}
E.~Cremmer and B.~Julia, {\it {The N=8 supergravity theory. 1. The
  Lagrangian}},  {\em Phys. Lett.} {\bf B80} (1978) 48.
%%CITATION = PHLTA,B80,48;%%

\bibitem{Gunaydin:1983rk}
M.~Gunaydin, G.~Sierra and P.~K. Townsend, {\it {Exceptional supergravity
  theories and the magic square}},  {\em Phys. Lett.} {\bf B133} (1983) 72.
%%CITATION = PHLTA,B133,72;%%

\bibitem{Gunaydin:1983bi}
M.~Gunaydin, G.~Sierra and P.~K. Townsend, {\it {The geometry of N=2
  Maxwell-Einstein supergravity and Jordan algebras}},  {\em Nucl. Phys.} {\bf
  B242} (1984) 244.
%%CITATION = NUPHA,B242,244;%%

\bibitem{Cecotti:1988ad}
S.~Cecotti, {\it {Homogeneous K\"ahler manifolds and T algebras in N=2
  supergravity and superstrings}},  {\em Commun. Math. Phys.} {\bf 124} (1989)
  23--55.
%%CITATION = CMPHA,124,23;%%

\bibitem{deWit:1993rr}
B.~de~Wit and A.~Van~Proeyen, {\it {Hidden symmetries, special geometry and
  quaternionic manifolds}},  {\em Int. J. Mod. Phys.} {\bf D3} (1994) 31--48
  [\href{http://arXiv.org/abs/hep-th/9310067}{{\tt hep-th/9310067}}].
%%CITATION = HEP-TH/9310067;%%

\bibitem{Swann:1990}
A.~Swann, {\it {Hyper-K\"ahler and quaternionic-K\"ahler geometry}},  {\em
  Math. Ann.} {\bf 289} (1991) 421.
%%CITATION = 0805.1310;%%

\bibitem{Wolf:1965zz}
J.~A. Wolf, {\it {Complex homogeneous contact manifolds and quaternionic
  symmetric spaces}},  {\em J. of Math. Mech.} {\bf 14} (1965) 1033.

\bibitem{Alekseevskii:1975zz}
D.~V. Alekseevskii, {\it {Classification of quaternionic spaces with transitive
  solvable group of motions}},  {\em Math. USSR Izvestija} {\bf 9} (1975) 297.

\bibitem{Kobak:2000mj}
P.~Kobak and A.~Swann, {\it {The hyper-K\"ahler geometry associated to Wolf
  spaces}},  \href{http://arXiv.org/abs/math/0001025}{{\tt math/0001025}}.
%%CITATION = MATH/0001025;%%

\bibitem{Berglund:2005dm}
P.~Berglund and P.~Mayr, {\it {Non-perturbative superpotentials in F-theory and
  string duality}},  \href{http://arXiv.org/abs/hep-th/0504058}{{\tt
  hep-th/0504058}}.
%%CITATION = HEP-TH/0504058;%%

\bibitem{Dall'Agata:2006nr}
G.~Dall'Agata, {\it {Non-K\"ahler attracting manifolds}},  {\em JHEP} {\bf 04}
  (2006) 001 [\href{http://arXiv.org/abs/hep-th/0602045}{{\tt
  hep-th/0602045}}].
%%CITATION = HEP-TH/0602045;%%

\bibitem{Strathdee:1986jr}
J.~A. Strathdee, {\it {Extended Poincar\'e supersymmetry}},  {\em Int. J. Mod.
  Phys.} {\bf A2} (1987) 273.
%%CITATION = IMPAE,A2,273;%%

\bibitem{deWit:1984pk}
B.~de~Wit and A.~Van~Proeyen, {\it {Potentials and symmetries of general gauged
  N=2 supergravity: Yang-Mills models}},  {\em Nucl. Phys.} {\bf B245} (1984)
  89.
%%CITATION = NUPHA,B245,89;%%

\bibitem{Craps:1997gp}
B.~Craps, F.~Roose, W.~Troost and A.~Van~Proeyen, {\it {What is
  special-K\"ahler geometry?}},  {\em Nucl. Phys.} {\bf B503} (1997) 565--613
  [\href{http://arXiv.org/abs/hep-th/9703082}{{\tt hep-th/9703082}}].
%%CITATION = HEP-TH/9703082;%%

\bibitem{Ceresole:1995ca}
A.~Ceresole, R.~D'Auria and S.~Ferrara, {\it {The symplectic structure of N=2
  supergravity and its central extension}},  {\em Nucl. Phys. Proc. Suppl.}
  {\bf 46} (1996) 67--74 [\href{http://arXiv.org/abs/hep-th/9509160}{{\tt
  hep-th/9509160}}].
%%CITATION = HEP-TH/9509160;%%

\bibitem{Cremmer:1984hj}
E.~Cremmer {\em et.~al.}, {\it {Vector multiplets coupled to N=2 supergravity:
  Super-Higgs effect, flat potentials and geometric structure}},  {\em Nucl.
  Phys.} {\bf B250} (1985) 385.
%%CITATION = NUPHA,B250,385;%%

\bibitem{Bagger:1983tt}
J.~Bagger and E.~Witten, {\it {Matter couplings in N=2 supergravity}},  {\em
  Nucl. Phys.} {\bf B222} (1983) 1.
%%CITATION = NUPHA,B222,1;%%

\bibitem{deWit:1984px}
B.~de~Wit, P.~G. Lauwers and A.~Van~Proeyen, {\it {Lagrangians of N=2
  supergravity -- matter systems}},  {\em Nucl. Phys.} {\bf B255} (1985) 569.
%%CITATION = NUPHA,B255,569;%%

\bibitem{Galicki:1986ja}
K.~Galicki, {\it {A generalization of the momentum mapping construction for
  quaternionic-K\"ahler manifolds}},  {\em Commun. Math. Phys.} {\bf 108}
  (1987) 117.
%%CITATION = CMPHA,108,117;%%

\bibitem{D'Auria:1990fj}
R.~D'Auria, S.~Ferrara and P.~Fre, {\it {Special and quaternionic isometries:
  General couplings in N=2 supergravity and the scalar potential}},  {\em Nucl.
  Phys.} {\bf B359} (1991) 705--740.
%%CITATION = NUPHA,B359,705;%%

\bibitem{deVroome:2007zd}
M.~de~Vroome and B.~de~Wit, {\it {Lagrangians with electric and magnetic
  charges of N=2 supersymmetric gauge theories}},  {\em JHEP} {\bf 08} (2007)
  064 [\href{http://arXiv.org/abs/0707.2717}{{\tt 0707.2717}}].
%%CITATION = 0707.2717;%%

\bibitem{D'Auria:2001kv}
R.~D'Auria and S.~Ferrara, {\it {On fermion masses, gradient flows and
  potential in supersymmetric theories}},  {\em JHEP} {\bf 05} (2001) 034
  [\href{http://arXiv.org/abs/hep-th/0103153}{{\tt hep-th/0103153}}].
%%CITATION = HEP-TH/0103153;%%

\bibitem{RoblesLlana:2006ez}
D.~Robles-Llana, F.~Saueressig and S.~Vandoren, {\it {String loop corrected
  hypermultiplet moduli spaces}},  {\em JHEP} {\bf 03} (2006) 081
  [\href{http://arXiv.org/abs/hep-th/0602164}{{\tt hep-th/0602164}}].
%%CITATION = HEP-TH/0602164;%%

\bibitem{House:2005yc}
  T.~House and E.~Palti,
  {\it {Effective action of (massive) IIA on manifolds with SU(3) structure}},
  {\em Phys. Rev.}   {\bf D72} (2005) 026004
  [\href{http://arXiv.org/abs/hep-th/0505177}{{\tt hep-th/0505177}}].
  %%CITATION = PHRVA,D72,026004;%%

\bibitem{D'Auria:2007ay}
R.~D'Auria, S.~Ferrara and M.~Trigiante, {\it {On the supergravity formulation
  of mirror symmetry in generalized Calabi-Yau manifolds}},  {\em Nucl. Phys.}
  {\bf B780} (2007) 28--39 [\href{http://arXiv.org/abs/hep-th/0701247}{{\tt
  hep-th/0701247}}].
%%CITATION = HEP-TH/0701247;%%

\bibitem{Ferrara:1983gn}
S.~Ferrara and P.~van Nieuwenhuizen, {\it {Noether coupling of massive
  gravitinos to N=1 supergravity}},  {\em Phys. Lett.} {\bf B127} (1983) 70.
%%CITATION = PHLTA,B127,70;%%

\bibitem{Ferrara:1985gj}
S.~Ferrara and L.~Maiani, {\it {An introduction to supersymmetry breaking in
  extended supergravity}},  {\em Bariloche SILARG Symp.} (1985) 349. Based on
  lectures given at SILARG V, 5th Latin American Symp. on Relativity and
  Gravitation.

\bibitem{Wess:1992cp}
J.~Wess and J.~Bagger, {\it {Supersymmetry and supergravity}}, . Princeton,
  USA: Univ. Pr. (1992) 259 p.

\bibitem{Gates:1983nr}
S.~J. Gates, M.~T. Grisaru, M.~Rocek and W.~Siegel, {\it {Superspace, or one
  thousand and one lessons in supersymmetry}},  {\em Front. Phys.} {\bf 58}
  (1983) 1--548 [\href{http://arXiv.org/abs/hep-th/0108200}{{\tt
  hep-th/0108200}}].
%%CITATION = HEP-TH/0108200;%%

\bibitem{D'Auria:2005yg}
R.~D'Auria, S.~Ferrara, M.~Trigiante and S.~Vaula, {\it {N = 1 reductions of N
  = 2 supergravity in the presence of tensor multiplets}},  {\em JHEP} {\bf 03}
  (2005) 052 [\href{http://arXiv.org/abs/hep-th/0502219}{{\tt
  hep-th/0502219}}].
%%CITATION = HEP-TH/0502219;%%

\bibitem{Frey:2003sd}
A.~R. Frey and M.~Grana, {\it {Type IIB solutions with interpolating
  supersymmetries}},  {\em Phys. Rev.} {\bf D68} (2003) 106002
  [\href{http://arXiv.org/abs/hep-th/0307142}{{\tt hep-th/0307142}}].
%%CITATION = HEP-TH/0307142;%%

\bibitem{Lust:2004ig}
D.~Lust and D.~Tsimpis, {\it {Supersymmetric AdS(4) compactifications of IIA
  supergravity}},  {\em JHEP} {\bf 02} (2005) 027
  [\href{http://arXiv.org/abs/hep-th/0412250}{{\tt hep-th/0412250}}].
%%CITATION = HEP-TH/0412250;%%

\bibitem{Behrndt:2005bv}
K.~Behrndt, M.~Cvetic and P.~Gao, {\it {General type IIB fluxes with SU(3)
  structures}},  {\em Nucl. Phys.} {\bf B721} (2005) 287--308
  [\href{http://arXiv.org/abs/hep-th/0502154}{{\tt hep-th/0502154}}].
%%CITATION = HEP-TH/0502154;%%

\bibitem{Micu:2006ey}
  A.~Micu, E.~Palti and P.~M.~Saffin,
  {\it {M-theory on seven-dimensional manifolds with SU(3) structure}},
  {\em JHEP} {\bf 0605} (2006) 048
  [\href{http://arXiv.org/abs/hep-th/0602163}{{\tt hep-th/0602163}}].

\bibitem{Micu:2007rd}
A.~Micu, E.~Palti and G.~Tasinato, {\it {Towards Minkowski vacua in type II
  string compactifications}},  {\em JHEP} {\bf 03} (2007) 104
  [\href{http://arXiv.org/abs/hep-th/0701173}{{\tt hep-th/0701173}}].
%%CITATION = HEP-TH/0701173;%%

\bibitem{KashaniPoor:2007tr}
A.-K. Kashani-Poor, {\it {Nearly-K\"ahler reduction}},  {\em JHEP} {\bf 11}
  (2007) 026 [\href{http://arXiv.org/abs/0709.4482}{{\tt 0709.4482}}].
%%CITATION = 0709.4482;%%

\bibitem{Andriot:2008va}
D.~Andriot, {\it {New supersymmetric flux vacua with intermediate SU(2)
  structure}},  {\em JHEP} {\bf 08} (2008) 096
  [\href{http://arXiv.org/abs/0804.1769}{{\tt 0804.1769}}].
%%CITATION = 0804.1769;%%

\bibitem{Anguelova:2008fm}
L.~Anguelova, {\it {Flux vacua attractors and generalized compactifications}},
  {\em JHEP} {\bf 01} (2009) 017 [\href{http://arXiv.org/abs/0806.3820}{{\tt
  0806.3820}}].
%%CITATION = 0806.3820;%%

\bibitem{Cassani:2009ck}
D.~Cassani and A.-K. Kashani-Poor, {\it {Exploiting N=2 in consistent coset
  reductions of type IIA}},  {\em Nucl. Phys.} {\bf B817} (2009) 25--57
  [\href{http://arXiv.org/abs/0901.4251}{{\tt 0901.4251}}].
%%CITATION = 0901.4251;%%

\bibitem{Cassani:2009na}
D.~Cassani, S.~Ferrara, A.~Marrani, J.~F. Morales and H.~Samtleben, {\it {A
  special road to AdS vacua}},  {\em JHEP} {\bf 02} (2010) 027
  [\href{http://arXiv.org/abs/0911.2708}{{\tt 0911.2708}}].
%%CITATION = 0911.2708;%%

\bibitem{deCarlos:2009qm}
B.~de~Carlos, A.~Guarino and J.~M. Moreno, {\it {Complete classification of
  Minkowski vacua in generalised flux models}},  {\em JHEP} {\bf 02} (2010) 076
  [\href{http://arXiv.org/abs/0911.2876}{{\tt 0911.2876}}].
%%CITATION = 0911.2876;%%

\bibitem{KashaniPoor:2005si}
A.-K. Kashani-Poor and A.~Tomasiello, {\it {A stringy test of flux-induced
  isometry gauging}},  {\em Nucl. Phys.} {\bf B728} (2005) 135--147
  [\href{http://arXiv.org/abs/hep-th/0505208}{{\tt hep-th/0505208}}].
%%CITATION = HEP-TH/0505208;%%

\bibitem{VanProeyen:1999ni}
A.~Van~Proeyen, {\it {Tools for supersymmetry}},
  \href{http://arXiv.org/abs/hep-th/9910030}{{\tt hep-th/9910030}}.
%%CITATION = HEP-TH/9910030;%%

\end{thebibliography}
\bibliographystyle{JHEP-2}

\providecommand{\href}[2]{#2}\begingroup\raggedright\endgroup

%%%%%%%%%%%%%%%%%%%%%%%%%%%%%%%%%%%%%%%%%%%%%%%%%%%%%%%%%%%

\end{document}